\documentclass[
headsepline, %Linie zwischen Kopfzeile und Hauptteil der Seite
12pt, %Schriftgröße 12
a4paper, %A4-Format
parskip=on,%parskip=on verhindert Einzug bei neuem Absatz
liststotoc] %Nimmt Verzeichnisse ins Inhaltsverzeichnis auf
{scrreprt}	%scrreprt-für mittlere Schriftwerke meist mit Titelblatt und einseitigem Druck und unterstützt europäische Typographierichtlinien
\usepackage[utf8]{inputenc} %Nutzung deutscher Umlaute ä, ö, ö und ß
	\usepackage{cite}						%Dateien mit UTF8 kodieren!
\usepackage{footmisc}						
\usepackage[automark]{scrlayer-scrpage} %Paket zur Nutzung von Kopf- und Fußzeilen
\usepackage[latin,english]{babel}
\usepackage[T1]{fontenc}
\usepackage{amssymb}
\usepackage{mathtools}
\usepackage{tikz}
\usetikzlibrary{angles, quotes, babel}
\usepackage{commath}
\usepackage{extarrows}
\usepackage{esvect}
\usepackage{lmodern} %Schriftart Latin Modern
\usepackage{textcmds}
\usepackage{dsfont}
\usepackage[left=30mm,right=25mm,bottom=25mm,top=30mm]{geometry}
\renewcommand{\arraystretch}{1.2} %Verändert Zeilenabstand in einer Tabelle
\usepackage{subcaption}
\usepackage[onehalfspacing]{setspace} %Zeilenabstand 1,5

\usepackage{graphicx}	%Ermöglicht das Eindbinden externer Bilder
\usepackage{xspace}

\usepackage{textpos} %Fixierung von Texten, Bildern usw. an einer bestimmten Stelle einer Seite

\usepackage[colorlinks,bookmarks]{hyperref} %hyperref zur Erstellung von PDF-Dokumenten

\usepackage{tabularx}	%Paket, um Tabellen zu erstellen. Hier können Breiten und Zeilenumbrüche vorgegeben werden

\usepackage{supertabular} %Paket für Tabellen, die über mehrere Seiten gehen; im Gegensatz zu longtable werden in diesem Paket die Spaltenbreiten auf jeder Seite neu berechnet

\usepackage{longtable} %Paket zum Erstellen von mehrseitigen Tabellen, unterschiedlich zu supertabular

\usepackage{ragged2e}	%Paket zur Veränderung der Textanordnung und Textausrichtung
	\newcolumntype{L}[1]{>{\RaggedRight\arraybackslash\hspace{0pt}}p{#1}} 	
	\newcolumntype{R}[1]{>{\RaggedLeft\arraybackslash\hspace{0pt}}p{#1}} 		
	\newcolumntype{C}[1]{>{\centering\arraybackslash\hspace{0pt}}p{#1}} 		
\usepackage{endnotes} %Paket zum Setzen von Fußnoten

\usepackage{hyperref}	%Verlinkung der Kapitel im Inhaltsverzeichnis; schwarze Schrift 
\hypersetup{colorlinks = true, linkcolor  = black, urlcolor=black, linkcolor=black}

\usepackage{booktabs} %Paket für diverse Tabellen-Befehle
\usepackage{multirow} %Wie kann man das vergessen.

\usepackage[binary-units = true]{siunitx}
\DeclareSIUnit\px{px}
\usepackage{amsmath} %Paket für mathematische Befehle

\usepackage{float} %Bilder an bestimmte Stelle zwingen mit H
\usepackage{placeins} %Mit \FloatBarrier obiges wirklich erreichen.
\usepackage{makecell} %Tabellenumbruch

\usepackage{caption} %Zum Einrichten von Bildunterschriften und Tabellenüberschriften
\usepackage{appendix} %Paket für den Anhang

\usepackage{pdfpages} %Paket zum Einfügen von PDF-Dateien

\usepackage[version=4]{mhchem}

\usepackage{lscape} %Querformat

\usepackage{rotating} %Paket zum Rotieren von Tabellen und Texten

\usepackage{upgreek} %Nichtkursive griechische Buchstaben

\usepackage{wrapfig} %Bilder neben Fließtext einfügen

\usepackage{icomma}

\makeatletter
\begingroup
\catcode`\_=\active
\protected\gdef_{\@ifnextchar|\subtextit\subtextup }
\endgroup
\def\subtextit|#1|{\sb{#1}}
\def\subtextup#1{\sb{\mathrm{#1}}}
\AtBeginDocument{\catcode`\_=12 \mathcode`\_=32768}
\makeatother

\RequirePackage{ifthen}

\usepackage{courier}
\usepackage{cancel}
\usepackage{chngcntr}
\counterwithout{footnote}{chapter}
\usepackage{caption}
\usepackage{pgfplots}
\usepackage{listings} %Matlab-Codes reinziehen
\lstdefinestyle{Matlab}{
  frame=L,
  language=Matlab,
  showstringspaces=false,
}
\lstdefinestyle{c}{
  frame=L,
  language=C++,
  showstringspaces=false,
}
\begin{document}

\setcounter{tocdepth}{3}	%Festlegung der Tiefe des Inhaltsverzeichnisses
\setcounter{secnumdepth}{3}	%Festlegung der Tiefe der Sektionen

\hypersetup{
pdftitle={Efficient calculation of moments of runaway electron distribution functions},
pdfsubject={Studienarbeit},
pdfauthor={Benjamin Buchholz},
citecolor=black}
%Einstellung von Dokumenteninformationen

\pagestyle{scrheadings} %Kopfzeile zentriert das aktuelle Kapitel, Seitenzahl in Fußzeile
\clearscrheadings	%Die Vorgaben im "scrheadings"-Stil werden gelöscht
\clearscrplain		%Die Vorgaben im "plain"-Stil werden gelöscht
\rohead{\pagemark}	%Einfügen der Seitenzahl rechts in der Kopfzeile
\lohead{\headmark}	%Einfügen des aktuellen Kapitelnamens links in der Kopfzeile

\setlength\abovedisplayshortskip{0.4cm}
\setlength\belowdisplayshortskip{0.4cm}
\setlength\abovedisplayskip{0.4cm}
\setlength\belowdisplayskip{0.4cm}
%Abstände der Formeln zum Text (Oben und Unten)

\thispagestyle{empty}
\begin{titlepage}
	
\begin{textblock}{65}(106,-16)
\includegraphics[width=0.5\textwidth]{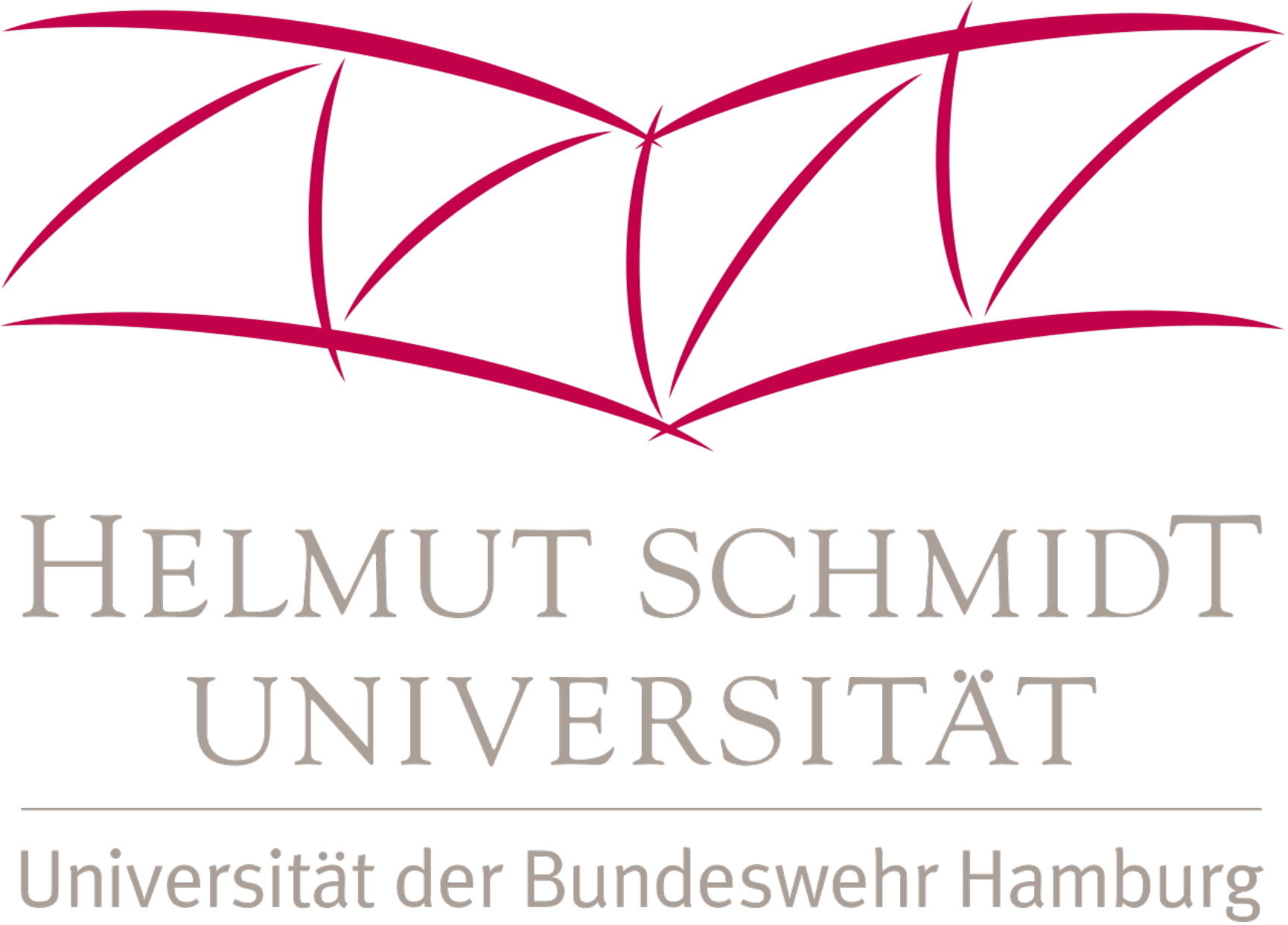}
\end{textblock}
					
\begin{textblock}{65}(80,-1.5)
\includegraphics[width=1.33\textwidth]{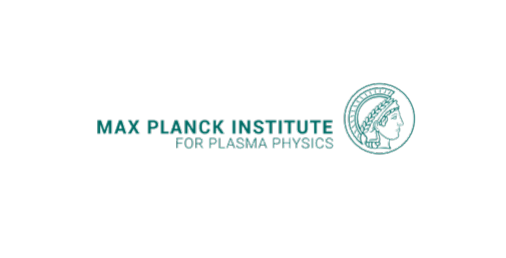}
\end{textblock}

\textcolor[RGB]{140,130,121}{	
\begin{textblock}{80}(0,-23)
\renewcommand{\baselinestretch}{1.1}\footnotesize
  \textsf{Helmut-Schmidt-University/ 
University of the \\German Federal Armed Forces Hamburg\\
  	Faculty of Mechanical and Civil Engineering\\ 
 	Institute for Applied Mathematics\\
	Prof. Dr. rer. nat. Thomas Carraro}
\end{textblock}}

\textcolor[RGB]{140,130,121}{	
\begin{textblock}{80}(0,-1)
\renewcommand{\baselinestretch}{1.1}\footnotesize
\textsf{Max Planck Institute for Plasma Physics, Garching\\
Tokamak Theory Division\\Dr. Gergely Papp}
\end{textblock}}

\textcolor[RGB]{140,130,121}{
\begin{textblock}{65}(0,6)										{\large\textsf{\underline{~~~~~~~~~~~~~~~~~~~~~~~~~~~~~~~~~~~~~~~~~~~~~~~~~~~~~~~~~~~~~~~~~~~~~~~~~~~~~~~~~~~~~~~~~~~~~}}}            
\end{textblock}}
	
\begin{textblock}{65}(44,29.5)
\begin{color}{black}											{\Huge \textbf{\textsf {Study Thesis}}}            
\end{color}
\end{textblock}																	
\begin{textblock}{120}(0,60.5)									
\textcolor[RGB]{165,0,52}{				
\textbf{{\Large Benjamin Buchholz}}}
\end{textblock}	

\begin{textblock}{130}(0,83.5)
\textcolor[RGB]{140,130,121}{
\textbf{{\Large Efficient calculation of the moments of\\ runaway electron distribution functions \\[1.2cm] 
Effiziente Berechnung der Momente von \\Runaway-Elektronen-Verteilungsfunktionen}} \\ [1.75cm]
\begin{tabular}{@{}cl@{}}
{\Large Supervisors:} & {\Large Prof. Dr. rer. nat. Thomas Carraro}\\
&{\Large Dr. Gergely Papp}\\
\end{tabular}\\[1.75cm]
{\Large Hamburg, March 30th, 2023}}

\end{textblock}

\begin{textblock}{100}(-30,135)												
 \includegraphics{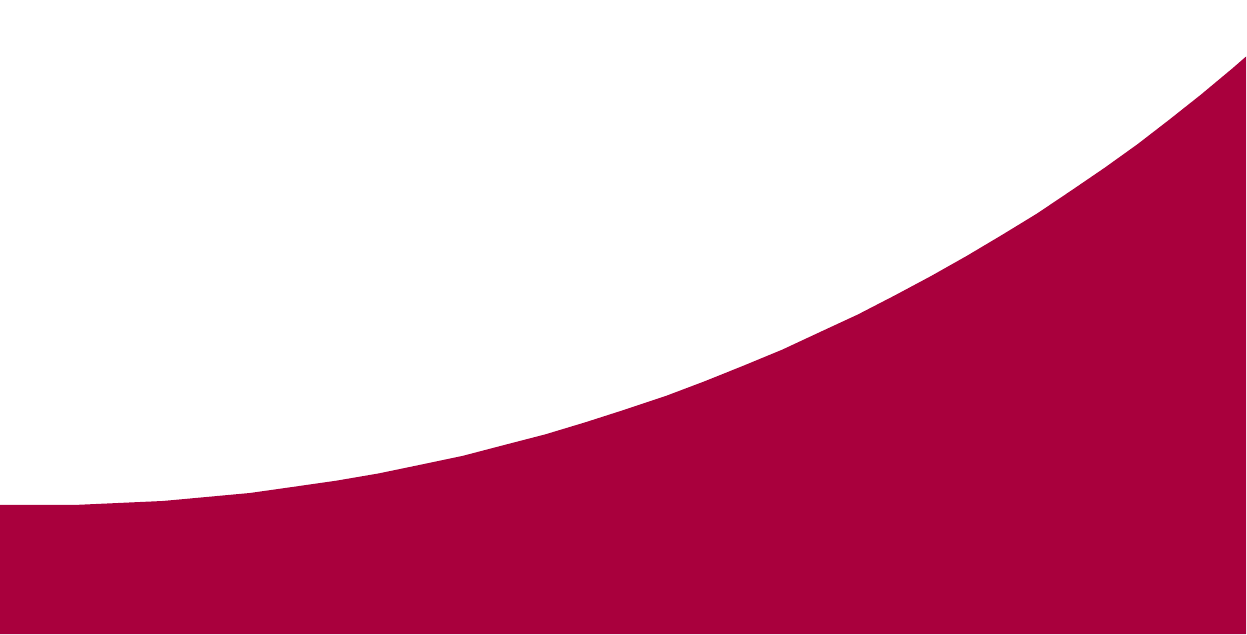}
\end{textblock}

\end{titlepage}
\newpage\thispagestyle{empty}\mbox{}\newpage %Leere Seite
\begin{titlepage}
\newgeometry{left=30mm,right=25mm,bottom=20mm,top=15mm}

\begin{center}
\begin{tabular}{p{\textwidth}}
\vspace{-2cm}
\centering
\begin{minipage}[c]{0.4\textwidth}
\hspace{-2.1cm}\includegraphics[scale=1.3]{IPP_eng.pdf}
\end{minipage}
\hspace{2.8cm}
\begin{minipage}[c]{0.2\textwidth}
\includegraphics[width=\textwidth]{Intellus-Partner-HSU.pdf}
\end{minipage}

\vspace{-0.8cm}

\large{Helmut-Schmidt-University/ \\
University of the German Federal Armed Forces Hamburg}\\[3pt]
\normalsize{Faculty of Mechanical and Civil Engineering\\
Institute for Applied Mathematics}

\vspace{0.85cm}

\LARGE{\textsc{
Efficient calculation of the moments\\of runaway electron\\distribution functions}}

\vspace{0.85cm}

\textbf{\LARGE{Study Thesis}}

\vspace{0.80cm}

\normalsize{submitted by}

\vspace{0.5cm}

\large{\textbf{Benjamin Buchholz}} \\[1pt]
\normalsize{born on June 24th, 1998 in Dresden }

\vspace{0.55cm}

\normalsize{supervised by}

\vspace{0.5cm}

\large{\textbf{Dr. Gergely Papp}} 

\vspace{0.1cm}

Max Planck Institute for Plasma Physics, Garching\\[1pt]
\normalsize{Tokamak Theory Division }

\vspace{2cm}

\begin{tabular}{lllll}
\textbf{Student ID Number:} & & & 00893529\\ 
\textbf{Date of submission:} & & & March 30th, 2023 \\
\textbf{First examiner:} & & & Prof. Dr. rer. nat. Thomas Carraro\\
\textbf{Second examiner:} & & & Dr. Gergely Papp \\
\end{tabular}

\end{tabular}
\end{center}
\end{titlepage}

\thispagestyle{empty}

\begin{titlepage}
\newgeometry{left=30mm,right=25mm,bottom=20mm,top=17mm}

\begin{center}
\begin{tabular}{p{\textwidth}}

\centering
\begin{minipage}[c]{0.4\textwidth}
\includegraphics[scale=0.4]{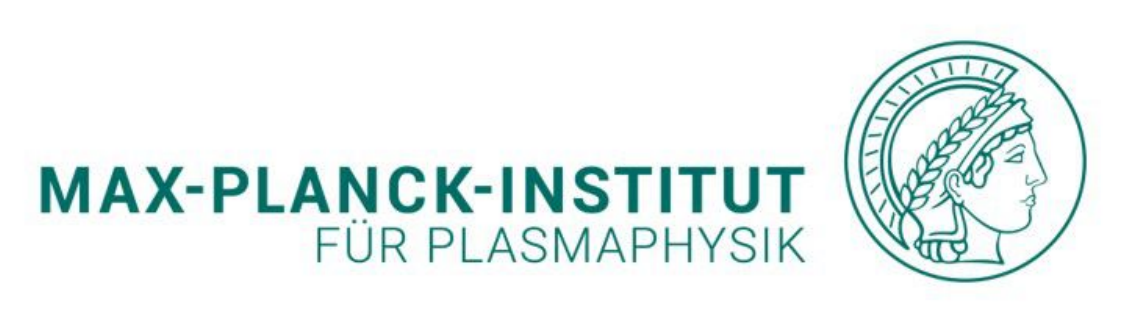}
\end{minipage}
\hspace{3cm}
\begin{minipage}[c]{0.2\textwidth}
\includegraphics[width=\textwidth]{Intellus-Partner-HSU.pdf}
\end{minipage}

\vspace{1cm}

\large{Helmut-Schmidt-Universit\"at/ \\
Universit\"at der Bundeswehr Hamburg}\\[3pt]
\normalsize{Fakult\"at f\"ur Maschinenbau und Bauingenieurwesen \\
Institut f\"ur Angewandte Mathematik }

\vspace{0.85cm}

\LARGE{\textsc{
Effiziente Berechnung der Momente von Runaway-Elektronen-Verteilungsfunktionen}}

\vspace{0.85cm}

\textbf{\LARGE{Studienarbeit}}

\vspace{0.8cm}

\normalsize{vorgelegt von}

\vspace{0.5cm}

\large{\textbf{Benjamin Buchholz}} \\[1pt]
\normalsize{geboren am 24.\hspace{1.4mm}Juni\hspace{1.5mm}1998 in Dresden}

\vspace{0.55cm}

betreut von

\vspace{0.5cm}

\large{\textbf{Dr. Gergely Papp}} 

\vspace{0.1cm}

Max-Planck-Institut f\"ur Plasmaphysik, Garching\\[1pt]
\normalsize{Bereich Tokamak-Theorie }

\vspace{2cm}

\begin{tabular}{lllll}
\textbf{Martikelnummer:} & & & 00893529\\ 
\textbf{Abgabedatum:} & & & 30.\hspace{1.4mm}März\hspace{1.5mm}2023\\
\textbf{Erstgutachter:} & & & Prof. Dr. rer. nat. Thomas Carraro\\
\textbf{Zweitgutachter:} & & & Dr. Gergely Papp\\
\end{tabular}

\end{tabular}
\end{center}
\end{titlepage}

\thispagestyle{empty}

\addchap*{Eidesstattliche Erklärung}
\label{cap:Erklaerung}

Hiermit erkläre ich, Benjamin Buchholz, die vorliegende Arbeit selbstständig angefertigt zu haben. Die Erstellung erfolgte ohne das unerlaubte Zutun Dritter. Alle Hilfs-mittel, die für die Erstellung der vorliegenden Arbeit benutzt wurden, befinden sich ausschließlich im Literaturverzeichnis. Alles, was aus anderen Arbeiten unverändert oder mit Abänderungen übernommen wurde, ist kenntlich gemacht.

Die Arbeit wurde bisher keiner anderen Prüfungsbehörde vorgelegt.

\vspace{2em}
Hamburg, 30.\hspace{1.4mm}März\hspace{1.5mm}2023

\vspace{1em}

\rule[-1cm]{5cm}{0.1mm}\\[1em]
Benjamin Buchholz

\clearpage

\pagenumbering{Roman} %Römische Seitenzahlen

\phantomsection % für link auf engl. Abstract Seite von Contents aus (ohne landet man bei der Seite für die Erklärung)

\addcontentsline{toc}{chapter}{Abstract}

\setcounter{page}{6}
\begin{abstract}
\vspace*{6.8cm}
{\Large \textbf{Abstract}}
\noindent\\\vspace*{-4mm}\\ \noindent 
In view of an increasing energy demand, research in the field of the thermonuclear toroidal \textit{tokamak} fusion reactor concept, which is characterized by a toroidal plasma current \cite{wesson} is of importance. Plasma current instabilities can destabilize the plasma discharge and cool the plasma rapidly \cite{Hender_2007,REsimulation}. In such \textit{disruptions} \cite{Hoppe_2021} or in the start-up phase of the reactor \cite{Hoppe_2022}, inductive electric fields are generated which accelerate electrons to relativistic velocities, resulting in a beam of \textit{runaway electrons} \cite{stahl}. This can potentially damage the reactor vessel and must be avoided in future reactors such as ITER \cite{Hoppe_2021,REsimulation,Hoppe_2022}. Thus, the efficient simulation of the evolution of the runaway electron current is motivated for prediction, avoidance and attenuation of disruptions \cite{Hoppe_2022}. In order to improve simulations based on a self-consistent calculation of the runaway electron current, the efficient computation of the moments of analytical runaway electron distribution functions is of interest. \\
In this respect, the general procedure is carried out through the example of the distribution function of the \textit{avalanche} generation of runaway electrons according to \mbox{\textit{Fülöp et al.}} \cite{REdistfuncderivation}. At this the runaway electron number density, the current density and the mean mass-related kinetic energy density, which result from the zeroth, first and second moment are considered. Their analysis is carried out analytically and numerically. By means of a \textsc{MATLAB} implementation, suitable calculation rules are derived and analyzed with regard to runtime efficiency. Finally, a physical evaluation of the components and the magnitude of the current density vector as well as the kinetic energy density for the plasma parameter space constructed from electric field, electron density and electron temperature is carried out, applying the derived efficient calculation rules. In addition, the applicability of the selected distribution function is discussed on the basis of graphical depictions of the results. 
\end{abstract}
\begin{abstract}
\vspace*{4.8cm}
{\Large \textbf{Kurzzusammenfassung}}
\noindent\\\vspace*{-4mm}\\ \noindent 
Angesichts des steigenden Energiebedarfs ist die Forschung im Bereich des thermonuklearen toroidalen \textit{Tokamak}-Fusionsreaktorkonzeptes von Bedeutung, welches durch einen toroidalen Plasmastrom \cite{wesson} gekennzeichnet ist. Hierbei können Plasmastrominstabilitäten die Plasmaentladung stören und das Plasma schnell abkühlen lassen \cite{Hender_2007,REsimulation}. Bei solchen \textit{Disruptionen} \cite{Hoppe_2021} oder in der Anlaufphase des Reaktors \cite{Hoppe_2022} werden induktive elektrische Felder erzeugt, die Elektronen auf relativistische Geschwindigkeiten beschleunigen, was zu einem Strahl von \textit{Runaway-Elektronen} \cite{stahl} führt. Dieser kann potentiell das Reaktorgefäß beschädigen und muss in zukünftigen Reaktoren wie ITER vermieden werden \cite{Hoppe_2021,REsimulation,Hoppe_2022}. Damit wird die effiziente Simulation der Zeitentwicklung des Runaway-Elektronenstroms zur Vorhersage, Vermeidung und Abschwächung von Disruptionen motiviert \cite{Hoppe_2022}. Zur Verbesserung von Simulationen, die auf einer selbstkonsistenten Berechnung des Runaway-Elektronenstromes beruhen, ist die effiziente Berechnung der Momente analytischer Runaway-Elektronen-Verteilungs-funktionen von Interesse. \\
Diesbezüglich wird die allgemeine Vorgehensweise exemplarisch für die Verteilungsfunktion der lawinenartigen Erzeugung von Runaway-Elektronen nach \mbox{\textit{Fülöp et al.}} \cite{REdistfuncderivation} durchgeführt. Dabei werden die Runaway-Elektronen-Anzahldichte, die Stromdichte und die mittlere massenbezogene kinetische Energiedichte, welche sich aus dem nullten, ersten und zweiten Moment ergeben, betrachtet. Ihre Auswertung wird analytisch und numerisch vorgenommen. Mittels einer \textsc{MATLAB}-Implementierung werden geeignete Berechnungsregeln abgeleitet und auf Laufzeit-Effizienz analysiert. Schließlich findet eine physikalische Bewertung der Komponenten und des Betrages des Stromdichtevektors sowie der kinetischen Energiedichte für den Plasmaparameterraum aus elektrischem Felde, Elektronendichte und Elektronentemperatur statt, wobei die abgeleiteten effizienten Berechnungsregeln Anwendung finden. Darüber hinaus wird die Anwendbarkeit der gewählten Verteilungsfunktion anhand grafischer Darstellungen der Ergebnisse diskutiert.
\end{abstract}

\pdfbookmark[chapter]{Contents}{contents}

\tableofcontents

\clearpage

\addchap{Nomenclature}

\textbf{Latin letters}
\begin{table}[H]
\renewcommand{\arraystretch}{1.1}
\newcolumntype{s}{>{\hsize=.4\hsize}X}
\newcolumntype{S}{>{\hsize=.7\hsize}X}
\begin{tabularx}{\textwidth}{sSX}
\toprule
Symbol				&				Unit				& 				Denotation \\    
\midrule
$t$		&				\si{\second}		&	time \\
$T$		&				\si{\kelvin}		&	temperature \\
$K$		&				\si{\joule\per\kilogram}\,=\,\si{\meter\squared\per\second\squared}		&	mass-related kinetic energy density \\
$q$		&				\si{\ampere\second}		&	electric charge \\
 
$n_{\alpha}$		&				\si{\per\cubic\meter}		&	particle density of particle species $\alpha$ \\
$N_{\alpha}$		&				-	&	number of particles of a species $\alpha$ \\
$m_{\alpha0}$ &				\si{\kilogram}		&	rest mass of the particle species $\alpha$ \\
$\mathbf{r}$		&					\si{\meter} &position vector\\
$v$\,=\,$\abs{\mathbf{v}}$		&					\si{\meter\per\second} &velocity\\
$p$\,=\,$\vert\mathbf{p}\vert$&					\si{\kilogram\meter\per\second} & (relativistic) momentum\\
$F$\,=\,$\vert\mathbf{F}\vert$&					\si{\kilogram\meter\square\per\second} & force\\
$E$\,=\,$\vert\mathbf{E}\vert$		&					\si{\volt\per\meter}\,=\,\si{\kilogram\meter\per\second\per\cubed\ampere}		&	electric field strength \\
$B$\,=\,$\vert\mathbf{B}\vert$		&				\si{\tesla}\,=\,\si{\kilogram\per\ampere\per\square\second}		&	magnetic flux density \\
$j$\,=\,$\vert\mathbf{j}\vert$		&				\si{\ampere\per\square\meter}		&	current density \\
$Z_{eff}$ &				-		&	effective ion charge  \\
$f_{\alpha}$ &				-		&	distribution function of the particle species $\alpha$ \\
$M_{n}$ & -& $n$th moment of a distribution function\\
\bottomrule						
\end{tabularx}
\end{table}

\textbf{Greek letters}
\begin{table}[H]
\renewcommand{\arraystretch}{1}
\newcolumntype{s}{>{\hsize=.5\hsize}X}
\newcolumntype{S}{>{\hsize=.8\hsize}X}
\begin{tabularx}{\textwidth}{sSX}
\toprule
Symbol				&				Unit				& 				Denotation \\      
\midrule

$\tau$		&			\si{\second}	&	characteristic time scale \\
\bottomrule						
\end{tabularx}
\end{table}

\clearpage

\textbf{Physical and mathematical constants}
\begin{table}[H]
\renewcommand{\arraystretch}{1.24}
\newcolumntype{s}{>{\hsize=.26\hsize}X}
\newcolumntype{S}{>{\hsize=0.95\hsize}X}
\newcolumntype{P}{>{\hsize=0.93\hsize}X}
\begin{tabularx}{\textwidth}{sSP}
\toprule
Symbol		&				Value and Unit				& 				Denotation \\    
\midrule

$\pi$ & $3.14159265358979323846264$&	ratio of a circle's circumference to its diameter \cite{WolframPI} \\

e & $2.71828182845904523536028$&	\textit{Euler's} number \cite{OEISe} \\

$c$  &				$2.99792458 \cdot 10^{8}$ \si{\meter\per\second}		&	speed of light in vacuum \cite{NISTc}\\

$e $ &				$1.602176634 \cdot 10^{-19}$ \si{\ampere\second}		&	elementary charge \cite{NISTe} \\

$m_{e0}$  &				$9.1093837015 \cdot 10^{-31}$ \si{\kilogram}		&	electron rest mass \cite{NISTme}\\

$\mu_{0} $&				$1.25663706212\cdot 10^{-6}$ \si{\newton\per\square\ampere}		& vacuum magnetic permeability \cite{NISTmue0}  \\

$k_{B} $&				$8.617333262\cdot 10^{-5}$ \si{\electronvolt\per\kelvin}		& \textit{Boltzmann} constant \cite{NISTkB}  \\

$\varepsilon_{0}$  &				\mbox{$8.8541878128\cdot10^{-12}$  \si{\ampere\second\per\volt\per\meter}}		&	
vacuum electric permittivity \cite{NISTeps0}\\

$\textup{eV}$  &	$e\cdot 1\textup{V}=1.602176634\cdot10^{-19}$ \si{\ampere\volt\second}			 	&	
electron volt \cite{NISTeV}\\

\bottomrule						
\end{tabularx}
\end{table}
\clearpage

\textbf{Mathematical and physical symbols and operators}
\begin{table}[H]
\renewcommand{\arraystretch}{1.25}
\newcolumntype{s}{>{\hsize=.3\hsize}X}
\newcolumntype{S}{>{\hsize=.98\hsize}X}
\begin{tabularx}{\textwidth}{sXS}
\toprule
Symbol				&				representation				& 				Denotation \\      
\midrule

$ \mathbf{e}_{i}$			&	 $ \mathbf{e}_{i}:=( a_{j})_{j=\left\lbrace 1,2,...,d\right\rbrace};  $		$\,\vert\mathbf{e}_{i}\vert=1$&	 $i$th unit vector in $d$ dimensions \\[9pt]

$ \vec{\nabla}	$			&		$ \vec{\nabla}=\displaystyle{\sum\limits_{i=1}^{d}}\,\mathbf{e}_{i}\cdot\dfrac{\partial}{\partial x_{i}}$			&	\mbox{cartesian \textit{Nabla}-Operator in $d$} dimensions \\

$ \vec{\nabla}	$			&		$ \vec{\nabla}=\displaystyle{\sum\limits_{i=1}^{d}}\,\vec{\nabla}u_{i}\cdot\dfrac{\partial}{\partial u_{i}}$			&	\textit{Nabla}-Operator in $d$ dimensions for curvilinear coordinates $u_{i}$\\

$ \textup{erf}(z)	$			&		$ \textup{erf}(z)=\dfrac{2}{\sqrt{\pi}}\,\displaystyle{\int\limits^{z}_{x=0}} \,\textup{e}^{-x^2}\, \mathrm{d}x$			&	error function \cite{helander}\\[20pt]

$ \textup{erfc}(z)	$			&		$ \textup{erfc}(z)\hspace{-1mm}=\hspace{-1mm}1\hspace{-0.9mm}-\hspace{-0.9mm}\textup{erf}(z)\hspace{-0.9mm}=\hspace{-1mm}\dfrac{2}{\sqrt{\pi}}\displaystyle{\int\limits^{\infty}_{x=z}} \hspace{-0.4mm} \textup{e}^{-x^2}  \mathrm{d}x$			&	\mbox{complementary error function \cite{NISTerfc}}\\[20pt]

$ \Gamma(z,\,a)	$			&		$ \Gamma(z,\,a)=  \displaystyle{\int\limits^{\infty}_{x=a}} \,x^{z-1}\textup{e}^{-x}\, \mathrm{d}x$			&	upper incomplete gamma\newline function \cite{incgammafunc}\\[20pt]

$ E_{1}(z)	$			&		$ E_{1}(z)=\displaystyle{\int\limits^{\infty}_{x=z}}x^{-1}\textup{e}^{-x}\,\mathrm{d}x$ 			&	exponential integral \cite{NISTexpint}  \\[20pt]

$ \gamma	$			&		$ \gamma=\gamma(v)=\dfrac{1}{\sqrt{1-\left(\dfrac{v}{c}\right)^2}}$			&	\textit{Lorentz} or gamma factor \\[27pt] 

$\ln{\left(\Lambda\right)}	$	&	 $ \ln{\left(\Lambda\right)}= 14.9-0.5\cdot\ln{\left(\dfrac{n_{e}}{10^{20}\,\textup{m}^{-3}}\right)} $			&	\mbox{relation for the \textit{Coulomb} logarithm} \\[8pt]
 &	 \hspace{1.55cm}$  +\; \ln{\left(\dfrac{k_{B}T_e}{10^3\,\textup{eV}}\right)}$			& \mbox{for electron-electron collisions \cite{wesson}} \\ [15pt]

\bottomrule						
\end{tabularx}
\end{table}

\clearpage

\textbf{Indices and abbreviations}
\begin{table}[H]
\renewcommand{\arraystretch}{0.9}
\newcolumntype{S}{>{\hsize=0.45\hsize}X}
\begin{tabularx}{\textwidth}{SX}
\toprule
Symbol				&				meaning \\      
\midrule

analyt & analytical \\

ava & avalanche generation mechanism \\

C & \textit{Coulomb}\\

c & critical \\

D & \textit{Dreicer}\\

DREAM & Disruption Runaway Electron Analysis Model\\

eff & effective \\

E & electric field\\

e & electron \\

e.\hspace{0.9mm}g. & exempli gratia \\

et.\hspace{1mm}al. &  et alia  \\

g & gyro \\

max & maximal \\

min & minimal \\

n& neutral particle\\

norm & normalized\\

num & numerical \\

i.\hspace{0.95mm}a. & inter alia \\

%I.b.P. & integration by parts \\

i & ion \\

ITER & International Thermonuclear Experimental Reactor\\

JET & Joint European Torus\\

L & \textit{Lorentz}\\

RE & runaway electrons \\

sa & slide-away\\

SI &  International System of Units \\

th & thermal \\

Z & effective ion charge $Z_{eff}$\\

$\|$ & parallel \\

$\perp$ & orthogonal \\

\bottomrule						
\end{tabularx}
\end{table}
%\begin{table}[H]
%\renewcommand{\arraystretch}{0.9}
%\newcolumntype{S}{>{\hsize=0.45\hsize}X}
%\begin{tabularx}{\textwidth}{SX}     
%\midrule
%
%i.a. & inter alia \\
%
%
%\bottomrule						
%\end{tabularx}
%\end{table}

\clearpage

\pagenumbering{arabic}	

\chapter{Introduction}

\begin{figure}[H]
\begin{center}
\includegraphics[trim=150 526 151 73,width=0.591\linewidth,clip]{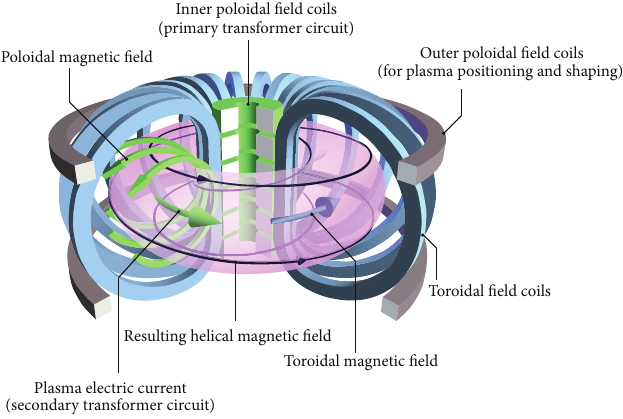} 
\caption[Schematic depiction of a tokamak chamber with the magnetic field and coil configuration \cite{Li_2014}]{Schematic depiction of a tokamak chamber with the magnetic field and coil configuration \cite{Li_2014}}
\label{tokamak_fig}
\end{center}
\end{figure}

In the face of a growing world population and a progressive rise in living standards, humanity needs an increasing amount of energy. Additionally, one averts from conventional energy production based on fossil fuels, due to the finiteness of their reserves, their current economical extraction range and negative influence on the sensitive balance of the environment of our planet. Thus particulary industrial and emerging countries with high energy demands orient themself towards the utilization of renewable energies and nuclear energy. Therefore, the current research in energy production and technology, for instance in Europe, China and the United States of America, focuses on the further development of these areas. At the moment, the development of and the research on nuclear fusion reactors is pronounced. At that, the currently most promising reactor design is as a toroidal thermonuclear fusion device relying on the magnetic confinement of a deuterium-tritium plasma, the so-called \textit{tokamak} \cite{wesson}. Hereinafter, the functional principles of \textit{tokamak} fusion reactors, like for example the JET-reactor or the under construction ITER-reactor, shall be explained by reference to the book from \textit{J.\hspace{0.9mm}Wesson} \cite{wesson}. In the course of this, one is referred to figure \ref{tokamak_fig} \cite{Li_2014}, which shows a schematic depiction of a tokamak.\vspace{1mm}
\vspace{-8.5mm}
In general, the magnetic confinement represents a state, in which the charged particles in the plasma approximately move along the mainly toroidal magnetic field lines, by performing a gyration motion around the field lines as a consequence of the \textit{Lorentz} force acting on them. This implies, that the plasma pressure is sufficiently compensated by the magnetic pressure. In a tokamak fusion reactor the magnetic confinement also requires a poloidal component of the magnetic field, which arises from a transformer-induced toroidal plasma current. In the consequence, the magnetic field lines are helically wound and the plasma in this pinch fusion reactor is confined in an unstable equilibrium. Moreover, only pulsed plasma discharges are possible in a tokamak, the electric current within the transformer cannot be increased unlimitedly.

The tokamak-intrinsic plasma
current can abruptly change as a consequence of plasma instabilities, during so-called \textit{disruptions}, in which the magnetic equilibrium and thus the plasma discharge becomes unstable, the plasma cools down and eventually terminates \cite{Hender_2007,wesson}. While a disruption occurs \cite{Hoppe_2021}, during disruption mitigation \cite{REsimulation} or as well in the start-up phase of the reactor \cite{Hoppe_2022} inductive electric fields are produced that can accelerate electrons to relativistic velocities, generating a beam of \textit{runaway electrons}, if a certain critical electric field is exceeded \cite{Dreicer_1959}. This beam has the potential to damage the components of the reactor vessel \cite{Hoppe_2021} and is a major concern for future reactors such as ITER \cite{Hoppe_2021,REsimulation,Hoppe_2022}. In plasma fusion physics the study of runaway electron generation is therefore important in the research area of disruption prediction, avoidance and mitigation \cite{Hoppe_2022}. \\
On this occasion, it should be remarked, that the runaway electron phenomenon is typical in plasma physics and not restricted to thermonuclear fusion plasma in tokamak reactors. In fact, runaway electrons also form e.\hspace{0.8mm}g.\hspace{1.6mm}in atmospheric
plasmas during lightning discharges or in astrophysical plasmas like solar flares, where the references related to the named examples can be found in the PhD thesis \cite{stahl}.

Useful runaway electron simulations are required to be as efficient as possible, in order to develop optimal control mechanisms to avoid damages to the reactor. Hence, the for instance computational expensive solution of the kinetic equation for the runaway electron distribution function is not suitable for certain studies. Typical simulation software like the DREAM-code \cite{Hoppe_2022} self-consistently simulate the plasma evolution and use approximations and lower-dimensional plasma descriptions. Similarly, a self-consistent evolution of the runaway current is possible \cite{pappPHD}, which could become responsible for damages to the reactor wall and is an eligible quantity for control of disruption mitigation. For this, the calculation of the moments of analytical distribution functions for the runaway electrons, where e.\hspace{0.8mm}g.\hspace{1.6mm}the runaway electron current density is determined by the first moment, can be an enrichment for a fast simulation software. Here, it should be remarked that runaway electrons are produced by different generation mechanisms, which contribute differently to the runaway current. 

Therefore the focus of this study thesis is the efficient calculation of the moments of an analytic runaway electron distribution function, for the purpose of presenting the general procedure, which can be transferred to other distribution functions in the future. \\
First, the runaway electron phenomenon is explained and framed by a theoretical context with a focus on the avalanche generation mechanism. At this, the distribution function from \mbox{\textit{Fülöp et al.}} \cite{REdistfuncderivation} is used, which is based on the growth rate proposed by \textit{Rosenbluth} \& \textit{Putvinski} \cite{Rosenbluth_1997}. 
\\
Second, the calculation of the moments is approached analytically and numerically with the goal to find calculation rules for the integrals in the definition of the moments, which require the shortest runtime and are therefore the most efficient computation rules applicable in future simulation codes. Nevertheless, one only expects programming language and implementation dependent statements, which is why this thesis is limited to \textsc{MATLAB}-implementations and further solely analyses the avalanche runaway electron number density, the current density and the mean mass-related kinetic energy density, related to the zeroth, the first and the second moment of the distribution function. \\
Third, a physical evaluation of the efficiently computed quantities is necessary to assess the applicability of the implementations based on the assumed approximations and the used analytic distribution function.

\clearpage

\chapter[Theoretical framework for the description of runaway electrons]{Theoretical framework for the description of runaway electrons in tokamak disruptions}\label{theory_chapter}

\section{Generation mechanisms for runaway electrons in tokamak fusion reactors}\label{mechanisms_section}

In the start-up phase of nuclear fusion reactors of the tokamak type low electron densities $n_{e}\approx 10^{18}$ \cite{Hoppe_2022} are prevalent and strong electric fields are induced. This is necessary to fastly create a sufficiently high plasma current and temperature, in order to achieve a fully ionized plasma and optimal magnetic confinement within the tokamak reactor. For continuative information about the start-up of a tokamak one is directed to the publication from \mbox{\textit{Hoppe et al.}} \cite{Hoppe_2022}. 

In addition, one can observe unusually high induced electric fields up to more than $100\,\mathrm{V/m}$ during plasma-terminating disruptions, if for instance impurity injections (i.e. pellet or gas injections) are used to avoid disruption-related damages \cite{REsimulation,Hender_2007}. This mitigation method is a planned method for future tokamak reactors, because it cools down the plasma via isotropic radiation, which efficiently distributes the energy and avoids high localised thermal loads to the walls \cite{Hender_2007}. However, the temperature drop, as a consequence of the diruption mitigation, occurs on the short time scale of the thermal quench. Thereupon an electric field is induced, because the toroidal current in the plasma cannot decay on the same timescale as the temperature and conductivity does, due to inductance. It shall be remarked, that this paragraph orients itself on the paper of \mbox{\textit{Fehér et al.}} \cite{REsimulation}, which discusses the simulation of runaway electron generation in scenarios, where a plasma shutdown is forced by an impurity injection. 

In both of the mentioned scenarios the electrons within the plasma are accelerated by the force $F_{E}=-eE$ orginating from the induced electric field $E$, where the negative elementary charge $q=-e$ is the charge of an electron. Roughly above the thermal velocity $v_{th}=\sqrt{(2ek_{B}T_{e})/m_{e0}}$ \cite{wesson,stahl}, this acceleration is mainly compensated by the friction force $F_{C} \sim v^{-2}$, due to electron-electron \textit{Coulomb} collisions, which decreases with the square of the electron velocity for $v\rightarrow c$, where $c$ is the speed of light in vacuum. From that, it is clear that electrons with a velocity $v$ higher than a certain critical velocity $v_{c}$ experience a continuous net acceleration as a consequence of the resulting accelerating force, if the accelerating electric field $E$ is larger than a critical electric field $E_{c}$. A more detailed description of the considered processes can be found in the PhD thesis of \textit{A. Stahl} \cite{stahl}, whereby the reference \cite{REsimulation} from the preceding paragraph allows further understanding. The representation for electric field associated to the minimum of the friction force - the \textit{critical} electric field - follows from the collision frequency for a highly relativistic electron and was found by \textit{Connor} and \textit{Hastie} in $1975$ \cite{stahl,Connor_1975} and reads:\vspace*{-0.5mm}
\begin{equation}\label{E_crit}
E_{c} = \dfrac{n_{e}\,e^3\ln{(\Lambda)}}{4\pi\,\varepsilon_{0}^2 \,m_{e0}\,c^2}\,.
\end{equation}
At this, the factor by which small-angle collisions are more effective than large-angle collisions, the \textit{Coulomb} logarithm, appears. It can be calculated for instance for electron-electron collisions from the relation \cite{wesson}:\vspace*{-0.5mm}
\begin{equation}\label{CoulombLog}
\ln{\left(\Lambda\right)}= 14.9-0.5\cdot\ln{\left(\dfrac{n_{e}}{10^{20}\,\textup{m}^{-3}}\right)}+ \ln{\left(\dfrac{k_{B}T_e}{10^3\,\textup{eV}}\right)}\,,
\end{equation}
with the \textit{Boltzmann} constant expressed as $k_{B}=8.617333262\cdot 10^{-5} \,\si{\electronvolt\per\kelvin}$ \cite{NISTkB}.
The mentioned continuously accelerated electrons with $v\gg v_{th}=\sqrt{(2ek_{B}T_{e})/m_{e0}}$ \cite{wesson,stahl} are superthermal in relation to the thermal electrons of a distriubtion of electrons in the phase space with $v\approx v_{th}$. Moreover, they reach relativistic velocities and are referred to as \textit{runaway electrons}. Their maximum relativisitic normalized momentum $p=(m_{e0}\gamma v)/(m_{e0}c)=(\gamma v)/c$ is bound by synchrotron radiation which (primarily) results from the gyro motion of the electrons around the magnetic field lines (see also \cite{stahl}), introducing the \textit{Lorentz} factor $\gamma=\gamma(v)$. Where in the case of a tokamak disruption, the total energy from the plasma current bounds the acceleration of the electrons, before the synchrotron radiation limit is reached. The critical relativistic momentum and thus the critical velocity is related to the critical electric field strength from (\ref{E_crit}) by the following expressions \cite{stahl}:\vspace*{-0.5mm}
\begin{equation}\label{p_crit}
p_{c} = \dfrac{\gamma_{c}v_{c}}{c}= \dfrac{1}{\sqrt{\dfrac{E}{E_{c}}-1}} \;\;;\;\; \gamma_{c}=\gamma(v_{c})=\dfrac{1}{\sqrt{1-\left(\dfrac{v_{c}}{c}\right)^2}}=\dfrac{1}{\sqrt{1-\dfrac{E_{c}}{E}}}\,.
\end{equation}
Typically, one finds the character of the runaway electron population in a tokamak disruption to be of a highly energetic beam along magnetic field lines or flux surfaces with a radially centered localization \cite{REdistfuncderivation}. If this beam comes into contact with the wall, this localised energy input can lead to severe damages \cite{Matthews2016,Reux2015}.
 
\begin{figure}[H]
\begin{center}
\includegraphics[trim=125 42 139 50,width=0.9\textwidth,clip]{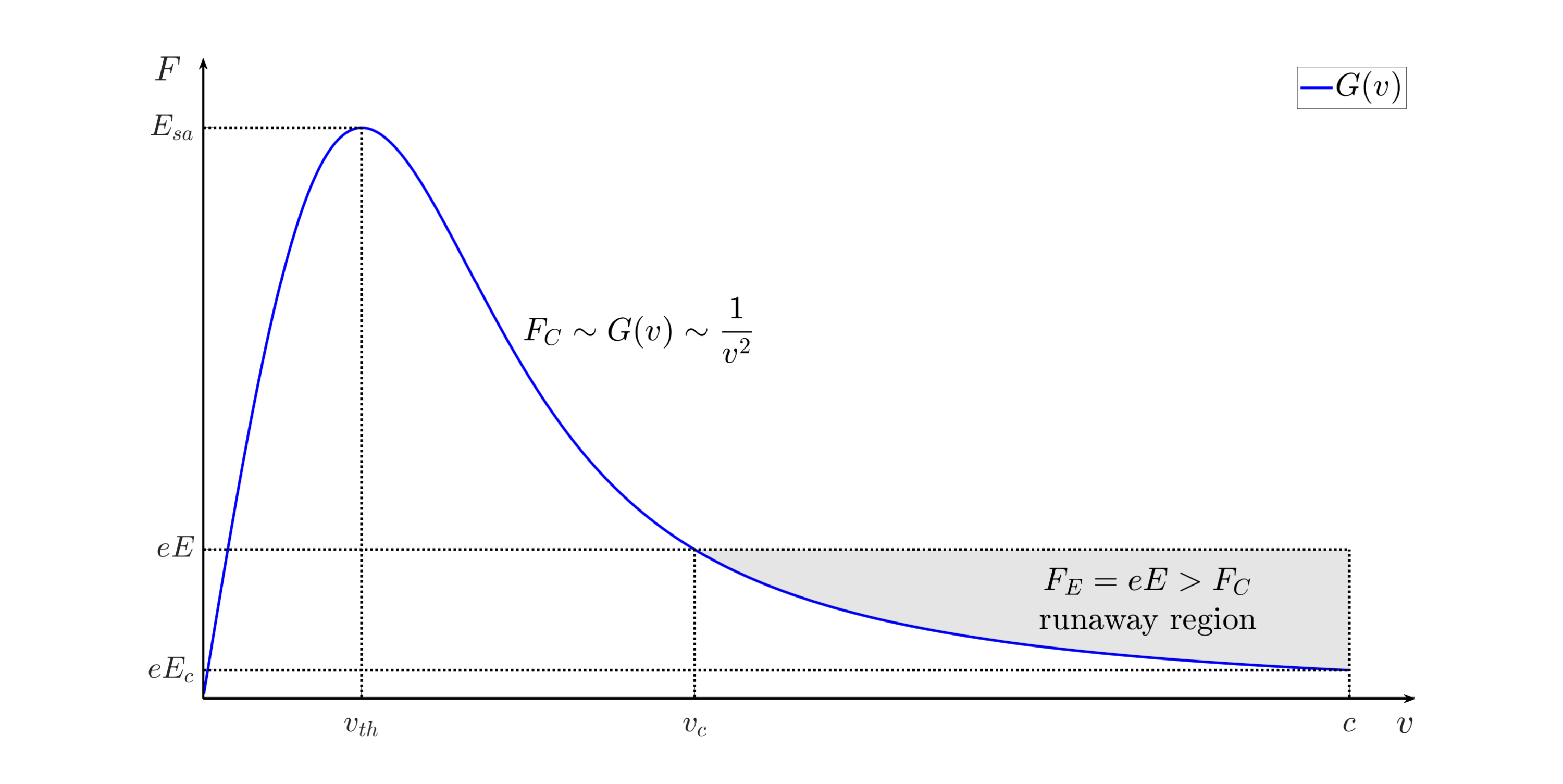}
\caption[Depiction of the forces effecting the electron dynamics for an induced electric field with $E_{c} < E \ll E_{sa}$ and schematic representation of the region for runaway-electron generation (see also \cite{stahl})]{Depiction\protect\footnotemark{} of the forces effecting the electron dynamics for an induced electric field with $E_{c} < E \ll E_{sa}$ and schematic representation of the region for runaway-electron generation (see also \cite{stahl})}
\label{fig_RE_region}
\end{center}
\end{figure}

The velocity dependence of the friction force $F_{C}$, due to the electron-electron \textit{Coulomb} collisions, can be expressed by the \mbox{\textit{Chandrasekhar}} function \cite{helander}:
\begin{equation}\label{Chadrasekhar}
G(v) = \dfrac{1}{2v^2}\cdot\left( \textup{erf}(v)-v\cdot\dfrac{\mathrm{d}}{\mathrm{d}v}\left[ \textup{erf}(v)\right] \right)= \dfrac{1}{2v^2}\cdot\left( \textup{erf}(v)-v\cdot\dfrac{2}{\sqrt{\pi}}\cdot e^{-v^2} \right)\,,
\end{equation}
where i.a. \hspace{-0.7mm}the error function $\textup{erf}(v)$ and its first derivative were used. The generation region for runaway electrons based on the velocity dependence of the friction force can then be looked upon schematically in figure \ref{fig_RE_region}.

For electric fields with $E>E_{sa}\cong0.214E_{D}$ \cite{Dreicer_1959}, where $E_{D}$ is the \textit{Dreicer} field, the whole electron population becomes a runaway electron population \cite{stahl}. The \textit{Dreicer} field is given as follows \cite{stahl}:\vspace*{-3mm}
\begin{equation}\label{Dreicerfield}
E_{D} = \dfrac{n_{e}\,e^3\ln{(\Lambda)}}{4\pi\,\varepsilon_{0}^{2}\,e\,k_{B}\,T_{e}}\,,
\end{equation}
with the \textit{Boltzmann} constant $k_{B}=8.617333262\cdot 10^{-5} \,\si{\electronvolt\per\kelvin}$ \cite{NISTkB}. 
However, this so-called \textit{slide-away} phenomenon \cite{Coppi_1976} does almost never appear in tokamak reactors, because it often holds $E\ll E_{sa}$ for fusion plasmas (compare \cite{stahl}). Hence, the important interval for the study of runaway electron dynamics is $E_{c} < E \ll E_{sa}$. The primary generation of runaway electrons through momentum space transport processes into the runaway region shown in figure \ref{fig_RE_region} is referred to as the \textit{Dreicer} generation mechanism. The PhD thesis of \textit{A. Stahl} \cite{stahl} is hereof suitable as a comparative reference.
\footnotetext{\label{fig_RE_region_footnote} The diagram in figure \ref{fig_RE_region} was produced with the \textsc{MATLAB}-file "plot_runaway_region.m",\\ \hspace*{8.7mm}which can be found in the digital appendix.}
\vspace{-7mm}
As mentioned above, there exists another important primary mechanism  for the production of runaway electrons, which has to be considered. This so-called \textit{hot-tail} generation mechanism relies on rapid plasma cooling as it might occur in scenarios of disruption mitigation. At this, one might also review the explanations in the previously mentioned PhD thesis \cite{stahl} and the publication of \textit{H.\hspace{1mm}Smith\hspace{1mm}et\hspace{0.8mm}al.} \hspace{-1.3mm}\cite{Smith_2005}. If the cooling time scale is shorter than the typical collision time 
\begin{equation}\label{tau_coll}
\tau=\dfrac{4\pi\,\varepsilon_{0}^{2}\,m_{e0}^{2}\,c^{3}}{n_{e}\,e^4\ln{(\Lambda)}}
\end{equation} 
according to \cite{REdistfuncderivation}, at which particles evolve towards equilibrium, the high energy tail of the particle distribution function behaves stagnant, because the collision time of those highly energetic particles is so small, that their evolution towards the equilibrium is significantly slower than those of the main part of the distribution. Hence, particles with a larger momentum than the critical momentum remain longer at their high energy level, even if the plasma temperature has dropped. In combination with an induced electric field, this can consequently lead to an acceleration of those superthermal electrons into the runaway region (compare \cite{stahl}). Here, one should remark, that a runaway electron population can also exist, as part of the electron state distribution in momentum space, whithout an acceleration by an induced electric field.

The primary generation mechanisms explain the presence and occurance of runaway populations in plasmas. Once a seed of runaway electrons exists in a plasma, secondary generation, such as the avalanche mechanism, can take place, exponentially increasing the runaway current \cite{Rosenbluth_1997}. However, for certain future tokamak scenarios the avalanche-like production mechanism as a secondary generation mechanism dominates the primary generation mechanisms, if there is an existing runaway electron population \cite{Hender_2007,Rosenbluth_1997}. At this, it is responsible for a runaway electron current, which could harm the reactor walls, which is one reason why this study thesis focuses on the development of efficient computation methods for the moments related to a avalanche runaway population. 
In the course of the avalanche generation of runaway electrons, a \textit{knock-on} or \textit{large-angle} collision of a runaway electron and a thermal electron takes place, at which the runaway electrons remain in the runaway region and the thermal electron is transferred into the runaway region (please refer \cite{stahl}). The growth rate of the runaway electron density was first calculated by \textit{Rosenbluth} \& \textit{Putvinski} in reference \cite{Rosenbluth_1997}, implying an exponential growth. Thus one can understand the sudden creation of a large number of runaway electrons. Since the acceleration of the electrons in a tokamak mainly results from the parallel component of an electric field, because the charged particles in magnetically confined fusion plasmas predominantly move parallel to the magnetic field lines.

\clearpage

\section{Kinetic description of plasmas}\label{kinetic_theory_section}

The kinetic theory of plasmas is based on the description of the particles within a plasma by means of a distribution function $f_{\alpha}(\mathbf{r},\,\mathbf{p},\,t)$ for a particle species $\alpha$. This function indicates how many particles of one kind per unit phase space volume $\mathrm{d}^3r\,\mathrm{d}^3p$ are near the state vector $(\mathbf{r},\,\mathbf{p})$ at the time $t$ (see also \cite{helander}). This means any given configuration of particles corresponds to a state distribution in the phase space and allowing to draw inferences from the dynamics of the phase space points about the dynamics of the real particle population. This dynamics is determined by the \textit{kinetic} equation \cite{stahl}:\vspace{-2mm}
\begin{equation}\label{kinetic_equation}
\dfrac{\partial f_{\alpha}}{\partial t} + \vec{\nabla}_{\mathbf{r}} \left[\dot{\mathbf{r}}\,f_{\alpha}\right]+ \vec{\nabla}_{\mathbf{p}} \left[\dot{\mathbf{p}}\,f_{\alpha}\right]= \displaystyle{\sum\limits_{\beta}} \,C_{\alpha\beta}\left\lbrace f_{\alpha},\,f_{\beta} \right\rbrace +S\,.
\end{equation}
The kinetic equation (\ref{kinetic_equation}) involves the \textit{Nabla}-operator with respect to position $\mathbf{r}$ and momentum $\mathbf{p}$ as well as its total derivative $\dot{\mathbf{p}}$, which can be replaced by a macroscopic equation of motion, like for instance the \textit{Lorentz}-force $\mathbf{F}_{L}=q_{\alpha}\,\left(\mathbf{E}+\mathbf{v}\times\mathbf{B}\,\right)$ generated by a macroscopic electric and magnetic field acting on the particles of the species $\alpha$ with electric charge $q_{\alpha}$. Furthermore, the collision operator $C_{\alpha}=\displaystyle{\sum\limits_{\beta}} \,C_{\alpha\beta}\left\lbrace f_{\alpha},\,f_{\beta} \right\rbrace$ appears on the right-hand-side of equation (\ref{kinetic_equation}), which sums over all particle species $\beta$ in the plasma and treats all microscopic interactions between the plasma
particles such as elastic and inelastic \textit{Coulomb} collisions \cite{stahl}. It is as well as the operator $S$, denoting sources or sinks of particles, due to ionization, recombination, fueling or loss processes, an inhomogeneity within the partial differential equation (\ref{kinetic_equation}) \cite{stahl}. The kinetic equation and especially the collision operator is expressed differently for each combination of approximations \cite{helander}. But in general, one can state that it is often not possible to find analytic solutions for the kinetic equation. As well it is numerically challenging, runtime- or computation-power-expensive to solve for the distribution function $f_{\alpha}$. Therefore, different approaches like the fluid description of a plasma, are used to overcome those difficulties \cite{helander}. For the fluid description of a plasma the zeroth, first and second moment of the kinetic equation and the balance equations of the number of particles, momentum and energy are used, in order to receive a set of partial differential equations containing the continuity, a momentum and an energy balance \cite{helander,wesson}. Those balance equations can be solved by existing solvers in finite runtime. However, they are not suitable for rapid predictions, for which simpler models are used. Due to the fact, that the moments of the distribution function are related to physical quantities, simple models often make use of them, which motivates the following paragraph.  
 
Let $f_{\alpha}(\mathbf{r},\,\mathbf{p},\,t)$ be a given distribution function of the particle species $\alpha$ (e.g.\hspace{1.5mm}\mbox{$\alpha=e,\,i,\,n$} for electrons, ions or neutral particles) then one can define the $m$th moment of the distribution function with respect to the vector $\mathbf{v}\in\mathbb{R}^3$ in an analogous form as it is shown in the lecture slides from \textit{Y. Mizuno} \cite{slides}:\vspace{-1mm}
\begin{equation}\label{mth_moment}
M_{m}(\mathbf{r},\,t)= \displaystyle{\iiint\limits_{\mathbb{R}^3}} \mathbf{v}^m\, f_{\alpha}(\mathbf{r},\,\mathbf{p},\,t)\,\mathrm{d}^3p\,.
\end{equation}
Conventionally, the zeroth, first and second moment are associated with physical interpretations. Hence, the zeroth moment is identified as the particle density \cite{slides,Bellan_2006}:
\begin{equation}\label{zeroth_moment}
n_{\alpha}(\mathbf{r},\,t):=M_{0}(\mathbf{r},\,t)= \displaystyle{\iiint\limits_{\mathbb{R}^3}} f_{\alpha}(\mathbf{r},\,\mathbf{p},\,t)\,\mathrm{d}^3p\,.
\end{equation}
The first moment normalized with the particle density can be interpreted as the mean or bulk velocity of the particle population within the momentum space: 
\begin{equation}\label{first_moment}
\mathbf{u}_{\alpha}(\mathbf{r},\,t):=\dfrac{M_{1}(\mathbf{r},\,t)}{n_{\alpha}(\mathbf{r},\,t)}=\dfrac{1}{n_{\alpha}(\mathbf{r},\,t)}\, \displaystyle{\iiint\limits_{\mathbb{R}^3}} \mathbf{v}\,f_{\alpha}(\mathbf{r},\,\mathbf{p},\,t)\,\mathrm{d}^3p\,.
\end{equation}
This can be understood in detail in the reference \cite{slides} and the book from \mbox{\textit{P. Bellan} \cite{Bellan_2006}.}

The second moment can be used to define the scalar pressure or the stress tensor. In addition, one can define the mean mass-related kinetic energy density of the particle population by means of the second moment normalized with two times the particle density \cite{slides}:\vspace{-1mm}
\begin{equation}\label{second_moment}
K_{\alpha}(\mathbf{r},\,t):=\dfrac{M_{2}(\mathbf{r},\,t)}{2\,n_{\alpha}(\mathbf{r},\,t)}= \dfrac{1}{2\,n_{\alpha}(\mathbf{r},\,t)}\,\displaystyle{\iiint\limits_{\mathbb{R}^3}} \mathbf{v}\cdot\mathbf{v}\, f_{\alpha}(\mathbf{r},\,\mathbf{p},\,t)\,\mathrm{d}^3p\,.
\end{equation}
One of the mentioned simple models, which allows faster calculations than by solving the kinetic equation, is the runaway electron generation computation with self-consistent electric field \cite{Papp2013,Hoppe_2021}. At that, the time evolution of the runaway electron current density is determined from ordinary differential equations for the growth rate of the runaway electron density and the diffusion equation for the electric field deduced from the parallel component of the induction equation \cite{pappPHD}. In addition, simplified balance equations are solved i.\hspace{0.7mm}a.\hspace{1.4mm}for the electron temperature and the effective ion charge. An implementation of this model together with the governing equations and the used approximation is for instance discussed in the PhD-thesis of \mbox{\textit{G.\hspace{1.2mm}Papp}} \cite{pappPHD} and the publication \cite{Papp2013}. 

\clearpage

\section{Gyro-radius-averaged momentum space coordinates}\label{mom_space_coord_section}

The magnetic confinement of particles of a species $\alpha$ with electric charge $q_{\alpha}$ and velocity $\mathbf{v}$ is based on a magnetic field $\mathbf{B}$ leading to a resulting \textit{Lorentz}-force \mbox{$\mathbf{F}_{L}=q_{\alpha}\,\left(\mathbf{v}\times\mathbf{B}\,\right)$} in absence of an external electric field $\mathbf{E}$. As a result, all particles are forced into a motion along the magnetic field lines superimposed with a circular motion orthogonal to the magnetic field at the so-called \textit{Larmor} or gyro radius \cite{wesson}
\begin{equation}\label{gyroradius}
r_{g}=\frac{m_{\alpha}v_{\perp}}{\vert q_{\alpha}\vert \vert\mathbf{B}\vert}\,,
\end{equation}
where $v_{\perp}$ is the velocity component perpendicular to the magnetic field. For the following discussion, the gyro radius is neglectable compared to the typical length scale of
the gradients and the typical time scale of the gyration is much smaller than those of other processes within the plasma \cite{stahl,helander}. This holds in an approximative sense for the considered fusion plasma scenarios. Consequently, one can average over the gyro motion, meaning that the particles mainly follow the magnetic field lines, allowing a simplified plasma description in two instead of three momentum space dimensions, implying a reduced phase space from the original six-dimensional phase space with three position and three momentum space dimensions. Based on that, it is possible to describe the momentum of the particles in a two-dimensional momentum space coordinate system instead of the usage of the three-dimensional momentum space. Note, that this is the first simplification improving the efficiency of the calculation of the moments of the distribution function.

Usually, the three-dimensional velocity and momentum space is described by a local spherical or cylindrical coordinate system moving with the particle. In case of the spherical coordinate system the length of the normalized momentum vector \mbox{$p=( \gamma \,\vert\mathbf{v}\vert)/c$} with \mbox{$p \in[0,\,\infty)$}, the azimuthal angle $\varphi\in[0,\,2\pi]$ measured orthogonally to the magnetic field and the polar or pitch angle $\theta\in[0,\,\pi]$ measured from the local magnetic field direction, are defining the coordinate triplet $(p,\,\varphi,\,\theta)$. Thus the neglection of the gyro motion corresponds to dropping the dependence on the azimuthal angle $\varphi$. The volume element in the momentum space in spherical coordinates is $\mathrm{d}^{3}p=p^2\sin{(\theta)}\mathrm{d}p\,\mathrm{d}\varphi\,\mathrm{d}\theta$ and by averaging over the gyro radius it becomes:\vspace{-1mm}
\begin{equation}\label{volelem_sphere_2D}
\mathrm{d}^{3}p=\int\limits_{0}^{2\pi}d\varphi\,p^2\sin{(\theta)} \,\mathrm{d}p\,\mathrm{d}\theta=2\pi\,p^2\sin{(\theta)}\,\mathrm{d}p\,\mathrm{d}\theta\,.
\end{equation}
At that, a point in the momentum space is characterised by the coordinate pair $(p,\,\theta)$. Another representation can be obtained by means of the substitution of the pitch coordinate $\xi:=\cos{(\theta)}$, by which one finds $\mathrm{d}^{3}p=-2\pi\,p^2\,\mathrm{d}p\,\mathrm{d}\xi$, where $\xi\in[1,-1]$. By interchanging the boundaries to $\xi\in[-1,\,1]$ and dropping the negative sign from the pitch-coordinate substitution one receives:\vspace{-1mm}
\begin{equation}\label{pitch_angle_volelem_sphere_2D}
\mathrm{d}^{3}p=2\pi\,p^2\,\mathrm{d}p\,\mathrm{d}\xi\,.
\end{equation} 
Where this leads to a coordinate description with the duplet $(p,\,\xi)$ for \mbox{$p\in[0,\,\infty)$} and \mbox{$\xi\in[-1,\,1]$}.

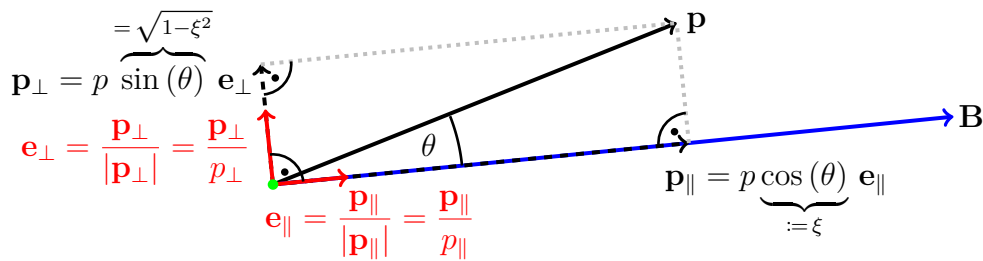
\begin{figure}[H]
\centering
\begin{tikzpicture}
\begin{axis}[hide axis,x=1cm,y=1cm, ymin=-1.3,ymax=2.4, xmin=-3.5,xmax=9.5]
\addplot+ [solid,->,mark=none,line width=1.5,black] coordinates { (0,0) (-1.6*0.1+5.5*0.9949,1.6*0.9949+5.5*0.1)};
\addplot+ [solid,->,mark=none,line width=1.5pt,blue] coordinates { (0,0) (9*0.9949,9*0.1)}; 
\addplot+ [dashed,->,mark=none,line width=1.5pt, black] coordinates { (0,0) (-1.6*0.1,1.6*0.9949)}; 
\addplot+ [dashed,->,mark=none,line width=1.5pt, black] coordinates { (0,0) (5.5*0.9949,5.5*0.1)}; 
\addplot+ [solid,->,mark=none,line width=1.5pt,red] coordinates { (0,0) (0.9949,0.1)}; 
\addplot+ [solid,->,mark=none,line width=1.5pt,red] coordinates { (0,0) (-0.1,0.9949)}; 
\addplot+ [dotted,mark=none,line width=1.5,black!25] coordinates { (-1.6*0.1,1.6*0.9949) (-1.6*0.1+5.5*0.9949,1.6*0.9949+5.5*0.1)};
\addplot+ [dotted,mark=none,line width=1.5,black!25] coordinates { (5.5*0.9949,5.5*0.1) (-1.6*0.1+5.5*0.9949,1.6*0.9949+5.5*0.1)};
\addplot+[solid,mark=none,line width=1,black,domain=2.33:2.48725] {(2.5^2-x^2)^0.5};
\node at (axis cs:9.5,9*0.1) [anchor=east] {$\mathbf{B}$};
\node at (axis cs:5.85,1.6*0.9949+5.5*0.1) [anchor=east] {$\mathbf{p}$};
\node at (axis cs:-3.3,2.5) [anchor=north east] {$\mathrlap{\mathbf{p}_{\perp}=p\overbrace{\sin{(\theta)}}^{=\,\sqrt{1-\xi^2}}\mathbf{e}_{\perp}}$};
\node at (axis cs:5.3,0.45) [anchor=north east] {$\mathrlap{\mathbf{p}_{\|}=p\underbrace{\cos{(\theta)}}_{\coloneqq\,\xi}\,\mathbf{e}_{\|}}$};
\node at (axis cs:-0.16,1) [anchor=north east] {\textcolor{red}{$\mathbf{e}_{\perp}=\dfrac{\mathbf{p}_{\perp}}{\vert\mathbf{p}_{\perp}\vert}=\dfrac{\mathbf{p}_{\perp}}{p_{\perp}}$}};
\node at (axis cs:2.8,0.06) [anchor=north east] {\textcolor{red}{$\mathbf{e}_{\|}=\dfrac{\mathbf{p}_{\|}}{\vert\mathbf{p}_{\|}\vert}=\dfrac{\mathbf{p}_{\|}}{p_{\|}}$}};
\node at (axis cs:2.3,0.81) [anchor=north east] {$\theta$};
\coordinate[] (A) at (axis cs:0.9949,0.1); 
\coordinate[] (B) at (axis cs:0,0); 
\coordinate[] (C) at (axis cs:-0.1,0.9949); 
\draw pic [draw,angle radius=4mm,line width=1pt]{angle =A--B--C};
\coordinate[] (D) at (axis cs:-1.6*0.1+5.5*0.9949,1.6*0.9949+5.5*0.1); 
\coordinate[] (E) at (axis cs:5.5*0.9949,5.5*0.1); 
\coordinate[] (F) at (axis cs:-0.9949,-0.1); 
\draw pic [draw,angle radius=4mm,line width=1pt]{angle =D--E--F};
\coordinate[] (G) at (axis cs:0,0); 
\coordinate[] (H) at (axis cs:-1.6*0.1,1.6*0.9949); 
\coordinate[] (I) at (axis cs:-1.6*0.1+5.5*0.9949,1.6*0.9949+5.5*0.1); 
\draw pic [draw,angle radius=4mm,line width=1pt]{angle =G--H--I};
\node at (axis cs:0.155,0.165)[circle,fill,inner sep=1pt]{};
\node at (axis cs:0.03,1.42*0.9949)[circle,fill,inner sep=1pt]{};
\node at (axis cs:5.31*0.9949,0.7)[circle,fill,inner sep=1pt]{};
\node at (axis cs:0,0)[circle,fill,inner sep=1.4pt,green]{};
\end{axis}
\end{tikzpicture}
\captionsetup{format=hang,indention=0cm}
\caption[Two-dimensional moving momentum coordinate system for a particle (light green) with respect to the local magnetic field $\mathbf{B}$ and depiction of the relations to the pitch-coordinate $\xi\hspace{-0.2mm}\coloneqq\hspace{-0.2mm}\cos(\theta)\hspace{-0.2mm}\in\hspace{-0.2mm}\lbrack -1,\,1 \rbrack $ for $\theta\hspace{-0.2mm}\in\hspace{-0.2mm}\lbrack -\pi,\,0 \rbrack$]{Two-dimensional moving momentum coordinate system\protect\footnotemark{} for a particle (light green) with respect to the local magnetic field $\mathbf{B}$ and depiction of the relations to the pitch-coordinate $\xi:=\cos{(\theta)}\in[-1,\,1]$ for $\theta\in[-\pi,\,0]\,$}
\label{fig_mom_coord}
\end{figure}

As well, it is possible to express the volume element in the three-dimensional momentum space in the cylindrical coordinates triplet $(p_{\|},\,p_{\perp},\,\varphi)$, using the component of the momentum parallel to the magnetic field $p_{\|}=\vert\mathbf{p}_{\|}\vert=(\gamma\,\vert\mathbf{v}_{\|}\vert)/c \in(-\infty,\,\infty)$, the component of the momentum perpendicular to the magnetic field direction \mbox{$p_{\perp}=\vert\mathbf{p}_{\perp}\vert$} with \mbox{$p_{\perp}=( \gamma \,\vert\mathbf{v}_{\perp}\vert)/c\in[0,\,\infty)$} and the azimuthal angle $\varphi\in[0,\,2\pi]$ measured in the plane orthogonal to the magnetic field. This is graphically displayed in figure \ref{fig_mom_coord}. \vspace{2mm}
\footnotetext{\label{coord_sys_fig_footnote} The graphic in figure \ref{fig_mom_coord} was created with \LaTeX-internal routines.}
\vspace*{-7mm}
Therewith, the volume element in the three-dimensional momentum space reads\linebreak\mbox{$\mathrm{d}^{3}p=p_{\perp}\mathrm{d}p_{\perp}\mathrm{d}p_{\|}\,\mathrm{d}\varphi$}. By eliminating the dependence on the azimuthal angle it \mbox{becomes:} 
\begin{equation}\label{volelem_cyl_2D}
\mathrm{d}^{3}p=\int\limits_{0}^{2\pi}\mathrm{d}\varphi\,p_{\perp}\mathrm{d}p_{\perp}\mathrm{d}p_{\|}=2\pi\,p_{\perp}\mathrm{d}p_{\perp}\mathrm{d}p_{\|}\,.
\end{equation}
Moreover, the equality $2\pi\,p_{\perp}\mathrm{d}p_{\perp}\mathrm{d}p_{\|}=2\pi\,p^2\,\mathrm{d}p\,\mathrm{d}\xi$ shows the equivalence of the volume elements in the two dicussed coordinate descriptions. It is gained by the relations \mbox{$p_{\|}=p\,\xi$} and $p_{\perp}=p\,\sqrt{1-\xi^2}$ between the coordinates $(p,\,\xi)$ and $(p_{\|},\,p_{\perp})$, which can be derived from figure \ref{fig_mom_coord} using the trigonometric pythagoras, where $\mathbf{e}_{\|}$ and $\mathbf{e}_{\perp}$ are orthonormal basis vectors. Furthermore, the \textit{Jacobian} determinant is needed, in order to perform the transformation between the coordinate systems. The according calculation is shown in detail in the appendix subsection \ref{subsection_mom_deriv_vol_elem_appendix}.

\clearpage

\chapter[Calculation of the moments of avalanche runaway electrons]{\mbox{Calculation of the moments of the} avalanche runaway electron distribution function}\label{avalanche_chapter}

\section{Distribution function for the avalanche runaway electron generation}\label{avalanche_dist_section}

The avalanche runaway electron generation, as described in section \ref{mechanisms_section}, leads to an exponential growth of the runaway electron density, modeled by the growth rate proposed by \textit{Rosenbluth} \& \textit{Putvinski} in their publication \cite{Rosenbluth_1997}, if the avalanche generation mechanism dominates the plasma dynamics. From this the momentum space distribution function $f_{RE}^{\textup{ava}}(\mathbf{p},\,t)$ for runaway electrons produced by avalanching was derived by \mbox{\textit{Fülöp et al.}} in reference \cite{REdistfuncderivation}. Therein the momentum space was described by the two-dimensional momentum coordinate system depicted in figure \ref{fig_mom_coord} from section \ref{mom_space_coord_section} resulting in the following function: \vspace{-1.5mm}
\begin{equation}\label{f_RE_ava}
f_{RE}^{\textup{ava}}(p_{\|},\,p_{\perp},\,t)=\dfrac{C}{p_{\|}}\cdot\exp{\left(\dfrac{(E-1)}{c_{Z} \ln{(\Lambda)}}\cdot\dfrac{t}{\tau}-\dfrac{p_{\|}}{c_{Z} \ln{(\Lambda)}}-\tilde{E}\cdot\dfrac{p_{\perp}^{2}}{p_{\|}}\right)}\,.
\end{equation} 
In (\ref{f_RE_ava}) $\ln{\Lambda}$ is the \textit{Coulomb} logarithm, $E=\vert E_{\|}\vert/E_{c}$ is the component of the electric field parallel to the magnetic field lines normalized with the critical 
electric field from (\ref{E_crit}), $\tau$ is the collision time from (\ref{tau_coll}) and the normalization factor\vspace{-1mm}
\begin{equation}\label{C_f_RE_ava}
C=\frac{n_{RE}\,\tilde{E}}{\pi\,c_{Z}\ln{(\Lambda)}}\cdot\exp{\left(\frac{(E-1) }{c_{Z}\ln{(\Lambda)} }\cdot\dfrac{t}{\tau}\right)}
\end{equation} 
includes the runaway electron density $n_{RE}$. Moreover, the abreviating constants\vspace{-1mm}
\begin{equation}\label{E_cZ_f_RE_ava}
\tilde{E}=\frac{E-1}{2\,(Z_{eff}+1)}\;\;\;\;;\;\;\;\;c_{Z}=\sqrt{\frac{3\,(Z_{eff}+5)}{\pi}}
\end{equation}
were used, which are i.a. related to the effective ion charge $Z_{eff}$. At that, the effective ion charge is defined as a density-weighted charge square, with the help of the book by Stroth \cite{Stroth_2018}:\vspace{-3mm}
\begin{equation}\label{Zeff}
Z_{eff}= \dfrac{1}{n_{e}}\,\displaystyle{\sum\limits_{k=1}^{N_{i}}} \,n_{i,k}\,q^{2}_{i,k}\,,
\end{equation}
where $n_{i,k}$ is the ion density of the $k$th ion, $n_{e}$ is the total electron density, $q_{i,k}^2$ is the squared charge of the $k$th ion and $N_{i}$ is the total number of ion species in the plasma.
   
\begin{figure}[H]
\begin{center}
\includegraphics[trim=108 170 110 200,width=\textwidth,clip]{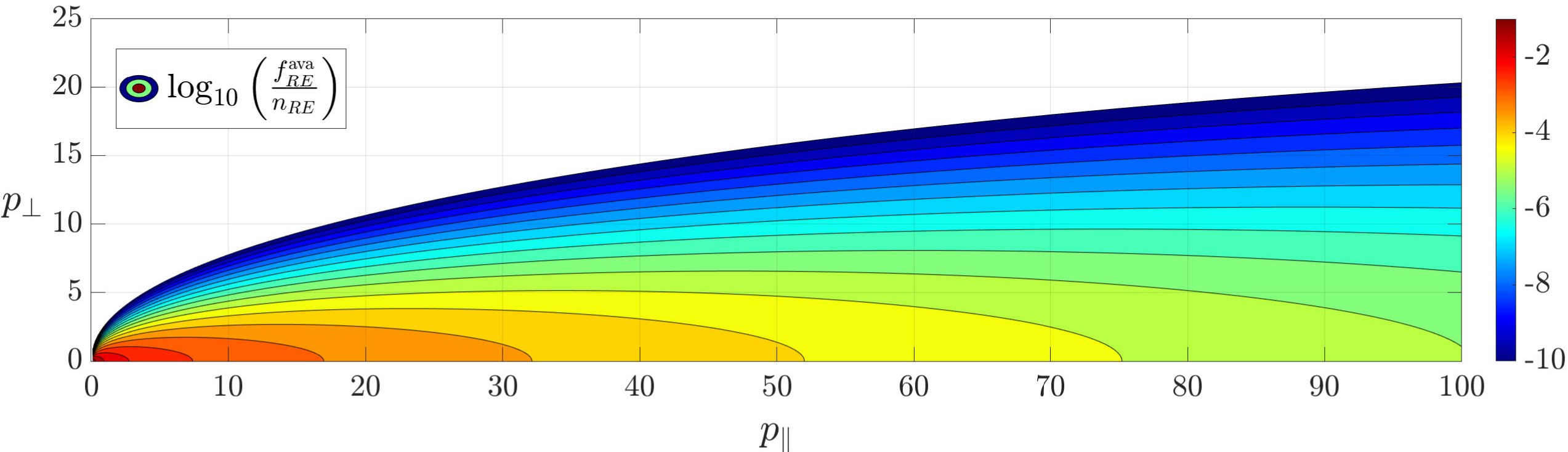}
\caption[Contour plot of the analytical quasi-steady-state avalanche runaway electron distribution function for $k_{B}\,T_{e}=50 \,\textup{eV}$, $n_{e}=10^{20}\,\textup{m}^{-3}$, $Z_{eff}=1.5$ and $E/E_{crit}=15$ (for comparison see also \cite{stahl})]{Contour plot\protect\footnotemark{} of the analytical quasi-steady-state avalanche runaway electron distribution function for $k_{B}\,T_{e}=50 \,\textup{eV}$, $n_{e}=10^{20}\,\textup{m}^{-3}$, $Z_{eff}=1.5$ and $E/E_{crit}=15$ (for comparison see also \cite{stahl})}
\label{fig_ava_dist_func}
\end{center}
\end{figure}

The quasi-steady-state runaway distribution
function follows from (\ref{f_RE_ava}) with $t=0$ and is therefore time-independent and in coherence with \cite{stahl,RunawayPositrons} becomes:
\begin{equation}\label{steady_f_RE_ava}
f_{RE}^{\textup{ava}}(p_{\|},\,p_{\perp})=\frac{n_{RE}\,\tilde{E}}{\pi\,c_{Z}\ln{(\Lambda)}\cdot p_{\|}} \cdot\exp{\left( -\dfrac{p_{\|}}{c_{Z} \ln{(\Lambda)}}-\tilde{E}\cdot\dfrac{p_{\perp}^{2}}{p_{\|}}\right)}\,.
\end{equation} 
In order to visualize the distribution function from (\ref{steady_f_RE_ava}), it is computed and displayed for typical tokamak plasma parameters and can be viewed in figure \ref{fig_ava_dist_func}. 
\footnotetext{\label{ava_dist_func_fig_footnote} The diagram in figure \ref{fig_ava_dist_func} was generated by means of the \textsc{MATLAB}-script\\\hspace*{8.5mm}"plot_quasi_steady_state_avalanche_runaway_electron_distribution.m", \\ \hspace*{8.7mm}which can be viewed in the digital appendix.}
\vspace{-7mm}
Note that in figure \ref{fig_ava_dist_func} all quantities are dimensionless and hence no units are given. The \textit{Coulomb} logarithm \mbox{$\ln{(\Lambda)}=14.9$}, the critical electric field strength \mbox{$E_{c} = 0.07598 \,\si{\volt\per\meter}$}, the \textit{Dreicer} field \mbox{$E_{D} = 38.82395 \,\si{\volt\per\meter}$}, the \textit{slide-away} field \mbox{$E_{sa} = 8.30833 \,\si{\volt\per\meter}$} and the electric field strength $E = 1.13965\,\si{\volt\per\meter}$ associated to this configuration were calculated, by means of a \textsc{MATLAB}-script$^{\ref{ava_dist_func_fig_footnote}}$. \\Furthermore, the characteristic of a runaway electron beam in momentum space can be observed in figure \ref{fig_ava_dist_func}, due to the fact that the distribution is localised along the $p_{\|}$-axis for $p_{\|}\in[0,\infty)$. Therefore, for the most electrons $p_{\|}\gg p_{\perp}$ holds. As well, it is important to mention, that the distribution function was originally derived for a homogeneous magnetic field, which can not be found in the magnetic field configuration of a tokamak, because of its toroidal geometry and its therefore higher coil density and thus stronger magnetic field on its inner side. Nevertheless, the magnetic field is mainly inhomogeneous along the minor radius of the tokamak torus, in order to confine the plasma. In poloidal direction, one finds helically coiled field lines. Runaway electron populations form beam-like structures following those field lines and are primarily generated in the core of the plasma. There, near the magnetic axis of the tokamak, the inhomogeneity in the magnetic field is less distinctiv and therefore approximately constant. This finally allows the use of the distribution function proposed by \mbox{\textit{Fülöp et al.}}.

\clearpage

\section{Avalanche runaway electron density - zeroth moment}\label{ava_zero_moment_section}

The avalanche runaway electron density is equivalent to the zeroth moment of the distribution function for the secondary generation mechanism of avalanche runaway electrons as presented in section \ref{kinetic_theory_section}. Hence, one has to evaluate the integral from equation (\ref{zeroth_moment}) for the distribution function from (\ref{f_RE_ava}) with the integration boundaries $p_{\perp}\in[0,\,\infty)$ and $p_{\|}\in(-\infty,\,\infty)$ as well as with the volume element $\mathrm{d}^3p=2\pi\,p_{\perp}\mathrm{d}p_{\perp}\mathrm{d}p_{\|}$ from (\ref{volelem_cyl_2D}) according to the coordinate system discussed in section \ref{mom_space_coord_section}:\vspace{-1mm}
\begin{equation}\label{RE_density_integral}
\begin{split}
\begin{gathered}
n_{RE}(t)=\displaystyle{\int\limits_{p_{\|}=-\infty}^{\infty}\int\limits_{p_{\perp}=0}^{\infty}} f_{RE}^{\textup{ava}}(p_{\|},\,p_{\perp},\,t)\,2\pi\,p_{\perp}\mathrm{d}p_{\perp}\mathrm{d}p_{\|}
\\[10pt]
=\displaystyle{\int\limits_{p_{\|}=-\infty}^{\infty}\int\limits_{p_{\perp}=0}^{\infty}} \dfrac{2\pi\,C}{p_{\|}}\cdot\exp{\left(\dfrac{(E-1)}{c_{Z} \ln{(\Lambda)}}\cdot\dfrac{t}{\tau}-\dfrac{p_{\|}}{c_{Z} \ln{(\Lambda)}}-\tilde{E}\cdot\dfrac{p_{\perp}^{2}}{p_{\|}}\right)}\,p_{\perp}\mathrm{d}p_{\perp}\mathrm{d}p_{\|}\,.
\end{gathered}
\end{split}
\end{equation}
In the following, the integral from (\ref{RE_density_integral}) is solved analytically. Detailed calculations are shown in subsection \ref{RE_dens_int_appendix_subsection} of the appendix.

First, one solves the $p_{\perp}$-integral from (\ref{RE_density_integral}), by introducing the substitution:
\begin{equation}\label{substitution1_RE_dens_integral}
\begin{split}
\begin{gathered}
\lambda(p_{\perp}):=\dfrac{(E-1)}{c_{Z} \ln{(\Lambda)}}\cdot\dfrac{t}{\tau}-\dfrac{p_{\|}}{c_{Z} \ln{(\Lambda)}}-\tilde{E}\cdot\dfrac{p_{\perp}^{2}}{p_{\|}}\;\;;\;\;\dfrac{\mathrm{d}\lambda}{\mathrm{d}p_{\perp}}=-\dfrac{2\,\tilde{E}\,p_{\perp}}{p_{\|}}\,.
\end{gathered}
\end{split}
\end{equation} 
With this, one rewrites the equation (\ref{RE_density_integral}):\vspace{-1mm}
\begin{equation}\label{V2_RE_density_integral}
\begin{split}
\begin{gathered}
n_{RE}(t)=-\hspace{-1mm}\displaystyle{\int\limits_{p_{\|}=-\infty}^{\infty}}\hspace{-2mm} \pi\,C  \,\displaystyle{\int\limits_{\lambda(0)}^{\lambda(\infty)}} \, \dfrac{ \textup{e}^{\lambda}}{\tilde{E}}\;\mathrm{d}\lambda\,\mathrm{d}p_{\|}
=-\dfrac{\pi\,C}{\tilde{E}} \,\displaystyle{\int\limits_{p_{\|}=-\infty}^{\infty}}  \left[\textup{e}^{\,\lambda}\right]^{\lambda(\infty)=-\infty}_{\lambda(0)=\frac{(E-1)}{c_{Z} \ln{(\Lambda)}}\cdot\frac{t}{\tau}-\frac{p_{\|}}{c_{Z} \ln{(\Lambda)}}} \,\mathrm{d}p_{\|}
\\[10pt] 
=-\dfrac{\pi\,C}{\tilde{E}} \,\displaystyle{\int\limits_{p_{\|}=-\infty}^{\infty}}   \underbrace{\textup{e}^{\,\lambda\rightarrow -\infty}}_{=\,0}-\textup{e}^{\,\frac{(E-1)}{c_{Z} \ln{(\Lambda)}}\cdot\frac{t}{\tau}-\frac{p_{\|}}{c_{Z} \ln{(\Lambda)}}} \,\mathrm{d}p_{\|}
=\dfrac{\pi\,C}{\tilde{E}} \,\displaystyle{\int\limits_{p_{\|}=-\infty}^{\infty}}    \textup{e}^{\,\frac{(E-1)}{c_{Z} \ln{(\Lambda)}}\cdot\frac{t}{\tau}-\frac{p_{\|}}{c_{Z} \ln{(\Lambda)}}} \,\mathrm{d}p_{\|}\,.
\end{gathered}
\end{split}
\end{equation}
Second, one solves the $p_{\|}$-integral from (\ref{V2_RE_density_integral}), by another substitution:\vspace{-1mm}
\begin{equation}\label{substitution2_RE_dens_integral}
\begin{split}
\begin{gathered}
\kappa(p_{\|}):=\dfrac{(E-1)}{c_{Z} \ln{(\Lambda)}}\cdot\dfrac{t}{\tau}-\dfrac{p_{\|}}{c_{Z} \ln{(\Lambda)}} \;\;;\;\;\dfrac{\mathrm{d}\kappa}{\mathrm{d}p_{\|}}=-\dfrac{1}{c_{Z} \ln{(\Lambda)}}\,.
\end{gathered}
\end{split}
\end{equation} 
With the help of (\ref{substitution2_RE_dens_integral}) one can evaluate (\ref{V2_RE_density_integral}), whereby the lower integration bound is set to $p_{\|,min}\rightarrow -\infty$:\vspace{-4mm}
\begin{equation*} 
\begin{split}
\begin{gathered}
n_{RE}(t)
=-\dfrac{\pi\,C\,c_{Z} \ln{(\Lambda)}}{\tilde{E}} \hspace{-1mm}\displaystyle{\int\limits_{\kappa(-\infty)}^{\kappa(\infty)}}    \hspace{-1mm}\textup{e}^{\,\kappa} \,\mathrm{d}\kappa=-\dfrac{\pi\,C\,c_{Z} \ln{(\Lambda)}}{\tilde{E}} \,\bigl[\textup{e}^{\,\kappa}\bigr]^{\kappa(\infty)=-\infty}_{\kappa(p_{\|,min})=\frac{(E-1)}{c_{Z} \ln{(\Lambda)}}\cdot\frac{t}{\tau}-\frac{p_{\|,min}}{c_{Z} \ln{(\Lambda)}}}
\end{gathered}
\end{split}
\end{equation*}
\begin{equation}\label{V3_RE_density_integral}
\begin{split}
\begin{gathered} 
=\hspace{-0.5mm}-\dfrac{\pi\,C\,c_{Z} \ln{(\Lambda)}}{\tilde{E}} \hspace{-1mm}\left(\underbrace{\textup{e}^{\,\kappa\rightarrow -\infty}}_{=\,0}-\textup{e}^{\,\frac{(E-1)}{c_{Z} \ln{(\Lambda)}}\cdot\frac{t}{\tau}-\frac{p_{\|,min}}{c_{Z} \ln{(\Lambda)}}}\hspace{-1mm} \right)\hspace{-1mm}
=\hspace{-0.5mm}\dfrac{\pi\,C\,c_{Z} \ln{(\Lambda)}}{\tilde{E}} \, \textup{e}^{\,\frac{(E-1)}{c_{Z} \ln{(\Lambda)}}\cdot\frac{t}{\tau}-\frac{p_{\|,min}}{c_{Z} \ln{(\Lambda)}}} \,.
\end{gathered}
\end{split}
\end{equation}
Finally, one inserts the expressions for $C$ and $\tilde{E}$ from (\ref{C_f_RE_ava}) and (\ref{E_cZ_f_RE_ava}) into the result from (\ref{V3_RE_density_integral}), which yields:
\vspace{-2.5mm}
\begin{equation}\label{n_RE_dens_int}
\underline{\underline{ n_{RE}(t)  =n_{RE}\cdot\exp{\left( \dfrac{2\,(E-1)}{c_{Z} \ln{(\Lambda)}}\cdot\dfrac{t}{\tau}-\dfrac{p_{\|,min}}{c_{Z} \ln{(\Lambda)}} \right)}  }}\,.
\end{equation}
By analyzing equation (\ref{n_RE_dens_int}), one recognizes that the zeroth moment of the avalanche runaway electron distribution function is only equal to the runaway electron density for one value of $p_{\|,min}$. This particular lower bound is computable and can be derived from (\ref{n_RE_dens_int}) as stated in equation (\ref{p_par_min_appendix}): \vspace{-1mm}
\begin{equation}\label{p_par_min_ava_RE}
p_{\|,min}=2\,(E-1)\cdot\dfrac{t}{\tau} \,.
\end{equation}
This lower boundary for the parallel momentum component can be expressed as a dimensionless time-dependent momentum difference, which represents the effective momentum transfer to the electrons originating from the electric field parallel to the magnetic field lines, by analysis of the units. If one inserts the relations $E=\vert E_{\|}\vert/E_{c}$ from section \ref{avalanche_dist_section} and $E_{c}\,\tau=(m_{e0}\,c)/e$, which follows from (\ref{E_crit}) and (\ref{tau_coll}), into equation (\ref{p_par_min_ava_RE}), one finds:\vspace{-2.5mm}
\begin{equation}\label{p_par_min_ava_RE_new}
p_{\|,min}=2\cdot\dfrac{(\vert E_{\|}\vert-E_{c})\,t}{E_{c}\,\tau}= 2\cdot\dfrac{e\,(\vert E_{\|}\vert-E_{c})\,t}{m_{e0}\,c}=2\cdot\dfrac{\Delta p (t)}{p_{norm}}\,.
\end{equation}
In (\ref{p_par_min_ava_RE_new}) $p_{norm}:=m_{e0}\,c$ is the momentum associated to the rest mass, which is the used normalization constant for the relativistic momentum. Moreover, it shall be remarked that the momentum difference transferred to the electrons $\Delta p (t)$ increases linearly for a time-independent electric field as time passes. Although, this is not expected in reality, since the electric field is time-dependent especially during plasma disruptions.  

Hereinafter, the zeroth moment is used as a benchmark for the \textsc{MATLAB}-implemen- tation, because the integral (\ref{RE_density_integral}) can be solved analytically. For this benchmark case, one defines the following integral from (\ref{RE_density_integral}) for $t=0\,\mathrm{s}$ and a finite lower integration bound $p_{\|,min}=0$:\vspace{-2.5mm}
\begin{equation}\label{benchmark_integral}
\begin{split}
\begin{gathered}
\textup{I}^{\,n_{\mathrm{RE}}}_{\,RE,ava}:=\dfrac{1}{n_{RE}}\;\displaystyle{\int\limits_{p_{\|}=p_{\|,min}}^{\infty}\int\limits_{p_{\perp}=0}^{\infty}} \dfrac{2\pi\,C}{p_{\|}}\cdot\exp{\left(-\dfrac{p_{\|}}{c_{Z} \ln{(\Lambda)}}-\tilde{E}\cdot\dfrac{p_{\perp}^{2}}{p_{\|}}\right)}\,p_{\perp}\mathrm{d}p_{\perp}\mathrm{d}p_{\|}
\\[6pt]
\underset{t\,=\,0\,\mathrm{s}}{\overset{(\ref{C_f_RE_ava})}{=}}\dfrac{2\,\tilde{E}}{c_{Z} \ln{(\Lambda)}}\underbrace{\displaystyle{\int\limits_{p_{\|}=p_{\|,min}}^{\infty}\int\limits_{p_{\perp}=0}^{\infty}} \dfrac{p_{\perp}}{p_{\|}}\cdot\exp{\left(-\dfrac{p_{\|}}{c_{Z} \ln{(\Lambda)}}-\tilde{E}\cdot\dfrac{p_{\perp}^{2}}{p_{\|}}\right)}\mathrm{d}p_{\perp}\mathrm{d}p_{\|}}_{=:\,\textup{I}^{\,n_{\mathrm{RE}}}_{\,num}}\overset{!}{=}\,1\,,
\end{gathered}
\end{split}
\end{equation}
where the constant $C$ from (\ref{C_f_RE_ava}) for $t=0\,\mathrm{s}$ was inserted and it was used, that the double integral in the first line is equal to $n_{RE}$ according to (\ref{n_RE_dens_int}) for $t=0\,\mathrm{s}$ and $p_{\|,min}=0$. Thus, the value of the quantity $\textup{I}^{\,n_{\mathrm{RE}}}_{\,RE,ava}$ should be one, which can be used as a validation for the \textsc{MATLAB}-implementation. Furthermore, one can compare the mean runtime duration for the two-dimensional numerical integration of $\textup{I}^{\,n_{\mathrm{RE}}}_{\,num}$ with the \textsc{MATLAB}-routines "\texttt{integral2}" and "\texttt{trapz}". The second routine requires finite integration bounds. Hence, one uses the substitutions:\vspace{-0.5mm}
\begin{equation}\label{substitutions_RE_n_2Dint}
\begin{split}
\begin{gathered}
p_{\|}=p_{\|,min}+\dfrac{w}{1-w} \;;\;\dfrac{\mathrm{d}p_{\|}}{\mathrm{d}w}= \dfrac{1}{(1-w)^2}\;;\; w(p_{\|}=p_{\|,min})=0\;,\;w(p_{\|}\rightarrow\infty)=1
\\[8pt]
p_{\perp}=\dfrac{z}{1-z} \;;\;\dfrac{\mathrm{d}p_{\perp}}{\mathrm{d}z}= \dfrac{1}{(1-z)^2}\;;\; z(p_{\perp}=0)=0\;,\;z(p_{\perp}\rightarrow\infty)=1\,.
\end{gathered}
\end{split}
\end{equation}
The mean runtime is then measured for different parameters $\tilde{E}$, $c_{Z}$ and $\ln{(\Lambda)}$ for the numerical integration of:\vspace{-3mm}
\begin{equation}\label{benchmark_integral_num}
\begin{split}
\begin{gathered}
\textup{I}^{\,n_{\mathrm{RE}}}_{\,num}\,=\displaystyle{\int\limits_{w=0}^{1}\int\limits_{z=0}^{1}} \,\frac{z\cdot\exp{\left(-\frac{p_{\|,min}+\frac{w}{1-w}}{c_{Z} \ln{(\Lambda)}}-\frac{\tilde{E}\left(\frac{z}{1-z}\right)^2}{p_{\|,min}+\frac{w}{1-w}}\right)}}{(p_{\|,min}(1-w)+w)(1-w)(1-z)^3} \;\mathrm{d}z\,\mathrm{d}w
\end{gathered}
\end{split}
\end{equation}
By means of the \textsc{MATLAB}-script "efficiency_analysis_for_moment_calculations.m", which is stored in the digital appendix, one can therefore compute the quantities $\textup{I}^{\,n_{\mathrm{RE}}}_{\,num}$ and $\textup{I}^{\,n_{\mathrm{RE}}}_{\,num}$. At that, the absolute error tolerance for the "\texttt{integral2}"-routine is set to $10^{-6}$. The "\texttt{trapz}"-method uses the trapezoidal rule with an error estimation proportional to the square of the uniform spacing-width of the sampling points $(\Delta x)^2$. Therefore, one uses $10^3$ sampling points for the integration variables $w,z\in[0,1]$, which leads to a grid with $10^6$ sampling points and an estimated absolute error tolerance of $10^{-6}$. Consequently, one can compare the two integration routines, although it should be mentioned that the grid for the trapezoidal rule requires a grid with $w,z\in(0,1)$ respectively excluded integration boundaries, in order to converge.

The measured mean runtime for the calculation of the integral $\textup{I}^{\,n_{\mathrm{RE}}}_{\,num}$ with the different computation rules can then be used to derive statements about the efficiency. The corresponding output for the values of the integrals and the mean runtime duration is displayed in listing \ref{outMATLABeffcalcanalysisbench}. Note, that the complete output of the \textsc{MATLAB}-script is given in the subsection \ref{subsection_num_efficiency_appendix} of the appendix and only an excerpt, is shown in the listing \ref{outMATLABeffcalcanalysisbench}. However, the runtime duration for each computation rule was calculated as the mean value of the runtime of $729$ computations for different parameter combinations for $\tilde{E}\in[5,\,45]$, $c_{Z}\in[5,\,45]$ and $\ln{(\Lambda)}\in[5,\,45]$. 
\begin{lstlisting}[language=Matlab,keywordstyle=\empty,frame=single, caption={Output connected to the benchmark quantity $\textup{I}^{\,n_{\mathrm{RE}}}_{\,num}$ of the \textsc{MATLAB}-script "efficiency_\\analysis_for_moment_calculations.m" \vspace{1mm}},label={outMATLABeffcalcanalysisbench} ]
set of parameters:

EoverEcrit = 150; n_e = 1e+20 m^-3; k_B*T_e = 100 eV; Z_eff = 1.5;

calculated quantities:

lnLambda = 12.59741; E_crit = 0.06424 V/m; p_crit = 0.08192;
E_D = 328.24256 V/m; E_sa = 70.24391 V/m; E = 9.63532 V/m; 

I_num_ava_nRE = 1.00000002 
(2D-integration with "integral2", from 0 to 1, runtime benchmark, abs_tol~10^-6);
mean runtime duration = 0.0094 s (for 729 parameter combinations);

I_num_ava_nRE = 0.99998290 
(2D-integration with "trapz", from 0 to 1, runtime benchmark, 10^6 grid points);
mean runtime duration = 0.0165 s (for 729 parameter combinations);
\end{lstlisting}
From the output one validates the correct integration and deduces, that the \mbox{\textsc{MATLAB}}-routine "\texttt{integral2}" is more efficient than the "\texttt{trapz}"-method. In addition, one notices that the estimated absolute error for the trapezoidal rule was not reproduced for the chosen grid. This means, that an even finer grid resolution and thus a higher runtime is needed for the "\texttt{trapz}"-method, if the accuracy of the "\texttt{integral2}"-routine has to be achieved.

\clearpage

\section{Avalanche runaway electron current density - first moment}\label{ava_first_moment_section}

The current density carried by a runaway electron population resulting from the avalanche generation mechanism is defined as:\vspace{-1mm}
\begin{equation}\label{j_ava_RE_def}
\mathbf{j}_{\,RE}^{\,\textup{ava}}(\mathbf{r},\,t)\hspace{-0.2mm}=\hspace{-0.2mm}q_{e}\,n_{RE}(\mathbf{r},\,t)\,\mathbf{u}_{\,RE}^{\,\textup{ava}}(\mathbf{r},\,t)\hspace{-1mm}\overset{(\ref{first_moment})}{=}\hspace{-0.8mm}q_{e}\,M_{1,RE}^{\textup{ava}}(\mathbf{r},\,t)=-e\,\displaystyle{\iiint\limits_{\mathbb{R}^3}} \mathbf{v}\,f_{\,RE}^{\,\textup{ava}}(\mathbf{r},\,\mathbf{p},\,t)\,\mathrm{d}^3p\,,
\end{equation}
where the definiton of the first moment of the distribution function from (\ref{first_moment}) and the charge $q_{e}=-e$ of an electron were used. Since the avalanche runaway electron distribution function from (\ref{f_RE_ava}) is expressed in the momentum coordinate system from section \ref{mom_space_coord_section}, one writes the current density as $\mathbf{j}_{\,RE}^{\,\textup{ava}}=(j_{\,\|,RE}^{\,\textup{ava}}\;j_{\,\perp,RE}^{\,\textup{ava}})^{\top}$, in order to be coherent with this coordinate system. Similarly, one can rewrite the velocity vector $\mathbf{v}$ in terms of the momentum components parallel and orthogonal to the magnetic field. For this, the relations between the components of the velocity vector and the normalized momentum components $p_{\|}$ and $p_{\perp}$, involving the definiton of the \textit{Lorentz}-factor $\gamma=\sqrt{1+p^2}$ with $p=\sqrt{p_{\|}^2+p_{\perp}^2}$, are helpful:
\begin{equation}\label{v_p_relation}
v_{\|}\hspace{-0.5mm}=\hspace{-0.5mm}\frac{p_{\|}\,c}{\gamma}\hspace{-0.5mm}=\hspace{-0.5mm}\frac{p_{\|}\,c}{\sqrt{1+p^2}}\hspace{-0.5mm}=\hspace{-0.5mm}\frac{p_{\|}\,c}{\sqrt{1+p_{\|}^2+p_{\perp}^2}}\;;\;v_{\perp}=\frac{p_{\perp}\,c}{\gamma}\hspace{-0.5mm}=\hspace{-0.5mm}\frac{p_{\perp}\,c}{\sqrt{1+p^2}}\hspace{-0.5mm}=\hspace{-0.5mm}\frac{p_{\perp}\,c}{\sqrt{1+p_{\|}^2+p_{\perp}^2}}\,.
\end{equation}
Hereinafter, the calculation and computation of the parallel and orthogonal component of the avalanche runaway electron current density are treated seperately.
\vspace{2mm}

\subsection{Parallel component of the avalanche runaway electron current density}\label{j_RE_ava_par_subsection}

In the consequence of relation (\ref{v_p_relation}), one would receive for example the parallel component $j_{\,\|,RE}^{\,\textup{ava}}$ of the current density by the calculation of:\vspace{-1mm}
\begin{equation}\label{j_ava_RE_parallel_def}
\begin{split}
\begin{gathered}
j_{\,\|,RE}^{\,\textup{ava}}=-e\,\displaystyle{\int\limits_{p_{\|}=-\infty}^{\infty}\int\limits_{p_{\perp}=0}^{\infty}} v_{\|}\,f_{RE}^{\textup{ava}}(p_{\|},\,p_{\perp},\,t)\,2\pi\,p_{\perp}\,\mathrm{d}p_{\perp}\,\mathrm{d}p_{\|}
\\[8pt]
=-2\pi\,c\,e\,\displaystyle{\int\limits_{p_{\|}=-\infty}^{\infty}\int\limits_{p_{\perp}=0}^{\infty}} \frac{p_{\perp}\,p_{\|}}{\sqrt{1+p_{\|}^{2}+p_{\perp}^2}}\,f_{RE}^{\textup{ava}}(p_{\|},\,p_{\perp},\,t)\,\mathrm{d}p_{\perp}\,\mathrm{d}p_{\|}\,.
\end{gathered}
\end{split}
\end{equation}
Note that in equation (\ref{j_ava_RE_parallel_def}) the volume element $\mathrm{d}^3p$ in (\ref{j_ava_RE_def}) was replaced by the relation from (\ref{volelem_cyl_2D}) in the two-dimensional momentum coordinate system from figure \ref{fig_mom_coord} in section \ref{mom_space_coord_section}. Moreover, the parallel component of velocity was expressed by the use of the according relation from (\ref{v_p_relation}).

It is expected, that the efficiency of the calculation of the first moment will increase, if the two-dimensional integral in (\ref{j_ava_RE_parallel_def}) can at least be partly evaluated analytically. Otherwise, a two-dimensional numerical integration is needed, which is computational more expensive than a one-dimensional numerical integration, if only one integration in (\ref{j_ava_RE_parallel_def}) can be carried out or a function evaluation, if both integrations in (\ref{j_ava_RE_parallel_def}) can be performed.

First, one deduces a calculation rule, which is useful for a full two-dimensional numerical integration of (\ref{j_ava_RE_parallel_def}). At that, a finite lower integration bound $p_{\|,min}$ is needed to ensure a finite result of the integration, which is reasoned in the analytic analysis of the integral carried out in a second step. Hence, for the two-dimensional numerical integration, the following substitutions are set up:
\begin{equation}\label{substitutions_RE_j_para_2Dint}
\begin{split}
\begin{gathered}
p_{\|}=p_{\|,min}+\dfrac{w}{1-w} \;;\;\dfrac{\mathrm{d}p_{\|}}{\mathrm{d}w}= \dfrac{1}{(1-w)^2}\;;\; w(p_{\|}=p_{\|,min})=0\;,\;w(p_{\|}\rightarrow\infty)=1
\\[8pt]
p_{\perp}=\dfrac{z}{1-z} \;;\;\dfrac{\mathrm{d}p_{\perp}}{\mathrm{d}z}= \dfrac{1}{(1-z)^2}\;;\; z(p_{\perp}=0)=0\;,\;z(p_{\perp}\rightarrow\infty)=1\,.
\end{gathered}
\end{split}
\end{equation}
With those substitutions one rewrites the equation (\ref{j_ava_RE_parallel_def}) as shown in (\ref{j_ava_RE_par_def_appendix}), while inserting the constant $C$ from  (\ref{C_f_RE_ava}) and the distribution function from (\ref{f_RE_ava}). In the consequence, one finds two representations for the computation of the parallel component $j_{\,\|,RE}^{\,\textup{ava}}$ of the runaway electron current density:\vspace{-1mm}
\begin{equation}\label{j_ava_RE_parallel_2Dnumeffy}
\begin{split}
\begin{gathered}
j_{\,\|,RE}^{\,\textup{ava}}\hspace{-1.5mm}\underset{(\ref{C_f_RE_ava})}{\overset{(\ref{j_ava_RE_par_def_appendix})}{=}}
\hspace{-1.5mm}-\dfrac{2\,c\,e\,n_{RE}\,\tilde{E}}{c_{Z} \ln{(\Lambda)}}\,\textup{e}^{\,\frac{2\,(E-1)}{c_{Z} \ln{(\Lambda)}}\,\frac{t}{\tau}}\hspace{-4mm}\displaystyle{\int\limits_{p_{\|}=p_{\|,min}}^{\infty}\int\limits_{p_{\perp}=0}^{\infty}} \hspace{-1mm}\frac{p_{\perp} }{\sqrt{1+p_{\|}^{2}+p_{\perp}^2}}\,\textup{e}^{\,-\frac{p_{\|}}{c_{Z} \ln{(\Lambda)}}-\tilde{E}\frac{p_{\perp}^2}{p_{\|}}}\,\mathrm{d}p_{\perp}\mathrm{d}p_{\|}
\\[6pt]
\overset{(\ref{substitutions_RE_j_para_2Dint})}{=}\hspace{-2mm}-\dfrac{2\,c\,e\,n_{RE}\,\tilde{E}}{c_{Z} \ln{(\Lambda)}}\,\textup{e}^{\,\frac{2\,(E-1)}{c_{Z} \ln{(\Lambda)}}\,\frac{t}{\tau}}\hspace{-2mm}\underbrace{\displaystyle{\int\limits_{w=0}^{1}\int\limits_{z=0}^{1}} \frac{\frac{z}{(1-w)^2(1-z)^3}\,\textup{e}^{\,-\frac{p_{\|,min}+\frac{w}{1-w}}{c_{Z} \ln{(\Lambda)}}-\frac{\tilde{E}\left(\frac{z}{1-z}\right)^2}{p_{\|,min}+\frac{w}{1-w}}}}{\sqrt{1+\left(p_{\|,min}+\frac{w}{1-w}\right)^{2}+\left(\frac{z}{1-z}\right)^2}} \,\mathrm{d}z\,\mathrm{d}w}_{:=\,\textup{I}_{\,\textup{num,ava}}^{\,j_{\|} }}\,.
\end{gathered}
\end{split}
\end{equation} 
Those expressions can be implemented and involve a numerical integration routine for two-dimensional integrals as it is for instance provided by \textsc{MATLAB}.

\begin{figure}[H]
\begin{center}
\includegraphics[trim=49 45 120 41,width=1\textwidth,clip]{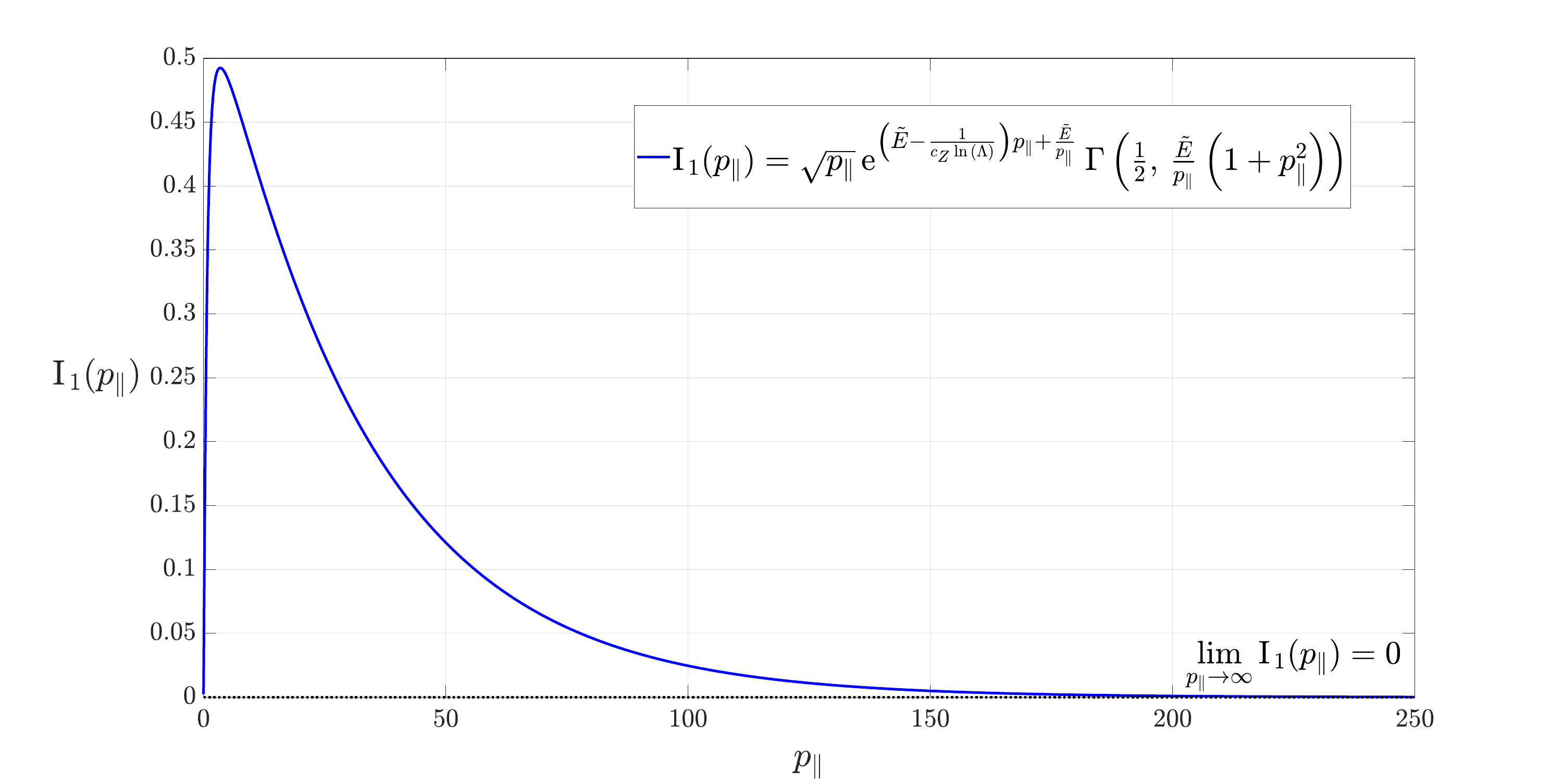}
\caption[Characteristic and representation of the limits at infinity for the integrand function $\textup{I}_{\,1}(p_{\|})$ from (\ref{j_ava_RE_par_calc1forgamma}) for $c_{Z}\approx 2.49$, $\ln{(\Lambda)}=12.6$ and\linebreak$\tilde{E}=2.8$]{Characteristic and representation of the limits at infinity\protect\footnotemark{} for the integrand function $\textup{I}_{\,1}(p_{\|})$ from (\ref{j_ava_RE_par_calc1forgamma}) for $c_{Z}\approx 2.49$, $\ln{(\Lambda)}=12.6$ and $\tilde{E}=2.8$}
\label{fig_j_ava_par_integrand_1}
\end{center}
\end{figure}

Second, an analytic analysis of the integral in (\ref{j_ava_RE_parallel_def}) shows, that the $p_{\perp}$-integral in (\ref{j_ava_RE_parallel_def}) is connected to the upper incomplete gamma function. Note, that the detailed analytic evaluation is shown in subsection \ref{j_ava_RE_parallel_int_appendix_subsection} of the appendix).\\
The starting point is the integral, which yields from (\ref{j_ava_RE_parallel_def}) by insertion of the distribution function from (\ref{f_RE_ava}):\vspace{-1mm}
\begin{equation}\label{j_ava_RE_par_def_aux}
\begin{split}
\begin{gathered}
j_{\,\|,RE}^{\,\textup{ava}}=\hspace{-0.5mm}-2\pi\,c\,e\hspace{-3.5mm}\displaystyle{\int\limits_{p_{\|}=-\infty}^{\infty}\int\limits_{p_{\perp}=0}^{\infty}} \frac{C\,p_{\perp}}{\sqrt{1+p_{\|}^{2}+p_{\perp}^2}} \,\textup{e}^{\frac{(E-1)}{c_{Z} \ln{(\Lambda)}}\,\frac{t}{\tau}-\frac{p_{\|}}{c_{Z} \ln{(\Lambda)}}-\tilde{E}\,\frac{p_{\perp}^{2}}{p_{\|}}} \,\mathrm{d}p_{\perp} \mathrm{d}p_{\|}\,.
\end{gathered}
\end{split}
\end{equation}
For the $p_{\perp}$-integration one can use the following substitution with the goal to rewrite the integral, in terms of an upper incomplete gamma function $\Gamma(z,\,a)$ \cite{incgammafunc}:\vspace{-1mm}
\begin{equation}\label{substitution_eta_j_par}
\begin{split}
\begin{gathered}
\eta(p_{\perp}):=\dfrac{\tilde{E}}{p_{\|} }\left(1+p_{\|}^{2}+p_{\perp}^2\right)\;;\;\dfrac{\mathrm{d}\eta}{\mathrm{d}p_{\perp}}=\dfrac{2\,\tilde{E}\,p_{\perp}}{p_{\|}}\;;
\\[4pt]
 \eta(p_{\perp}=0)=\dfrac{\tilde{E}}{p_{\|} }\left(1+p_{\|}^{2} \right)\;,\;\eta(p_{\perp}\rightarrow\infty)=\infty\,.
\end{gathered}
\end{split}
\end{equation}
Hence, the integral from (\ref{j_ava_RE_par_def_aux}) together with the relations for $C$ from (\ref{C_f_RE_ava}) and the substitution from (\ref{substitution_eta_j_par}) becomes:\vspace{-1mm}
\begin{equation}\label{j_ava_RE_par_calc1forgamma}
\begin{split}
\begin{gathered}
j_{\,\|,RE}^{\,\textup{ava}}=\hspace{-0.5mm}
- \frac{\pi\,c\,e\,C\,\sqrt{\tilde{E}}}{\tilde{E} }\hspace{-1.5mm}\displaystyle{\int\limits_{p_{\|}=-\infty}^{\infty}}\hspace{-2.5mm} \sqrt{p_{\|}}\,\textup{e}^{\,\frac{(E-1)}{c_{Z} \ln{(\Lambda)}}\,\frac{t}{\tau}+\left(\tilde{E}-\frac{1}{c_{Z} \ln{(\Lambda)}}\right)p_{\|}+\frac{\tilde{E}}{p_{\|}}}\hspace{-4mm}\displaystyle{\int\limits_{\eta=\frac{\tilde{E}}{p_{\|} }(1+p_{\|}^{2} )}^{\infty}}\hspace{-3.5mm} \eta^{-\frac{1}{2}}\,\textup{e}^{\, -\eta}  \, \mathrm{d}\eta  \,\mathrm{d}p_{\|}
\\[8pt]
=- \frac{ c\,e\,n_{RE}\, \sqrt{\tilde{E}} }{c_{Z}\ln{(\Lambda)}}\,\textup{e}^{\,\frac{2\,(E-1)}{c_{Z} \ln{(\Lambda)}}\,\frac{t}{\tau}} \hspace{-2mm}\displaystyle{\int\limits_{p_{\|}=-\infty}^{\infty}} \hspace{-2mm} \underbrace{\sqrt{p_{\|}}\,\textup{e}^{  \left(\tilde{E}-\frac{1}{c_{Z} \ln{(\Lambda)}}\right)p_{\|}+\frac{\tilde{E}}{p_{\|}}}\,\Gamma\hspace{-0.5mm}\left(\dfrac{1}{2},\,\frac{\tilde{E}}{p_{\|} }\hspace{-0.5mm}\left(1+p_{\|}^{2}   \right)\hspace{-0.5mm}\right)}_{=:\,\textup{I}_{\,1}(p_{\|})}\hspace{-0.5mm}\mathrm{d}p_{\|}\,.
\end{gathered}
\end{split}
\end{equation} 
Note, that the appearing special value of the incomplete gamma function is related to the complementary error function \cite{NISTincompleteGammafunction} and in the consequence as to the error function \cite{NISTerfc} by:\vspace{-2.5mm}
\begin{equation}\label{incgammafunc_erfc}
\Gamma\hspace{-0.5mm}\left(\dfrac{1}{2},\,\frac{\tilde{E}}{p_{\|} }\hspace{-0.5mm}\left(\hspace{-0.5mm}1\hspace{-0.5mm}+\hspace{-0.5mm}p_{\|}^{2}   \right)\hspace{-0.5mm}\right)\hspace{-0.8mm}=\hspace{-0.5mm}\sqrt{\pi}\,\mathrm{erfc}\hspace{-0.5mm}\left(\hspace{-0.5mm}\sqrt{\frac{\tilde{E}}{p_{\|} }\hspace{-0.5mm}\left(1+p_{\|}^{2}   \right)}\hspace*{0.4mm}\right)\hspace{-0.8mm}=\hspace{-0.5mm}\sqrt{\pi} \left(\hspace{-1mm}1\hspace{-0.5mm}-\hspace{-0.5mm}\mathrm{erf}\hspace{-0.5mm}\left(\hspace{-0.5mm}\sqrt{\frac{\tilde{E}}{p_{\|} }\hspace{-0.5mm}\left(1+p_{\|}^{2}   \right)}\hspace*{0.4mm}\right)\hspace{-0.7mm}\right).
\end{equation}
The integral in (\ref{j_ava_RE_par_calc1forgamma}) is convergent for a finite and positive lower integration bound $0\leq p_{\|,min}\leq\infty$, since one has $\displaystyle \lim_{p_{\|} \to \infty}\textup{I}_{\,1}(p_{\|})=0$ and $p_{\|}\geq 0$, so that the upper incomplete gamma function is defined. This can be validated as well by the plot in figure \ref{fig_j_ava_par_integrand_1}. 

Consequently, the integral in (\ref{j_ava_RE_par_calc1forgamma}) is exclusively solvable by means of numerical integration for $p_{\|}\in[p_{\|,min},\,\infty)$. Therefore another substitution is introduced:\vspace{-2mm}
\begin{equation}\label{substitutionNUM_RE_j_para}
\begin{split}
\begin{gathered}
p_{\|}=p_{\|,min}+\dfrac{w}{1-w} \;;\;\dfrac{dp_{\|}}{dw}= \dfrac{1}{(1-w)^2}\;;\; w(p_{\|}=p_{\|,min})=0\;,\;w(p_{\|}\rightarrow\infty)=1\,.
\end{gathered}
\end{split}
\end{equation} 
Thus, one has to numerically evaluate the integral $\textup{I}_{\,num,ava}^{\,j_{\|}}\,$, which is defined through (\ref{j_ava_RE_par_calc1forgamma}) in combination with (\ref{substitutionNUM_RE_j_para}):\vspace{-1.5mm}
\begin{equation*} 
\begin{split}
\begin{gathered}
\underline{\underline{j_{\,\|,RE}^{\,\textup{ava}}
}}= - \frac{ c\,e\,n_{RE}\,\sqrt{\tilde{E}}\,\textup{e}^{\,\frac{2\,(E-1)}{c_{Z} \ln{(\Lambda)}}\,\frac{t}{\tau}} }{ c_{Z}\ln{(\Lambda)}}\,\times  
\end{gathered}
\end{split}
\end{equation*}
\begin{equation}\label{j_ava_RE_par_I_num}
\begin{split}
\begin{gathered}
\times\hspace{-0.4mm}\displaystyle{\int\limits_{w=0 }^{1} } \hspace{-0.9mm}  \Gamma\hspace{-0.7mm}\left(\hspace{-0.5mm}\dfrac{1}{2},\frac{\tilde{E}\left(1\hspace{-0.5mm}+\hspace{-0.5mm}\left(p_{\|,min}\hspace{-0.5mm}+\hspace{-0.5mm}\frac{w}{w-1}\right)^{2}   \right)}{\left(p_{\|,min}+\frac{w}{w-1}\right) }\hspace{-0.5mm}\right) \hspace{-0.7mm}
 \frac{\sqrt{p_{\|,min}\hspace{-0.5mm}+\hspace{-0.5mm}\frac{w}{w-1}}}{(1-w)^2}\,\textup{e}^{  \left(\hspace{-0.5mm}\tilde{E}-\frac{1}{c_{Z} \ln{(\Lambda)}}\hspace{-0.5mm}\right)\left(p_{\|,min} + \frac{w}{w-1}\right) +\frac{\tilde{E}}{p_{\|}}} \, \mathrm{d}w
\\[8pt]
=:\underline{\underline{ - \frac{ c\,e\,n_{RE}\,\sqrt{\tilde{E}}\,\textup{e}^{\,\frac{2\,(E-1)}{c_{Z} \ln{(\Lambda)}}\,\frac{t}{\tau}} }{ c_{Z}\ln{(\Lambda)}}\cdot\textup{I}_{\,\textup{num,ava}}^{\,j_{\|}\,1\textup{D}} }}\,.
\end{gathered}
\end{split}
\end{equation}
Regarding (\ref{j_ava_RE_par_I_num}) it should be mentioned, that the value of the integral $\textup{I}_{\,\textup{num,ava}}^{\,j_{\|}\,1\textup{D}}$ depends on $p_{\|,min}$, $c_Z$, $\ln{(\Lambda)}$ and $\tilde{E}$. 
\footnotetext{\label{fig_j_ava_par_integrand1_footnote} The graph in figure \ref{fig_j_ava_par_integrand_1} was plotted with the \textsc{MATLAB}-file "j_ava_par_integrand_1.m",\\ \hspace*{8.7mm}which can be found in the digital appendix.}
\vspace{-7mm}
An evaluation of the efficiency of the two solution methods, requiring a one- or a two-dimensional numerical integration routine can now be carried out. By means of a \textsc{MATLAB}-script\footnote{\label{footnote_calculations_MATLAB} "efficiency_analysis_for_moment_calculations.m"}, which is stored in the digital appendix, one can therefore compute the parallel component of the current density from the two-dimensional integral expressions given in (\ref{j_ava_RE_parallel_2Dnumeffy}) and from the two one-dimensional integrals in (\ref{j_ava_RE_par_calc1forgamma}) and (\ref{j_ava_RE_par_I_num}). At that, the high-precision integration routine "\texttt{vpaintegral}" is used for the integrals from (\ref{j_ava_RE_par_calc1forgamma}) and (\ref{j_ava_RE_par_I_num}), due to the fact that the evaluations of their integrands exceed the default floating point precision. The measured runtime for the different computation rules can then be used to derive statements about the efficiency. The corresponding output for the values of the integrals and the parallel component of the current density for a certain set of parameters is shown for $p_{\|,min}=0$ in listing \ref{outMATLABeffcalcanalysisJPar}. Note, that the complete output of the \textsc{MATLAB}-script$^{\text{\ref{footnote_calculations_MATLAB}}}$ is given in the subsection \ref{subsection_num_efficiency_appendix} of the appendix and only an excerpt, related to the parallel component of the avalanche runaway electron current density, is displayed in the listing \ref{outMATLABeffcalcanalysisJPar}. However, the runtime duration for each computation rule was calculated as the mean value of the runtime of $729$ computations for different parameter combinations for $\tilde{E}\in[5,\,45]$, $c_{Z}\in[5,\,45]$ and $\ln{(\Lambda)}\in[5,\,45]$,
\begin{lstlisting}[language=Matlab,keywordstyle=\empty,frame=single, caption={Output connected to the parallel component of the avalanche runaway electron current density $j_{\,\|,RE}^{\,\textup{ava}}$ of the \textsc{MATLAB}-script "efficiency_\\analysis_for_moment_calculations.m" \vspace{1mm}},label={outMATLABeffcalcanalysisJPar} ]
set of parameters:

EoverEcrit = 150; n_e = 1e+20 m^-3; k_B*T_e = 100 eV; Z_eff = 1.5;

calculated quantities:

lnLambda = 12.59741; E_crit = 0.06424 V/m; p_crit = 0.08192;
E_D = 328.24256 V/m; E_sa = 70.24391 V/m; E = 9.63532 V/m; 

I_num_ava_nRE = 1.00000002 
(2D-integration with "integral2", from 0 to 1, runtime benchmark, abs_tol~10^-6);
mean runtime duration = 0.0094 s (for 729 parameter combinations);

I_num_ava_nRE = 0.99998290 
(2D-integration with "trapz", from 0 to 1, runtime benchmark, 10^6 grid points);
mean runtime duration = 0.0165 s (for 729 parameter combinations);

I_num_ava_j_par_1D = 5.5701 (1D, from p_par_min to infinity);
mean runtime duration = 0.2563 s (for 729 parameter combinations);

I_num_ava_j_par_1D = 5.5701 (1D, from 0 to 1);
mean runtime duration = 0.2584 s (for 729 parameter combinations);

I_num_ava_j_par = 0.5102 (2D, from p_par_min to infinity);
mean runtime duration = 0.0380 s (for 729 parameter combinations);

I_num_ava_j_par = 0.5102 (2D, from 0 to 1);
mean runtime duration = 0.0060 s (for 729 parameter combinations);

j_RE_ava_par_1D/(c*e*n_RE*exp(2*(E-1)/(c_Z*lnLambda)*t/tau)) 
		= u_RE_ava_par/c = -0.9688 (from 1D-integration);

j_RE_ava_par/(c*e*n_RE*exp(2*(E-1)/(c_Z*lnLambda)*t/tau)) 
		= u_RE_ava_par/c = -0.9688 (from 2D-integration);
\end{lstlisting}
in order to achieve reliable results. From this, one can deduce that the most efficient method for the calculation of the parallel component of the avalanche runaway electron current density is the two-dimensional numerical integration of the integral in (\ref{j_ava_RE_parallel_def}) with the substitutions suggested in (\ref{substitutions_RE_j_para_2Dint}). This computation method leads to the shortest runtime according to listing \ref{outMATLABeffcalcanalysisJPar} and it is therefore attested to be the most efficient procedure for this \textsc{MATLAB}-implementation.

In order to enable a physical validation of the results for $j_{\,\|,RE}^{\,\textup{ava}}$, the \textsc{MATLAB}-script displays the value of the parallel component of the mean velocity normalized with the speed of light in vacuum, for which $0\leq |u_{\,\|,RE}^{\,\textup{ava}}|/c\leq 1$ has to be fulfilled. With (\ref{j_ava_RE_def}) one finds:\vspace{-3.5mm}
\begin{equation}\label{normalized_u_mean_par}
\dfrac{u_{\,\|,RE}^{\,\textup{ava}}}{c}\overset{(\ref{j_ava_RE_def})}{=}\dfrac{j_{\,\|,RE}^{\,\textup{ava}}}{c\,e\,n_{RE}}\overset{(\ref{j_ava_RE_parallel_2Dnumeffy})}{=}-\dfrac{2\,\tilde{E}}{c_{Z} \ln{(\Lambda)}}\cdot\textup{e}^{\,\frac{2\,(E-1)}{c_{Z} \ln{(\Lambda)}}\,\frac{t}{\tau}}\cdot\textup{I}_{\,\textup{num,ava}}^{\,j_{\|}}\,.
\end{equation}
Note, that in (\ref{normalized_u_mean_par}) the time dependent part of (\ref{j_ava_RE_parallel_2Dnumeffy}) can be either set to one for $t=0\textup{ s}$ or it appears as a normalization factor for $u_{\,\|,RE}^{\,\textup{ava}}/c$, which only acts decreasing for $t\rightarrow\infty$ and is always greater than zero. In the consequence, it can not be resonsible for unphysical values of $u_{\,\|,RE}^{\,\textup{ava}}/c$. For the case presented in listing \ref{outMATLABeffcalcanalysisJPar} only physically valid values occur, depicting a large parallel mean velocity component at approximately $97\%$ of the speed of light. 
\vspace{2mm}

\subsection{Orthogonal component of the avalanche runaway electron current density}\label{j_RE_ava_perp_subsection}

For the orthogonal component of the current density arising from the avalanche runaway electron generation mechanism, one has to calculate the following integral, which results from (\ref{j_ava_RE_def}) with the help of (\ref{v_p_relation}) and by insertion of the volume element from equation (\ref{volelem_cyl_2D}) in analogy to (\ref{j_ava_RE_parallel_def}):
\vspace{-2mm}
\begin{equation}\label{j_ava_RE_perp_def}
\begin{split}
\begin{gathered}
j_{\,\perp,RE}^{\,\textup{ava}}=-e\,\displaystyle{\int\limits_{p_{\|}=-\infty}^{\infty}\int\limits_{p_{\perp}=0}^{\infty}} v_{\perp}\,f_{RE}^{\textup{ava}}(p_{\|},\,p_{\perp},\,t)\,2\pi\,p_{\perp}\,\mathrm{d}p_{\perp}\,\mathrm{d}p_{\|}
\\[8pt]
=-2\pi\,c\,e\,\displaystyle{\int\limits_{p_{\|}=-\infty}^{\infty}\int\limits_{p_{\perp}=0}^{\infty}} \frac{p_{\perp}^{2} }{\sqrt{1+p_{\|}^2+p_{\perp}^{2}}}\,f_{RE}^{\textup{ava}}(p_{\|},\,p_{\perp},\,t)\,\mathrm{d}p_{\perp}\,\mathrm{d}p_{\|}\,.
\end{gathered}
\end{split}
\end{equation}
One now inserts the distribution function from (\ref{f_RE_ava}) and the expression for $C$ according to (\ref{C_f_RE_ava}) in the integral in (\ref{j_ava_RE_perp_def}). The detailed conversions are carried out in the subsection \ref{j_ava_RE_perp_int_appendix_subsection} of the appendix, leading to the result: 
\begin{equation}\label{j_ava_RE_perp_def2}
\begin{split}
\begin{gathered}
j_{\,\perp,RE}^{\,\textup{ava}}=- \frac{2 \, c\,e\,n_{RE}\,\tilde{E}\,\textup{e}^{\,\frac{2\,(E-1)}{c_{Z} \ln{(\Lambda)}}\,\frac{t}{\tau}} }{ c_{Z}\ln{(\Lambda)}}\hspace{-0.7mm}\displaystyle{\int\limits_{p_{\|}=-\infty}^{\infty}\int\limits_{p_{\perp}=0}^{\infty}} \hspace{-0.4mm}\underbrace{\frac{p_{\perp}^{2}}{p_{\|}\sqrt{1\hspace{-0.4mm}+\hspace{-0.4mm}p_{\|}^2\hspace{-0.4mm}+\hspace{-0.4mm}p_{\perp}^{2}}}\,\textup{e}^{\,-\frac{p_{\|}}{c_{Z} \ln{(\Lambda)}}-\tilde{E}\frac{p_{\perp}^2}{p_{\|}}}}_{:=\,\textup{I}_{\,2}(p_{\|},\,p_{\perp})}\mathrm{d}p_{\perp}\mathrm{d}p_{\|}
\end{gathered}
\end{split}
\end{equation}
The resulting integral in (\ref{j_ava_RE_perp_def2}) is convergent for a finite lower integration bound $\vert p_{\|,min}\vert\leq\infty$, since one has $\displaystyle \lim_{p_{\|} \to \infty}\textup{I}_{\,2}(p_{\|})=0$ and $\displaystyle \lim_{p_{\|} \to -\infty}\textup{I}_{\,2}(p_{\|})=-\infty$ for constant finite values of $p_{\perp}$, due to the dominant behaviour of the exponential in $\,\textup{I}_{\,2}(p_{\|},\,p_{\perp})$. In the course of this, it is remarked, that the $p_{\perp}$-integration leads to a finite result as well, for the same reason of the dominant exponential function. One finds that $\displaystyle \lim_{p_{\perp} \to \infty}\textup{I}_{\,2}(p_{\perp})=\displaystyle \lim_{p_{\perp} \to 0}\textup{I}_{\,2}(p_{\perp})=0$ for constant finite values of $p_{\|}$, which does not require a change of the boundaries of the $p_{\perp}$-integration.

An analytic simplification of the integral in (\ref{j_ava_RE_perp_def2}) was not found, consequently requiring a two-dimensional numerical solving procedure. Therefore, one again utilizes the substitutions: 
\begin{equation}\label{substitutions_num_jperpagain}
\begin{split}
\begin{gathered}
p_{\|}=p_{\|,min}+\dfrac{w}{1-w} \;;\;\dfrac{\mathrm{d}p_{\|}}{\mathrm{d}w}= \dfrac{1}{(1-w)^2}\;;\; w(p_{\|}=p_{\|,min})=0\;,\;w(p_{\|}\rightarrow\infty)=1
\\[6pt]
p_{\perp}=\dfrac{z}{1-z} \;;\;\dfrac{\mathrm{d}p_{\perp}}{\mathrm{d}z}= \dfrac{1}{(1-z)^2}\;;\; z(p_{\perp}=0)=0\;,\;z(p_{\perp}\rightarrow\infty)=1\,,
\end{gathered}
\end{split}
\end{equation}
which were already defined in (\ref{substitutions_RE_j_para_2Dint}). The resulting two integrals, following from (\ref{j_ava_RE_perp_def2}) for $p_{\|}\in[p_{\|,min},\,\infty)$, then allow a numerical computation and are given by:\vspace{-1mm}
\begin{equation}\label{j_ava_RE_perp_2Dnum_result}
\begin{split}
\begin{gathered}
j_{\,\perp,RE}^{\,\textup{ava}}\hspace{-1.3mm}\overset{(\ref{j_ava_RE_perp_def2})}{=}\hspace{-1.8mm}- \frac{2 \, c\,e\,n_{RE}\,\tilde{E} }{ c_{Z}\ln{(\Lambda)}}\,\textup{e}^{\,\frac{2\,(E-1)}{c_{Z} \ln{(\Lambda)}}\,\frac{t}{\tau}}\hspace{-4.5mm}\underbrace{\displaystyle{\int\limits_{p_{\|}=p_{\|,min}}^{\infty}\int\limits_{p_{\perp}=0}^{\infty}}  \frac{p_{\perp}^{2}\,\textup{e}^{\,-\frac{p_{\|}}{c_{Z} \ln{(\Lambda)}}-\tilde{E}\frac{p_{\perp}^2}{p_{\|}}}}{p_{\|}\sqrt{1\hspace{-0.4mm}+\hspace{-0.4mm}p_{\|}^2\hspace{-0.4mm}+\hspace{-0.4mm}p_{\perp}^{2}}}\,\mathrm{d}p_{\perp}\mathrm{d}p_{\|} }_{=:\,\textup{I}_{\,\textup{num,ava}}^{\,j_{\perp}}}
\end{gathered}
\end{split}
\end{equation}
on the onehandside and by utilization of (\ref{substitutions_RE_j_para_2Dint}) by
\begin{equation}\label{j_ava_RE_perp_2Dnum_substituted_result}
\begin{split}
\begin{gathered}
\underline{\underline{j_{\,\perp,RE}^{\,\textup{ava}}}} \underset{(\ref{substitutions_RE_j_para_2Dint})}{\overset{(\ref{j_ava_RE_perp_def2})}{=}} - \frac{2 \, c\,e\,n_{RE}\,\tilde{E} }{ c_{Z}\ln{(\Lambda)}}\,\textup{e}^{\,\frac{2\,(E-1)}{c_{Z} \ln{(\Lambda)}}\,\frac{t}{\tau}}  
\\[6pt]
\times\,\displaystyle{\int\limits_{w=0}^{1}\int\limits_{z=0}^{1}}  \frac{ z^2\cdot\exp{\left(-\frac{ p_{\|,min}+\frac{w}{1-w} }{c_{Z} \ln{(\Lambda)}}-\frac{\tilde{E}\left(\frac{z}{1-z}\right)^2}{p_{\|,min}+\frac{w}{1-w}}\right)}}{(1-w)^2\,(1-z)^4\left(p_{\|,min}+\frac{w}{1-w}\right)\sqrt{1+\left(p_{\|,min}+\frac{w}{1-w}\right)^{2}+\left(\frac{z}{1-z}\right)^2}}\,\mathrm{d}z\,\mathrm{d}w
\\[7pt]
=:\underline{\underline{- \frac{2 \, c\,e\,n_{RE}\,\tilde{E} }{ c_{Z}\ln{(\Lambda)}}\,\textup{e}^{\,\frac{2\,(E-1)}{c_{Z} \ln{(\Lambda)}}\,\frac{t}{\tau}}\cdot \textup{I}_{\,\textup{num,ava}}^{\,j_{\perp}}}}
\end{gathered}
\end{split}
\end{equation}
on the otherhandside. By comparing the computation rules in (\ref{j_ava_RE_perp_2Dnum_result}) and (\ref{j_ava_RE_perp_2Dnum_substituted_result}) one notices, that they only differ in the calculation of $\textup{I}_{\,\textup{num,ava}}^{\,j_{\perp}}$. 

An evaluation of the efficiency of the two calculation methods is again possible with the \textsc{MATLAB}-script$^{\text{\ref{footnote_calculations_MATLAB}}}$. Hence a runtime-based efficiency analysis for the computation of the perpendicular component of the avalanche runaway electron current density based on the two-dimensional integral expressions given in (\ref{j_ava_RE_perp_2Dnum_result}) and (\ref{j_ava_RE_perp_2Dnum_substituted_result}) can be performed. The measured runtime for the different computation rules is again used to derive statements about the efficiency and calculated as the mean value of the runtime of $729$ computations for different parameter combinations for $\tilde{E}\in[5,\,45]$, $c_{Z}\in[5,\,45]$ and $\ln{(\Lambda)}\in[5,\,45]$, in order to achieve reliable results. The corresponding output for the values of the integrals and the orthogonal component of the current density for a certain set of parameters is shown for $p_{\|,min}=0$ in listing \ref{outMATLABeffcalcanalysisJPerp}.  
Note, that the complete output of the \textsc{MATLAB}-script$^{\text{\ref{footnote_calculations_MATLAB}}}$ is given in the subsection \ref{subsection_num_efficiency_appendix} of the appendix and only an excerpt, related to the perpendicular component of the avalanche runaway electron current density, is displayed in the listing \ref{outMATLABeffcalcanalysisJPerp}.
 
\begin{lstlisting}[language=Matlab,keywordstyle=\empty,frame=single, caption={Output connected to the orthogonal component of the avalanche runaway electron current density $j_{\,\perp,RE}^{\,\textup{ava}}$ of the \textsc{MATLAB}-script "efficiency_\\analysis_for_moment_calculations.m" \vspace{1mm}},label={outMATLABeffcalcanalysisJPerp} ]
set of parameters:

EoverEcrit = 150; n_e = 1e+20 m^-3; k_B*T_e = 100 eV; Z_eff = 1.5;

calculated quantities:

lnLambda = 12.59741; E_crit = 0.06424 V/m; p_crit = 0.08192;
E_D = 328.24256 V/m; E_sa = 70.24391 V/m; E = 9.63532 V/m; 

I_num_ava_nRE = 1.00000002 
(2D-integration with "integral2", from 0 to 1, runtime benchmark, abs_tol~10^-6);
mean runtime duration = 0.0094 s (for 729 parameter combinations);

I_num_ava_nRE = 0.99998290 
(2D-integration with "trapz", from 0 to 1, runtime benchmark, 10^6 grid points);
mean runtime duration = 0.0165 s (for 729 parameter combinations); 

I_num_ava_j_perp = 0.0224 (2D, from p_par_min to infinity);
mean runtime duration = 0.0357 s (for 729 parameter combinations);

I_num_ava_j_perp = 0.0224 (2D, from 0 to 1);
mean runtime duration = 0.0041 s (for 729 parameter combinations);

j_RE_ava_perp/(c*e*n_RE*exp(2*(E-1)/(c_Z*lnLambda)*t/tau))
		= u_RE_ava_perp/c = -0.0426;
\end{lstlisting}
As a result of the analysis, one can state that the most efficient method for the calculation of the orthogonal component of the avalanche runaway electron current density is the two-dimensional numerical integration of the integral in (\ref{j_ava_RE_perp_2Dnum_substituted_result}), which makes use of the substitutions from (\ref{substitutions_num_jperpagain}), allowing the integration between finite integration boundaries. This statement is reasoned, because the computation method involving the substitutions needs a significantly shorter mean runtime, according to listing \ref{outMATLABeffcalcanalysisJPerp}, compared to the calculations rule from (\ref{j_ava_RE_perp_2Dnum_result}).

A physical validation of the result for $j_{\,\perp,RE}^{\,\textup{ava}}$ is also observed in listing \ref{outMATLABeffcalcanalysisJPerp}, due to the fact that the normalized orthogonal component of the mean velocity 
\begin{equation}\label{normalized_u_mean_perp}
\dfrac{u_{\,\perp,RE}^{\,\textup{ava}}}{c}\overset{(\ref{j_ava_RE_def})}{=}\dfrac{j_{\,\perp,RE}^{\,\textup{ava}}}{c\,e\,n_{RE}}\overset{(\ref{j_ava_RE_perp_2Dnum_substituted_result})}{=}-\dfrac{2\,\tilde{E}}{c_{Z} \ln{(\Lambda)}}\cdot\textup{e}^{\,\frac{2\,(E-1)}{c_{Z} \ln{(\Lambda)}}\,\frac{t}{\tau}}\cdot\textup{I}_{\,\textup{num,ava}}^{\,j_{\perp}}\,.
\end{equation}
fulfills the requirement $0\leq |u_{\,\perp,RE}^{\,\textup{ava}}|/c\leq 1$ for the case presented in listing \ref{outMATLABeffcalcanalysisJPerp}. Note, that in contrast to the parallel component of the mean velocity from subsection \ref{j_RE_ava_par_subsection} the perpendicular component is noticeably smaller as it only reaches $4\%$ of the speed of light. In addition, it should be mentioned that the result for $u_{\,\perp,RE}^{\,\textup{ava}}/c$ is the same for both analysed calculation rules, because they generate the same value for $\textup{I}_{\,\textup{num,ava}}^{\,j_{\perp}}$.
\vspace{2mm}

\subsection{Efficient computation of the avalanche runaway electron current density}\label{j_RE_ava_full_subsection}

The magnitude of the avalanche runaway electron current density in SI-units is efficiently calculable if the discussion from the previous subsection \ref{j_RE_ava_par_subsection} and \ref{j_RE_ava_perp_subsection} is taken into account. Hence, the computation rule (\ref{j_ava_RE_parallel_2Dnumeffy}) is chosen for the parallel component of the current density and (\ref{j_ava_RE_perp_2Dnum_substituted_result}) is used for the perpendicular component. In the course of this, both methods include a two-dimensional numerical integration and make use of the variable substitutions in (\ref{substitutions_RE_j_para_2Dint}). The magnitude of the avalanche runaway electron current density then yields from:\vspace{-1mm}
\begin{equation}\label{j_ava_RE_num_magnitude}
\begin{split}
\begin{gathered}
\underline{\underline{j_{\,RE}^{\,\textup{ava}}}}=\vert\mathbf{j}_{\,RE}^{\,\textup{ava}}\vert=\sqrt{\left(j_{\,\|,RE}^{\,\textup{ava}}\right)^2+\left(j_{\,\perp,RE}^{\,\textup{ava}}\right)^2}
\\[9pt]
\underset{(\ref{j_ava_RE_perp_2Dnum_substituted_result})}{\overset{(\ref{j_ava_RE_parallel_2Dnumeffy})}{=}}\sqrt{\left(-\dfrac{2\,c\,e\,n_{RE}\,\tilde{E}}{c_{Z} \ln{(\Lambda)}}\,\textup{e}^{\,\frac{2\,(E-1)}{c_{Z} \ln{(\Lambda)}}\,\frac{t}{\tau}}\,\textup{I}_{\,\textup{num,ava}}^{\,j_{\|} } \right)^2+\left(- \frac{2 \, c\,e\,n_{RE}\,\tilde{E} }{ c_{Z}\ln{(\Lambda)}}\,\textup{e}^{\,\frac{2\,(E-1)}{c_{Z} \ln{(\Lambda)}}\,\frac{t}{\tau}}\,\textup{I}_{\,\textup{num,ava}}^{\,j_{\perp}}\right)^2}
\\[9pt]
=\underline{\underline{\frac{ 2\,c\,e\,n_{RE}\,\tilde{E} }{ c_{Z} \ln{(\Lambda)}}\,\textup{e}^{\,\frac{2\,(E-1)}{c_{Z} \ln{(\Lambda)}}\,\frac{t}{\tau} }\,\sqrt{\bigl( \textup{I}_{\,\textup{num,ava}}^{\,j_{\|}}\bigr)^2+ \bigl(  \textup{I}_{\,\textup{num,ava}}^{\,j_{\perp}}\bigr)^2} }}\,.
\end{gathered}
\end{split}
\end{equation}
The \textsc{MATLAB}-script$^{\text{\ref{footnote_calculations_MATLAB}}}$ uses the last equation in (\ref{j_ava_RE_num_magnitude}) and verifies the result with the values resulting from the first equation in (\ref{j_ava_RE_num_magnitude}). The corresponding output for the magnitude of the current density for a certain set of parameters is shown for $p_{\|,min}=0$ in listing \ref{outMATLABeffcalcanalysisJ}. Note, that the complete output of the \textsc{MATLAB}-script$^{\text{\ref{footnote_calculations_MATLAB}}}$ is given in the subsection \ref{subsection_num_efficiency_appendix} of the appendix and only an excerpt is displayed in the listing \ref{outMATLABeffcalcanalysisJPerp}.

The physical validity of the result for $j_{\,RE}^{\,\textup{ava}}$ is observed in listing \ref{outMATLABeffcalcanalysisJ}, due to the fact that the normalized magnitude of the mean velocity 
\begin{equation}\label{normalized_u_mean_magnitude}
\dfrac{u_{\,RE}^{\,\textup{ava}}}{c}\overset{(\ref{j_ava_RE_def})}{=}\dfrac{j_{\,RE}^{\,\textup{ava}}}{c\,e\,n_{RE}}\overset{(\ref{j_ava_RE_num_magnitude})}{=}\dfrac{2\,\tilde{E}}{c_{Z} \ln{(\Lambda)}}\cdot\textup{e}^{\,\frac{2\,(E-1)}{c_{Z} \ln{(\Lambda)}}\,\frac{t}{\tau}}\cdot\sqrt{\bigl( \textup{I}_{\,\textup{num,ava}}^{\,j_{\|}}\bigr)^2+ \bigl(  \textup{I}_{\,\textup{num,ava}}^{\,j_{\perp}}\bigr)^2}
\end{equation}
fulfills the requirement $0\leq |u_{\,RE}^{\,\textup{ava}}|/c\leq 1$ for the case shown in listing \ref{outMATLABeffcalcanalysisJ}. 
   
\begin{lstlisting}[language=Matlab,keywordstyle=\empty,frame=single, caption={Output connected to the magnitude of the avalanche runaway electron current density $j_{\,RE}^{\,\textup{ava}}$ of the \textsc{MATLAB}-script "efficiency_analysis_for_\\moment_calculations.m" \vspace{1mm}},label={outMATLABeffcalcanalysisJ} ]
set of parameters:

EoverEcrit = 150; n_e = 1e+20 m^-3; k_B*T_e = 100 eV; Z_eff = 1.5;

calculated quantities:

lnLambda = 12.59741; E_crit = 0.06424 V/m; p_crit = 0.08192; 
E_D = 328.24256 V/m; E_sa = 70.24391 V/m; E = 9.63532 V/m;  

j_RE_ava/(c*e*n_RE*exp(2*(E-1)/(c_Z*lnLambda)*t/tau)) 
		= u_RE_ava/c = 0.9698; (check: 0.9698)
\end{lstlisting}

\clearpage

\section{Mean mass-related kinetic energy density of an avalanche runaway electron population - second moment}\label{ava_second_moment_section}

As described in section \ref{kinetic_theory_section}, it is possible to utilize the second moment of a distribution function can be used to calculate the mean mass-related kinetic energy density of a particle population. Hence, the mean mass-related kinetic energy density of an avalanche runaway electron population can be evaluated, by means of equation (\ref{second_moment}):
\begin{equation}\label{ava_RE_second_moment}
K_{\,RE}^{\,\textup{ava}}(\mathbf{r},\,t)\overset{(\ref{second_moment})}{=}\dfrac{M^{\textup{ava}}_{2,RE}(\mathbf{r},\,t)}{2\,n_{RE}(\mathbf{r},\,t)}= \dfrac{1}{2\,n_{RE}(\mathbf{r},\,t)}\,\displaystyle{\iiint\limits_{\mathbb{R}^3}} \mathbf{v}\cdot\mathbf{v}\, f_{\,RE}^{\,\textup{ava}}(\mathbf{r},\,\mathbf{p},\,t)\,\mathrm{d}^3p\,.
\end{equation}
For the scalar product of the velocity vector $\mathbf{v}$ with itself it holds:\vspace{-1mm}
\begin{equation}\label{v_scalar_product}
\mathbf{v}\cdot\mathbf{v}=\vert\mathbf{v}\vert^2=v^2= v_{\|}^2+v_{\perp}^2 = c^2\hspace{-0.5mm}\left(\hspace{-0.5mm}\dfrac{p_{\|}^2}{1+p_{\|}^2+p_{\perp}^2}+\dfrac{p_{\perp}^2  }{1+p_{\|}^2+p_{\perp}^2}\hspace{-0.5mm}\right)\hspace{-0.5mm}= c^2\,\dfrac{p_{\|}^2+p_{\perp}^2}{1+p_{\|}^2+p_{\perp}^2} \,.
\end{equation}
Whereby, it should be remarked, that the magnitude of the velocity vector $v$ was expressed by its components parallel and perpendicular to the magnetic field (compare figure \ref{fig_mom_coord}). Additionally the relations between the velocity components and the associated components of the normalized momentum from (\ref{v_p_relation}) were applied. In the consequence, the equation (\ref{ava_RE_second_moment}) is rewritten, by means of (\ref{v_scalar_product}) and the volume element from (\ref{volelem_cyl_2D}):\vspace{-1mm}
\begin{equation}\label{K_ava_RE_def}
\begin{split}
\begin{gathered}
K_{\,RE}^{\,\textup{ava}}\underset{(\ref{volelem_cyl_2D})}{\overset{(\ref{ava_RE_second_moment})}{=}}\dfrac{1}{2\,n_{RE}}\,\displaystyle{\int\limits_{p_{\|}=-\infty}^{\infty}\int\limits_{p_{\perp}=0}^{\infty}} v^2\,f_{RE}^{\textup{ava}}(p_{\|},\,p_{\perp},\,t)\,2\pi\,p_{\perp}\mathrm{d}p_{\perp}\mathrm{d}p_{\|}
\\[8pt]
\overset{(\ref{v_scalar_product})}{=}\dfrac{\pi\,c^2}{n_{RE}}\,\displaystyle{\int\limits_{p_{\|}=-\infty}^{\infty}\int\limits_{p_{\perp}=0}^{\infty}}\dfrac{p_{\perp}p_{\|}^2+p_{\perp}^3}{1+p_{\|}^2+p_{\perp}^2} \,f_{RE}^{\textup{ava}}(p_{\|},\,p_{\perp},\,t)\,\mathrm{d}p_{\perp}\mathrm{d}p_{\|}\,.
\end{gathered}
\end{split}
\end{equation}
The expression for the mean mass-related kinetic energy density of an avalanche runaway electron population from (\ref{K_ava_RE_def}) together with the distribution function from (\ref{f_RE_ava}) yields:\vspace{-1mm}
\begin{equation}\label{K_ava_RE_V2}
K_{\,RE}^{\,\textup{ava}}\underset{(\ref{f_RE_ava})}{\overset{(\ref{K_ava_RE_def})}{=}}\hspace{-1mm}\dfrac{\pi\,c^2\,C}{n_{RE}}\hspace{-1mm}\displaystyle{\int\limits_{p_{\|}=-\infty}^{\infty}\int\limits_{p_{\perp}=0}^{\infty}}\underbrace{\dfrac{p_{\perp}p_{\|}^2+p_{\perp}^3}{p_{\|}(1+p_{\|}^2+p_{\perp}^2)} \,\textup{e}^{ \frac{(E-1)}{c_{Z} \ln{(\Lambda)}}\,\frac{t}{\tau}-\frac{p_{\|}}{c_{Z} \ln{(\Lambda)}}-\tilde{E}\,\frac{p_{\perp}^{2}}{p_{\|}} }}_{:=\,\textup{I}_{\,3}(p_{\|},\,p_{\perp})}\mathrm{d}p_{\perp}\mathrm{d}p_{\|}\,.
\end{equation}
First, the integral (\ref{K_ava_RE_V2}) is prepared for a two-dimensional computation using a numerical integration routine. Therefore, one analyses the behaviour of the integrand function $\textup{I}_{\,2}(p_{\|},\,p_{\perp})$ in (\ref{K_ava_RE_V2}), if $p_{\|}$ strives toward the lower or upper integration bound for constant finite values of $p_{\perp}$. Since one finds that $\displaystyle \lim_{p_{\|} \to \infty}\textup{I}_{\,3}(p_{\|})=0$ and $\displaystyle \lim_{p_{\|} \to -\infty}\textup{I}_{\,3}(p_{\|})=-\infty$ for constant finite values of $p_{\perp}$, due to the dominant behaviour of the exponential in $\,\textup{I}_{\,3}(p_{\|},\,p_{\perp})$. In the consequence, the integral from (\ref{K_ava_RE_V2}) has a finite result for a finite lower integration bound $\vert p_{\|,min}\vert\leq\infty$. Here it is remarked, that the $p_{\perp}$-integration is convergent as well, for the same reason of the dominant exponential function. One finds that $\displaystyle \lim_{p_{\perp} \to \infty}\textup{I}_{\,3}(p_{\perp})=\displaystyle \lim_{p_{\perp} \to 0}\textup{I}_{\,3}(p_{\perp})=0$ for constant finite values of $p_{\|}$, which does not require a change of the boundaries of the $p_{\perp}$-integration.
Based on the above discussion, one again utilizes the substitutions previously defined in (\ref{substitutions_RE_j_para_2Dint}):\vspace{-2mm} 
\begin{equation}\label{substitutions_num_K2Dpagain}
\begin{split}
\begin{gathered}
p_{\|}=p_{\|,min}+\dfrac{w}{1-w} \;;\;\dfrac{\mathrm{d}p_{\|}}{\mathrm{d}w}= \dfrac{1}{(1-w)^2}\;;\; w(p_{\|}=p_{\|,min})=0\;,\;w(p_{\|}\rightarrow\infty)=1
\\[6pt]
p_{\perp}=\dfrac{z}{1-z} \;;\;\dfrac{\mathrm{d}p_{\perp}}{\mathrm{d}z}= \dfrac{1}{(1-z)^2}\;;\; z(p_{\perp}=0)=0\;,\;z(p_{\perp}\rightarrow\infty)=1\,,
\end{gathered}
\end{split}
\end{equation}
in order to possibly support the efficiency of the numerical integration routine. Hence, two integrals with and without the substitutions from (\ref{substitutions_num_K2Dpagain}) result from (\ref{K_ava_RE_V2}) for $p_{\|}\in[p_{\|,min},\,\infty)$, allowing a numerical computation. By inserting the expression for $C$ from (\ref{C_f_RE_ava}), which is also shown in equation (\ref{K_ava_RE_V2_appendix}) of the subsection \ref{K_ava_RE_int_appendix_subsection}, the first two-dimensional integral reads:
\vspace{0mm}
\begin{equation}\label{K_ava_RE_V3_num2D}
K_{\,RE}^{\,\textup{ava}}\underset{(\ref{C_f_RE_ava}) }{\overset{(\ref{K_ava_RE_V2})}{=}}\hspace{-1mm}\dfrac{c^2\,\tilde{E}}{c_{Z} \ln{(\Lambda)}}\;\textup{e}^{ \frac{2\,(E-1)}{c_{Z} \ln{(\Lambda)}}\,\frac{t}{\tau}}\hspace{-4mm}\underbrace{\displaystyle{\int\limits_{p_{\|}=p_{\|,min}}^{\infty}\int\limits_{p_{\perp}=0}^{\infty}}\dfrac{p_{\perp}p_{\|}^2+p_{\perp}^3}{p_{\|}(1+p_{\|}^2+p_{\perp}^2)} \,\textup{e}^{ -\frac{p_{\|}}{c_{Z} \ln{(\Lambda)}}-\tilde{E}\,\frac{p_{\perp}^{2}}{p_{\|}} }\,\mathrm{d}p_{\perp}\mathrm{d}p_{\|}}_{:=\,\textup{I}_{\,num,ava}^{\,K}}\,.
\end{equation}
The second integral then follows from (\ref{K_ava_RE_V3_num2D}) and the substitutions from (\ref{substitutions_num_K2Dpagain}):
\begin{equation}\label{K_ava_RE_V3_num2D_sustituted}
\begin{split}
\begin{gathered}
\underline{\underline{ K_{\,RE}^{\,\textup{ava}} }}\underset{(\ref{substitutions_num_K2Dpagain}) }{\overset{(\ref{K_ava_RE_V3_num2D})}{=}} \dfrac{c^2\,\tilde{E}}{c_{Z} \ln{(\Lambda)}}\;\textup{e}^{ \frac{2\,(E-1)}{c_{Z} \ln{(\Lambda)}}\,\frac{t}{\tau}} 
\\[5pt]
\times \displaystyle{\int\limits_{w=0}^{1}\int\limits_{z=0}^{1}}\hspace{-0.5mm}\frac{\left(z\left(p_{\|,min}+\frac{w}{1-w}\right)^{2}+\frac{z^3}{(1-z)^2}\right) \cdot\exp{\left(-\frac{ p_{\|,min}+\frac{w}{1-w} }{c_{Z} \ln{(\Lambda)}}-\frac{\tilde{E}\left(\frac{z}{1-z}\right)^2}{p_{\|,min}+\frac{w}{1-w}}\right)}}{(1-w)^2\,(1-z)^3\left(p_{\|,min}+\frac{w}{1-w}\right)\hspace{-1mm}\left(1+\left(p_{\|,min}+\frac{w}{1-w}\right)^{2}+\left(\frac{z}{1-z}\right)^2\right)}\,\mathrm{d}z\,\mathrm{d}w
\\[6pt]
:=\underline{\underline{\dfrac{c^2\,\tilde{E}}{c_{Z} \ln{(\Lambda)}}\;\textup{e}^{ \frac{2\,(E-1)}{c_{Z} \ln{(\Lambda)}}\,\frac{t}{\tau}}\cdot\textup{I}_{\,num,ava}^{\,K}}}\,.
\end{gathered}
\end{split}
\end{equation}
Second, the integral (\ref{K_ava_RE_V2}) is partly solved in an analytic manner and results in an expression, which requires only a one-dimensional numerical integration. The detailed derivation is carried out in the subsection \ref{K_ava_RE_int_appendix_subsection} of the appendix. Hereinafter, the general procedure shall be outlined. 

\begin{figure}[H]
\begin{center}
\includegraphics[trim=62 52 120 41,width=1\textwidth,clip]{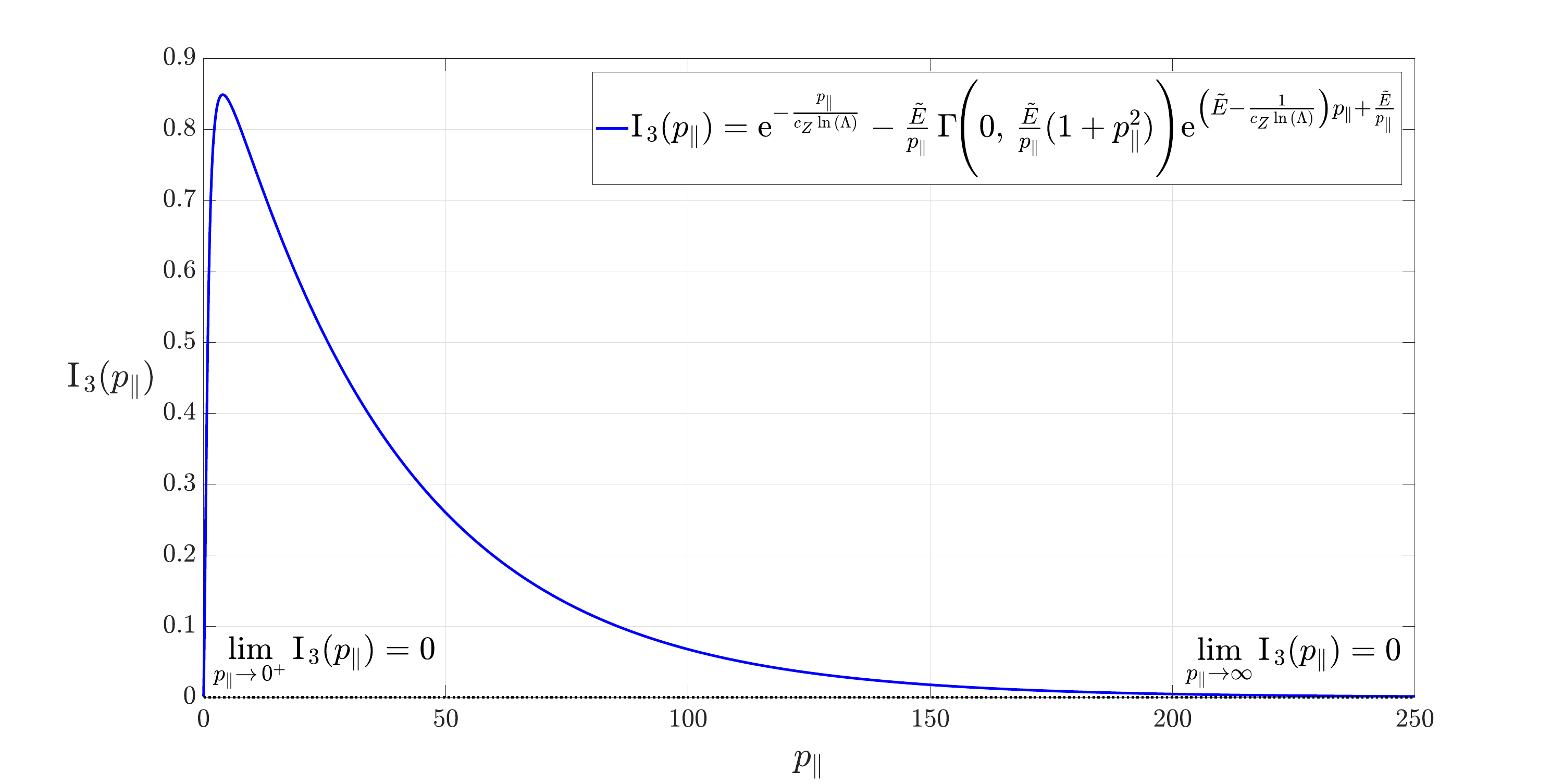}
\caption[Characteristic and representation of the limits at infinity for the integrand function $\textup{I}_{\,3}(p_{\|})$ from (\ref{K_ava_RE_V4_deriv}) for $c_{Z}\approx 2.49$, $\ln{(\Lambda)}=14.9$ and\linebreak$\tilde{E}=2.8$]{Characteristic and representation of the limits at infinity\protect\footnotemark{} for the integrand function $\textup{I}_{\,3}(p_{\|})$ from (\ref{K_ava_RE_V4_deriv}) for $c_{Z}\approx 2.49$, $\ln{(\Lambda)}=14.9$ and $\tilde{E}=2.8$}
\label{fig_K_ava_integrand_3}
\end{center}
\end{figure}

For that purpose, one starts from equation (\ref{K_ava_RE_V3_num2D}) and applies the substitution (\ref{substitution_eta_j_par}), which was already used in subsection \ref{j_RE_ava_par_subsection}, where it was defined as: \vspace{-2mm}
\begin{equation}\label{substitution_eta_K_ava}
\begin{split}
\begin{gathered}
\eta(p_{\perp})\hspace{-0.5mm}:=\hspace{-0.5mm}\dfrac{\tilde{E}}{p_{\|} }\hspace{-0.5mm}\left(\hspace{-0.5mm}1+p_{\|}^{2}+p_{\perp}^2\hspace{-0.5mm}\right)\,;\;\dfrac{\mathrm{d}\eta}{\mathrm{d}p_{\perp}}\hspace{-0.5mm}=\hspace{-0.5mm}\dfrac{2\,\tilde{E}\,p_{\perp}}{p_{\|}}\;;
\\[4pt]
 \eta(p_{\perp}\hspace{-0.5mm}=\hspace{-0.5mm}0)\hspace{-0.5mm}=\hspace{-0.5mm}\dfrac{\tilde{E}}{p_{\|} }\hspace{-0.5mm}\left(\hspace{-0.5mm}1+p_{\|}^{2}\hspace{-0.5mm} \right)\hspace{0.5mm},\;\eta(p_{\perp}\hspace{-0.5mm}\rightarrow\infty)\hspace{-0.5mm}=\hspace{-0.5mm}\infty\,.
\end{gathered}
\end{split}
\end{equation}
Hence, the integral from (\ref{K_ava_RE_V3_num2D}) becomes:\vspace{-2mm}
\begin{equation}\label{K_ava_RE_V3_deriv}
\begin{split}
\begin{gathered}
K_{\,RE}^{\,\textup{ava}}\underset{(\ref{substitution_eta_K_ava})}{\overset{(\ref{K_ava_RE_V3_num2D})}{=}} \dfrac{ c^2 \,\textup{e}^{ \frac{2\,(E-1)}{c_{Z} \ln{(\Lambda)}}\,\frac{t}{\tau}}}{ 2\,c_{Z} \ln{(\Lambda)}}
\\[4pt]
\times\, \displaystyle{\int\limits_{p_{\|}=-\infty}^{\infty}}\hspace{-3mm}\textup{e}^{  \left(\tilde{E}-\frac{1}{c_{Z} \ln{(\Lambda)}}\right)p_{\|}+\frac{\tilde{E}}{p_{\|}}}\hspace{-1mm}\left(\displaystyle{\int\limits_{\eta=\frac{\tilde{E}}{p_{\|} } ( 1+p_{\|}^{2})}^{\infty}} \hspace{-6.5mm} \textup{e}^{-\eta} \,\mathrm{d}\eta-\frac{\tilde{E}}{p_{\|}}\displaystyle{\int\limits_{\eta=\frac{\tilde{E}}{p_{\|} } ( 1+p_{\|}^{2})}^{\infty}}\hspace{-5.5mm}\eta^{-1} \textup{e}^{-\eta} \,\mathrm{d}\eta\hspace{-0.5mm}\right)\hspace{-0.5mm}\mathrm{d}p_{\|}\,.
\end{gathered}
\end{split}
\end{equation}
The first appearing integral with respect to $\eta$ can be solved analytically and the second integral is expressable in terms of the upper incomplete gamma function $\Gamma(z,\,a)$ \cite{incgammafunc}:
\vspace{-4mm}
\begin{equation}\label{K_ava_RE_V4_deriv}
\begin{split}
\begin{gathered}
K_{\,RE}^{\,\textup{ava}}\underset{ }{\overset{(\ref{K_ava_RE_V3_deriv})}{=}}\dfrac{ c^2 \,\textup{e}^{ \frac{2\,(E-1)}{c_{Z} \ln{(\Lambda)}}\,\frac{t}{\tau}}}{ 2\,c_{Z} \ln{(\Lambda)}}\hspace{-1mm} \displaystyle{\int\limits_{p_{\|}=-\infty}^{\infty}}\hspace{-3mm}\textup{e}^{  \left(\tilde{E}-\frac{1}{c_{Z} \ln{(\Lambda)}}\right)p_{\|}+\frac{\tilde{E}}{p_{\|}}}
\\[5pt]
 \times\,\Biggl\lgroup \, \underbrace{ \bigl[-\textup{e}^{-\eta}\bigr]_{\eta=\frac{\tilde{E}}{p_{\|} } ( 1+p_{\|}^{2})}^{\eta\rightarrow\infty}}_{=\,\textup{e}^{-\frac{\tilde{E}}{p_{\|} } ( 1+p_{\|}^{2})}}-\frac{\tilde{E}}{p_{\|}}\,\Gamma\hspace{-0.5mm}\Biggl(0,\,\frac{\tilde{E}}{p_{\|} } ( 1+p_{\|}^{2})\Biggr) \hspace{-0.5mm}\Biggr\rgroup \mathrm{d}p_{\|}
\\[5pt]
=\dfrac{ c^2 \,\textup{e}^{ \frac{2\,(E-1)}{c_{Z} \ln{(\Lambda)}}\,\frac{t}{\tau}}}{ 2\,c_{Z} \ln{(\Lambda)}}  \displaystyle{\int\limits_{p_{\|}-\infty}^{\infty}} \underbrace{\textup{e}^{ -\frac{p_{\|}}{c_{Z} \ln{(\Lambda)}}} - \frac{\tilde{E}}{p_{\|}} \,\Gamma\hspace{-0.5mm}\Biggl(0,\,\frac{\tilde{E}}{p_{\|} } ( 1+p_{\|}^{2})\Biggr)\textup{e}^{  \left(\tilde{E}-\frac{1}{c_{Z} \ln{(\Lambda)}}\right)p_{\|}+\frac{\tilde{E}}{p_{\|}}}}_{=:\,\textup{I}_{\,3}(p_{\|})}\,\mathrm{d}p_{\|}\,.
\end{gathered}
\end{split}
\end{equation}
Note, that a finite lower integration bound $p_{\|,min}$ with $ 0\hspace{-0.3mm}\leq\hspace{-0.3mm}p_{\|,min} \hspace{-0.3mm}<\hspace{-0.3mm}\infty$ is needed for the last integral in (\ref{K_ava_RE_V4_deriv}), so that the upper incomplete gamma function is defined and produces real values. Here, it should be remarked, that the appearing special value of the incomplete gamma function is related to the exponential integral \cite{NISTexpint} by the relation \cite{NISTincompleteGammafunction}:\vspace{-2.5mm}
\begin{equation}\label{incgammafunc_ExpInt}
\Gamma\hspace{-0.5mm}\left(0,\,\frac{\tilde{E}}{p_{\|} }\hspace{-0.5mm}\left(\hspace{-0.5mm}1\hspace{-0.5mm}+\hspace{-0.5mm}p_{\|}^{2}   \right)\hspace{-0.5mm}\right) =E_{1}\hspace{-0.5mm}\left(\frac{\tilde{E}}{p_{\|} }\hspace{-0.5mm}\left(\hspace{-0.5mm}1\hspace{-0.5mm}+\hspace{-0.5mm}p_{\|}^{2}   \right)\right)=\hspace{-1.5mm}\displaystyle{\int\limits^{\infty}_{x=\frac{\tilde{E}}{p_{\|} }\hspace{-0.5mm} (\hspace{-0.5mm}1\hspace{-0.5mm}+\hspace{-0.5mm}p_{\|}^{2}    )}}\hspace{-2.5mm}x^{-1}\,\textup{e}^{-x}\,\mathrm{d}x
\end{equation}
This leads to a finite total integration result for $K_{\,RE}^{\,\textup{ava}}$, where the integrand $\textup{I}_{\,3}(p_{\|})$ is plotted in figure \ref{fig_K_ava_integrand_3} for the possible interval for $p_{\|}$ and verifies that the integral has to converge, if a numerical integration routine is applied. 
\footnotetext{\label{fig_K_ava_integrand3_footnote} The diagram in figure \ref{fig_K_ava_integrand_3} was created with the \textsc{MATLAB}-file "K_ava_integrand_3.m",\\ \hspace*{8.7mm}which can be found in the digital appendix.}
\vspace{-8.5mm}
If the finite lower integration bound $p_{\|,min}$ is used, the last integral in (\ref{K_ava_RE_V4_deriv}) yields:
\vspace{-0.5mm}
\begin{equation}\label{K_ava_RE_V5_deriv}
\begin{split}
\begin{gathered}
\underline{\underline{K_{\,RE}^{\,\textup{ava}}}}\underset{ }{\overset{(\ref{K_ava_RE_V4_deriv})}{=}} \dfrac{ c^2 \,\textup{e}^{ \frac{2\,(E-1)}{c_{Z} \ln{(\Lambda)}}\,\frac{t}{\tau}}}{ 2\,c_{Z} \ln{(\Lambda)}} \Biggl(\;\, \displaystyle{\int\limits_{p_{\|}=p_{\|,min}}^{\infty}}\hspace{-3mm}\textup{e}^{ -\frac{p_{\|}}{c_{Z} \ln{(\Lambda)}}}\,\mathrm{d}p_{\|}   
  \\[6pt]
  -\,\tilde{E}\underbrace{\displaystyle{\int\limits_{p_{\|}=p_{\|,min}}^{\infty}}\hspace{-1mm} \frac{1}{p_{\|}} \,\Gamma\hspace{-0.5mm}\Biggl(0,\,\frac{\tilde{E}}{p_{\|} } ( 1+p_{\|}^{2})\Biggr)\textup{e}^{  \left(\tilde{E}-\frac{1}{c_{Z} \ln{(\Lambda)}}\right)p_{\|}+\frac{\tilde{E}}{p_{\|}}}\mathrm{d}p_{\|}}_{:=\,\textup{I}_{\,\textup{num,ava}}^{\,K,\,1\textup{D}}}\;\Biggr)
 \\[6pt]
= \dfrac{ c^2 \,\textup{e}^{ \frac{2\,(E-1)}{c_{Z} \ln{(\Lambda)}}\,\frac{t}{\tau}}}{ 2\,c_{Z} \ln{(\Lambda)}} \Biggl(\;\underbrace{ \left[-c_{Z} \ln{(\Lambda)}\,\textup{e}^{ -\frac{p_{\|}}{c_{Z} \ln{(\Lambda)}}}\right]_{p_{\|}=p_{\|,min}}^{p_{\|}\rightarrow\infty}}_{=\,c_{Z} \ln{(\Lambda)}\,\textup{e}^{ -\frac{p_{\|,min}}{c_{Z} \ln{(\Lambda)}}}}\hspace{-1.5mm} - \; \tilde{E}\cdot \textup{I}_{\,\textup{num,ava}}^{\,K,\,1\textup{D}}\Biggr)
\\[5pt]
=\underline{\underline{ \dfrac{ c^2 \,\textup{e}^{ \frac{2\,(E-1)}{c_{Z} \ln{(\Lambda)}}\,\frac{t}{\tau}}}{ 2\,c_{Z} \ln{(\Lambda)}} \biggl( c_{Z} \ln{(\Lambda)}\,\textup{e}^{ -\frac{p_{\|,min}}{c_{Z} \ln{(\Lambda)}}}  - \; \tilde{E}\cdot \textup{I}_{\,\textup{num,ava}}^{\,K,\,1\textup{D}}\biggr)}}\,.
\end{gathered}
\end{split}
\end{equation}
The boundaries of the integral $\textup{I}_{\,\textup{num,ava}}^{\,K,\,1\textup{D}}$ from (\ref{K_ava_RE_V5_deriv}) can be transformed, so that they are finite. Therefore, one recapitulates the substitution from (\ref{substitutionNUM_RE_j_para}):\vspace{-1.5mm}
\begin{equation}\label{substitution1again_RE_K_ava}
\begin{split}
\begin{gathered}
p_{\|}=p_{\|,min}+\dfrac{w}{1-w} \;;\;\dfrac{\mathrm{d}p_{\|}}{\mathrm{d}w}= \dfrac{1}{(1-w)^2}\;;\; w(p_{\|}=p_{\|,min})=0\;,\;w(p_{\|}\rightarrow\infty)=1\,.
\end{gathered}
\end{split}
\end{equation} 
Together with (\ref{K_ava_RE_V5_deriv}) it leads to:
\vspace{-1.5mm}
\begin{equation}\label{K_ava_RE_V6_deriv}
\begin{split}
\begin{gathered}
\underline{\underline{K_{\,RE}^{\,\textup{ava}}}}\underset{(\ref{substitution1again_RE_K_ava})}{\overset{(\ref{K_ava_RE_V5_deriv})}{=}} \dfrac{ c^2 \,\textup{e}^{ \frac{2\,(E-1)}{c_{Z} \ln{(\Lambda)}}\,\frac{t}{\tau}}}{ 2\,c_{Z} \ln{(\Lambda)}} \Biggl(c_{Z} \ln{(\Lambda)}\,\textup{e}^{ -\frac{p_{\|,min}}{c_{Z} \ln{(\Lambda)}}}   
 \\[7pt]
 -\,\tilde{E} \displaystyle{\int\limits_{w=0}^{1}}  \frac{\Gamma\hspace{-0.5mm}\left(\hspace{-0.5mm}0,\,\frac{\tilde{E}\,(1-w)}{p_{\|,min}\,(1-w)+w}\left(\hspace{-0.5mm} 1\hspace{-0.5mm}+\hspace{-0.5mm}\left(\hspace{-0.5mm}p_{\|,min}\hspace{-0.5mm}+\hspace{-0.5mm}\frac{w}{1-w}\hspace{-0.5mm}\right)^{2} \right)\hspace{-0.5mm}\right)\,}{(p_{\|,min}\,(1-w)+w)(1-w)} 
 \\[7pt]
 \times\,\exp{ \left( \left(\tilde{E}-\frac{1}{c_{Z} \ln{(\Lambda)}}\right)\left(p_{\|,min}+\frac{w}{1-w}\right)+\frac{\tilde{E}\,(1-w)}{p_{\|,min}\,(1-w)+w}\right)}\,\mathrm{d}w \;\Biggr)
  \\[9pt]
=:\underline{\underline{ \dfrac{ c^2 \,\textup{e}^{ \frac{2\,(E-1)}{c_{Z} \ln{(\Lambda)}}\,\frac{t}{\tau}}}{ 2\,c_{Z} \ln{(\Lambda)}} \biggl( c_{Z} \ln{(\Lambda)}\,\textup{e}^{ -\frac{p_{\|,min}}{c_{Z} \ln{(\Lambda)}}}  - \; \tilde{E}\cdot \textup{I}_{\,\textup{num,ava}}^{\,K,\,1\textup{D}}\biggr)}}\,.
\end{gathered}
\end{split}
\end{equation}
Consequently, one can compute the mean mass-related kinetic energy density of an avalanche runaway electron population $K_{\,RE}^{\,\textup{ava}}$ from a two-dimensional numerical integration, whereby the rules (\ref{K_ava_RE_V3_num2D}) and (\ref{K_ava_RE_V3_num2D_sustituted}) are usable or from a one-dimensional numerical integration, utilizing the rules stated in (\ref{K_ava_RE_V5_deriv}) and (\ref{K_ava_RE_V6_deriv}).

The efficiency of the presented solution methods, requiring a one- or a two-dimensional numerical integration routine can now be carried out, by means of the \textsc{MATLAB}-script$^{\text{\ref{footnote_calculations_MATLAB}}}$. It computes the calculation rules (\ref{K_ava_RE_V3_num2D}) with the help of two-dimensional numerical integration routine, while the two one-dimensional integrals stated in (\ref{K_ava_RE_V5_deriv}) and (\ref{K_ava_RE_V6_deriv}) require a high precision integration routine, because their integrand evaluation exceed the default floating point precision. The measured runtime for the different computation rules can then be used to derive statements about the efficiency. The corresponding output for the values of the integrals and the parallel component of the current density for a certain set of parameters is shown for $p_{\|,min}=0$ in listing \ref{outMATLABeffcalcanalysisK}.
Note, that the complete output of the \textsc{MATLAB}-script$^{\text{\ref{footnote_calculations_MATLAB}}}$ is given in the subsection \ref{subsection_num_efficiency_appendix} of the appendix and only an excerpt, related to the mean mass-related kinetic energy density, is displayed in the listing \ref{outMATLABeffcalcanalysisK}. However, the runtime duration for each computation rule was calculated as the mean value of the runtime of $729$ computations for different parameter combinations for $\tilde{E}\in[5,\,45]$, $c_{Z}\in[5,\,45]$ and $\ln{(\Lambda)}\in[5,\,45]$, in order to achieve reliable results. From this, one can deduce that the most efficient method for the calculation of mean mass-related kinetic energy density is the two-dimensional numerical integration of the integral in (\ref{K_ava_RE_V3_num2D_sustituted}). This computation method leads to the shortest runtime according to listing \ref{outMATLABeffcalcanalysisK} and it is therefore attested to be the most efficient procedure for this \textsc{MATLAB}-implementation.

In order to enable a physical validation of the results for $K_{\,RE}^{\,\textup{ava}}$, the \textsc{MATLAB}-script displays the value of the mean mass-related kinetic energy density normalized with the mass-related kinetic energy density related to the rest mass $c^2/2$, for which $0\leq K_{\,RE}^{\,\textup{ava}}/(c^2/2)\leq 1$ corresponds to energy densities, which are lower than the rest mass kinetic energy density. 
\begin{lstlisting}[language=Matlab,keywordstyle=\empty,frame=single, caption={Output connected to the mean mass-related kinetic energy density of an avalanche runaway electron population $K_{\,RE}^{\,\textup{ava}}$ of the \textsc{MATLAB}-script\\"efficiency_analysis_for_moment_calculations.m" \vspace{1mm}},label={outMATLABeffcalcanalysisK} ]
set of parameters:

EoverEcrit = 150; n_e = 1e+20 m^-3; k_B*T_e = 100 eV; Z_eff = 1.5;

calculated quantities:

lnLambda = 12.59741; E_crit = 0.06424 V/m; p_crit = 0.08192;
E_D = 328.24256 V/m; E_sa = 70.24391 V/m; E = 9.63532 V/m; 

I_num_ava_nRE = 1.00000002 
(2D-integration with "integral2", from 0 to 1, runtime benchmark, abs_tol~10^-6);
mean runtime duration = 0.0094 s (for 729 parameter combinations);

I_num_ava_nRE = 0.99998290 
(2D-integration with "trapz", from 0 to 1, runtime benchmark, 10^6 grid points);
mean runtime duration = 0.0165 s (for 729 parameter combinations);

I_num_ava_K = 0.5026 (2D, from p_par_min to infinity);
mean runtime duration = 0.0397 s (for 729 parameter combinations);

I_num_ava_K = 0.5026 (2D, from 0 to 1);
mean runtime duration = 0.0092 s (for 729 parameter combinations);

I_num_ava_K_p_par_1D = 0.0480 (1D, from p_par_min to infinity);
mean runtime duration = 0.1024 s (for 729 parameter combinations);

I_num_ava_K_p_par_1D = 0.0480 (1D, from 0 to 1);
mean runtime duration = 0.1088 s (for 729 parameter combinations);

K_RE_ava/((c^2/2)*exp(2*(E-1)/(c_Z*lnLambda)*t/tau)) = 0.9544 (from 1D-integration);

K_RE_ava/((c^2/2)*exp(2*(E-1)/(c_Z*lnLambda)*t/tau)) = 0.9544 (from 2D-integration);
\end{lstlisting}
With (\ref{K_ava_RE_V3_num2D_sustituted}) one finds:\vspace{-0.5mm}
\begin{equation}\label{normalized_K_mean}
\dfrac{K_{\,RE}^{\,\textup{ava}}}{c^2/2}\overset{(\ref{K_ava_RE_V3_num2D_sustituted})}{=} \dfrac{2\, \tilde{E}}{c_{Z} \ln{(\Lambda)}}\cdot\textup{e}^{\,\frac{2\,(E-1)}{c_{Z} \ln{(\Lambda)}}\,\frac{t}{\tau}}\cdot\textup{I}_{\,num,ava}^{\,K}\,.
\end{equation}
Note, that in (\ref{normalized_K_mean}) the time dependent part of (\ref{K_ava_RE_V3_num2D_sustituted}) can be either set to one for $t=0\textup{ s}$ or it appears as a normalization factor for $K_{\,RE}^{\,\textup{ava}}/(c^2/2)$ as it is observable in the listing \ref{outMATLABeffcalcanalysisK} from above. In addition, it is remarked, that this time dependent part only acts decreasing for $t\rightarrow\infty$ and is always greater than zero. In the consequence, it can not cause unphysical negative values of $K_{\,RE}^{\,\textup{ava}}/(c^2/2)$. That is why only physically valid values occur for the case presented in listing \ref{outMATLABeffcalcanalysisJPar}, at which the normalized mean mass-related kinetic energy density of the avalanche runaway population is approximately $95\%$ of the mass-related rest mass kinetic energy density.

\clearpage

\chapter{Physical evaluation and discussion}\label{evalution_chapter}

Efficient calculation rules i.a.\hspace{1mm}for the first and second moment of the analytical avalanche runaway electron distribution function were discussed in detail based on a \textsc{MATLAB}-implementation in chapter \ref{avalanche_chapter}. In the process, computation methods emerged, which allow the evaluation of the parallel and orthogonal component as well as the magnitude of the mean velocity $\mathbf{u}_{\,RE}^{\,\mathrm{ava}}$ and in addition the mean mass-related kinetic energy density $K_{\,RE}^{\,\mathrm{ava}}$ of avalanche runaway electrons. It was found, that an analytic approach in combination with a one-dimensional numerical integration was less efficient than two-dimensional numerical integration. Furthermore, the two-dimensional numerical integration together with substitutions, which avoid infinite integration bounds, turned out to be faster for a wide range of parameters. Hence, the important runtimes for the calculation of the moments of the avalanche runaway distribution function, represented by the duration of the integrations, are optimized for the used \textsc{MATLAB}-code, if the evaluation is based on a two-dimensional numerical integration over a finite domain. At that, the mentioned physical quantities related to the moments follow from the equations (\ref{j_ava_RE_parallel_def}), (\ref{j_ava_RE_perp_2Dnum_substituted_result}), (\ref{j_ava_RE_num_magnitude}) and (\ref{K_ava_RE_V3_num2D_sustituted}).\\
The variable parameters in those expressions are the \textit{Coulomb} logarithm $\ln{(\Lambda)}$ and the abbreviating parameter $\tilde{E}$. The \textit{Coulomb} logarithm can be written as a function of the electron density $n_{e}$ according to (\ref{CoulombLog}). The parameter $\tilde{E}$ from (\ref{E_cZ_f_RE_ava}) only varies for different values of the normalized component of the electric field parallel to the magnetic field lines $E=\vert E_{\|}\vert/E_{c}$ for a constant effective ion charge $Z_{eff}$. Where it should be remarked, that the critical 
electric field $E_{c}$ is calculated from the relation (\ref{E_crit}), by utilization of the electron density. In the consequence, one discovers the fundamental parameters $\vert E_{\|}\vert$ and $n_{e}$ of the computation expressions for the current density and the mean mass-related kinetic energy density of an avalanche runaway electron population with a given electron temperature $k_{B}T_{e}$ in electron volts and a constant effective ion charge $Z_{eff}$. Therewith, an evaluation and discussion of the physical quantities related to the first and second moment, in the space of those basic parameters, would allow to understand their value for different plasma configurations. On this occasion, a plasma configuration, e.g. within a tokamak fusion reactor, is defined by the electron density $n_{e}$, the electron temperature $k_{B}T_{e}$, the effective ion charge $Z_{eff}$ and the prevailing electric field represented by  its component parallel to the magnetic field lines $\vert E_{\|}\vert$. 

In the following, a general evaluation of the current density and the mean mass-related kinetic energy density of an avalanche runaway electron population shall take place. For that, the analyzed plasma configurations are solely varied in the electron density and the parallel electric field component, while the effective ion charge  and the electron temperatures are kept constant. For the purpose of the deduction of basic statements about the behaviour of the calculated quantities for tokamak fusion plasmas, the marginally changing effective ion charge can be set to $Z_{eff}=1.5$, meaning that it is not varied hereinafter. Moreover, the variation in the electron temperature $k_{B}T_{e}\in\lbrace 10\,\mathrm{eV},\,50\,\mathrm{eV},\,100\,\mathrm{eV},\,1000\,\mathrm{eV}\rbrace$ is chosen to be coarse, assuming that it will show possible changes in the results for increasing thermal energy, indicated by the electron temperature. A finer resolution is used for the typical range of the  electron density $n_{e}\in[10^{19}\,\mathrm{m}^{-3},\,10^{21}\,\mathrm{m}^{-3}]$ and the parallel component of the electric field $E_{c}<\vert E_{\|}\vert<E_{sa}<E_{D}$. Note that, the electric field has to be greater than the critical electric field from (\ref{E_crit}) and smaller than the slide-away field $E_{sa}\approx0.214\,E_{D}$ and the \textit{Dreicer} field $E_{D}$ from (\ref{Dreicerfield}), because otherwise either no runaway electrons are produced, if $\vert E_{\|}\vert<E_{c}$ or all electrons are runaway electrons and the avalanche generation mechanism has no physical meaning, if $\vert E_{\|}\vert>E_{sa}$ or $\vert E_{\|}\vert>E_{D}$.    \\
A more natural parameter for the computation, by means of the \textsc{MATLAB}-implemen- tation, is the normalized parallel electric field component $E=\vert E_{\|}\vert/E_{c}$, which satisfies the possible range of $\vert E_{\|}\vert$ for:\vspace{-2mm}
\begin{equation}\label{EoverEcrit_range}
1<\dfrac{\vert E_{\|}\vert}{E_{c}}<\dfrac{E_{D}}{E_{c}}\underset{(\ref{Dreicerfield})}{\overset{(\ref{E_crit})}{=}}\dfrac{m_{e0}\,c^2}{e\,k_{B}\,T_{e}}\,.
\end{equation}
In the consequence, the \textsc{MATLAB}-codes\footnote{\hspace{1mm}\mbox{"generate_num_data_10eV", "generate_num_data_100eV" and "generate_num_data_1000eV"}} make use of the parameter space:
\begin{equation}\label{EoverEcrit_range}
\lbrace n_{e}\rbrace\,\times\,\lbrace\vert E_{\|}\vert/E_{c}\rbrace\,=\,[1.01,\,1.1\cdot E_{D}/E{c}]\,\times\,[10^{19}\,\mathrm{m}^{-3},\,10^{21}\,\mathrm{m}^{-3}]
\end{equation}
with $400\times400$ grid points. The parallel and orthogonal component as well as the magnitude of the mean velocity $\mathbf{u}_{\,RE}^{\,\mathrm{ava}}$ and the mean mass-related kinetic energy density $K_{\,RE}^{\,\mathrm{ava}}$ are then calculated on this parameter space for the electron temperatures $k_{B}T_{e}\in\lbrace 10\,\mathrm{eV},\,100\,\mathrm{eV},\,1000\,\mathrm{eV}\rbrace$ for $Z_{eff}=1.5$. In addition, the parallel component of the electric field $\vert E_{\|}\vert$ is evaluated from $\vert E_{\|}\vert/E_{c}$ and the critical electric field $E_{c}$, which is given by the equation (\ref{E_crit}). The produced data is then written to a \textit{txt.}-file and plotted with the \textsc{MATLAB}-scripts\footnote{\label{plotfootnote}\hspace{1mm}\mbox{"plot_num_data_10eV.m", "plot_num_data_100eV.m" and "plot_num_data_1000eV.m"}}. They generate contour plots for the $n_{e}\,\times\,E=\vert E_{\|}\vert/E_{c}$-grid and scatter plots for the $n_{e}\,\times\, \vert E_{\|}\vert $-grid for the same data points. 

\begin{figure}[H]
\centering
\begin{subfigure}{\textwidth}
\centering \vspace{-2mm}
\includegraphics[trim=87 13 91 43,width=0.805\textwidth,clip]{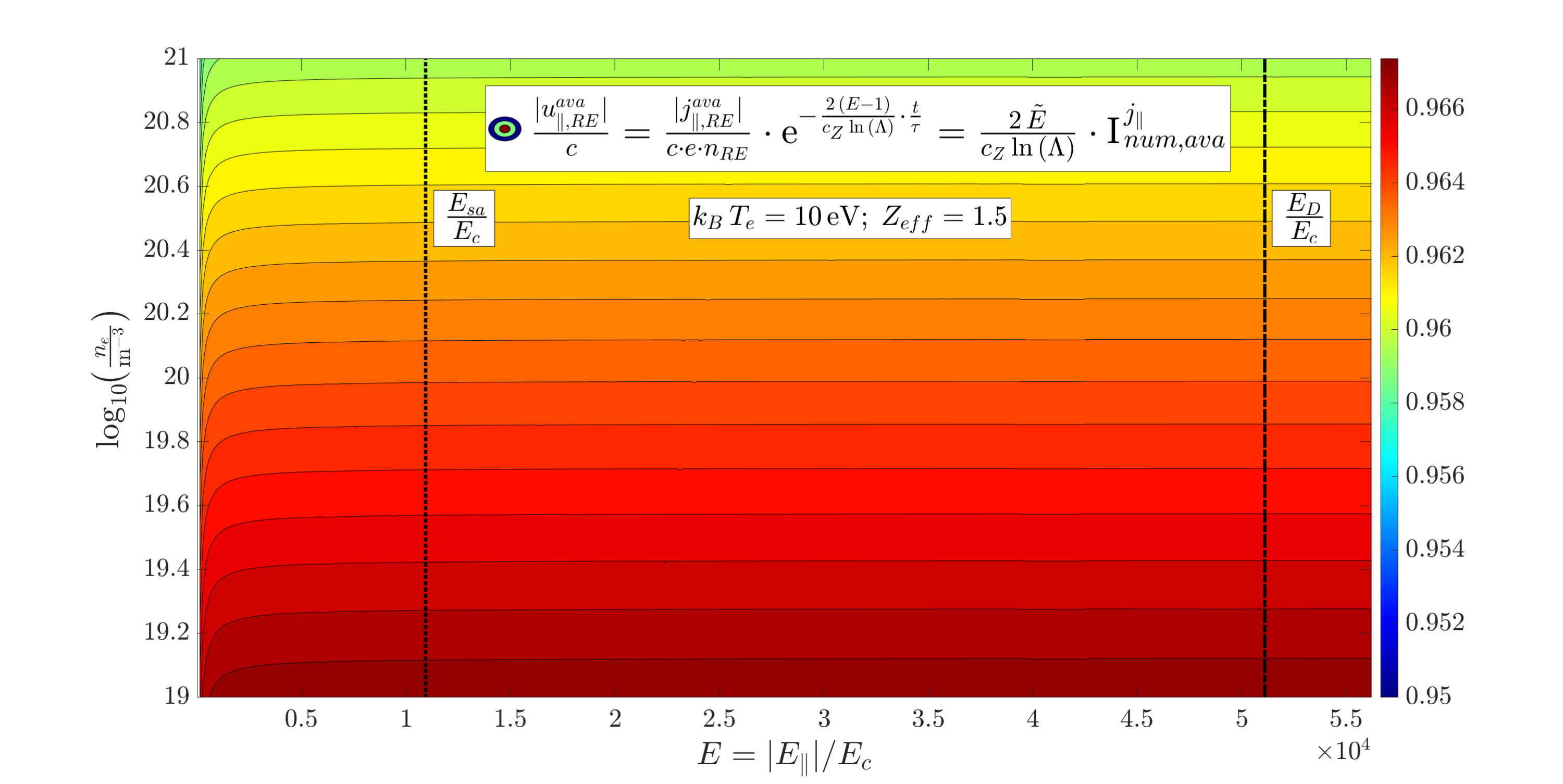} 
\caption{Avalanche runaway electron population with $k_{B}\,T_{e}=10\,\textup{eV}$ and $Z_{eff}=1.5$}
\label{fig_ava_u_par_EoverEc_n_e10eV}
\end{subfigure}
\begin{subfigure}{\textwidth}
\centering\vspace{2mm}
\includegraphics[trim=88 13 91 43,width=0.805\textwidth,clip]{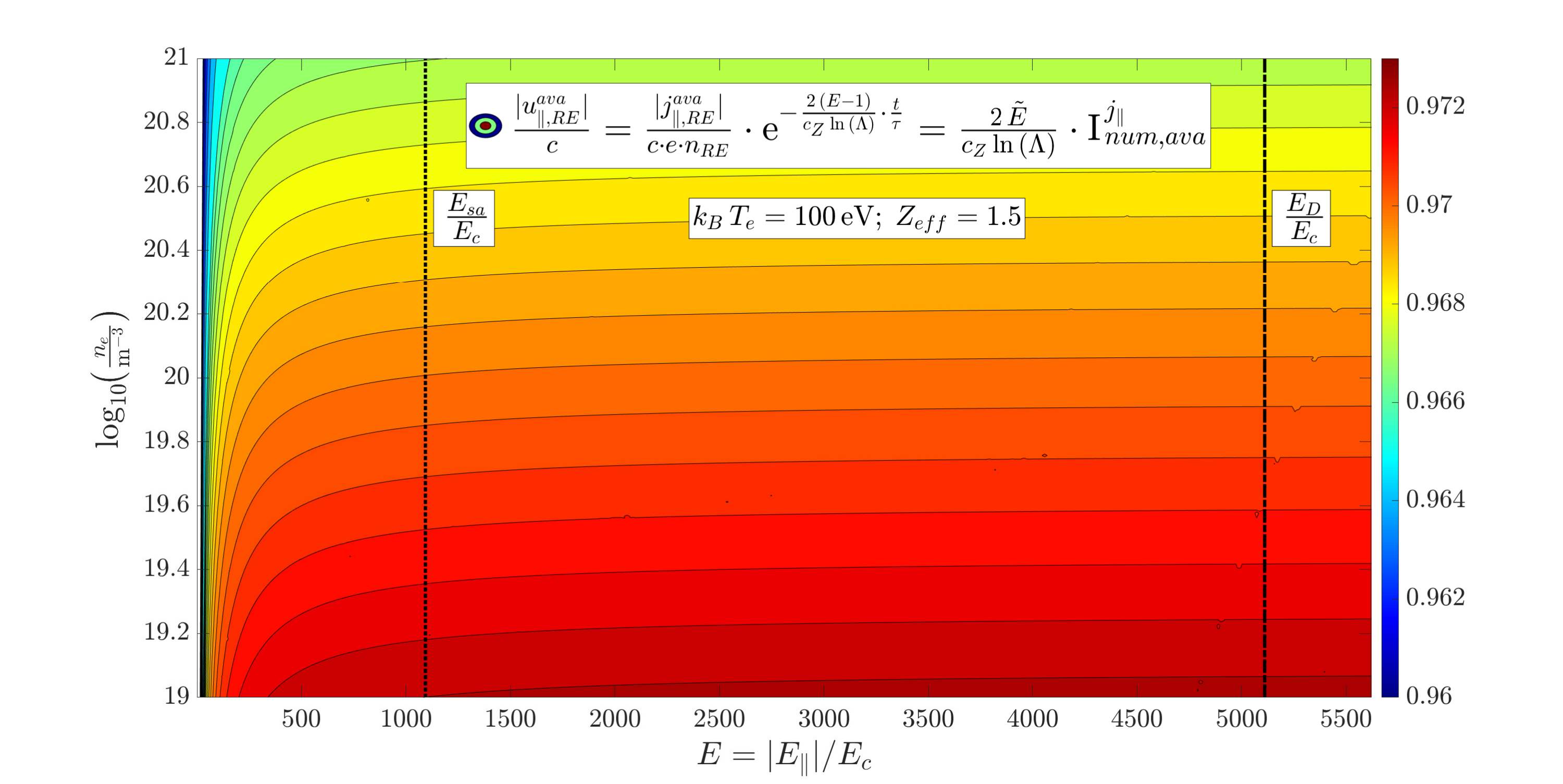}
\caption{Avalanche runaway electron population with $k_{B}\,T_{e}=100\,\textup{eV}$ and $Z_{eff}=1.5$}
\label{fig_ava_u_par_EoverEc_n_e100eV}
\end{subfigure}
\begin{subfigure}{\textwidth}
\centering\vspace{2mm}
\includegraphics[trim=85 13 91 43,width=0.805\textwidth,clip]{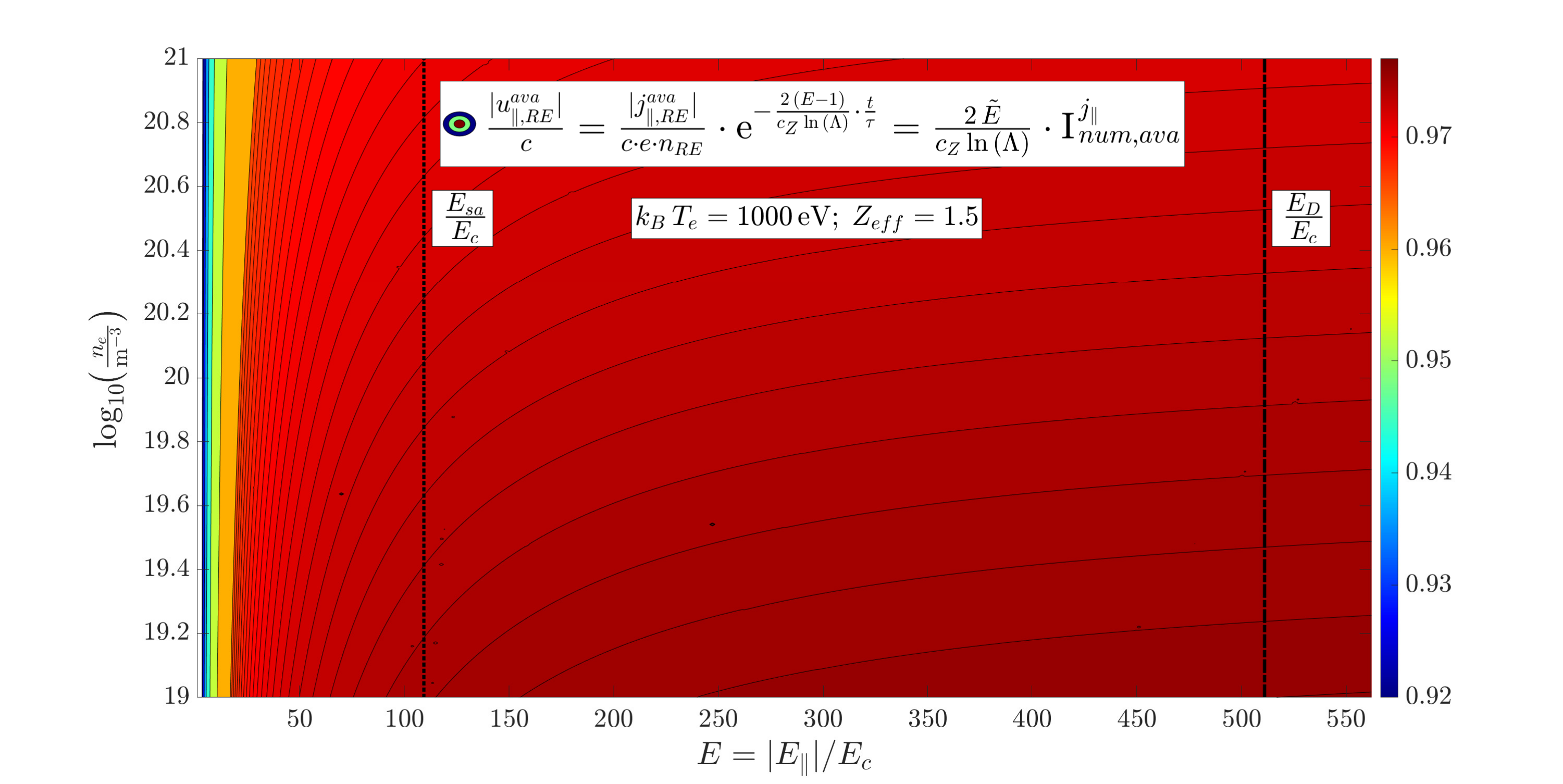}
\caption{Avalanche runaway electron population with $k_{B}\,T_{e}=1000\,\textup{eV}$ and $Z_{eff}=1.5$ }
\label{fig_ava_u_par_EoverEc_n_e1000eV}
\end{subfigure}
\captionsetup{format=hang,indention=0cm}
\caption[Contour plots of the absolute value of the normalized parallel component of the mean velocity $|u_{\,\|,RE}^{\,\textup{ava}}|/c$ for different avalanche runaway electron populations with respect to the electron density $n_{e}$ and the normalized electric field strength $E=\vert E_{\|}\vert/E_{c}$]{Contour plots$^{\ref{fig_eV_footnote}}$ of the absolute value of the normalized parallel component of the mean velocity $|u_{\,\|,RE}^{\,\textup{ava}}|/c$ for different avalanche runaway electron populations with respect to the electron density $n_{e}$ and the normalized electric field strength $E=\vert E_{\|}\vert/E_{c}$}
\label{fig_ava_u_par_EoverEc_n_e}
\end{figure}

\begin{figure}[H]
\centering\vspace{-2mm}
\begin{subfigure}{\textwidth}
\centering
\includegraphics[trim=80 12 90 40,width=0.805\textwidth,clip]{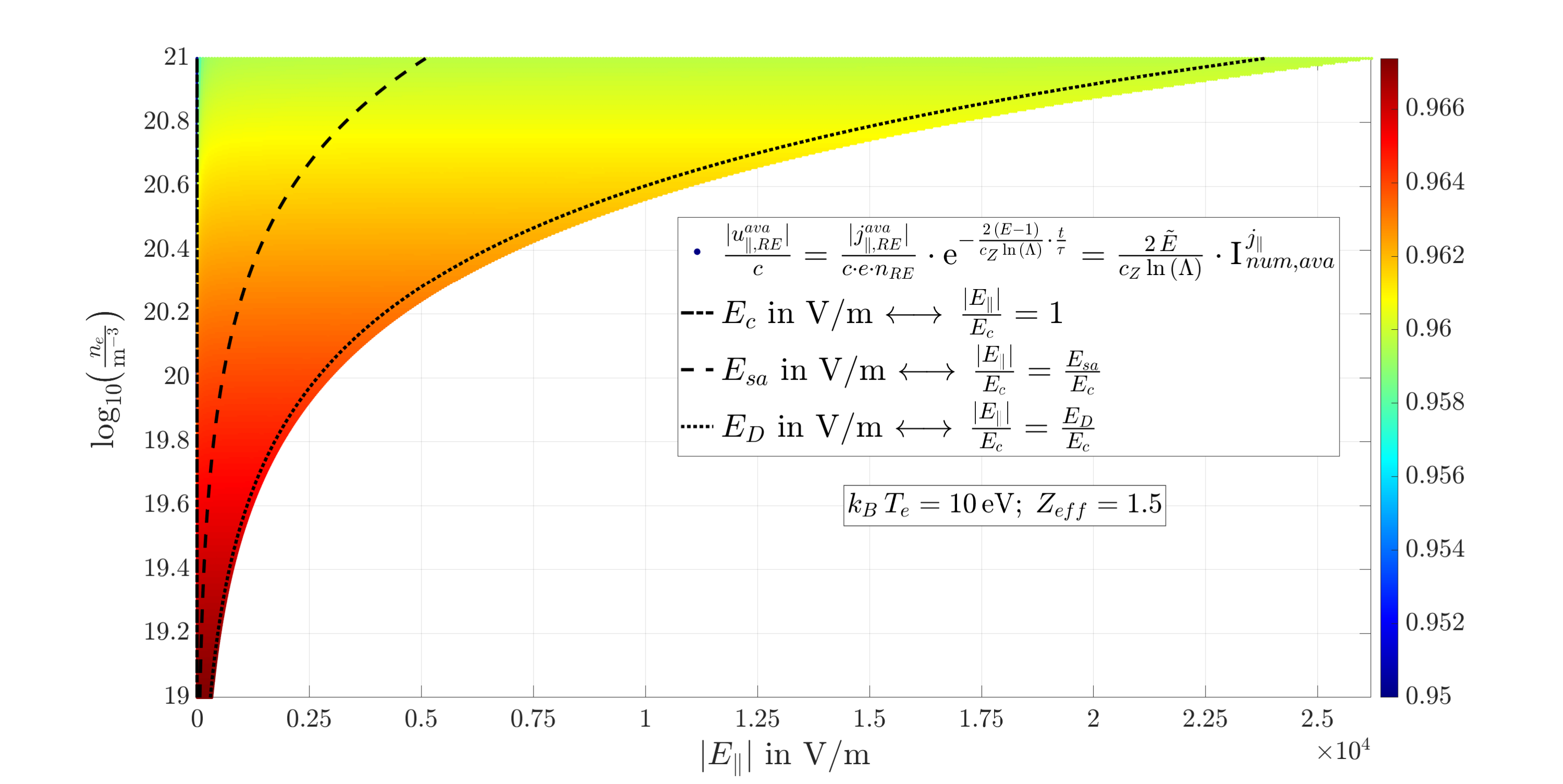}
\caption{Avalanche runaway electron population with $k_{B}\,T_{e}=10\,\textup{eV}$ and $Z_{eff}=1.5$}
\label{fig_ava_u_par_E_par_n_e10eV}
\end{subfigure}
\begin{subfigure}{\textwidth}
\centering\vspace{2mm}
\includegraphics[trim=87 12 90 40,width=0.805\textwidth,clip]{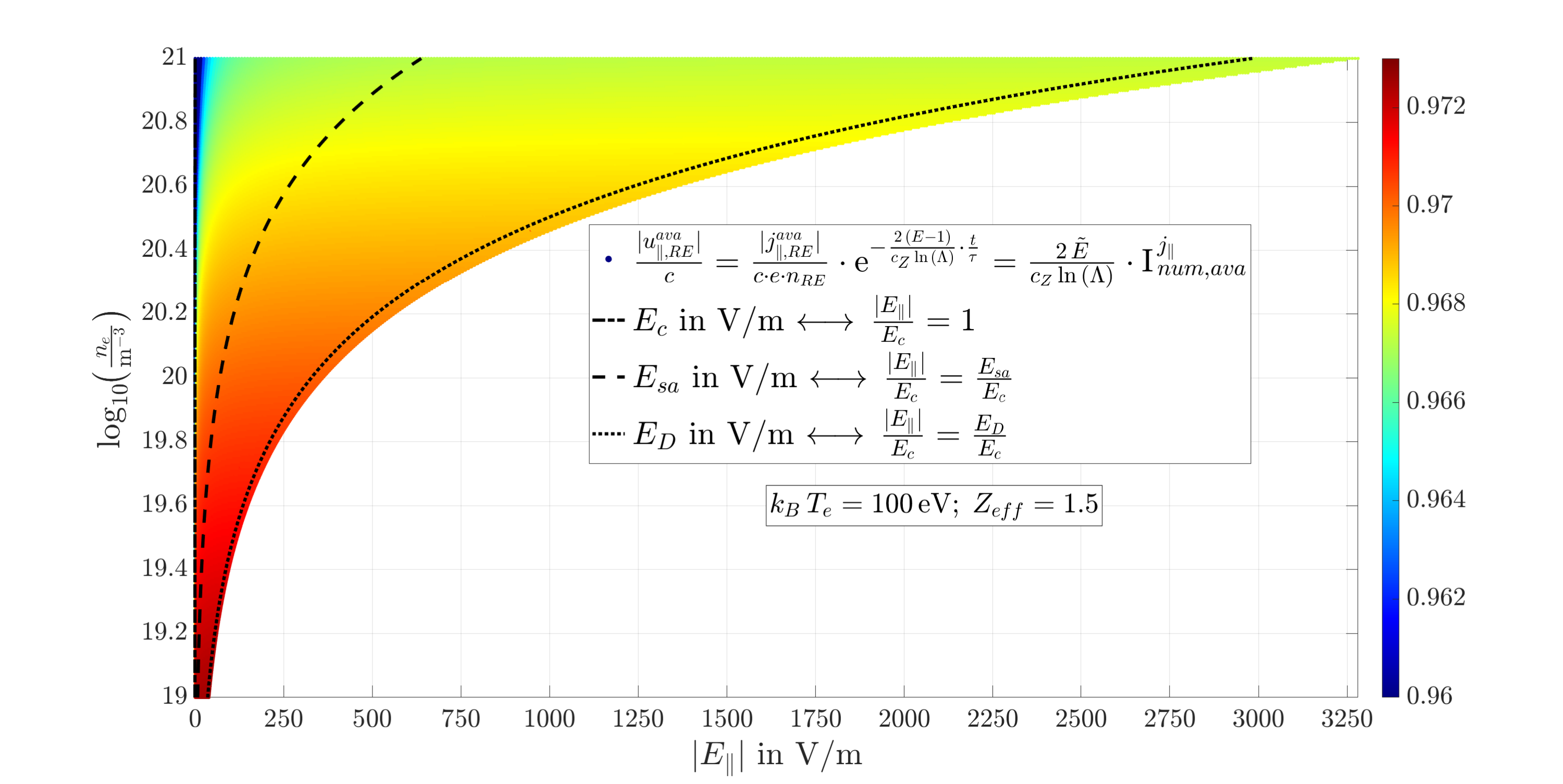}
\caption{Avalanche runaway electron population with $k_{B}\,T_{e}=100\,\textup{eV}$ and $Z_{eff}=1.5$}
\label{fig_ava_u_par_E_par_n_e100eV}
\end{subfigure}
\begin{subfigure}{\textwidth}
\centering\vspace{2mm}
\includegraphics[trim=87 12 90 40,width=0.805\textwidth,clip]{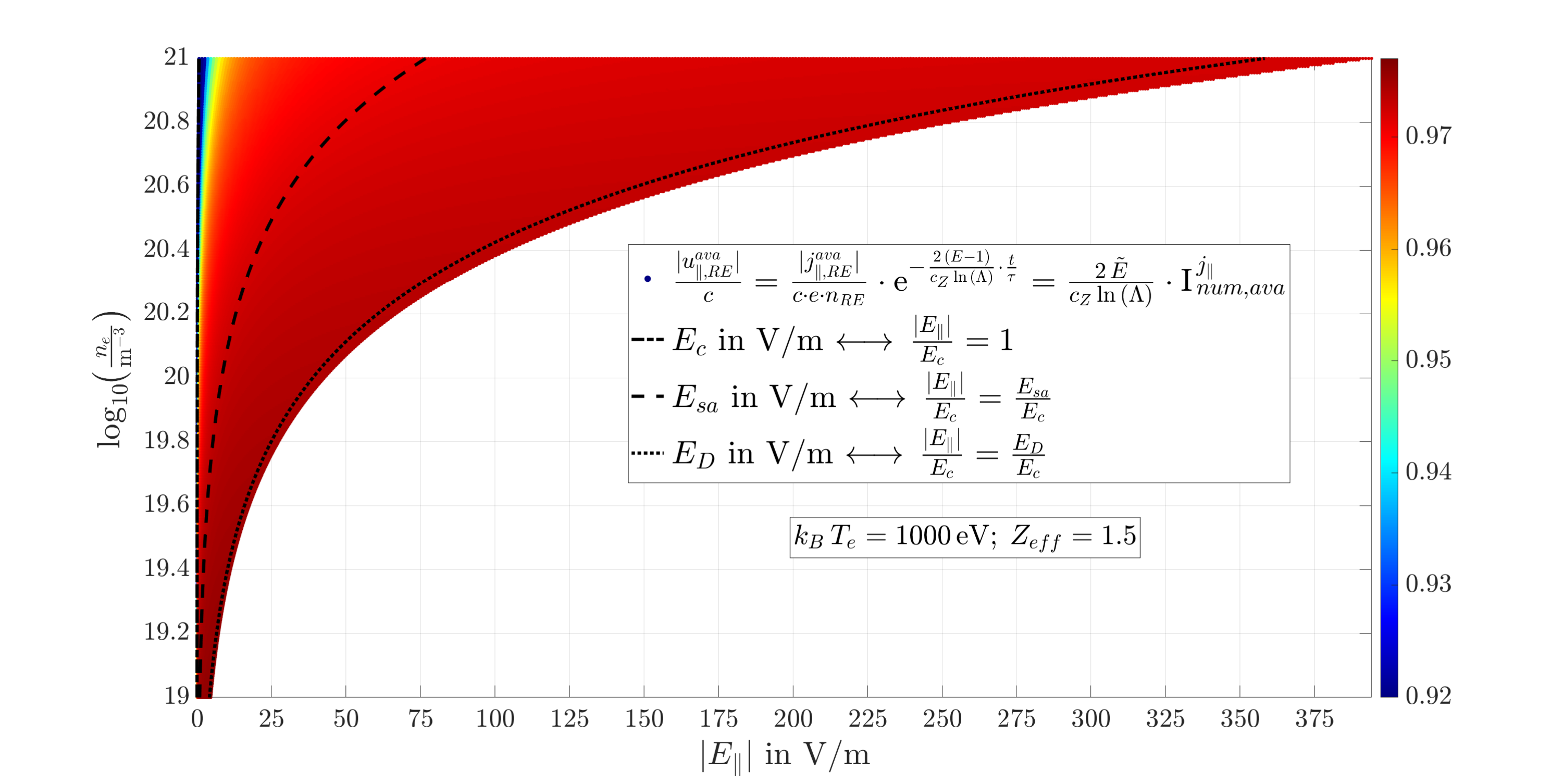}
\caption{Avalanche runaway electron population with $k_{B}\,T_{e}=1000\,\textup{eV}$ and $Z_{eff}=1.5$ }
\label{fig_ava_u_par_E_par_n_e1000eV}
\end{subfigure}
\captionsetup{format=hang,indention=0cm}
\caption[Contour plots of the absolute value of the normalized parallel component of the mean velocity $|u_{\,\|,RE}^{\,\textup{ava}}|/c$ for different avalanche runaway electron populations with respect to the electron density $n_{e}$ and the absolute value of the parallel component of the electric field $\vert E_{\|}\vert$]{Contour plots$^{\ref{fig_eV_footnote}}$ of the absolute value of the normalized parallel component of the mean velocity $|u_{\,\|,RE}^{\,\textup{ava}}|/c$ for different avalanche runaway electron populations with respect to the electron density $n_{e}$ and the absolute value of the parallel component of the electric field $\vert E_{\|}\vert$}
\label{fig_ava_u_par_E_par_n_e}
\end{figure}

In figure \ref{fig_ava_u_par_EoverEc_n_e} one can observe the contour plots of the absolute value of the normalized parallel component of the mean velocity $|u_{\,\|,RE}^{\,\textup{ava}}|/c$ for different  electron temperatures with respect to the electron density $n_{e}$ and the normalized electric field strength $E=\vert E_{\|}\vert/E_{c}$. At that, the resulting values are in the ranges $|u_{\,\|,RE}^{\,\textup{ava}}|/c\in[0.262,\,0.967]$ for $k_{B}\,T_{e}=10\,\textup{eV}$, $|u_{\,\|,RE}^{\,\textup{ava}}|/c\in[0.286,\,0.973]$ for $k_{B}\,T_{e}=100\,\textup{eV}$ and $|u_{\,\|,RE}^{\,\textup{ava}}|/c\in[0.306,\,0.981]$ for $k_{B}\,T_{e}=1000\,\textup{eV}$ according to the combined console output of the plotting scripts$^{\mathrm{\ref{plotfootnote}}}$ from listing \ref{outMATLABgeneratedataAppendix} in subsection \ref{subsection_generated_data} of the appendix. Note, that the color scales do not match this ranges, in order to enhance the visibility of gradients in the contour plot via different colors, which holds for all diagrams for the electron temperatures $k_{B}T_{e}\in\lbrace 10\,\mathrm{eV},\,100\,\mathrm{eV},\,1000\,\mathrm{eV}\rbrace$. In general, one notices that this component of the mean velocity is increases for high electric fields and low electron densities. As well, one regards an increase of this quantity with rising electron temperature. The same statements can be found in the figure \ref{fig_ava_u_par_E_par_n_e} in the scatter plots for the normalized parallel component of the mean velocity $|u_{\,\|,RE}^{\,\textup{ava}}|/c$ with respect to the electron density $n_{e}$ and the absolute value of the parallel component of the electric field strength $\vert E_{\|}\vert$. Further, one reminds, that the slide-away electric field $E_{sa}$ is basically never exceeded in tokamak disruption. Therefore, mainly the area for lower electric fields bounded by the slide-away electric field $E_{sa}$ is interesting for fusion plasmas in both figures.

\begin{figure}[H]
\centering
\begin{subfigure}{\textwidth}
\centering\vspace{-2mm}
\includegraphics[trim=87 13 91 43,width=0.805\textwidth,clip]{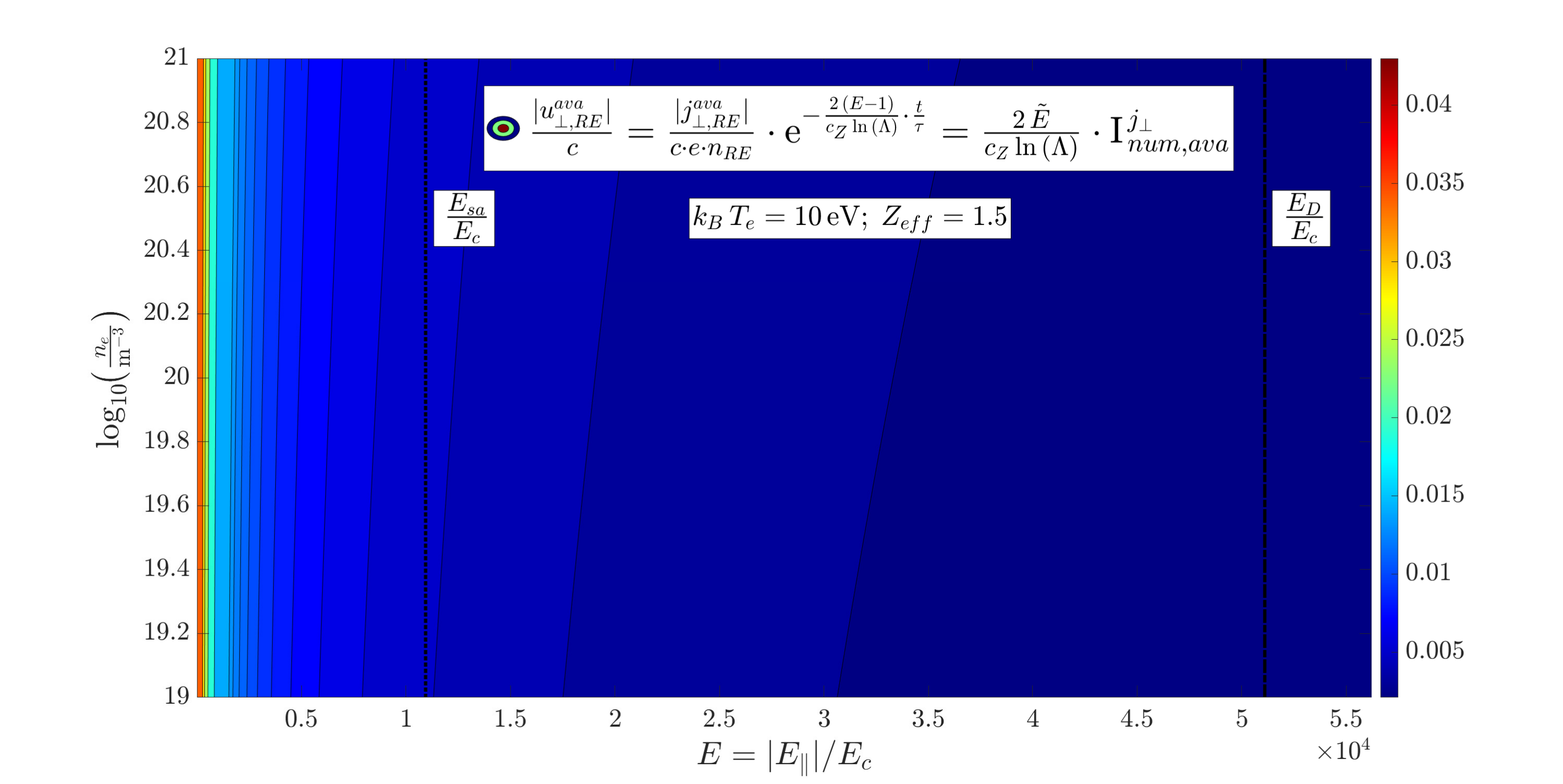}
\caption{Avalanche runaway electron population with $k_{B}\,T_{e}=10\,\textup{eV}$ and $Z_{eff}=1.5$}
\label{fig_ava_u_perp_EoverEc_n_e10eV}
\end{subfigure}
\begin{subfigure}{\textwidth}
\centering\vspace{2mm}
\includegraphics[trim=87 13 91 43,width=0.805\textwidth,clip]{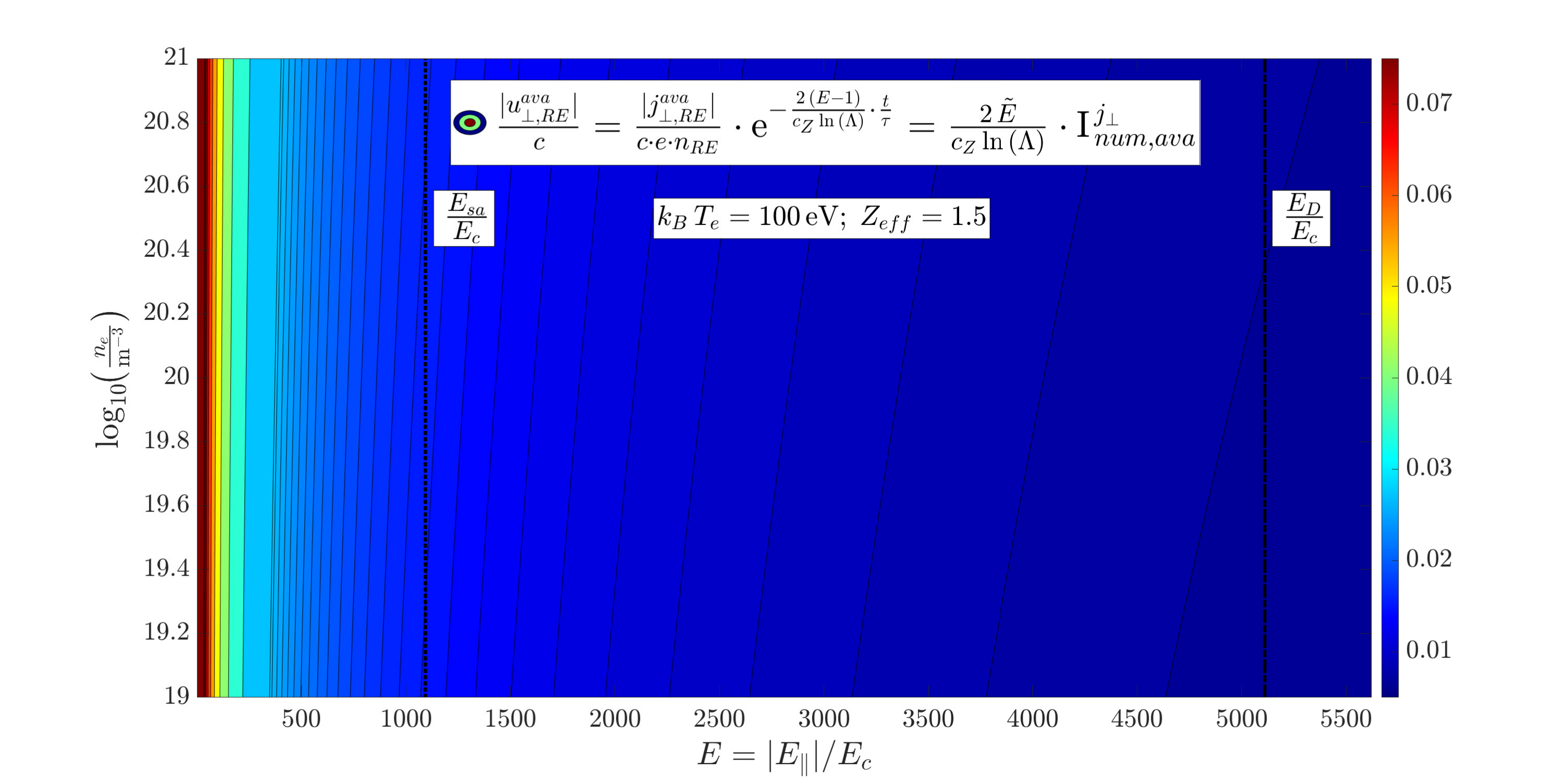}
\caption{Avalanche runaway electron population with $k_{B}\,T_{e}=100\,\textup{eV}$ and $Z_{eff}=1.5$}
\label{fig_ava_u_perp_EoverEc_n_e100eV}
\end{subfigure}
\begin{subfigure}{\textwidth}
\centering\vspace{2mm}
\includegraphics[trim=85 13 91 43,width=0.805\textwidth,clip]{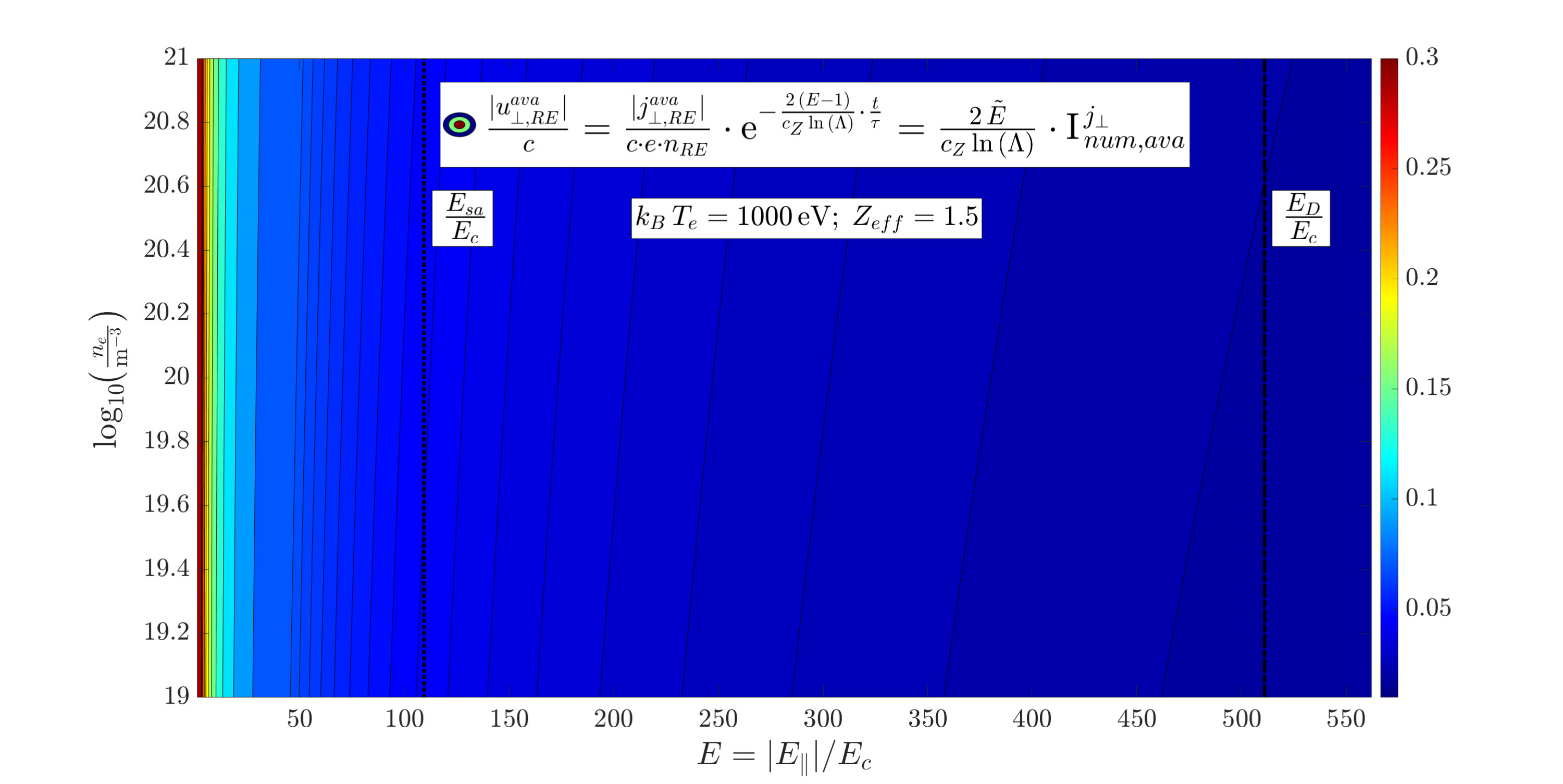}
\caption{Avalanche runaway electron population with $k_{B}\,T_{e}=1000\,\textup{eV}$ and $Z_{eff}=1.5$ }
\label{fig_ava_u_perp_EoverEc_n_e1000eV}
\end{subfigure}
\captionsetup{format=hang,indention=0cm}
\caption[Contour plots of the absolute value of the normalized orthogonal component of the mean velocity $|u_{\,\perp,RE}^{\,\textup{ava}}|/c$ for different avalanche runaway electron populations with respect to the electron density $n_{e}$ and the normalized electric field strength $E=\vert E_{\|}\vert/E_{c}$]{Contour plots$^{\ref{fig_eV_footnote}}$ of the absolute value of the normalized orthogonal component of the mean velocity $|u_{\,\perp,RE}^{\,\textup{ava}}|/c$ for different avalanche runaway electron populations with respect to the electron density $n_{e}$ and the normalized electric field strength $E=\vert E_{\|}\vert/E_{c}$}
\label{fig_ava_u_perp_EoverEc_n_e}
\end{figure}

\begin{figure}[H]
\centering
\begin{subfigure}{\textwidth}
\centering\vspace{-2mm}
\includegraphics[trim=85 13 91 43,width=0.805\textwidth,clip]{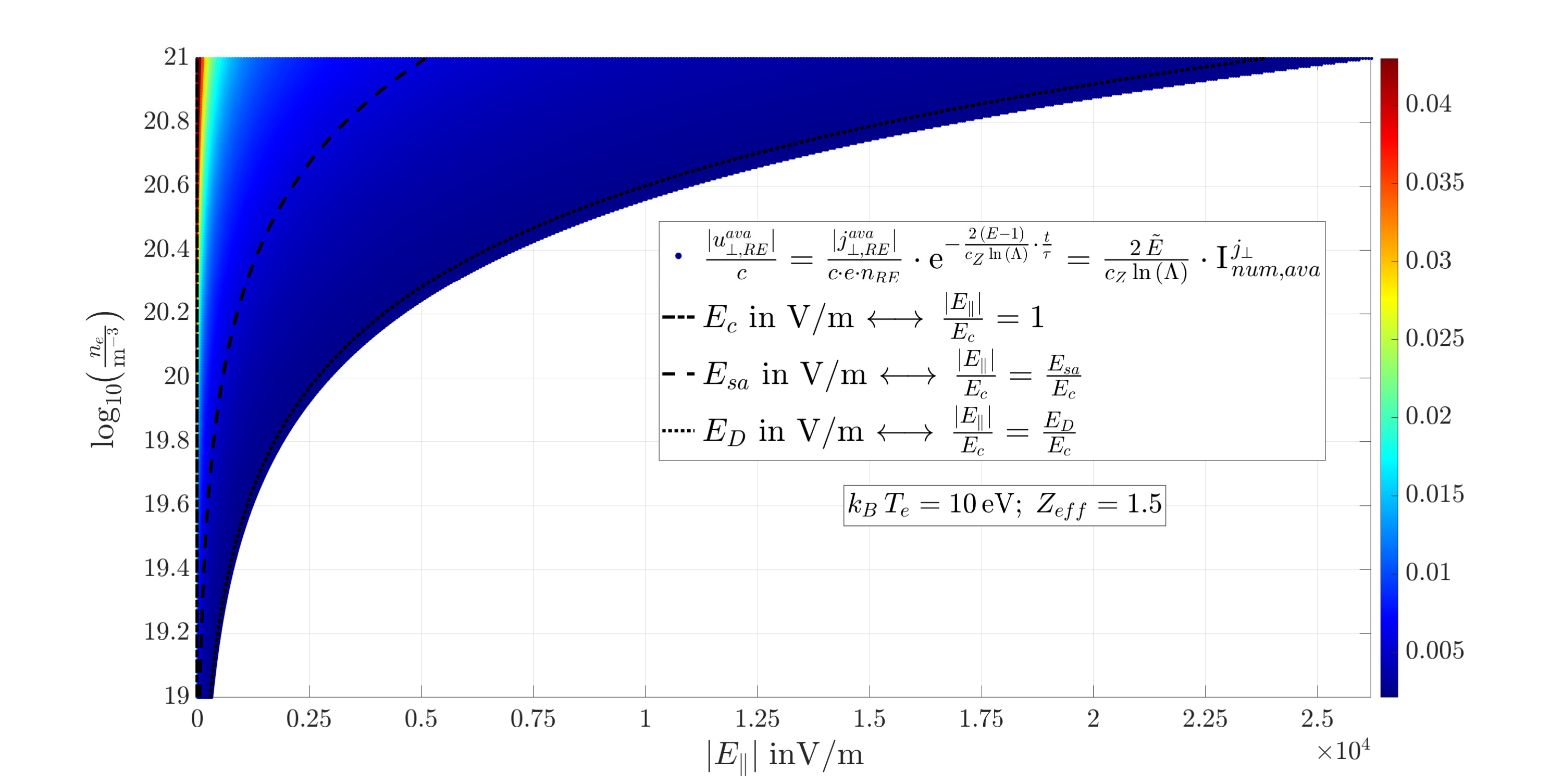}
\caption{Avalanche runaway electron population with $k_{B}\,T_{e}=10\,\textup{eV}$ and $Z_{eff}=1.5$}
\label{fig_ava_u_perp_E_par_n_e10eV}
\end{subfigure}
\begin{subfigure}{\textwidth}
\centering\vspace{2mm}
\includegraphics[trim=83 13 91 43,width=0.805\textwidth,clip]{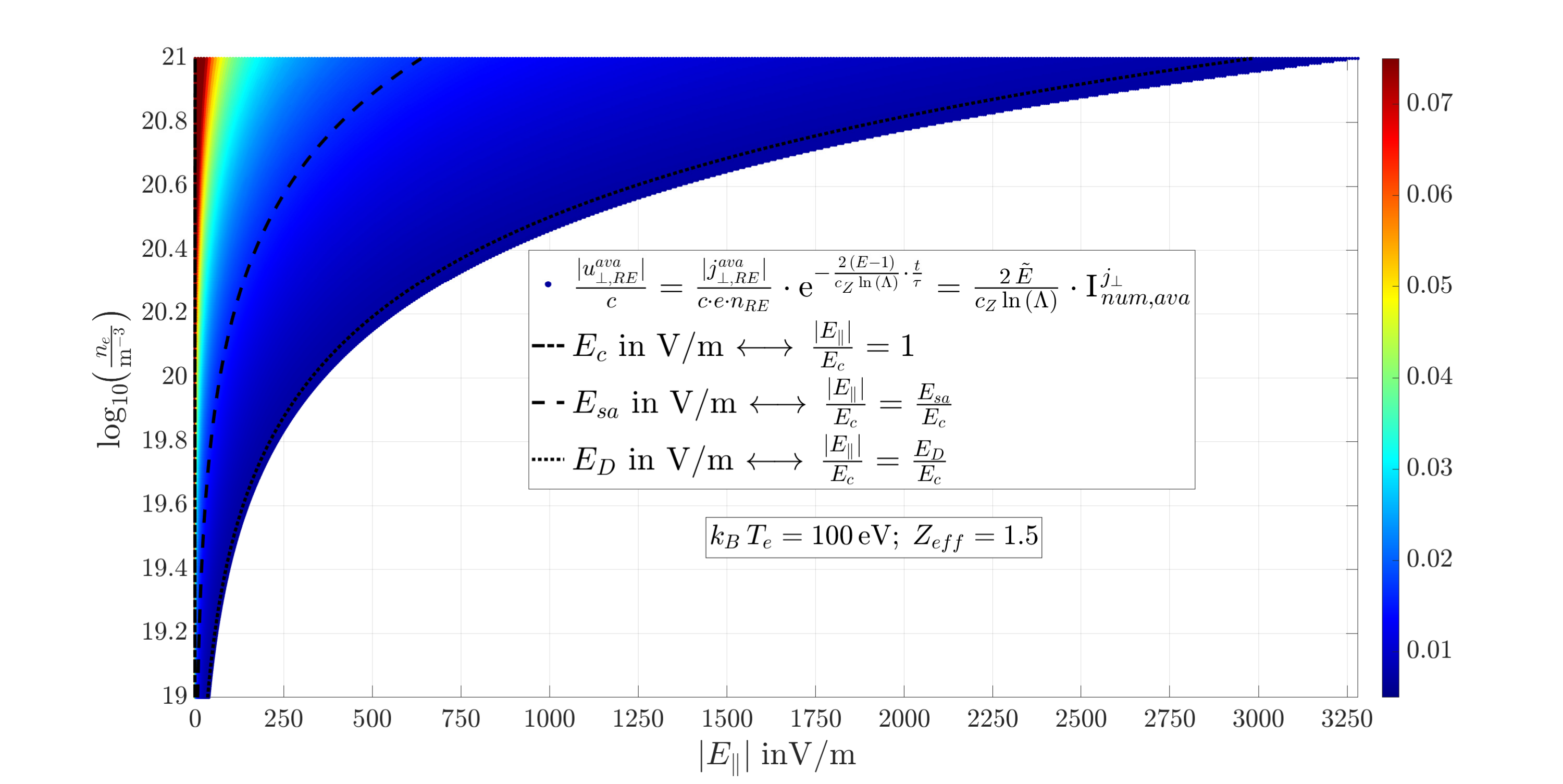}
\caption{Avalanche runaway electron population with $k_{B}\,T_{e}=100\,\textup{eV}$ and $Z_{eff}=1.5$}
\label{fig_ava_u_perp_E_par_n_e100eV}
\end{subfigure}
\begin{subfigure}{\textwidth}
\centering\vspace{2mm}
\includegraphics[trim=85 13 91 43,width=0.805\textwidth,clip]{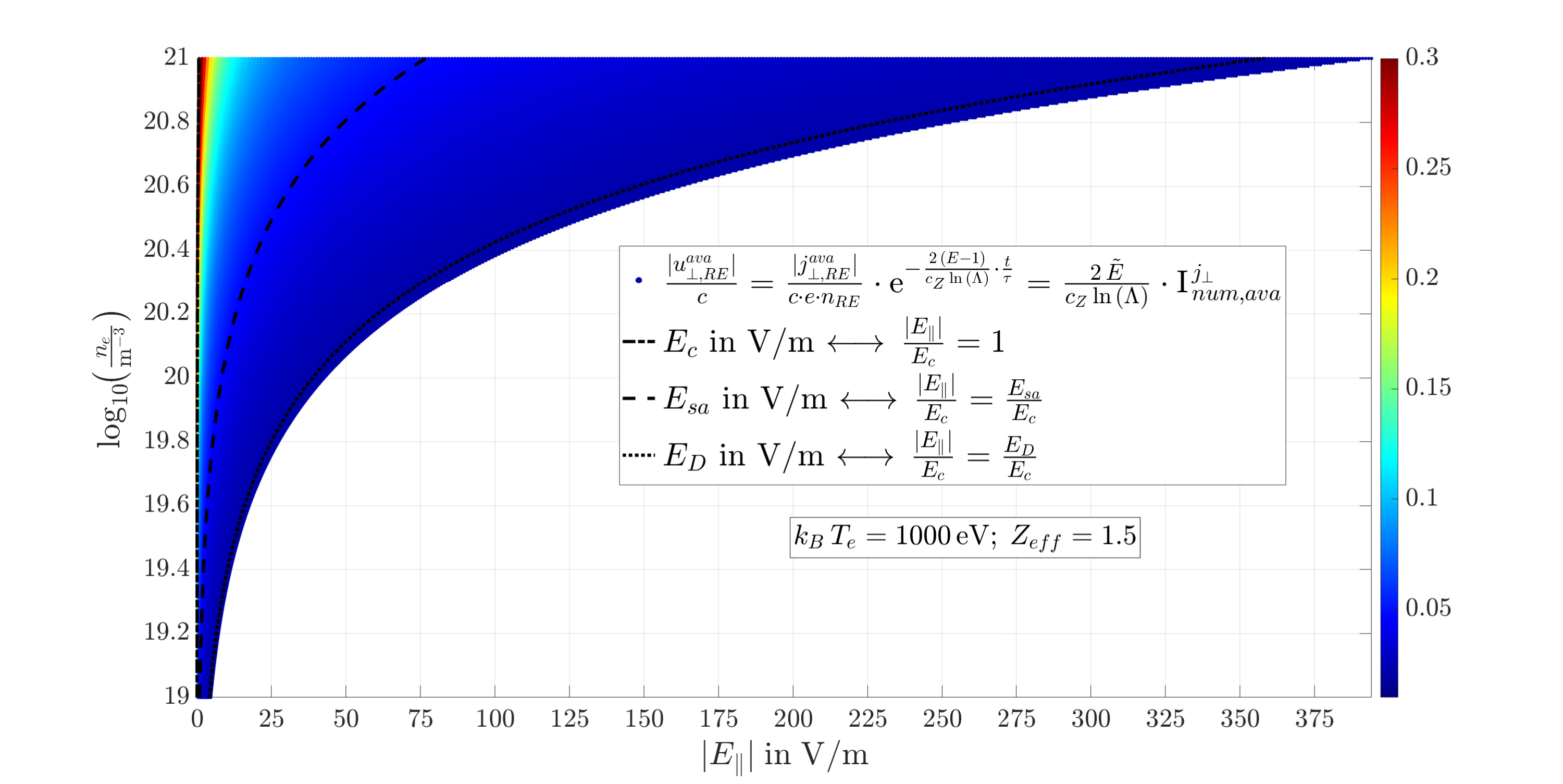}
\caption{Avalanche runaway electron population with $k_{B}\,T_{e}=1000\,\textup{eV}$ and $Z_{eff}=1.5$ }
\label{fig_ava_u_perp_E_par_n_e1000eV}
\end{subfigure}
\captionsetup{format=hang,indention=0cm}
\caption[Contour plots of the absolute value of the normalized orthogonal component of the mean velocity $|u_{\,\perp,RE}^{\,\textup{ava}}|/c$ for different avalanche runaway electron populations with respect to the electron density $n_{e}$ and the absolute value of the parallel component of the electric field $\vert E_{\|}\vert$]{Contour plots\protect\footnotemark{} of the absolute value of the normalized orthogonal component of the mean velocity $|u_{\,\perp,RE}^{\,\textup{ava}}|/c$ for different avalanche runaway electron populations with respect to the electron density $n_{e}$ and the absolute value of the parallel component of the electric field $\vert E_{\|}\vert$}
\label{fig_ava_u_perp_E_par_n_e}
\end{figure}

In order to study the absolute value of the normalized orthogonal component of the mean velocity $|u_{\,\perp,RE}^{\,\textup{ava}}|/c$ for different avalanche runaway electron populations with respect to the electron density $n_{e}$ and the normalized electric field strength $E=\vert E_{\|}\vert/E_{c}$, one can use the figure \ref{fig_ava_u_perp_EoverEc_n_e}. It shows values in the ranges $|u_{\,\perp,RE}^{\,\textup{ava}}|/c\in[0.002,\,0.938]$ for $k_{B}\,T_{e}=10\,\textup{eV}$, $|u_{\,\perp,RE}^{\,\textup{ava}}|/c\in[0.007,\,0.928]$ for $k_{B}\,T_{e}=100\,\textup{eV}$ and $|u_{\,\perp,RE}^{\,\textup{ava}}|/c\in[0.020,\,0.919]$ for $k_{B}\,T_{e}=1000\,\textup{eV}$ according to the combined console output of the plotting scripts$^{\mathrm{\ref{plotfootnote}}}$ from listing \ref{outMATLABgeneratedataAppendix} in subsection \ref{subsection_generated_data} of the appendix. Here, the maximum values are reached for low electric fields and high electron densities, while the minimum values appear for low electric fields and high electron densities. In particular, this can be seen in the scatter plots for the normalized orthogonal component of the mean velocity $|u_{\,\perp,RE}^{\,\textup{ava}}|/c$ with respect to the electron density $n_{e}$ and the absolute value of the parallel component of the electric field strength $\vert E_{\|}\vert$ from figure \ref{fig_ava_u_perp_E_par_n_e}. Thus the perpendicular component behaves in the opposite way to the parallel component with regard to the extreme values, while for the most parameter combinations the orthogonal component is much smaller than the orthogonal component of the normalized mean velocity. This is further emphasized by the output of the plotting scripts$^{\mathrm{\ref{plotfootnote}}}$ from listing \ref{outMATLABgeneratedataAppendix}, which shows that the perpendicular component of the mean velocity $|u_{\,\perp,RE}^{\,\textup{ava}}|/c$ can reach more than $300\,\%$ of the parallel component, while for the most parameters it approximately in the size of less than $5\,\%$ of the parallel component $|u_{\,\|,RE}^{\,\textup{ava}}|/c$. 

\begin{figure}[H]
\centering
\begin{subfigure}{\textwidth}
\centering\vspace{-2mm}
\includegraphics[trim=83 13 91 43,width=0.805\textwidth,clip]{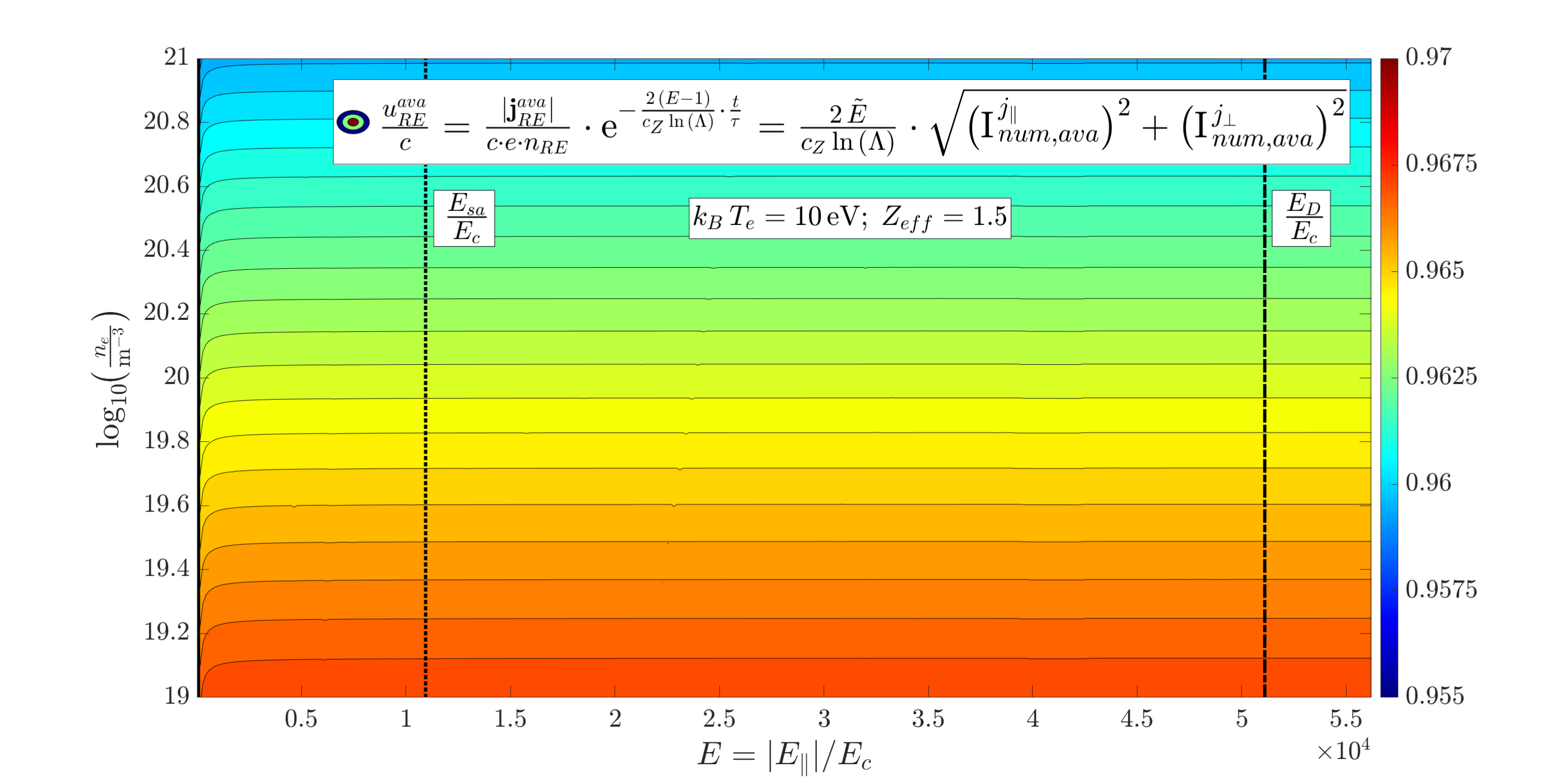}
\caption{Avalanche runaway electron population with $k_{B}\,T_{e}=10\,\textup{eV}$ and $Z_{eff}=1.5$}
\label{fig_ava_u_EoverEc_n_e10eV}
\end{subfigure}
\begin{subfigure}{\textwidth}
\centering\vspace{2mm}
\includegraphics[trim=83 13 91 43,width=0.805\textwidth,clip]{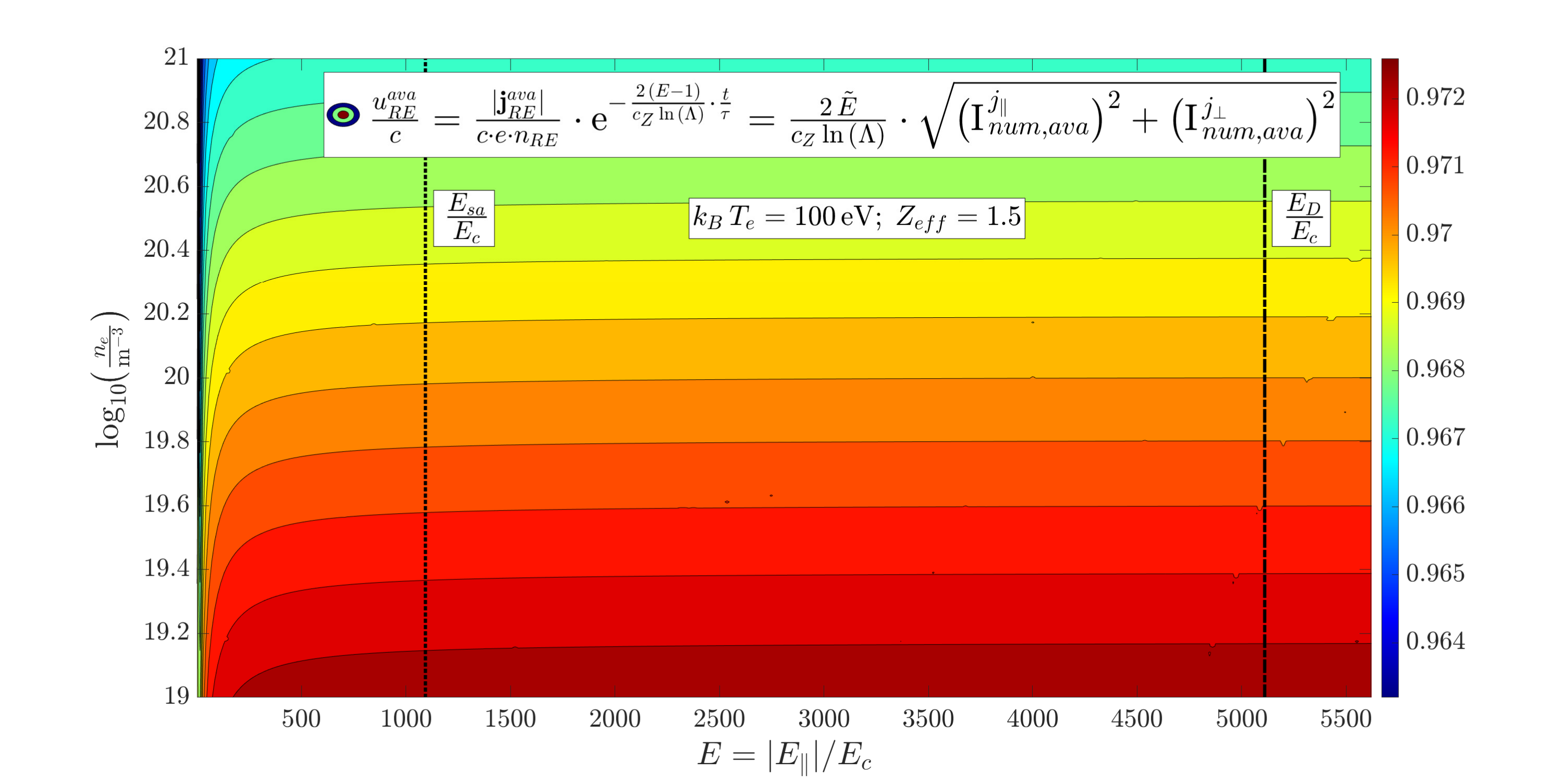}
\caption{Avalanche runaway electron population with $k_{B}\,T_{e}=100\,\textup{eV}$ and $Z_{eff}=1.5$}
\label{fig_ava_u_EoverEc_n_e100eV}
\end{subfigure}
\begin{subfigure}{\textwidth}
\centering\vspace{2mm}
\includegraphics[trim=83 13 91 43,width=0.805\textwidth,clip]{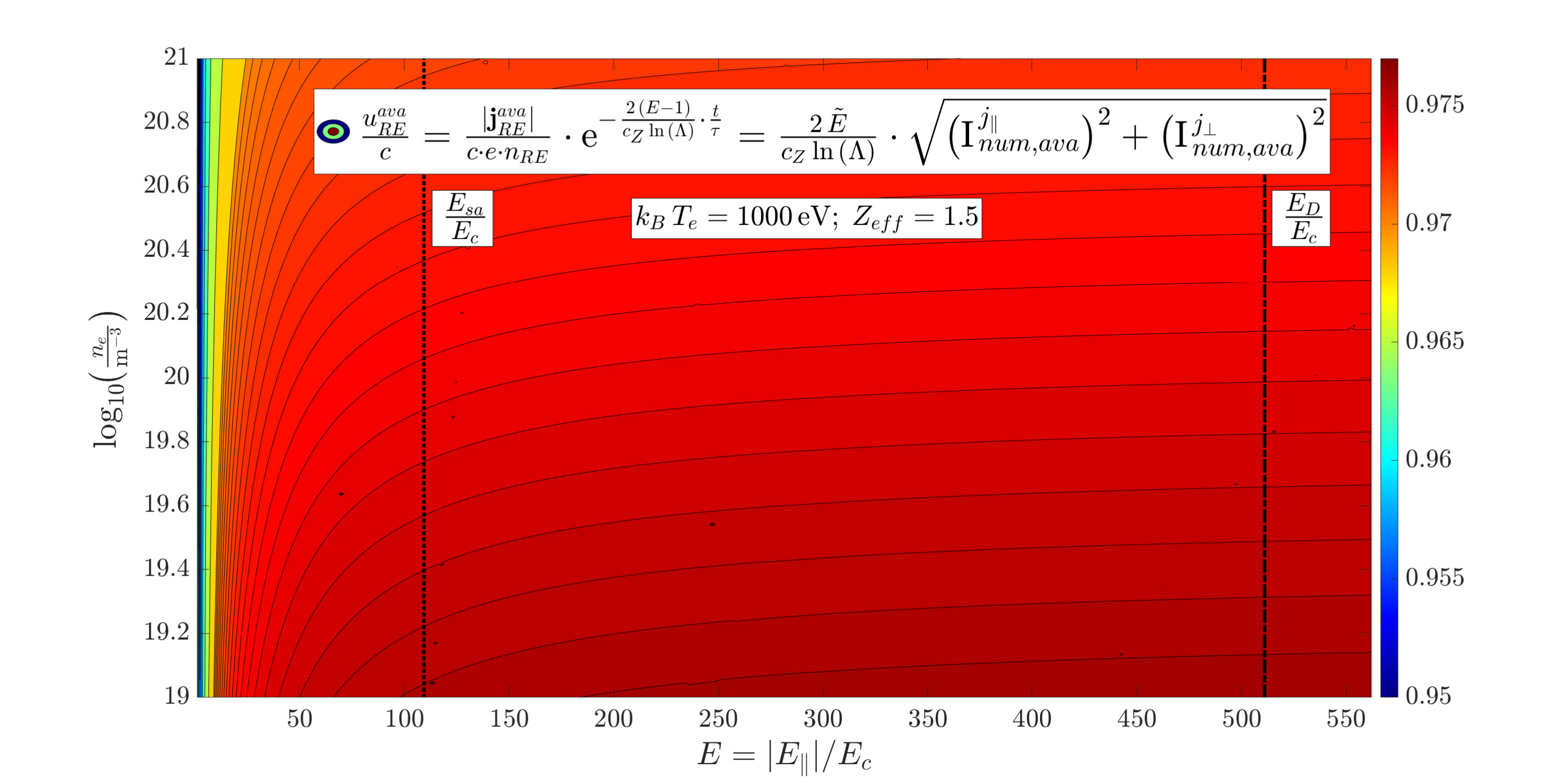}
\caption{Avalanche runaway electron population with $k_{B}\,T_{e}=1000\,\textup{eV}$ and $Z_{eff}=1.5$ }
\label{fig_ava_u_EoverEc_n_e1000eV}
\end{subfigure}
\captionsetup{format=hang,indention=0cm}
\caption[Contour plots of the normalized magnitude of the mean velocity $u_{\,RE}^{\,\textup{ava}}/c$ for different avalanche runaway electron populations with respect to the electron density $n_{e}$ and the normalized electric field strength\linebreak$E=\vert E_{\|}\vert/E_{c}$]{Contour plots$^{\ref{fig_eV_footnote}}$ of the normalized magnitude of the mean velocity $u_{\,RE}^{\,\textup{ava}}/c$ for different avalanche runaway electron populations with respect to the electron density $n_{e}$ and the normalized electric field strength $E=\vert E_{\|}\vert/E_{c}$}
\label{fig_ava_u_EoverEc_n_e}
\end{figure}

\begin{figure}[H]
\centering
\begin{subfigure}{\textwidth}
\centering\vspace{-2mm}
\includegraphics[trim=83 13 91 43,width=0.805\textwidth,clip]{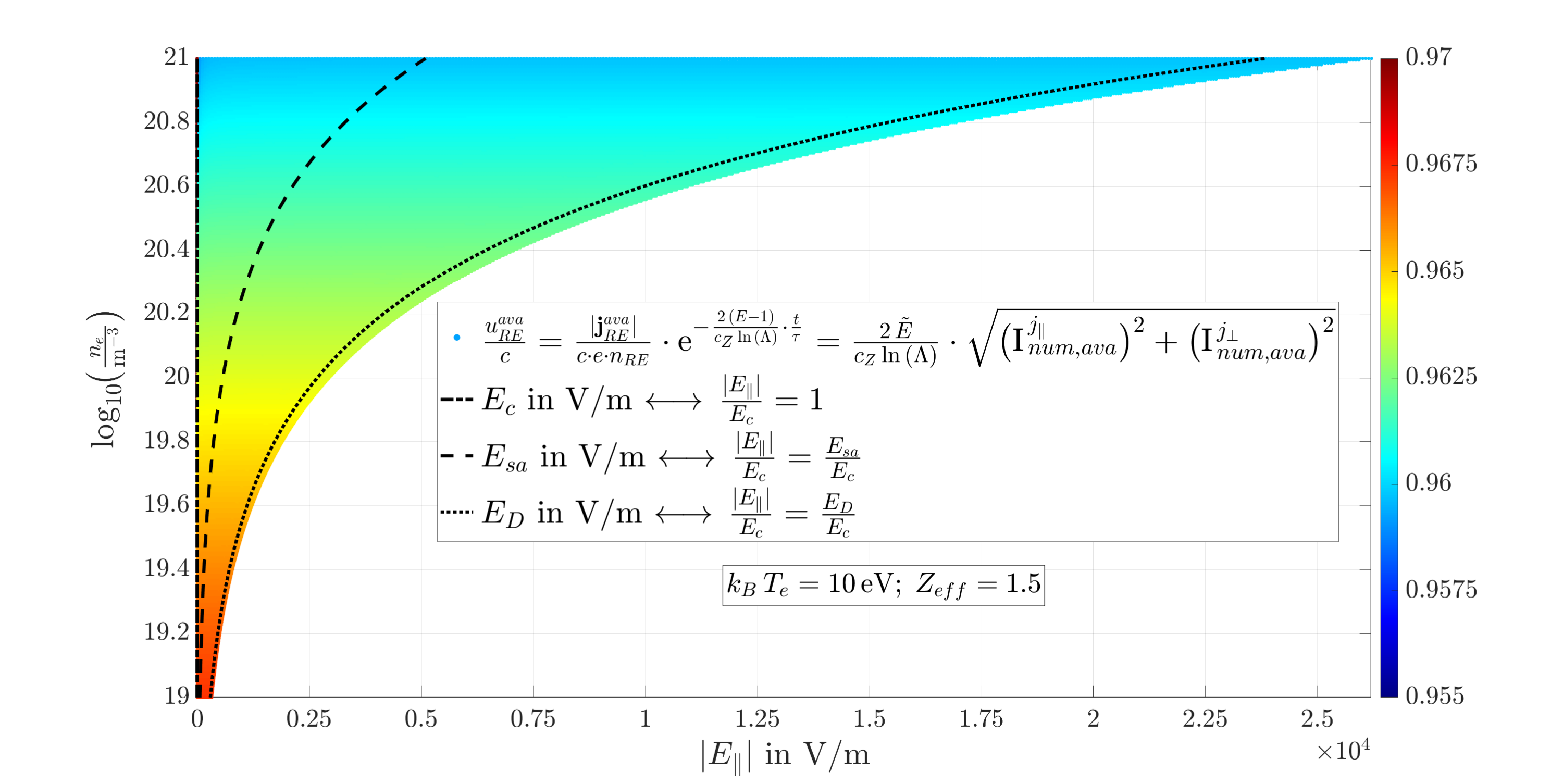}
\caption{Avalanche runaway electron population with $k_{B}\,T_{e}=10\,\textup{eV}$ and $Z_{eff}=1.5$}
\label{fig_ava_u_E_par_n_e10eV}
\end{subfigure}
\begin{subfigure}{\textwidth}
\centering\vspace{2mm}
\includegraphics[trim=83 13 91 43,width=0.805\textwidth,clip]{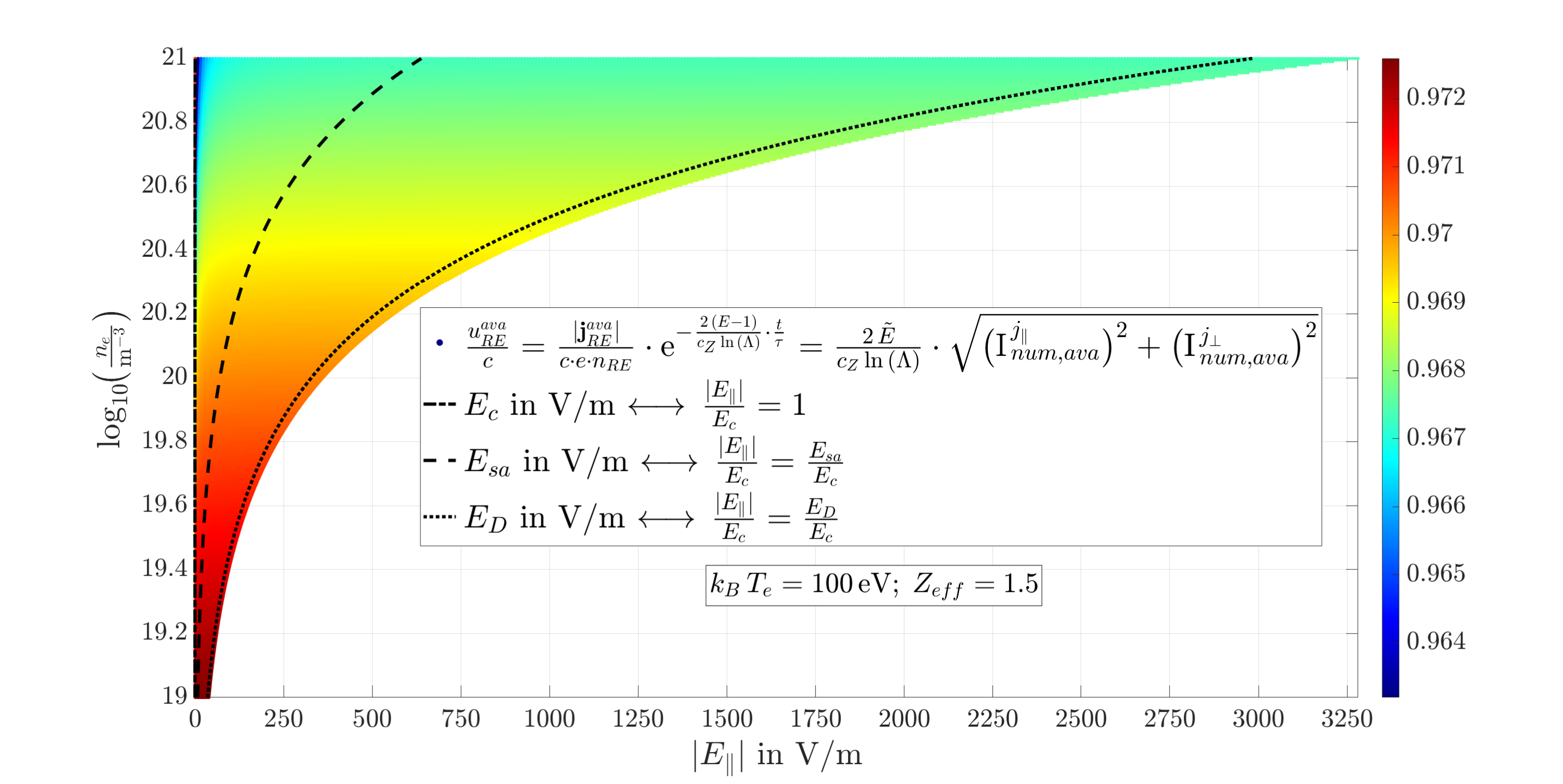}
\caption{Avalanche runaway electron population with $k_{B}\,T_{e}=100\,\textup{eV}$ and $Z_{eff}=1.5$}
\label{fig_ava_u_E_par_n_e100eV}
\end{subfigure}
\begin{subfigure}{\textwidth}
\centering\vspace{2mm}
\includegraphics[trim=83 13 91 43,width=0.805\textwidth,clip]{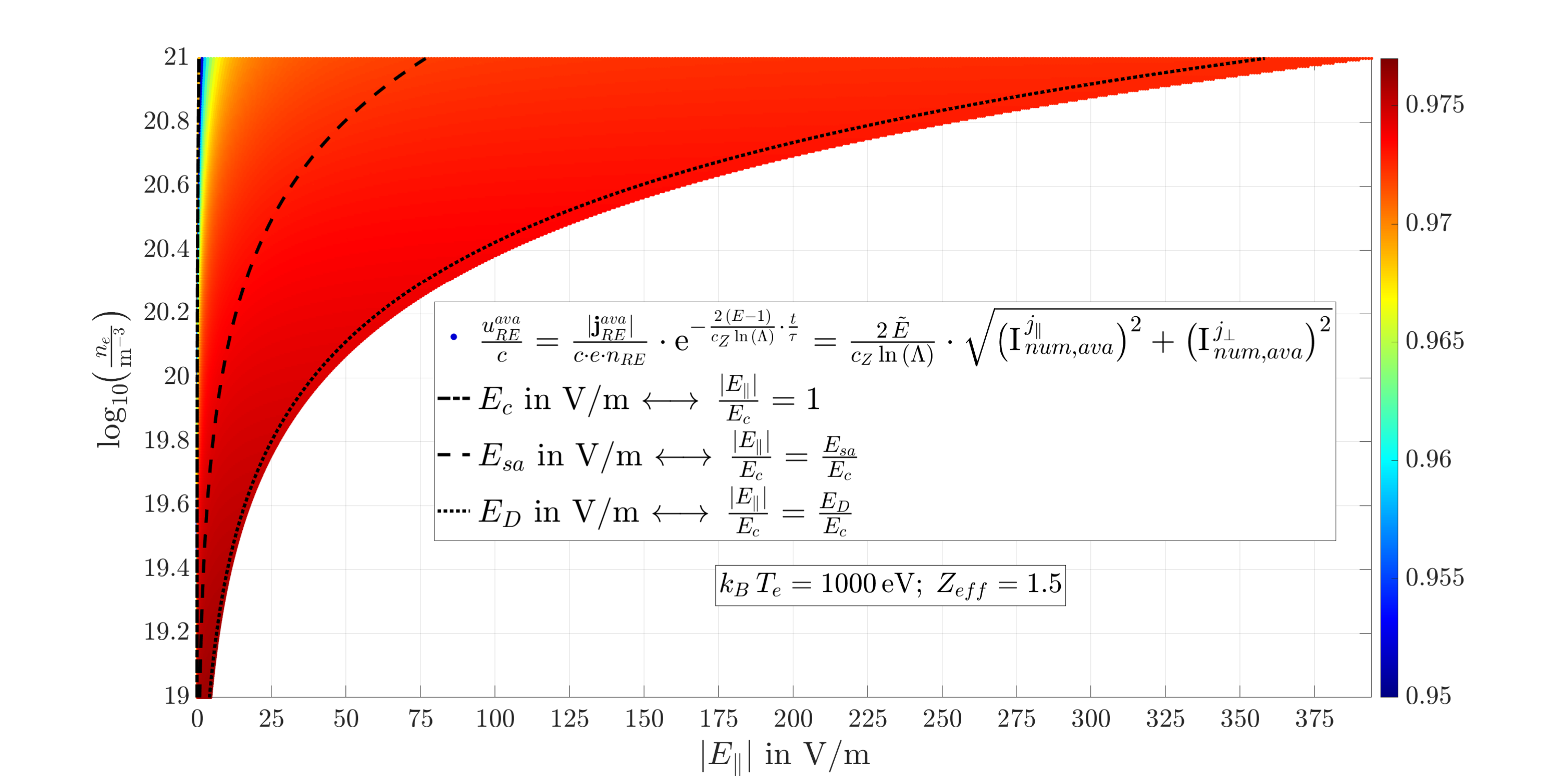}
\caption{Avalanche runaway electron population with $k_{B}\,T_{e}=1000\,\textup{eV}$ and $Z_{eff}=1.5$ }
\label{fig_ava_u_E_par_n_e1000eV}
\end{subfigure}
\captionsetup{format=hang,indention=0cm}
\caption[Contour plots of the normalized magnitude of the mean velocity $u_{\,RE}^{\,\textup{ava}}/c$ for different avalanche runaway electron populations with respect to the electron density $n_{e}$ and the absolute value of the parallel component of the electric field $\vert E_{\|}\vert$]{Contour plots$^{\ref{fig_eV_footnote}}$ of the normalized magnitude of the mean velocity $u_{\,RE}^{\,\textup{ava}}/c$ for different avalanche runaway electron populations with respect to the electron density $n_{e}$ and the absolute value of the parallel component of the electric field $\vert E_{\|}\vert$}
\label{fig_ava_u_E_par_n_e}
\end{figure}

\begin{figure}[H]
\centering
\begin{subfigure}{\textwidth}
\centering\vspace{-2mm}
\includegraphics[trim=83 13 91 43,width=0.805\textwidth,clip]{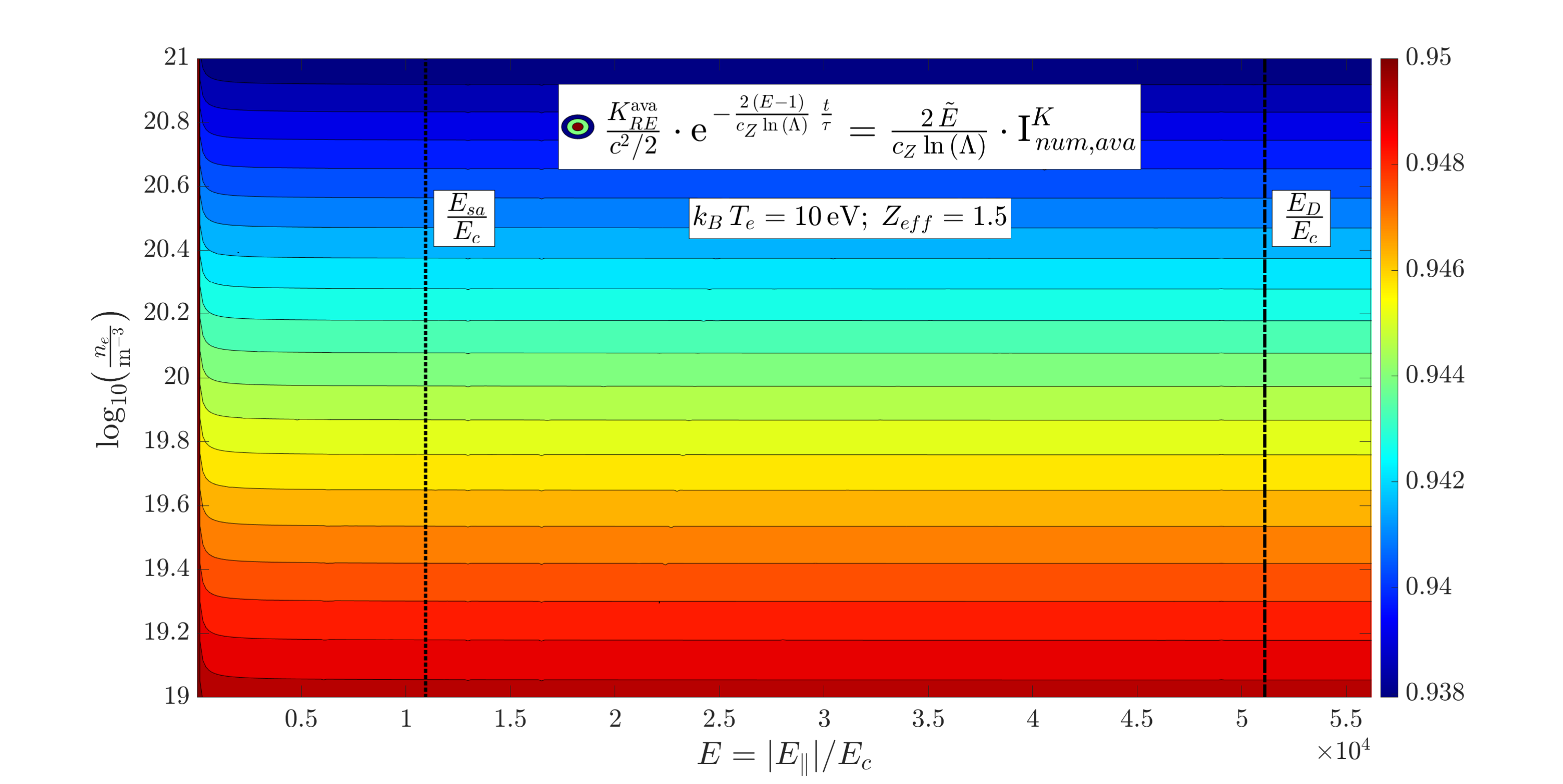}
\caption{Avalanche runaway electron population with $k_{B}\,T_{e}=10\,\textup{eV}$ and $Z_{eff}=1.5$}
\label{fig_ava_K_EoverEc_n_e10eV}
\end{subfigure}
\begin{subfigure}{\textwidth}
\centering\vspace{2mm}
\includegraphics[trim=83 13 91 43,width=0.805\textwidth,clip]{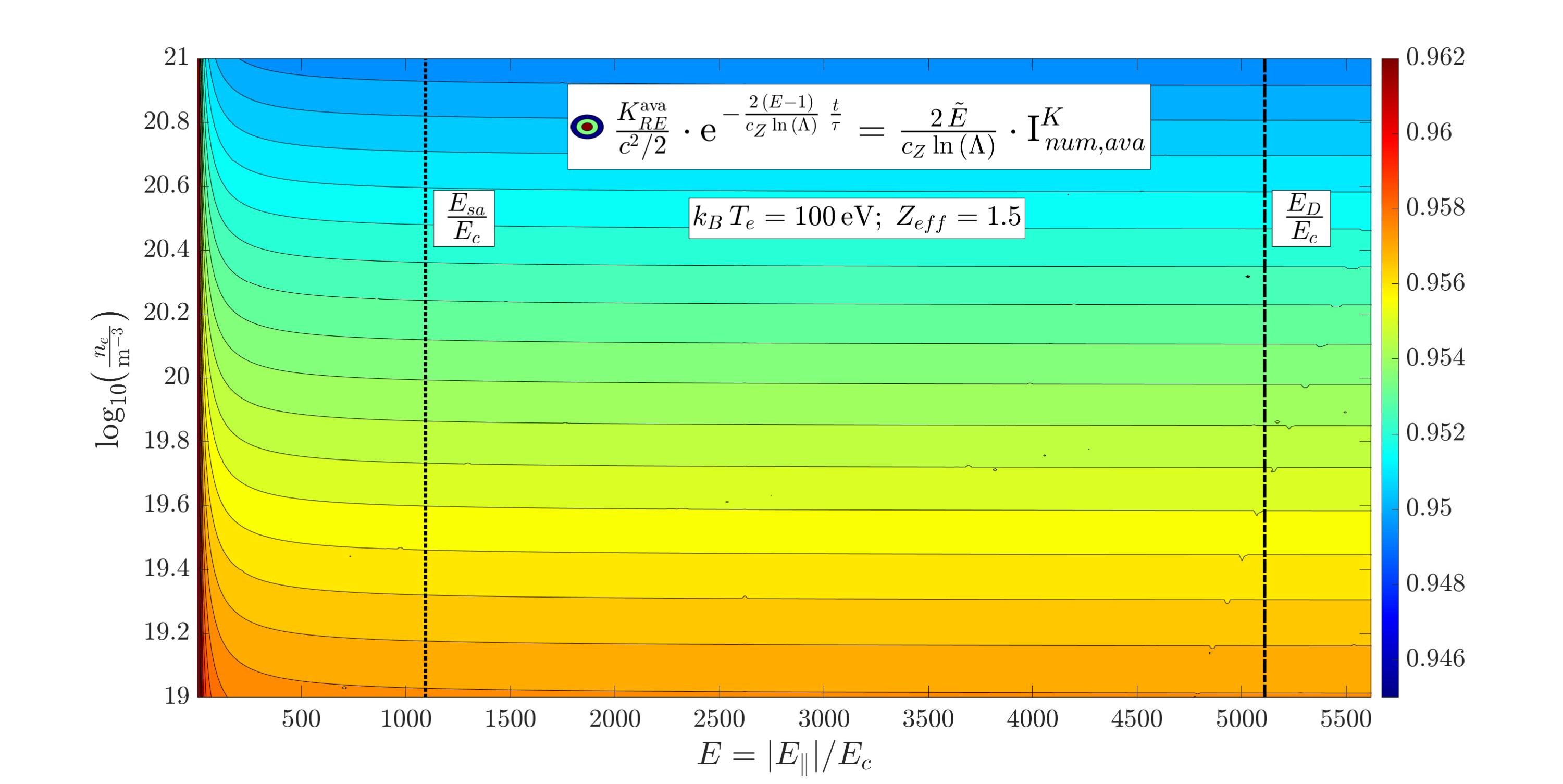}
\caption{Avalanche runaway electron population with $k_{B}\,T_{e}=100\,\textup{eV}$ and $Z_{eff}=1.5$}
\label{fig_ava_K_EoverEc_n_e100eV}
\end{subfigure}
\begin{subfigure}{\textwidth}
\centering\vspace{2mm}
\includegraphics[trim=83 13 91 43,width=0.805\textwidth,clip]{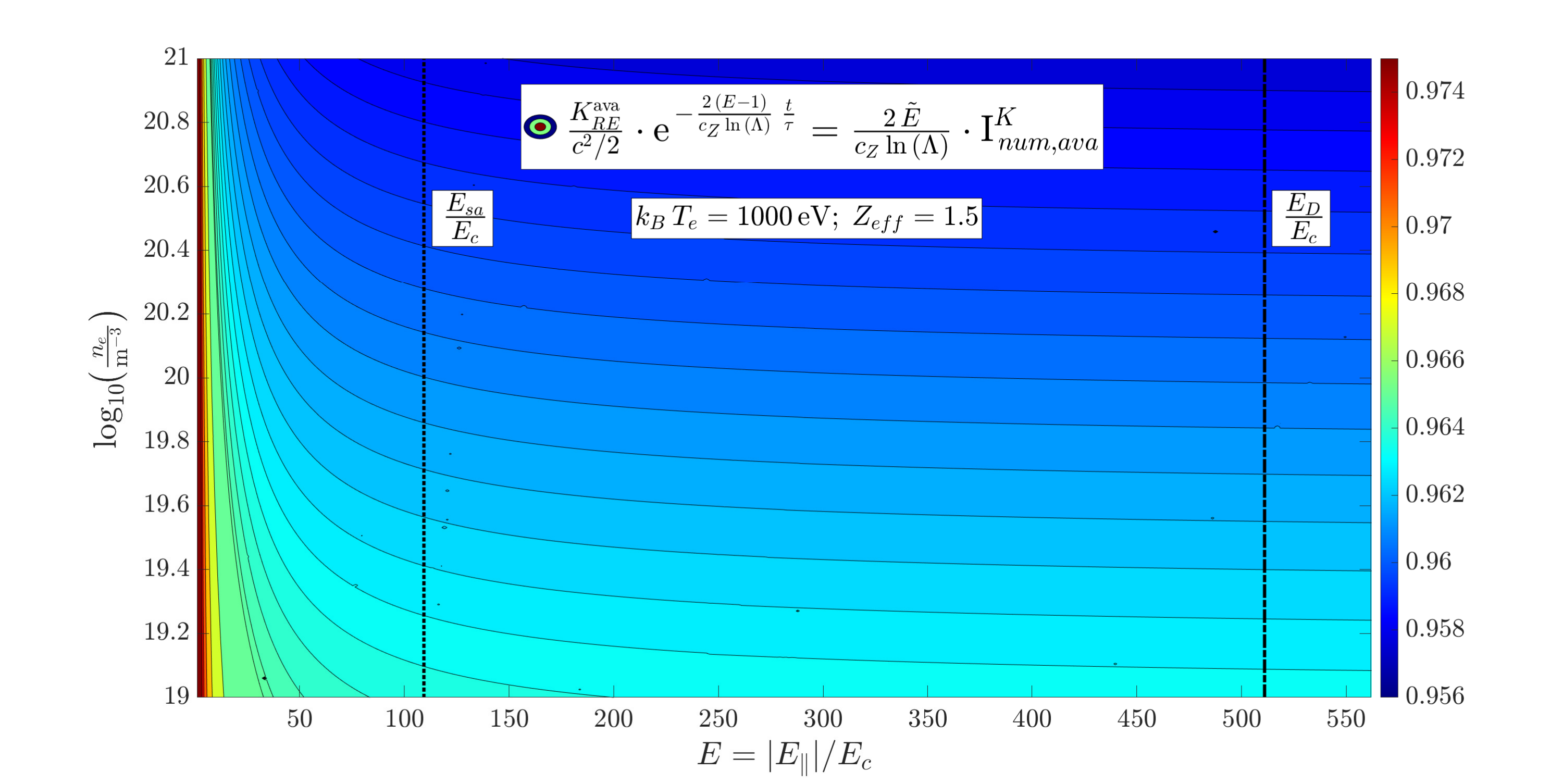}
\caption{Avalanche runaway electron population with $k_{B}\,T_{e}=1000\,\textup{eV}$ and $Z_{eff}=1.5$ }
    \label{fig_ava_K_EoverEc_n_e1000eV}
\end{subfigure}
\captionsetup{format=hang,indention=0cm}
\caption[Contour plots of the normalized mean mass-related kinetic energy density $K_{\,RE}^{\,\textup{ava}}/(c^2/2)$ for different avalanche runaway electron populations with respect to the electron density $n_{e}$ and the normalized electric field strength $E=\vert E_{\|}\vert/E_{c}$]{Contour plots$^{\ref{fig_eV_footnote}}$ of the normalized mean mass-related kinetic energy density $K_{\,RE}^{\,\textup{ava}}/(c^2/2)$ for different avalanche runaway electron populations with respect to the electron density $n_{e}$ and the normalized electric field strength $E=\vert E_{\|}\vert/E_{c}$}
\label{fig_ava_K_EoverEc_n_e}
\end{figure}

\begin{figure}[H]
\centering
\begin{subfigure}{\textwidth}
\centering\vspace{-2mm}
\includegraphics[trim=83 13 91 43,width=0.805\textwidth,clip]{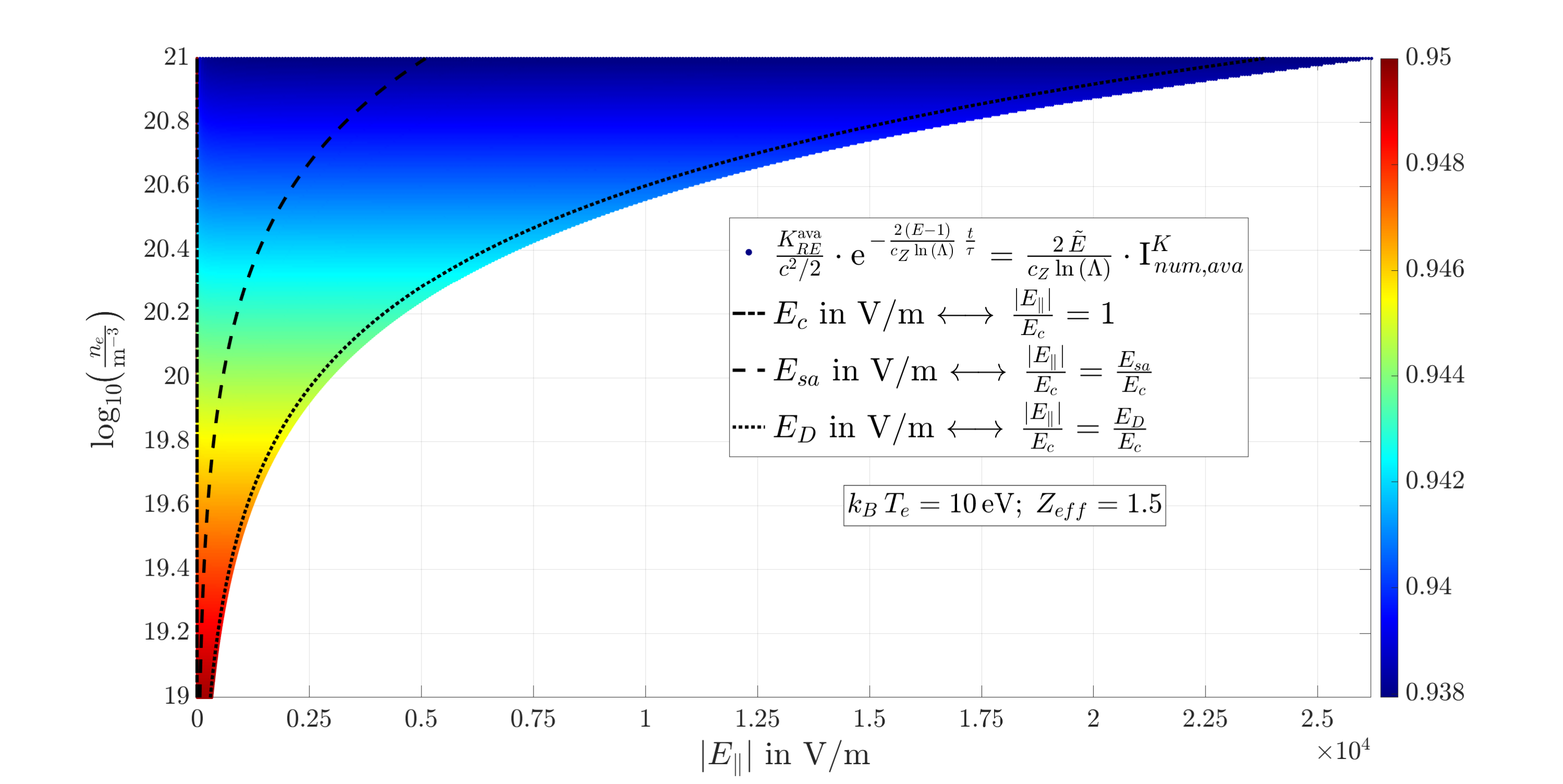}
\caption{Avalanche runaway electron population with $k_{B}\,T_{e}=10\,\textup{eV}$ and $Z_{eff}=1.5$}
\label{fig_ava_K_E_par_n_e10eV}
\end{subfigure}
\begin{subfigure}{\textwidth}
\centering\vspace{2mm}
\includegraphics[trim=83 13 91 43,width=0.805\textwidth,clip]{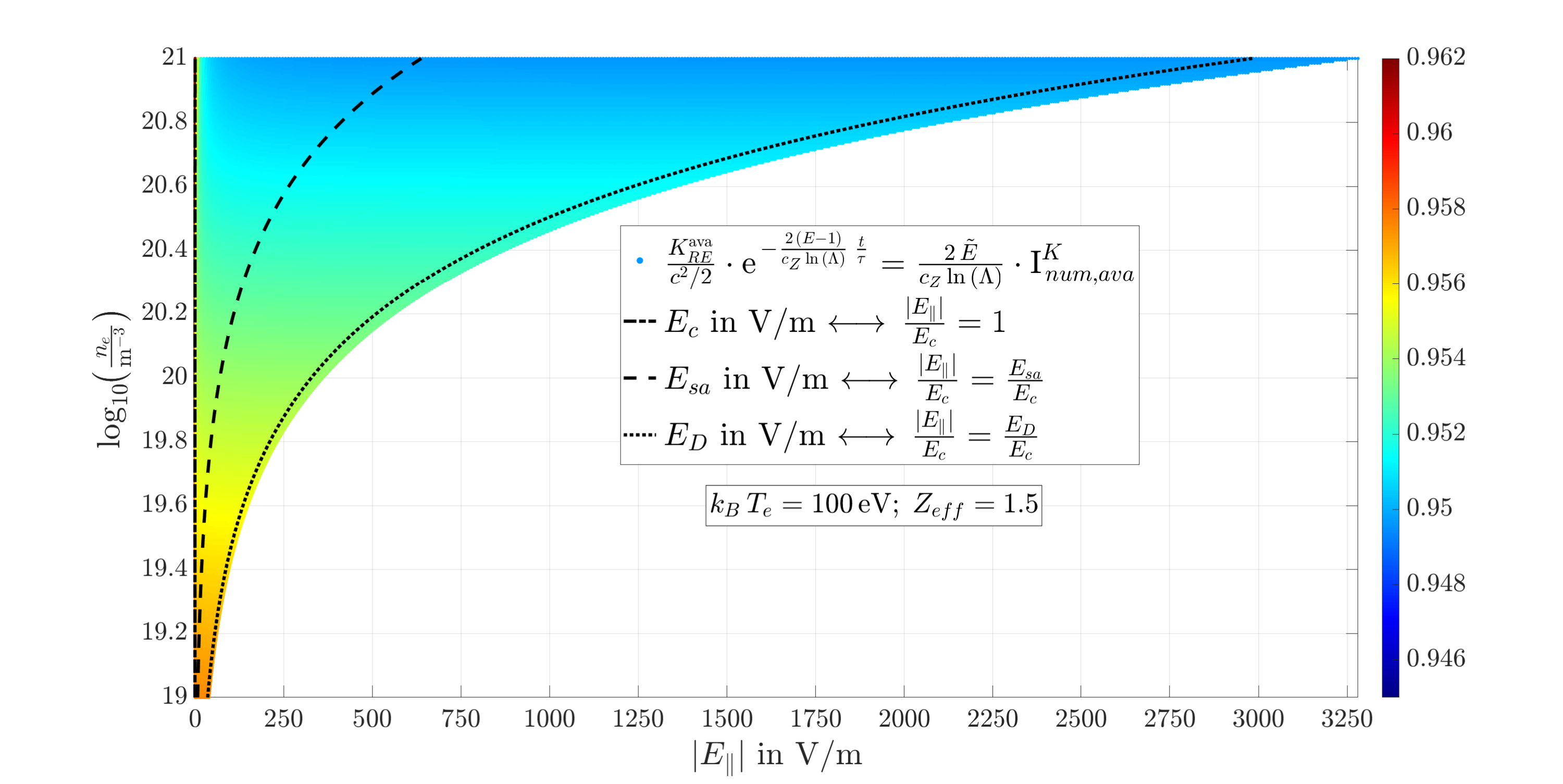}
\caption{Avalanche runaway electron population with $k_{B}\,T_{e}=100\,\textup{eV}$ and $Z_{eff}=1.5$}
\label{fig_ava_K_E_par_n_e100eV}
\end{subfigure}
\begin{subfigure}{\textwidth}
\centering\vspace{2mm}
\includegraphics[trim=83 13 91 43,width=0.805\textwidth,clip]{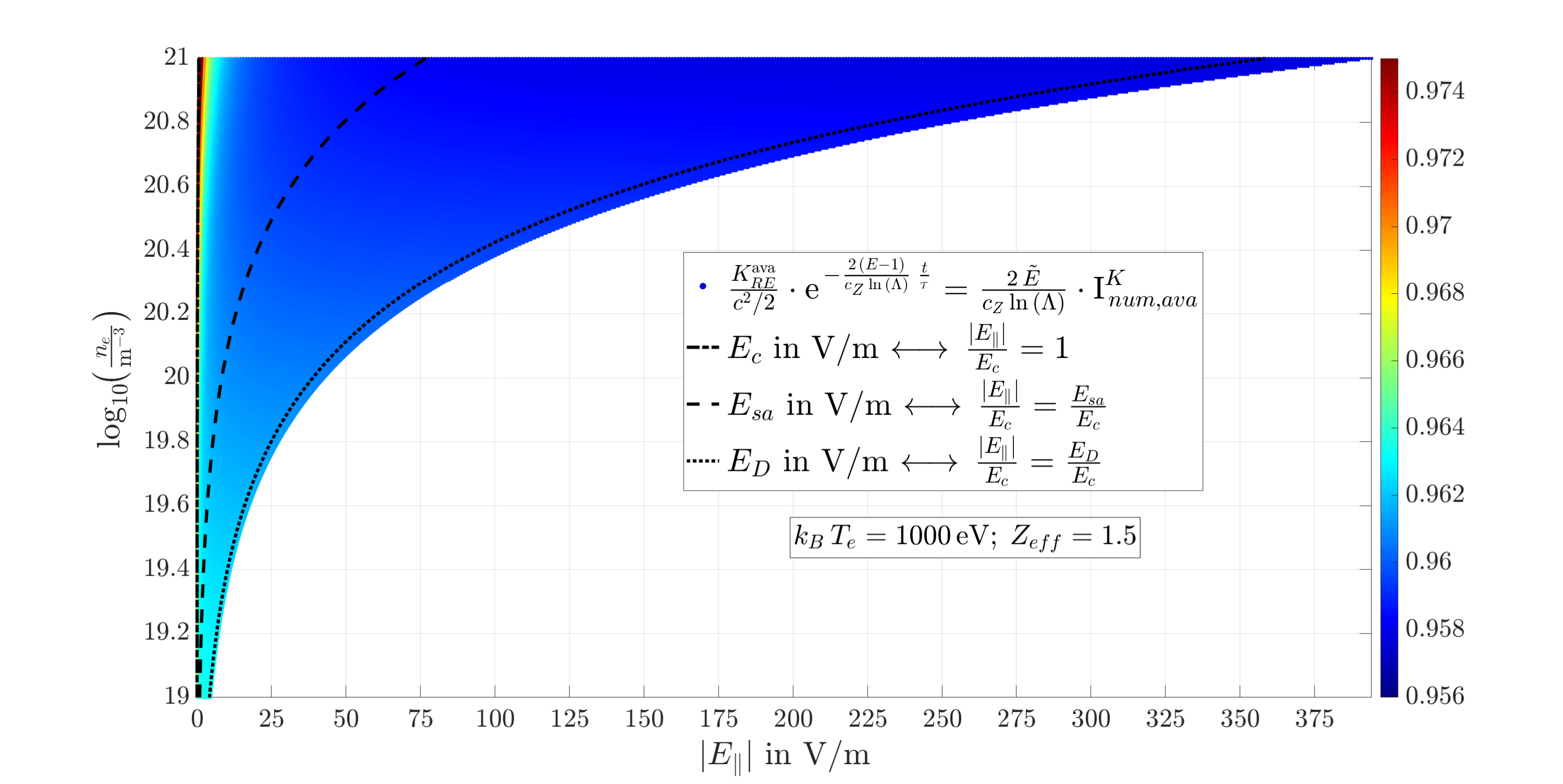}
\caption{Avalanche runaway electron population with $k_{B}\,T_{e}=1000\,\textup{eV}$ and $Z_{eff}=1.5$ }
\label{fig_ava_K_E_par_n_e1000eV}
\end{subfigure}
\captionsetup{format=hang,indention=0cm}
\caption[Contour plots of the normalized mean mass-related kinetic energy density $K_{\,RE}^{\,\textup{ava}}/(c^2/2)$ for different avalanche runaway electron populations with respect to the electron density $n_{e}$ and the absolute value of the parallel component of the electric field $\vert E_{\|}\vert$]{Contour plots$^{\ref{fig_eV_footnote}}$ of the normalized mean mass-related kinetic energy density $K_{\,RE}^{\,\textup{ava}}/(c^2/2)$ for different avalanche runaway electron populations with respect to the electron density $n_{e}$ and the absolute value of the parallel component of the electric field $\vert E_{\|}\vert$}
\label{fig_ava_K_E_par_n_e}
\end{figure}

The figures \ref{fig_ava_u_EoverEc_n_e} and \ref{fig_ava_u_perp_E_par_n_e} show the contour and scatter plots of the normalized magnitude of the mean velocity $u_{\,RE}^{\,\textup{ava}}/c$ for different electron temperatures with respect to the electron density $n_{e}$ either the normalized electric field strength $E=\vert E_{\|}\vert/E_{c}$ or the absolute value of the parallel component of the electric field $\vert E_{\|}\vert$. The displayed results are in the ranges $u_{\,RE}^{\,\textup{ava}}/c\in[0.959,\,0.974]$ for $k_{B}\,T_{e}=10\,\textup{eV}$, $u_{\,RE}^{\,\textup{ava}}/c\in[0.963,\,0.973]$ for $k_{B}\,T_{e}=100\,\textup{eV}$ and $u_{\,RE}^{\,\textup{ava}}/c\in[0.952,\,0.982]$ for $k_{B}\,T_{e}=1000\,\textup{eV}$ according to the\linebreak\nopagebreak\vspace{-7mm}
\footnotetext{\label{fig_eV_footnote} The contour plots in the figures \ref{fig_ava_u_par_EoverEc_n_e}, \ref{fig_ava_u_par_E_par_n_e}, \ref{fig_ava_u_perp_EoverEc_n_e}, \ref{fig_ava_u_perp_E_par_n_e}, \ref{fig_ava_u_EoverEc_n_e}, \ref{fig_ava_u_E_par_n_e}, \ref{fig_ava_K_EoverEc_n_e} and \ref{fig_ava_K_E_par_n_e} were computed with the\\\hspace*{8.5mm}\mbox{help of the \textsc{MATLAB}-scripts "generate_num_data_10eV.m", "plot_num_data_10eV.m", "gen-}\\\hspace*{8.5mm}\mbox{erate_num_data_100eV.m", "plot_num_data_100eV.m", and "generate_num_data_1000eV.m",}\\\hspace*{8.5mm}"plot_num_data_1000eV.m", which can be viewed in the digital appendix.}
\vspace{-7mm}
combined console output of the plotting scripts$^{\mathrm{\ref{plotfootnote}}}$ from listing \ref{outMATLABgeneratedataAppendix} in subsection \ref{subsection_generated_data} of the appendix. At this, the dominance of the parallel component of the mean velocity is noticeable, because the variation of the results is smaller than for the single components, due to the root mean square calculation $u_{\,RE}^{\,\textup{ava}}/c=\sqrt{(|u_{\,\|,RE}^{\,\textup{ava}}|/c)^2+(|u_{\,\perp,RE}^{\,\textup{ava}}|/c)^2}$ needed for the magnitude of the mean velocity vector $\mathbf{u}_{\,RE}^{\,\mathrm{ava}}$. Furthermore, one can deduce the same statements as for the parallel component, namely that the magnitude of the mean velocity increases for high electric fields and low electron densities, while its shows higher values for higher electron temperatures. \\ This can be explained with a more effective runaway generation via the avalanche mechanism, because for higher electron temperatures and higher electric fields more electrons are already in the runaway region and additionally more electrons are transferred into this region via acceleration by the electric field. However, the increase of the normalized magnitude of the mean velocity with lower electron density can be understood by the decrease of the friction force, which arises from \textit{Coulomb} collisions, which are less likely and less effective in decelerating the electrons for lower electron densities.

The discussed behaviour in the results for $|u_{\,\|,RE}^{\,\textup{ava}}|/c$ and $u_{\,RE}^{\,\textup{ava}}/c$ is also viewable in the figures \ref{fig_ava_K_EoverEc_n_e} and \ref{fig_ava_K_E_par_n_e}, which present the contour and scatter plots the normalized mean mass-related kinetic energy density $K_{\,RE}^{\,\textup{ava}}/(c^2/2)$ for different avalanche runaway electron populations with respect to the electron density $n_{e}$ and either the normalized electric field strength $E=\vert E_{\|}\vert/E_{c}$ or the absolute value of the parallel component of the electric field $\vert E_{\|}\vert$. The displayed results are in the ranges $K_{\,RE}^{\,\textup{ava}}/(c^2/2)\in[0.938,\,0.998]$ for $k_{B}\,T_{e}=10\,\textup{eV}$, $K_{\,RE}^{\,\textup{ava}}/(c^2/2)\in[0.950,\,0.998]$ for $k_{B}\,T_{e}=100\,\textup{eV}$ and $K_{\,RE}^{\,\textup{ava}}/(c^2/2)\in[0.958,\,0.999]$ for $k_{B}\,T_{e}=1000\,\textup{eV}$ according to the combined console output of the plotting scripts$^{\mathrm{\ref{plotfootnote}}}$ from listing \ref{outMATLABgeneratedataAppendix} in subsection \ref{subsection_generated_data} of the appendix. In general, one can deduce, that the range for the kinetic energy density increases for higher electron temperatures as expected. As well, it decreases for higher electron densities and high electric fields, where less runaway electrons are produced and the mean velocity is smaller than for low densities.  

\begin{figure}[H]
\centering
\begin{subfigure}{\textwidth}
\centering\vspace{-2mm}
\includegraphics[trim=83 13 91 43,width=0.59\textwidth,clip]{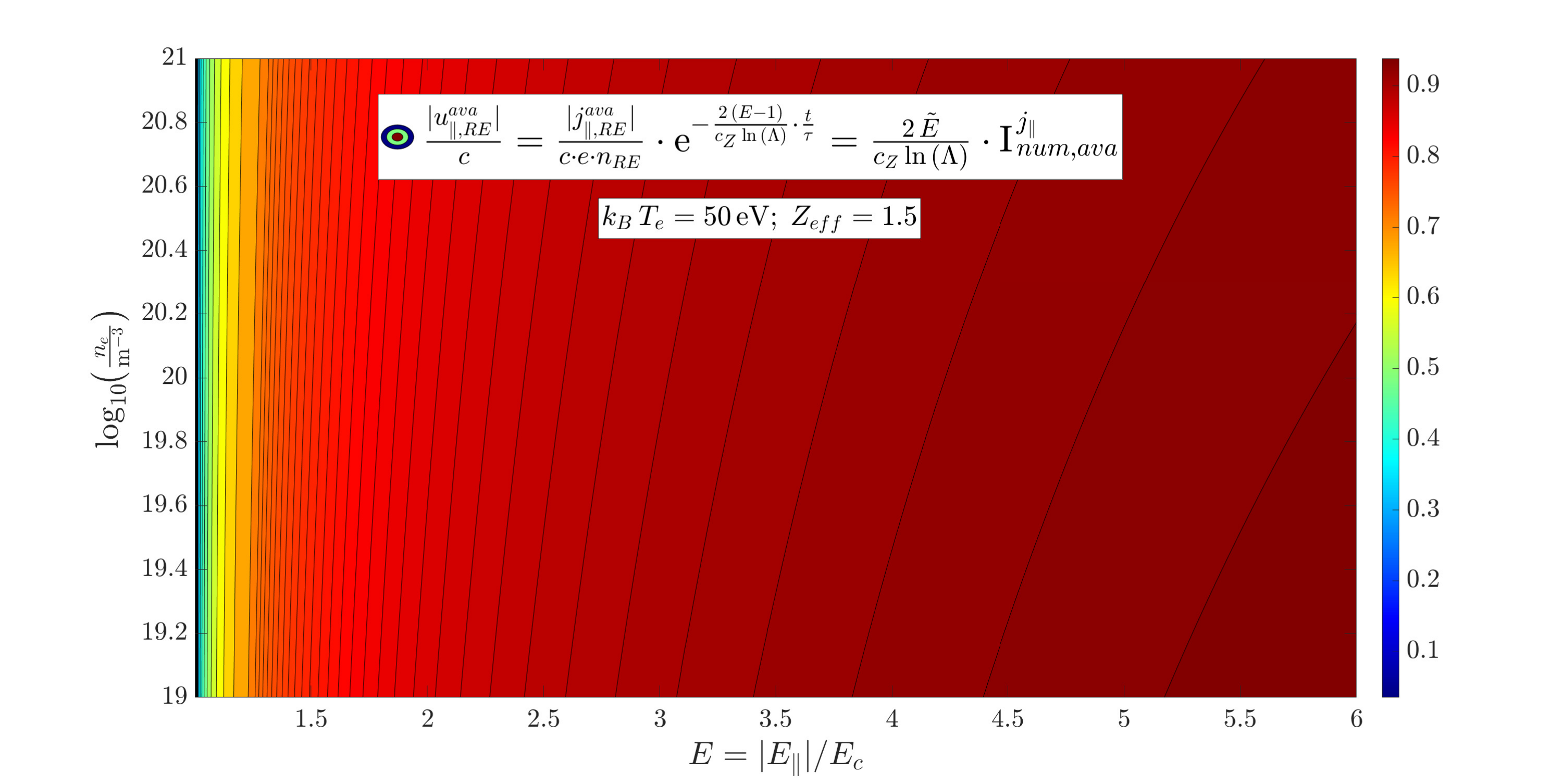}
\caption{Absolute value of the normalized parallel component of the mean velocity $|u_{\,\|,RE}^{\,\textup{ava}}|/c$}
\label{fig_ava_u_par_ava_EoverEc_n_e_50eV}
\end{subfigure}
\begin{subfigure}{\textwidth}
\centering\vspace{2mm}
\includegraphics[trim=83 13 91 43,width=0.59\textwidth,clip]{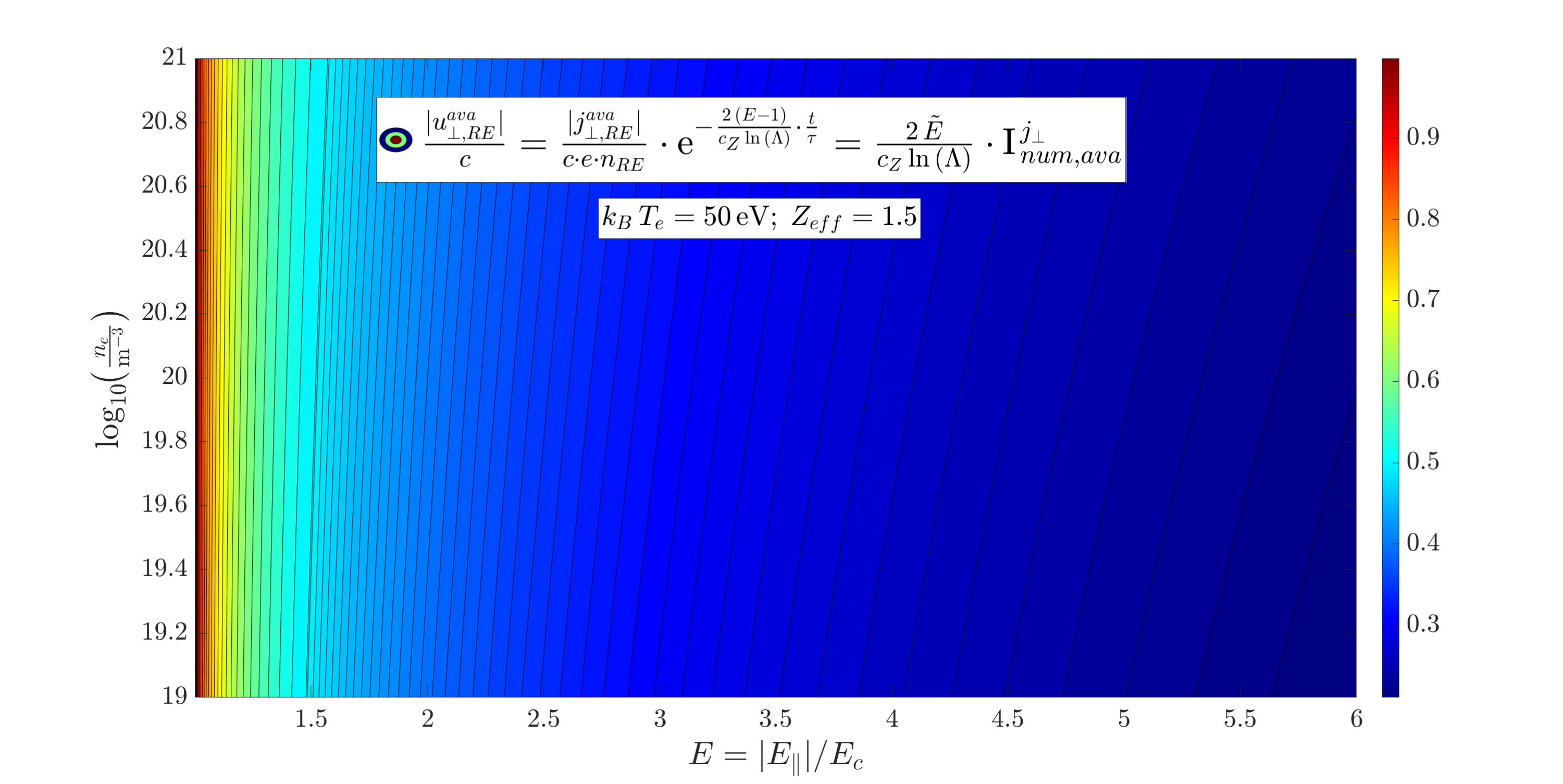}
\caption{Absolute value of the normalized orthogonal component of the mean velocity $|u_{\,\perp,RE}^{\,\textup{ava}}|/c$}
\label{fig_ava_u_perp_EoverEc_n_e50eVsub}
\end{subfigure}
\begin{subfigure}{\textwidth}
\centering\vspace{2mm}
\includegraphics[trim=83 13 91 43,width=0.59\textwidth,clip]{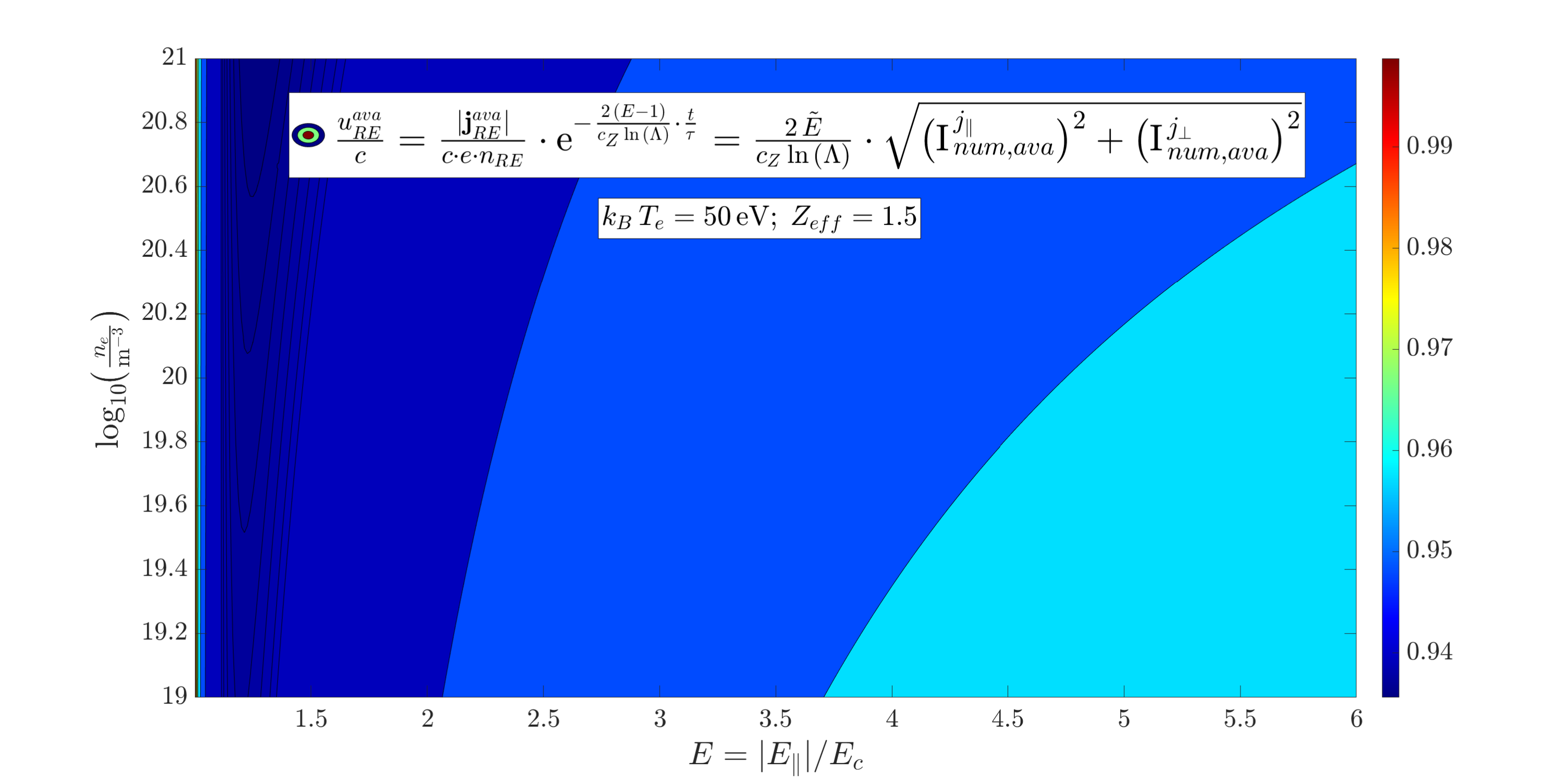}
\caption{Normalized magnitude of the mean velocity $|u_{\,RE}^{\,\textup{ava}}|/c$}
\label{fig_ava_u_EoverEc_n_e50eV}
\end{subfigure}
\begin{subfigure}{\textwidth}
\centering\vspace{2mm}
\includegraphics[trim=83 13 91 43,width=0.59\textwidth,clip]{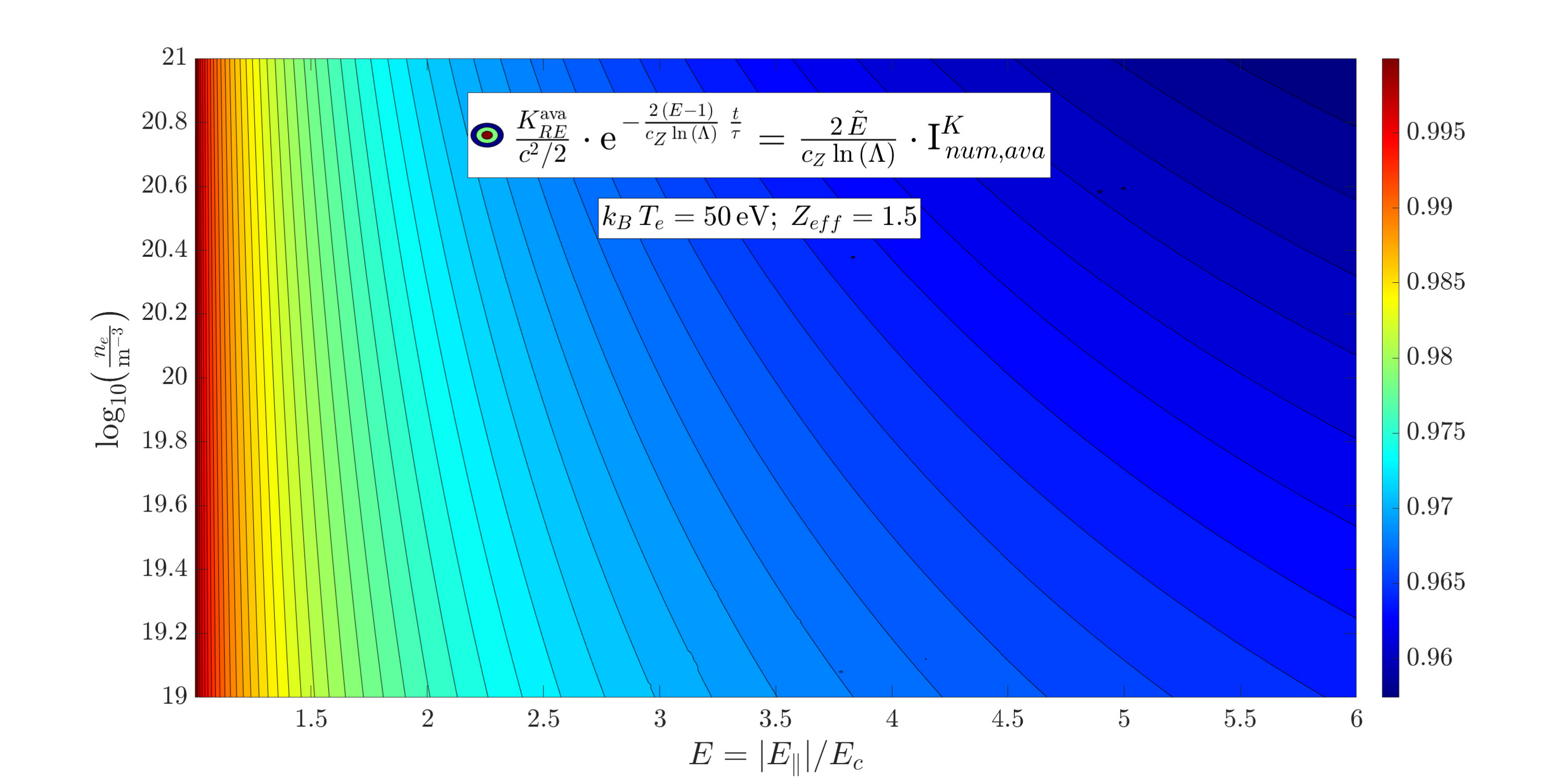}
\caption{Normalized mean mass-related kinetic energy density $K_{\,RE}^{\,\textup{ava}}/(c^2/2)$}
\label{fig_ava_K_EoverEc_n_e50eVsub}
\end{subfigure}
\captionsetup{format=hang,indention=0cm}
\caption[Contour plots with respect to the electron density $n_{e}$ and the normalized electric field strength $E=\vert E_{\|}\vert/E_{c}$ for an avalanche runaway electron population with $k_{B}\,T_{e}=50\,\textup{eV}$ and $Z_{eff}=1.5$]{Contour plots$^{\mathrm{\ref{fig_50eV_footnote}}}$ with respect to the electron density $n_{e}$ and the normalized electric field strength $E=\vert E_{\|}\vert/E_{c}$ for an avalanche runaway electron population with $k_{B}\,T_{e}=50\,\textup{eV}$ and $Z_{eff}=1.5$}
\label{fig_ava_EoverEc_n_e50eV}
\end{figure}

\begin{figure}[H]
\centering
\begin{subfigure}{\textwidth}
\centering\vspace{-2mm}
\includegraphics[trim=83 13 91 43,width=0.59\textwidth,clip]{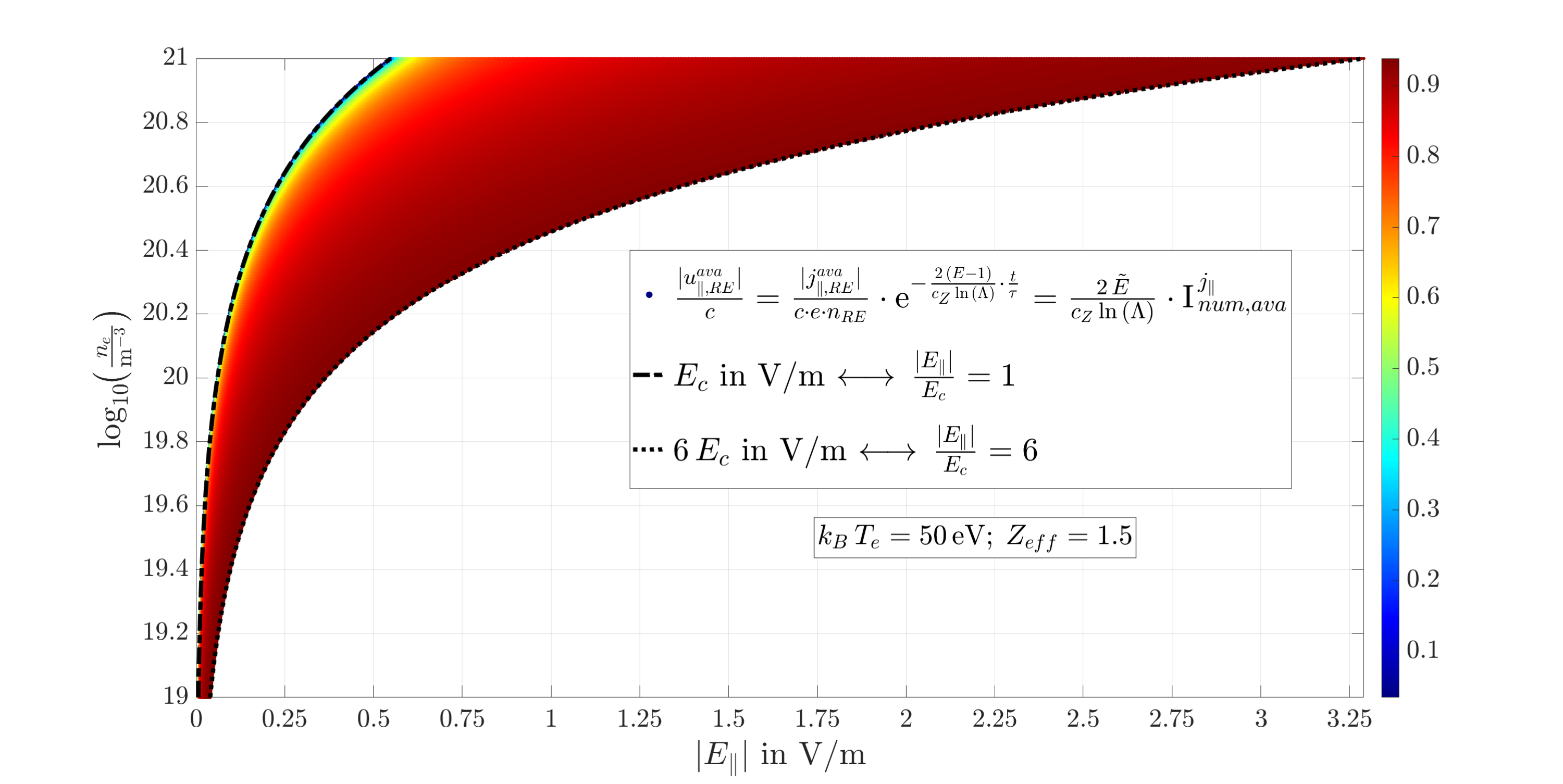}
\caption{Absolute value of the normalized parallel component of the mean velocity $|u_{\,\|,RE}^{\,\textup{ava}}|/c$}
\label{fig_ava_u_par_ava_E_par_n_e_50eV}
\end{subfigure}
\begin{subfigure}{\textwidth}
\centering\vspace{2mm}
\includegraphics[trim=83 13 91 43,width=0.59\textwidth,clip]{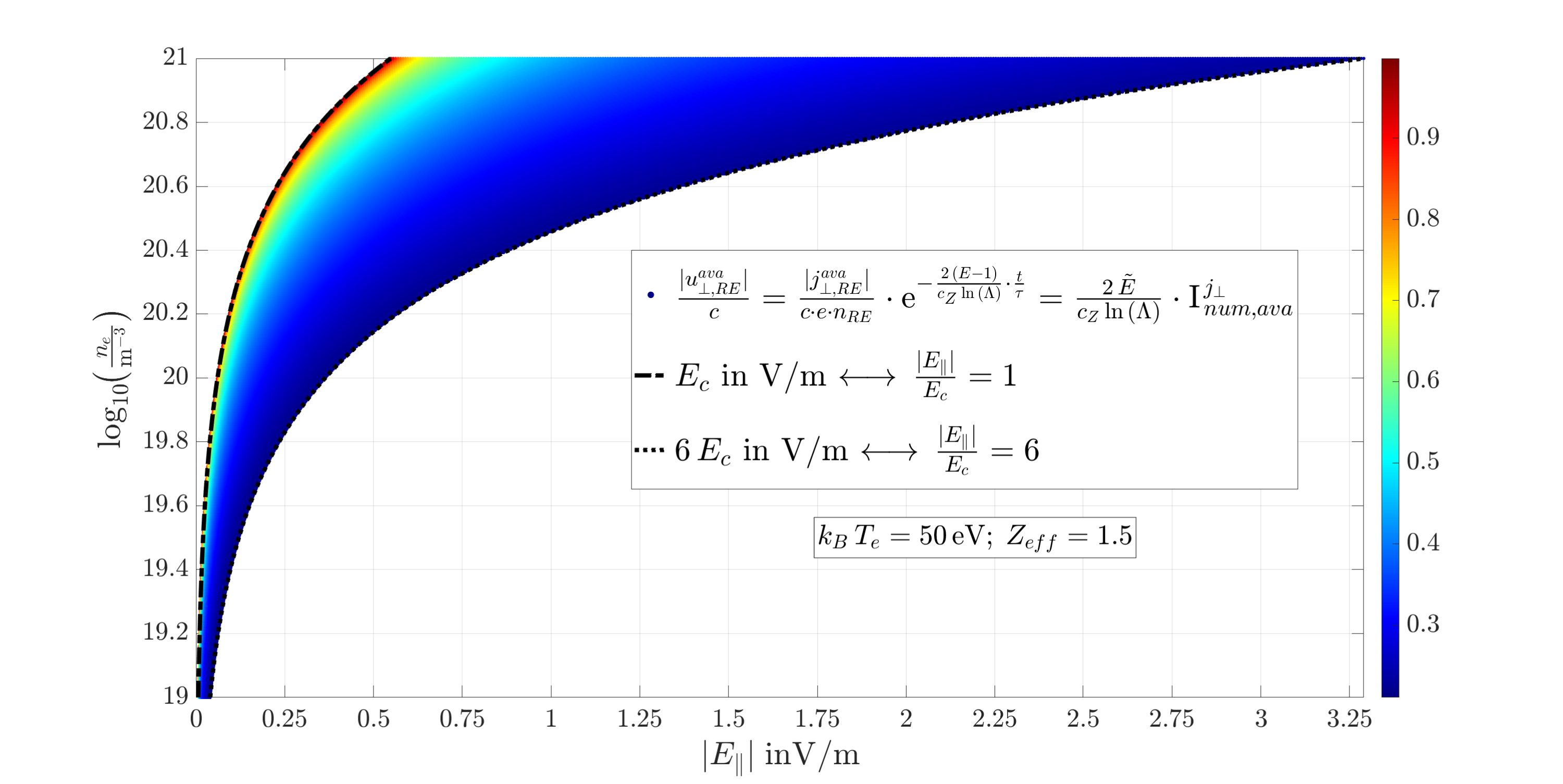}
\caption{Absolute value of the normalized orthogonal component of the mean velocity $|u_{\,\perp,RE}^{\,\textup{ava}}|/c$}
\label{fig_ava_u_perp_E_par_n_e50eV}
\end{subfigure}
\begin{subfigure}{\textwidth}
\centering\vspace{2mm}
\includegraphics[trim=83 13 91 43,width=0.59\textwidth,clip]{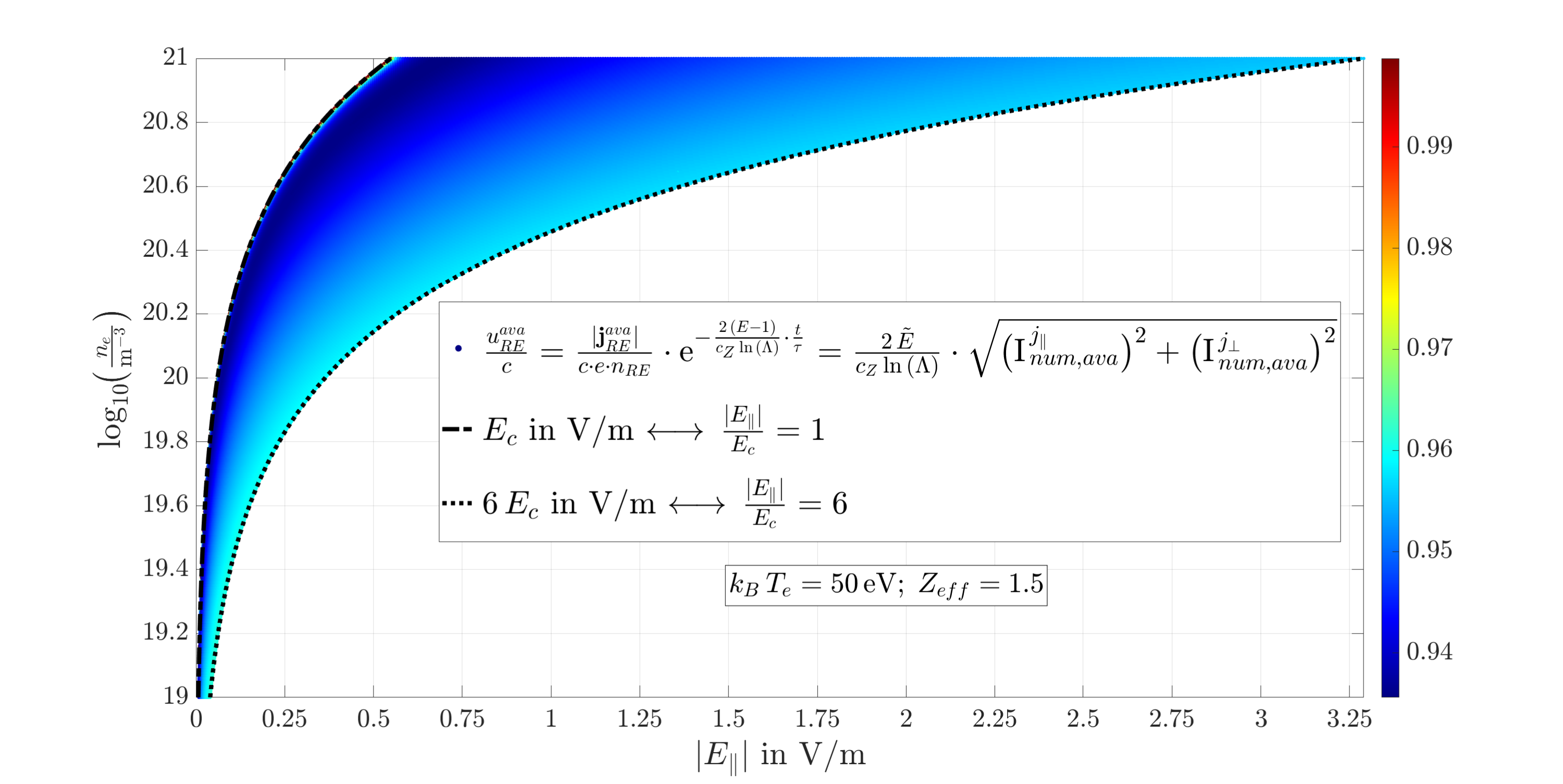}
\caption{Normalized magnitude of the mean velocity $|u_{\,RE}^{\,\textup{ava}}|/c$}
\label{fig_ava_u_E_par_n_e50eV}
\end{subfigure}
\begin{subfigure}{\textwidth}
\centering\vspace{2mm}
\includegraphics[trim=83 13 91 43,width=0.59\textwidth,clip]{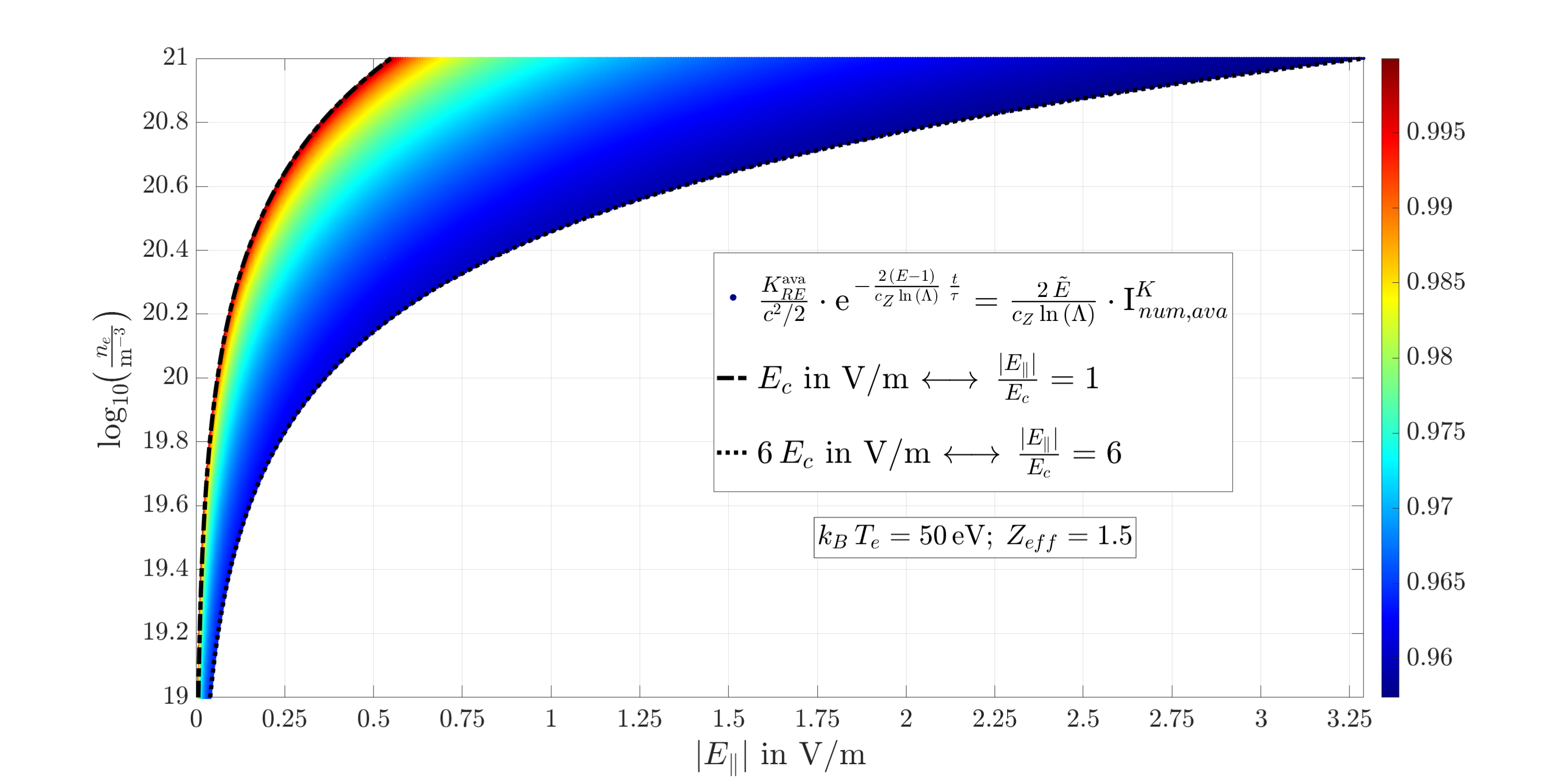}
\caption{Normalized mean mass-related kinetic energy density $K_{\,RE}^{\,\textup{ava}}/(c^2/2)$}
\label{fig_ava_K_E_par_n_e50eV}
\end{subfigure}
\captionsetup{format=hang,indention=0cm}
\caption[Contour plots with respect to the electron density $n_{e}$ and the absolute value of the parallel component of the electric field $\vert E_{\|}\vert$ for an avalanche runaway electron population with $k_{B}\,T_{e}=50\,\textup{eV}$ and $Z_{eff}=1.5$]{Contour plots$^{\ref{fig_50eV_footnote}}$ with respect to the electron density $n_{e}$ and the absolute value of the parallel component of the electric field $\vert E_{\|}\vert$ for an avalanche runaway electron population with $k_{B}\,T_{e}=50\,\textup{eV}$ and $Z_{eff}=1.5$}
\label{fig_ava_E_par_n_e50eV}
\end{figure}

Due to the fact, that the parameter area for electric fields near the critical field shows interesting gradients and a dominance of the perpendicular component of the mean velocity, an additional runaway electron population with an electron temperature of $k_{B}T_{e}=50\,\mathrm{eV}$ and an effective ion charge $Z_{eff}=1.5$ is analyzed for the smaller interval $\vert E_{\|}\vert/E_{c}\in[1.0001,\,6]$. The computation with the code "generate_num_data_50eV.m" and the plotting with the script "plot_num_data_50eV.m" then leads to the figures \ref{fig_ava_EoverEc_n_e50eV} and \ref{fig_ava_EoverEc_n_e50eV}. The displayed results are in the ranges $|u_{\,\|,RE}^{\,\textup{ava}}|/c\in[0.035,\,0.938]$, $|u_{\,\perp,RE}^{\,\textup{ava}}|/c\in[0.211,\,0.998]$, $u_{\,RE}^{\,\textup{ava}}/c\in[0.936,\,0.999]$ and $K_{\,RE}^{\,\textup{ava}}/(c^2/2)\in[0.957,\,1.000]$ according to the console output of the plotting script "plot_num_data_50eV.m" from listing \ref{outMATLABgeneratedataAppendix} in subsection \ref{subsection_generated_data} of the appendix. There one can also notice, that the orthogonal component is at least $23\,\%$ of the size of the parallel component of the mean velocity and can exceed it by up to $2826\,\%$ for low electric fields and high electron densities. Consequently, a rather unexpected result are the large values for the mean mass-related kinetic energy density for low electric fields near the critical field and high electron densities. Where, this is particularly evident in the contour and scatter plots\footnote{\label{fig_50eV_footnote} The contour plots in the subfigures \ref{fig_ava_u_par_ava_EoverEc_n_e_50eV}, \ref{fig_ava_u_perp_EoverEc_n_e50eVsub}, \ref{fig_ava_u_EoverEc_n_e50eV}, \ref{fig_ava_K_EoverEc_n_e50eVsub}, \ref{fig_ava_u_par_ava_E_par_n_e_50eV}, \ref{fig_ava_u_perp_E_par_n_e50eV}, \ref{fig_ava_u_E_par_n_e50eV}, \ref{fig_ava_K_E_par_n_e50eV} were\\\hspace*{8.5mm}calculated with the help of the \textsc{MATLAB}-scripts "generate_num_data_50eV.m" and\\ \hspace*{8.5mm}"plot_num_data_50eV.m", which are stored in the digital appendix.} in the figures \ref{fig_ava_EoverEc_n_e50eV} and \ref{fig_ava_EoverEc_n_e50eV} and could denote, that the analytic avalanche distribution function and the focus on the parallel component of the electric field are not applicable for electric fields with $\vert E_{\|}\vert < 1.2\,E_{c}$ or do not represent the main physical processes.

\vspace{-7mm}


\clearpage

\chapter{Summary}

The calculation of the moments of an analytical distribution function for the avalanche runaway electrons was motivated by the presentation of the procedure based on an example, in order to find efficient computation rules, for instance for the first moment determining the current density and the second moment, which is related to the kinetic energy density of a runaway electron density. Those efficient computation equations can then be an enrichment for a simulation software, which self-consistently calculate the evolution of the runaway current \cite{pappPHD} and can consequently contribute to the prediction of disruptions or the improvement of disruption mitigation control mechanism, for the purpose of avoiding damages to the reactor wall of future thermonuclear fusion reactors such as ITER \cite{REsimulation}.

Hence, the runaway electron phenomenon was first explained in general and with a focus on the avalanche and the hot-tail generation mechanism. After that, the kinetic description of plasmas in the phase space was utilized as a theoretical framework and the definitions of the moments of a distribution function were collated. On the background of an efficient simulation a two-dimensional phase space in the momentum coordinates parallel and orthogonal to the magnetic field was introduced in section \ref{mom_space_coord_section}. 

Second, the calculation of the moments was approached by presenting the analytic momentum space distribution function (\ref{f_RE_ava}) for runaway electrons produced by avalanching from \mbox{\textit{Fülöp et al.}} \cite{REdistfuncderivation}, which is based on the growth rate proposed by \textit{Rosenbluth} \& \textit{Putvinski} \cite{Rosenbluth_1997} and holds for plasma scenarios dominated by the avalanche generation mechanism.\\
The integral over this distribution function respectively the zeroth moment could be evaluated completely analytically. Therefore it was used as a benchmark for the runtime efficiency analysis for the \textsc{MATLAB}-integration routines in section \ref{ava_zero_moment_section}. At that it was shown, that the two-dimensional numerical integration with the programming command "\texttt{integral2}" is more efficient with regard to the runtime and more accurate than the "\texttt{trapz}"-method, which uses a trapezoidal rule. \\
The parallel and orthogonal component of the avalanche runaway current density were then related to first moments of the distribution function in section \ref{ava_first_moment_section}. Both quantities were computable with the two-dimensional numerical integration routine "\texttt{integral2}" in \textsc{MATLAB}, where the runtime was decreased by the utilization of substitutions, which transform the integration boundaries to finite values. In addition, an analytic solution for the integration over the perpendicular momentum component was found in subsection \ref{j_RE_ava_par_subsection} in case of the parallel component, leading to the calculation rule (\ref{j_ava_RE_par_I_num}), which only requires a one-dimensional integration. However, it was discovered that this integration needs more runtime than the alternative previously mentioned two-dimensional numerical integration. The reason of this was found to be the characteristic of the integrand, which only allowed a computation for a high-precision integration with the "\texttt{vpaintegral}"-command as a consequence of floating point arithmetics. Conclusively, one was able to argue that the components and therefore also the magnitude of the avalanche runaway current density are most efficiently computable with the routine \mbox{"\texttt{integral2}"} in combination with a transformation of the integration bounds to finite values by means of a substitution, yielding to the calculation rules (\ref{j_ava_RE_parallel_2Dnumeffy}), (\ref{j_ava_RE_perp_2Dnum_substituted_result}) and (\ref{j_ava_RE_num_magnitude}). This was found exclusively for the implementation in \mbox{\textsc{MATLAB}}, where the criterion for efficiency was the mean runtime for different parameters if the integrand functions. Further, it shall be remarked, that for instance an implementation in C\texttt{++} might yield a shorter runtime duration for the one-dimensional integration of the parallel component of the current density, if a solution for the difficulties with the floating point accuracy for the evaluation of the integrand is found and efficient algorithms for the evaluation of the appearing incomplete gamma function are used. This work is left for a future study.\\
The second moment thus the mean mass-related kinetic energy density of an avalanche runaway electron population was defined and evaluated in section \ref{ava_second_moment_section}. At this, it was again possible to analytically solve one integration of the two-dimensional integral defining the second moment, requiring a numerical solve of the integrals in the relations (\ref{K_ava_RE_V5_deriv}) and (\ref{K_ava_RE_V6_deriv}). However, the same difficulties with the floating point arithmetic arise as for the parallel component of the current density. A \mbox{\textsc{MATLAB}}-program then showed, that the second moment is numerically solvable based on the deduced one-dimensional integral and by computation of the full double integral over the momentum space without and with substitutions according to the equations (\ref{K_ava_RE_V3_num2D}) and (\ref{K_ava_RE_V3_num2D_sustituted}), which transform the integration boundaries to finite values. However, the lowest mean runtime was again reached, if the "\texttt{integral2}"-routine in combination with the substitutions from (\ref{substitutions_num_K2Dpagain}) was utilized as stated in the expression (\ref{K_ava_RE_V3_num2D_sustituted}). 

Third, a \mbox{\textsc{MATLAB}}-code was set up to physically evaluate the efficiently computed quantities. It used the "\texttt{integral2}"-routine in combination with the substitutions for the calculation of the parallel and orthogonal component as well as the magnitude of the mean velocity of a avalanche runaway population being the normalized current density. Moreover, the according normalized mean mass-related kinetic energy density was computed similarly. From the generated data for a parameter space in the electron density and the electric field for different electron temperatures one was able to assess the applicability of the implementations based on the used approximations and the used analytic distribution function. At that, one observed that the runaway electron generation due to avalanching is favoured for low electron densities, high electric fields and increasing electron temperatures as expected. Unexpectedly, large values for the mean mass-related kinetic energy density occur for low electric fields near the critical electric field and high electron densities, which seems to be related to dominance of the perpendicular component of the current density over the parallel component. However, it was not possible to determine, if the analytic avalanche distribution function and its focus on the parallel component of the electric field are not applicable for electric fields with $\vert E_{\|}\vert < 1.2\,E_{c}$ or do not represent the main physical processes. The reason therefore is, that the underlying distribution function becomes inaccurate for $\vert E_{\|}\vert \,\rightarrow\,E_{c}$, while no exact threshold is known, below which it is not applicable anymore. Where this is further discussed in the publications by \textit{A.\hspace{0.9mm}K{\'{o}}m{\'{a}}r} and \textit{G.\hspace{0.9mm}I.\hspace{0.9mm}Pokol} and \textit{T.\hspace{0.9mm}Fülöp} \cite{Komar2012,Komar2013}. Hence, one should consider to include a lower bound $\vert E_{\|,min}\vert \approx 1.2\,E_{c}$ for the absolute value of the parallel component of the electric field.

Finally, one was able to demonstrate the general procedure for the determination of runtime-efficient calculation rules for the moments of an analytic distribution function, using the example of the avalanche runaway distribution function. A rather coarse efficiency analysis was carried out for the programming language \mbox{\textsc{MATLAB}} based on the mean runtime for the single integrations needed for the calculations of the moments. Nevertheless, a physically validated efficient implementation could be presented. As well, its limitations and applicability were outlined with the help of graphical depictions of the computed results for a wide range of the plasma parameters electric field and electron density and for four different electron temperatures. It shall be noted that all statements concerning the efficiency of the computation are programming language and implementation dependent.\\ Future work could apply the procedure to the analytic distribution function for the hot-tail generation of runaway electrons as given by  \textit{H.\hspace{1mm}M.\hspace{1mm}Smith} \& \textit{E.\hspace{1mm}Verwichte} in reference \cite{hottailREdistfunc}. In addition, the deduced calculation rules could be implemented consistently in one suitable programming language and included in a self-consistent simulation code for the evolution of the runaway current. Thereupon, the code could be analysed with the focus on efficiency, possibly leading to an improved simulation tool for the runaway electron current. In the long term view, this might be an enrichment in the prediction of disruptions or the development of disruption mitigation control mechanisms.

\clearpage

\phantomsection % für Link zur Bibliography Seite von Contents aus (ohne landet man bei der Seite für die Conclusion

\addcontentsline{toc}{chapter}{References}

\bibliography{Ref}

\clearpage

\listoffigures

\appendix

\chapter{Appendix}

\section{Analytical calculations}

\subsection{Equivalence of the momentum space volume element in the $(p_{\|},\,p_{\perp})$ and $(p,\,\xi)$ coordinate basis}\label{subsection_mom_deriv_vol_elem_appendix}

The volume element in the coordinates $(p_{\|},\,p_{\perp})$ from (\ref{volelem_cyl_2D}) reads $\mathrm{d}^{3}p=2\pi\,p_{\perp}\mathrm{d}p_{\perp}\,\mathrm{d}p_{\|}$, where \mbox{the integration} over the azimuthal angle $\varphi$ was already carried out from $0$ to $2\pi$. In addition, it holds that $p_{\|}=p\,\xi$ and $p_{\perp}=p\,\sqrt{1-\xi^2}$ in accordance with figure \ref{fig_mom_coord}. From this one can transform the volume element from the coordinates $(p_{\|},\,p_{\perp})$ to the coordinate $(p,\,\xi)$, by evaluating absolute value of the \textit{Jacobian} determinant:\vspace{-1.5mm}
\begin{equation}\label{mom_deriv_vol_elem_appendix}
\begin{split}
\begin{gathered}
\underline{\underline{\mathrm{d}^{3}p=2\pi\,p_{\perp}\mathrm{d}p_{\perp}\,\mathrm{d}p_{\|}}} = 2\pi\,p\,\sqrt{1-\xi^2}\,\abs{\frac{\partial\mathbf{x}(p_{\|}(p,\,\xi),\,p_{\perp}(p,\,\xi))}{\partial(p,\,\xi)}}\mathrm{d}p\,\mathrm{d}\xi
\\[8pt]
= 2\pi\,p\,\sqrt{1-\xi^2}\,
\abs{\begin{pmatrix}
\dfrac{\partial\,p_{\|}}{\partial\,p}\hspace{-1mm}&\hspace{-1mm}\dfrac{\partial\,p_{\|}}{\partial\,\xi}\\
\dfrac{\partial\,p_{\perp}}{\partial\,p}\hspace{-1mm}&\hspace{-1mm}\dfrac{\partial\,p_{\perp}}{\partial\,\xi}
\end{pmatrix}}\mathrm{d}p\,\mathrm{d}\xi
\\[8pt]
= 2\pi\,p\,\sqrt{1-\xi^2}\,
\abs{\begin{pmatrix}
\xi\hspace{-1mm}&\hspace{-1mm}p\\
\sqrt{1-\xi^2}\hspace{-1mm}&\hspace{-1mm}\dfrac{-\cancel{2}\,p\,\xi}{\cancel{2}\,\sqrt{1-\xi^2}}
\end{pmatrix}}\mathrm{d}p\,\mathrm{d}\xi
\\[8pt]
= 2\pi\,p\,\sqrt{1-\xi^2}\,
\abs{-\dfrac{ p\,\xi^2}{ \sqrt{1-\xi^2}}-p\,\sqrt{1-\xi^2} }\mathrm{d}p\,\mathrm{d}\xi
\\[8pt]
\overset{\underbrace{{\scriptsize \text{$\vert\xi\vert\,\leq\,1$}}}}{=} 2\pi\,p\,\sqrt{1-\xi^2}\,
\left(\dfrac{ p\,\xi^2+p\,(1-\xi^2)}{ \sqrt{1-\xi^2}} \right)\mathrm{d}p\,\mathrm{d}\xi
\\[8pt]
=2\pi\,p\,\cancel{ \sqrt{1-\xi^2}}\,
 \dfrac{ p }{\cancel{ \sqrt{1-\xi^2}}} \, \mathrm{d}p\,\mathrm{d}\xi
=\underline{\underline{2\pi\,p^2\,\mathrm{d}p\,\mathrm{d}\xi}} 
\end{gathered}
\end{split}
\end{equation}   \vspace{1mm}\noindent
Note that the domain of definition $\xi\in[-1,\,1]$ was used, in order to ensure that $\vert\xi\vert\leq 1$ is valid. Thus, one has shown the validity of the equivalence of the azimuthal-angle-averaged volume elements $\mathrm{d}^{3}p=2\pi\,p_{\perp}\mathrm{d}p_{\perp}\,\mathrm{d}p_{\|}=2\pi\,p^2\,\mathrm{d}p\,\mathrm{d}\xi$ in (\ref{mom_deriv_vol_elem_appendix}).

\clearpage

\subsection{Evaluation of the integral for the avalanche runaway electron density from (\ref{RE_density_integral})}\label{RE_dens_int_appendix_subsection}

The avalanche runaway electron density can be calculated by evaluating the integral from (\ref{RE_density_integral}):\vspace{-2mm}
\begin{equation}\label{RE_dens_int_appendix}
\begin{split}
\begin{gathered}
n_{RE}(t)=\hspace{-2mm}\displaystyle{\int\limits_{p_{\|}=-\infty}^{\infty}\int\limits_{p_{\perp}=0}^{\infty}} \dfrac{2\pi\,C}{p_{\|}}\cdot\exp{\left(\dfrac{(E-1)}{c_{Z} \ln{(\Lambda)}}\cdot\dfrac{t}{\tau}-\dfrac{p_{\|}}{c_{Z} \ln{(\Lambda)}}-\tilde{E}\cdot\dfrac{p_{\perp}^{2}}{p_{\|}}\right)}\,p_{\perp}\mathrm{d}p_{\perp}\mathrm{d}p_{\|}\,.
\end{gathered}
\end{split}
\end{equation}
First, one solves the $p_{\perp}$-integral from (\ref{RE_dens_int_appendix}), by introducing the following substitution:
\begin{equation}\label{substitution1_RE_dens_int_appendix}
\begin{split}
\begin{gathered}
\lambda(p_{\perp}):=\dfrac{(E-1)}{c_{Z} \ln{(\Lambda)}}\cdot\dfrac{t}{\tau}-\dfrac{p_{\|}}{c_{Z} \ln{(\Lambda)}}-\tilde{E}\cdot\dfrac{p_{\perp}^{2}}{p_{\|}}\;\;;\;\;\dfrac{\mathrm{d}\lambda}{\mathrm{d}p_{\perp}}=-\dfrac{2\,\tilde{E}\,p_{\perp}}{p_{\|}}\,.
\end{gathered}
\end{split}
\end{equation} 
With this, one rewrites the equation (\ref{RE_dens_int_appendix}):\vspace{-1mm}
\begin{equation}\label{V2_RE_dens_int_appendix}
\begin{split}
\begin{gathered}
n_{RE}(t)=-\hspace{-4mm}\displaystyle{\int\limits_{p_{\|}=-\infty}^{\infty}}\hspace{-2mm} \dfrac{\cancel{2}\pi\,C}{\cancel{p_{\|}}} \displaystyle{\int\limits_{\lambda(0)}^{\lambda(\infty)}}\hspace{-1mm} \textup{e}^{\lambda} \,\cancel{p_{\perp}}\,\dfrac{\cancel{p_{\|}}}{\cancel{2}\,\tilde{E}\,\cancel{p_{\perp}}}\,\mathrm{d}\lambda\,\mathrm{d}p_{\|}
\\[7pt]
=\hspace{-0.5mm}-\dfrac{\pi\,C}{\tilde{E}} \displaystyle{\int\limits_{p_{\|}=-\infty}^{\infty}} \hspace{-.5mm} \left[\textup{e}^{\,\lambda}\right]^{\lambda(\infty)=-\infty}_{\lambda(0)=\frac{(E-1)}{c_{Z} \ln{(\Lambda)}}\cdot\frac{t}{\tau}-\frac{p_{\|}}{c_{Z} \ln{(\Lambda)}}}  \mathrm{d}p_{\|}
\\[7pt]
=-\dfrac{\pi\,C}{\tilde{E}} \,\displaystyle{\int\limits_{p_{\|}=-\infty}^{\infty}}  \hspace{-1.5mm}  \underbrace{\textup{e}^{\,\lambda\rightarrow -\infty}}_{=\,0}-\textup{e}^{\,\frac{(E-1)}{c_{Z} \ln{(\Lambda)}}\cdot\frac{t}{\tau}-\frac{p_{\|}}{c_{Z} \ln{(\Lambda)}}} \,\mathrm{d}p_{\|}
=\dfrac{\pi\,C}{\tilde{E}} \,\displaystyle{\int\limits_{p_{\|}=-\infty}^{\infty}}    \textup{e}^{\,\frac{(E-1)}{c_{Z} \ln{(\Lambda)}}\cdot\frac{t}{\tau}-\frac{p_{\|}}{c_{Z} \ln{(\Lambda)}}} \,\mathrm{d}p_{\|}\,.
\end{gathered}
\end{split}
\end{equation}
Second, one solves the $p_{\|}$-integral from (\ref{V2_RE_dens_int_appendix}), by another substitution:
\begin{equation}\label{substitution2_RE_dens_int_appendix}
\begin{split}
\begin{gathered}
\kappa(p_{\|}):=\dfrac{(E-1)}{c_{Z} \ln{(\Lambda)}}\cdot\dfrac{t}{\tau}-\dfrac{p_{\|}}{c_{Z} \ln{(\Lambda)}} \;\;;\;\;\dfrac{\mathrm{d}\kappa}{\mathrm{d}p_{\|}}=-\dfrac{1}{c_{Z} \ln{(\Lambda)}}\,.
\end{gathered}
\end{split}
\end{equation} 
With the help of (\ref{substitution2_RE_dens_int_appendix}) one can evaluate (\ref{V2_RE_dens_int_appendix}), whereby the lower integration bound is set to $p_{\|,min}\rightarrow -\infty$:\vspace{-4mm}
\begin{equation}\label{V3_RE_dens_int_appendix}
\begin{split}
\begin{gathered}
n_{RE}(t)
=-\dfrac{\pi\,C\,c_{Z} \ln{(\Lambda)}}{\tilde{E}}\hspace{-1.5mm} \displaystyle{\int\limits_{\kappa(-\infty)}^{\kappa(\infty)}}    \hspace{-1.5mm} \textup{e}^{\,\kappa} \,\mathrm{d}\kappa
\\[7pt]
=-\dfrac{\pi\,C\,c_{Z} \ln{(\Lambda)}}{\tilde{E}} \,\bigl[\textup{e}^{\,\kappa}\bigr]^{\kappa(\infty)=-\infty}_{\kappa(p_{\|,min})=\frac{(E-1)}{c_{Z} \ln{(\Lambda)}}\cdot\frac{t}{\tau}-\frac{p_{\|,min}}{c_{Z} \ln{(\Lambda)}}}
\\[7pt] 
=-\dfrac{\pi\,C\,c_{Z} \ln{(\Lambda)}}{\tilde{E}} \hspace{-0.5mm}\left(\underbrace{\textup{e}^{\,\kappa\rightarrow -\infty}}_{=\,0}-\textup{e}^{\,\frac{(E-1)}{c_{Z} \ln{(\Lambda)}}\cdot\frac{t}{\tau}-\frac{p_{\|,min}}{c_{Z} \ln{(\Lambda)}}} \hspace{-0.5mm} \right)
 \\[7pt] 
=\dfrac{\pi\,C\,c_{Z} \ln{(\Lambda)}}{\tilde{E}} \, \textup{e}^{\,\frac{(E-1)}{c_{Z} \ln{(\Lambda)}}\cdot\frac{t}{\tau}-\frac{p_{\|,min}}{c_{Z} \ln{(\Lambda)}}} \,.
\end{gathered}
\end{split}
\end{equation}
Finally, one inserts the expressions for $C$ and $\tilde{E}$ from (\ref{C_f_RE_ava}) and (\ref{E_cZ_f_RE_ava}) into the result from (\ref{V3_RE_dens_int_appendix}):\vspace{-2.5mm}
\begin{equation}\label{V4_RE_dens_int_appendix}
\begin{split}
\begin{gathered}
\underline{\underline{ n_{RE}(t)  }}
=\dfrac{\cancel{\pi}\,n_{RE}\,\cancel{\tilde{E}\,c_{Z} \ln{(\Lambda)}}}{\cancel{\tilde{E}\,\pi\,c_{Z}\ln{\Lambda}}} \, \textup{e}^{\,\frac{(E-1)}{c_{Z} \ln{(\Lambda)}}\cdot\frac{t}{\tau}}\, \textup{e}^{\,\frac{(E-1)}{c_{Z} \ln{(\Lambda)}}\cdot\frac{t}{\tau}-\frac{p_{\|,min}}{c_{Z} \ln{(\Lambda)}}} 
\\[10pt] 
=\underline{\underline{n_{RE}\exp{\left( \dfrac{2\,(E-1)}{c_{Z} \ln{(\Lambda)}}\cdot\dfrac{t}{\tau}-\dfrac{p_{\|,min}}{c_{Z} \ln{(\Lambda)}} \right)}  }}\,.
\end{gathered}
\end{split}
\end{equation}

The result of the integration in (\ref{V3_RE_dens_int_appendix}) is only finite, if $p_{\|,min}>-\infty$. Moreover, one can derive the lowest possible parallel momentum $p_{\|,min}$, for which the zeroth moment truly reproduces the runaway electron density from (\ref{V4_RE_dens_int_appendix}):
\begin{equation}\label{p_par_min_appendix}
\begin{split}
\begin{gathered}
n_{RE}(t)  \overset{!}{
=}n_{RE}\exp{\left( \dfrac{2\,(E-1)}{c_{Z} \ln{(\Lambda)}}\cdot\dfrac{t}{\tau}-\dfrac{p_{\|,min}}{c_{Z} \ln{(\Lambda)}} \right)}
\\[10pt]
\longleftrightarrow\;\dfrac{2\,(E-1)}{\cancel{c_{Z} \ln{(\Lambda)}}}\cdot\dfrac{t}{\tau} \overset{!}{
=}\dfrac{p_{\|,min}}{\cancel{c_{Z} \ln{(\Lambda)}}} 
\\[10pt] 
\underline{\underline{p_{\|,min}=2\,(E-1)\cdot\dfrac{t}{\tau}}}\,.
\end{gathered}
\end{split}
\end{equation}

\clearpage

\subsection{Evaluation of the integral for the parallel component of the avalanche runaway electron current density from (\ref{j_ava_RE_parallel_def})}\label{j_ava_RE_parallel_int_appendix_subsection}

The parallel component of the current density carried by a runaway electron population resulting from the avalanche generation mechanism can be calculated by evaluating the integral in (\ref{j_ava_RE_parallel_def}) from section \ref{ava_first_moment_section} by inserting the distribution function from (\ref{f_RE_ava}):\vspace{-1mm}
\begin{equation}\label{j_ava_RE_par_def_appendix}
\begin{split}
\begin{gathered}
j_{\,\|,RE}^{\,\textup{ava}}=-2\pi\,c\,e\hspace{-2mm}\displaystyle{\int\limits_{p_{\|}=-\infty}^{\infty}\int\limits_{p_{\perp}=0}^{\infty}} \frac{p_{\perp}\,p_{\|}}{\sqrt{1+p_{\|}^{2}+p_{\perp}^2}}\,f_{RE}^{\textup{ava}}(p_{\|},\,p_{\perp},\,t)\, \mathrm{d}p_{\perp} \mathrm{d}p_{\|}
\\[10pt]
\overset{(\ref{f_RE_ava})}{=}\hspace{-1.5mm}-2\pi\,c\,e\hspace{-3.5mm}\displaystyle{\int\limits_{p_{\|}=-\infty}^{\infty}\hspace{-0.5mm}\int\limits_{p_{\perp}=0}^{\infty}} \hspace{-0.5mm}\frac{C\,p_{\perp}\,\cancel{p_{\|}}}{\cancel{p_{\|}}\sqrt{1+p_{\|}^{2}+p_{\perp}^2}} \hspace{-0.2mm}\exp{\hspace{-0.8mm}\left(\dfrac{(E-1)}{c_{Z} \ln{(\Lambda)}}\hspace{0.5mm}\dfrac{t}{\tau}\hspace{-0.5mm}-\hspace{-0.5mm}\dfrac{p_{\|}}{c_{Z} \ln{(\Lambda)}}\hspace{-0.5mm}-\hspace{-0.5mm}\tilde{E}\,\dfrac{p_{\perp}^{2}}{p_{\|}}\hspace{-0.5mm}\right)} \mathrm{d}p_{\perp} \mathrm{d}p_{\|}
\\[10pt]
=\hspace{-0.5mm}-2\pi\,c\,e\hspace{-3.5mm}\displaystyle{\int\limits_{p_{\|}=-\infty}^{\infty}\int\limits_{p_{\perp}=0}^{\infty}} \frac{C\,p_{\perp}}{\sqrt{1+p_{\|}^{2}+p_{\perp}^2}} \exp{\left(\dfrac{(E-1)}{c_{Z} \ln{(\Lambda)}}\,\dfrac{t}{\tau}-\dfrac{p_{\|}}{c_{Z} \ln{(\Lambda)}}-\tilde{E}\,\dfrac{p_{\perp}^{2}}{p_{\|}}\hspace{-0.5mm}\right)} \mathrm{d}p_{\perp} \mathrm{d}p_{\|}\,.
\end{gathered}
\end{split}
\end{equation}
For the $p_{\perp}$-integration one can use the following substitution with the goal to rewrite the integral in terms of an upper incomplete gamma function $\Gamma(z,\,a)$ \cite{incgammafunc}:\vspace{-2mm}
\begin{equation}\label{substitution_eta_j_par_appendix}
\begin{split}
\begin{gathered}
\eta(p_{\perp}):=\dfrac{\tilde{E}}{p_{\|} }\left(1+p_{\|}^{2}+p_{\perp}^2\right)\;;\;\dfrac{\mathrm{d}\eta}{\mathrm{d}p_{\perp}}=\dfrac{2\,\tilde{E}\,p_{\perp}}{p_{\|}}\;;
\\[4pt]
 \eta(p_{\perp}=0)=\dfrac{\tilde{E}}{p_{\|} }\left(1+p_{\|}^{2} \right)\;,\;\eta(p_{\perp}\rightarrow\infty)=\infty\,.
\end{gathered}
\end{split}
\end{equation} 
Hence, the integral from (\ref{j_ava_RE_par_def_appendix}) together with the relations for $C$ from (\ref{C_f_RE_ava}) becomes:\vspace{-2.5mm}
\begin{equation}\label{j_ava_RE_par_calc1}
\begin{split}
\begin{gathered}
j_{\,\|,RE}^{\,\textup{ava}}=
-\cancel{2}\pi\,c\,e\hspace{-2mm}\displaystyle{\int\limits_{p_{\|}=-\infty}^{\infty}} \textup{e}^{\,\frac{(E-1)}{c_{Z} \ln{(\Lambda)}}\,\frac{t}{\tau}-\frac{p_{\|}}{c_{Z} \ln{(\Lambda)}}}\hspace{-2mm}\displaystyle{\int\limits_{\eta=\frac{\tilde{E}}{p_{\|} }(1+p_{\|}^{2} )}^{\infty}} \frac{C\,p_{\|}\,\cancel{p_{\perp}} \,\textup{e}^{\, -\eta+\frac{\tilde{E}}{p_{\|}}(1+p_{\|}^{2} )}}{\cancel{2} \,\tilde{E} \,\cancel{p_{\perp}}\,\sqrt{\frac{p_{\|}}{\tilde{E}}\,\eta}}  \, \mathrm{d}\eta  \,\mathrm{d}p_{\|}
\\[8pt]
=
- \frac{\pi\,c\,e\,C\,\sqrt{\tilde{E}}}{\tilde{E} }\hspace{-0.5mm}\displaystyle{\int\limits_{p_{\|}=-\infty}^{\infty}}\hspace{-0.5mm} \sqrt{p_{\|}}\,\textup{e}^{\,\frac{(E-1)}{c_{Z} \ln{(\Lambda)}}\,\frac{t}{\tau}+\left(\tilde{E}-\frac{1}{c_{Z} \ln{(\Lambda)}}\right)p_{\|}+\frac{\tilde{E}}{p_{\|}}}\hspace{-4mm}\displaystyle{\int\limits_{\eta=\frac{\tilde{E}}{p_{\|} }(1+p_{\|}^{2} )}^{\infty}}\hspace{-2mm} \eta^{-\frac{1}{2}}\,\textup{e}^{\, -\eta}  \, \mathrm{d}\eta  \,\mathrm{d}p_{\|}
\\[8pt]
\overset{(\ref{C_f_RE_ava})}{=}\hspace{-0.5mm}- \frac{\cancel{\pi}\,c\,e\,n_{RE}\,\cancel{\tilde{E}}\,\sqrt{\tilde{E}}}{\cancel{\tilde{E}}\,\cancel{\pi}\,c_{Z}\ln{(\Lambda)}}\,\textup{e}^{\,\frac{2\,(E-1)}{c_{Z} \ln{(\Lambda)}}\,\frac{t}{\tau}}\hspace{-3mm}\displaystyle{\int\limits_{p_{\|}=-\infty}^{\infty}} \hspace{-2mm}\sqrt{p_{\|}}\,\textup{e}^{  \left(\tilde{E}-\frac{1}{c_{Z} \ln{(\Lambda)}}\right)p_{\|}+\frac{\tilde{E}}{p_{\|}}}\,\Gamma\hspace{-0.5mm}\left(\dfrac{1}{2},\,\frac{\tilde{E}}{p_{\|} }\hspace{-0.5mm}\left(1+p_{\|}^{2} \right)\hspace{-0.5mm}\right) \hspace{-0.5mm}\mathrm{d}p_{\|}
\\[8pt]
=- \frac{ c\,e\,n_{RE}\, \sqrt{\tilde{E}} }{c_{Z}\ln{(\Lambda)}}\,\textup{e}^{\,\frac{2\,(E-1)}{c_{Z} \ln{(\Lambda)}}\,\frac{t}{\tau}} \hspace{-2mm}\displaystyle{\int\limits_{p_{\|}=-\infty}^{\infty}} \hspace{-2mm} \underbrace{\sqrt{p_{\|}}\,\textup{e}^{  \left(\tilde{E}-\frac{1}{c_{Z} \ln{(\Lambda)}}\right)p_{\|}+\frac{\tilde{E}}{p_{\|}}}\,\Gamma\hspace{-0.5mm}\left(\dfrac{1}{2},\,\frac{\tilde{E}}{p_{\|} }\hspace{-0.5mm}\left(1+p_{\|}^{2}   \right)\hspace{-0.5mm}\right)}_{=:\,\textup{I}_{\,1}(p_{\|})}\hspace{-0.5mm}\mathrm{d}p_{\|}\,.
\end{gathered}
\end{split}
\end{equation}
The integral in (\ref{j_ava_RE_par_calc2}) is convergent for a finite and positive lower integration bound $0\leq p_{\|,min}\leq\infty$, since one has $\displaystyle \lim_{p_{\|} \to \infty}\textup{I}_{\,1}(p_{\|})=0$ and $p_{\|}\geq 0$, so that the upper incomplete gamma function is defined. This can be validated as well by the plot in figure \ref{fig_j_ava_par_integrand_1_appendix}. 
\begin{figure}[H]
\begin{center}
\includegraphics[trim=49 45 120 41,width=1\textwidth,clip]{j_ava_par_integrand_1.pdf}
\captionsetup{format=hang,indention=0cm}
\caption[Characteristic and representation of the limits at infinity for the integrand function $\textup{I}_{\,1}(p_{\|})$ from (\ref{j_ava_RE_par_calc2}) for $c_{Z}\approx 2.49$, $\ln{(\Lambda)}=12.6$ and\linebreak$\tilde{E}=2.8$]{Characteristic and representation of the limits at infinity\protect\footnotemark{} for the integrand function $\textup{I}_{\,1}(p_{\|})$ from (\ref{j_ava_RE_par_calc2}) for $c_{Z}\approx 2.49$, $\ln{(\Lambda)}=12.6$ and $\tilde{E}=2.8$}
\label{fig_j_ava_par_integrand_1_appendix}
\end{center}
\end{figure}\footnotetext{\label{fig_j_ava_par_integrand1_footnote_appendix} The graph in figure \ref{fig_j_ava_par_integrand_1_appendix} was plotted with the \textsc{MATLAB}-file "j_ava_par_integrand_1.m",\\ \hspace*{8.7mm}which can be found in the digital appendix.}
\vspace{-7mm}
Consequently, the integral in (\ref{j_ava_RE_par_calc2}) is exclusively solvable by means of numerical integration for $p_{\|}\in[p_{\|,min},\,\infty)$. Therefore the following substitution is introduced:\vspace{1mm}
\begin{equation}\label{substitution1_RE_j_para_appendix}
\begin{split}
\begin{gathered}
p_{\|}=p_{\|,min}+\dfrac{w}{1-w} \;;\;\dfrac{\mathrm{d}p_{\|}}{\mathrm{d}w}= \dfrac{1}{(1-w)^2}\;;\; w(p_{\|}=p_{\|,min})=0\;,\;w(p_{\|}\rightarrow\infty)=1\,.
\end{gathered}
\end{split}
\end{equation} 
Thus, one has to numerically evaluate the integral $\textup{I}_{\,num,ava}^{\,j_{\|}}\,$, which is defined through (\ref{j_ava_RE_par_calc2}) in combination with (\ref{substitution1_RE_j_para_appendix}): \vspace{-2mm}
\begin{equation}\label{j_ava_RE_par_calc2}
\begin{split}
\begin{gathered}
\underline{\underline{j_{\,\|,RE}^{\,\textup{ava}}}}=\hspace{-1mm}
- \frac{ c\,e\,n_{RE}\, \sqrt{\tilde{E}} }{c_{Z}\ln{(\Lambda)}}\,\textup{e}^{\,\frac{2\,(E-1)}{c_{Z} \ln{(\Lambda)}}\,\frac{t}{\tau}} \hspace{-5mm}\displaystyle{\int\limits_{p_{\|}=p_{\|,min} }^{\infty}} \hspace{-3mm}  \sqrt{p_{\|}}\,\textup{e}^{  \left(\tilde{E}-\frac{1}{c_{Z} \ln{(\Lambda)}}\right)p_{\|}+\frac{\tilde{E}}{p_{\|}}}\,\Gamma\hspace{-0.5mm}\left(\dfrac{1}{2},\,\frac{\tilde{E}}{p_{\|} }\hspace{-0.5mm}\left(1+p_{\|}^{2}   \right)\hspace{-0.5mm}\right) \hspace{-0.5mm}\mathrm{d}p_{\|}
\\[5pt]
\overset{(\ref{substitution1_RE_j_para_appendix})}{=}\hspace{-1mm}- \frac{ c\,e\,n_{RE}\,\sqrt{\tilde{E}}\,\textup{e}^{\,\frac{2\,(E-1)}{c_{Z} \ln{(\Lambda)}}\,\frac{t}{\tau}} }{ c_{Z}\ln{(\Lambda)}} \hspace{-1mm}\displaystyle{\int\limits_{w=0 }^{1} }    \Gamma\hspace{-0.7mm}\left(\hspace{-0.5mm}\dfrac{1}{2},\,\frac{\tilde{E}\left(1\hspace{-0.5mm}+\hspace{-0.5mm}\left(p_{\|,min}\hspace{-0.5mm}+\hspace{-0.5mm}\frac{w}{w-1}\right)^{2}   \right)}{\left(p_{\|,min}+\frac{w}{w-1}\right) }\hspace{-0.5mm}\right)  \frac{\sqrt{p_{\|,min}\hspace{-0.5mm}+\hspace{-0.5mm}\frac{w}{w-1}}}{(1-w)^2}
\\[5pt]
\times\,\textup{e}^{  \left(\hspace{-0.5mm}\tilde{E}-\frac{1}{c_{Z} \ln{(\Lambda)}}\hspace{-0.5mm}\right)\left(p_{\|,min}+\frac{w}{w-1}\right) +\frac{\tilde{E}}{p_{\|}}} \, \mathrm{d}w
=: \underline{\underline{- \frac{ c\,e\,n_{RE}\,\sqrt{\tilde{E}}\,\textup{e}^{\,\frac{2\,(E-1)}{c_{Z} \ln{(\Lambda)}}\,\frac{t}{\tau}} }{ c_{Z}\ln{(\Lambda)}}\cdot\textup{I}_{\,num,ava}^{\,j_{\|}\,1\textup{D}}}}\,.
\end{gathered}
\end{split}
\end{equation}

\clearpage

\subsection{Evaluation of the integral for the orthogonal component of the avalanche runaway electron current density from (\ref{j_ava_RE_perp_def})}\label{j_ava_RE_perp_int_appendix_subsection}

The orthogonal component of the current density carried by a runaway electron population resulting from the avalanche generation mechanism can be calculated by computing the integral in (\ref{j_ava_RE_perp_def}) from section \ref{ava_first_moment_section} by inserting the distribution function from (\ref{f_RE_ava}) and the expression for $C$ according to (\ref{C_f_RE_ava}):\vspace{-2mm}
\begin{equation}\label{j_ava_RE_perp_def_appendix}
\begin{split}
\begin{gathered}
j_{\,\perp,RE}^{\,\textup{ava}}=-2\pi\,c\,e\hspace{-2mm}\displaystyle{\int\limits_{p_{\|}=-\infty}^{\infty}\int\limits_{p_{\perp}=0}^{\infty}} \frac{p_{\perp}^{2}}{\sqrt{1+p_{\|}^2+p_{\perp}^{2}}}\,f_{RE}^{ava}(p_{\|},\,p_{\perp},\,t) \mathrm{d}p_{\perp} \mathrm{d}p_{\|}
\\[10pt]
\overset{(\ref{f_RE_ava})}{=}\hspace{-1.2mm}-2\pi\,c\,e\hspace{-3mm}\displaystyle{\int\limits_{p_{\|}=-\infty}^{\infty}\int\limits_{p_{\perp}=0}^{\infty}} \hspace{-0.4mm}\frac{C\,p_{\perp}^{2}}{p_{\|}\sqrt{1\hspace{-0.4mm}+\hspace{-0.4mm}p_{\|}^2\hspace{-0.4mm}+\hspace{-0.4mm}p_{\perp}^{2}}}\exp{\hspace{-0.7mm}\left(\hspace{-0.6mm}\dfrac{(E-1)}{c_{Z} \ln{(\Lambda)}}\,\dfrac{t}{\tau}\hspace{-0.6mm}-\hspace{-0.6mm}\dfrac{p_{\|}}{c_{Z} \ln{(\Lambda)}}\hspace{-0.6mm}-\hspace{-0.6mm}\tilde{E}\,\dfrac{p_{\perp}^{2}}{p_{\|}}\hspace{-0.6mm}\right)}\mathrm{d}p_{\perp}\mathrm{d}p_{\|}
\\[10pt]
\overset{(\ref{C_f_RE_ava})}{=}- \frac{2\,\cancel{\pi}\, c\,e\,n_{RE}\,\tilde{E}\,\textup{e}^{\,\frac{2\,(E-1)}{c_{Z} \ln{(\Lambda)}}\,\frac{t}{\tau}} }{\cancel{\pi}\, c_{Z}\ln{(\Lambda)}}\hspace{-0.7mm}\displaystyle{\int\limits_{p_{\|}=-\infty}^{\infty}\int\limits_{p_{\perp}=0}^{\infty}} \hspace{-0.4mm}\frac{p_{\perp}^{2}}{p_{\|}\sqrt{1\hspace{-0.4mm}+\hspace{-0.4mm}p_{\|}^2\hspace{-0.4mm}+\hspace{-0.4mm}p_{\perp}^{2}}}\,\textup{e}^{\,-\frac{p_{\|}}{c_{Z} \ln{(\Lambda)}}-\tilde{E}\frac{p_{\perp}^2}{p_{\|}}}\,\mathrm{d}p_{\perp}\mathrm{d}p_{\|}
\\[10pt]
=- \frac{2 \, c\,e\,n_{RE}\,\tilde{E}\,\textup{e}^{\,\frac{2\,(E-1)}{c_{Z} \ln{(\Lambda)}}\,\frac{t}{\tau}} }{ c_{Z}\ln{(\Lambda)}}\hspace{-0.7mm}\displaystyle{\int\limits_{p_{\|}=-\infty}^{\infty}\int\limits_{p_{\perp}=0}^{\infty}} \hspace{-0.4mm}\underbrace{\frac{p_{\perp}^{2}}{p_{\|}\sqrt{1\hspace{-0.4mm}+\hspace{-0.4mm}p_{\|}^2\hspace{-0.4mm}+\hspace{-0.4mm}p_{\perp}^{2}}}\,\textup{e}^{\,-\frac{p_{\|}}{c_{Z} \ln{(\Lambda)}}-\tilde{E}\frac{p_{\perp}^2}{p_{\|}}}}_{:=\,\textup{I}_{\,2}(p_{\|},\,p_{\perp})}\mathrm{d}p_{\perp}\mathrm{d}p_{\|}
\end{gathered}
\end{split}
\end{equation}
The integral in (\ref{j_ava_RE_perp_def_appendix}) is convergent for a finite lower integration bound $\vert p_{\|,min}\vert\leq\infty$, since one has $\displaystyle \lim_{p_{\|} \to \infty}\textup{I}_{\,2}(p_{\|})=0$ and $\displaystyle \lim_{p_{\|} \to -\infty}\textup{I}_{\,2}(p_{\|})=-\infty$ for constant finite values of $p_{\perp}$, due to the dominant behaviour of the exponential in $\,\textup{I}_{\,2}(p_{\|},\,p_{\perp})$. Additionally, an analytic simplification of the integral in (\ref{j_ava_RE_perp_def_appendix}) was not found. This requires a two-dimensional numerical solving procedure. Therefore, one again utilizes the substitutions:\vspace{-1mm}
\begin{equation}\label{substitutions_num_Appendixx}
\begin{split}
\begin{gathered}
p_{\|}=p_{\|,min}+\dfrac{w}{1-w} \;;\;\dfrac{\mathrm{d}p_{\|}}{\mathrm{d}w}= \dfrac{1}{(1-w)^2}\;;\; w(p_{\|}=p_{\|,min})=0\;,\;w(p_{\|}\rightarrow\infty)=1
\\[6pt]
p_{\perp}=\dfrac{z}{1-z} \;;\;\dfrac{\mathrm{d}p_{\perp}}{\mathrm{d}z}= \dfrac{1}{(1-z)^2}\;;\; z(p_{\perp}=0)=0\;,\;z(p_{\perp}\rightarrow\infty)=1\,,
\end{gathered}
\end{split}
\end{equation}
which were already defined in (\ref{substitutions_RE_j_para_2Dint}). The resulting two integrals, following from (\ref{j_ava_RE_perp_def_appendix}) for $p_{\|}\in[p_{\|,min},\,\infty)$, then allow a numerical computation and are given by:\vspace{-1mm}
\begin{equation}\label{j_ava_RE_perp_2Dnum_appendix}
\begin{split}
\begin{gathered}
j_{\,\perp,RE}^{\,\textup{ava}}\hspace{-1.3mm}\overset{(\ref{j_ava_RE_perp_def_appendix})}{=}\hspace{-1.8mm}- \frac{2 \, c\,e\,n_{RE}\,\tilde{E} }{ c_{Z}\ln{(\Lambda)}}\,\textup{e}^{\,\frac{2\,(E-1)}{c_{Z} \ln{(\Lambda)}}\,\frac{t}{\tau}}\hspace{-4.5mm}\displaystyle{\int\limits_{p_{\|}=p_{\|,min}}^{\infty}\int\limits_{p_{\perp}=0}^{\infty}}  \frac{p_{\perp}^{2}\,\textup{e}^{\,-\frac{p_{\|}}{c_{Z} \ln{(\Lambda)}}-\tilde{E}\frac{p_{\perp}^2}{p_{\|}}}}{p_{\|}\sqrt{1\hspace{-0.4mm}+\hspace{-0.4mm}p_{\|}^2\hspace{-0.4mm}+\hspace{-0.4mm}p_{\perp}^{2}}}\,\mathrm{d}p_{\perp}\mathrm{d}p_{\|} 
\end{gathered}
\end{split}
\end{equation}
on the onehandside and by utilization of (\ref{substitutions_RE_j_para_2Dint}) by
\begin{equation}\label{j_ava_RE_perp_2Dnum_substituted_appendix}
\begin{split}
\begin{gathered}
\underline{\underline{j_{\,\perp,RE}^{\,\textup{ava}}}}\hspace{-1.3mm}\overset{(\ref{j_ava_RE_perp_def_appendix})}{=}\hspace{-1.8mm}- \frac{2 \, c\,e\,n_{RE}\,\tilde{E} }{ c_{Z}\ln{(\Lambda)}}\,\textup{e}^{\,\frac{2\,(E-1)}{c_{Z} \ln{(\Lambda)}}\,\frac{t}{\tau}}
\\[6pt]
\times\,\displaystyle{\int\limits_{w=0}^{1}\int\limits_{z=0}^{1}}  \frac{\frac{z^2}{(1-w)^2(1-z)^4}\,\textup{e}^{\,-\frac{ p_{\|,min}+\frac{w}{1-w} }{c_{Z} \ln{(\Lambda)}}-\frac{\tilde{E}\left(\frac{z}{1-z}\right)^2}{p_{\|,min}+\frac{w}{1-w}}}}{\left(p_{\|,min}+\frac{w}{1-w}\right)\sqrt{1+\left(p_{\|,min}+\frac{w}{1-w}\right)^{2}+\left(\frac{z}{1-z}\right)^2}}\,\mathrm{d}z\,\mathrm{d}w
\\[6pt]
=:\underline{\underline{- \frac{2 \, c\,e\,n_{RE}\,\tilde{E} }{ c_{Z}\ln{(\Lambda)}}\,\textup{e}^{\,\frac{2\,(E-1)}{c_{Z} \ln{(\Lambda)}}\,\frac{t}{\tau}}\cdot \textup{I}_{\,\textup{num,ava}}^{\,j_{\perp}}}}
\end{gathered}
\end{split}
\end{equation}
on the otherhandside.

\clearpage

\subsection{Evaluation of the integral for the mean mass-related kinetic energy density of an avalanche runaway electron population from (\ref{K_ava_RE_V2})}\label{K_ava_RE_int_appendix_subsection}

The mean mass-related kinetic energy density of an avalanche runaway electron population can be calculated by evaluating the integral in (\ref{K_ava_RE_V2}) from section \ref{ava_second_moment_section}. Together with the constant $C$ from (\ref{C_f_RE_ava}) it reads:\vspace{-1mm}
\begin{equation}\label{K_ava_RE_V2_appendix}
\begin{split}
\begin{gathered}
K_{\,RE}^{\,\textup{ava}}\overset{(\ref{K_ava_RE_V2})}{=}\hspace{-1mm}\dfrac{\pi\,c^2\,C}{n_{RE}}\hspace{-1mm}\displaystyle{\int\limits_{p_{\|}=-\infty}^{\infty}\int\limits_{p_{\perp}=0}^{\infty}} \dfrac{p_{\perp}p_{\|}^2+p_{\perp}^3}{p_{\|}(1+p_{\|}^2+p_{\perp}^2)} \,\textup{e}^{ \frac{(E-1)}{c_{Z} \ln{(\Lambda)}}\,\frac{t}{\tau}-\frac{p_{\|}}{c_{Z} \ln{(\Lambda)}}-\tilde{E}\,\frac{p_{\perp}^{2}}{p_{\|}} }\,\mathrm{d}p_{\perp}\mathrm{d}p_{\|}
\\[8pt]
\overset{(\ref{C_f_RE_ava})}{=} \hspace{-2mm}\dfrac{\cancel{\pi}\,c^2\,\cancel{n_{RE}}\,\tilde{E}}{\cancel{n_{RE}}\,\cancel{\pi}\,c_{Z} \ln{(\Lambda)}}\,\textup{e}^{ \frac{2\,(E-1)}{c_{Z} \ln{(\Lambda)}}\,\frac{t}{\tau}}\hspace{-4mm}\displaystyle{\int\limits_{p_{\|}=-\infty}^{\infty}\int\limits_{p_{\perp}=0}^{\infty}} \dfrac{p_{\perp}p_{\|}^2+p_{\perp}^3}{p_{\|}(1+p_{\|}^2+p_{\perp}^2)} \,\textup{e}^{  -\frac{p_{\|}}{c_{Z} \ln{(\Lambda)}}-\tilde{E}\,\frac{p_{\perp}^{2}}{p_{\|}} }\,\mathrm{d}p_{\perp}\mathrm{d}p_{\|}
\\[8pt]
=\dfrac{ c^2\, \tilde{E}}{ c_{Z} \ln{(\Lambda)}}\,\textup{e}^{ \frac{2\,(E-1)}{c_{Z} \ln{(\Lambda)}}\,\frac{t}{\tau}}\displaystyle{\int\limits_{p_{\|}=-\infty}^{\infty}\int\limits_{p_{\perp}=0}^{\infty}} \dfrac{p_{\perp}\hspace{-0.8mm}\left(p_{\|}^2+p_{\perp}^2\right)}{p_{\|}(1+p_{\|}^2+p_{\perp}^2)} \,\textup{e}^{  -\frac{p_{\|}}{c_{Z} \ln{(\Lambda)}}-\tilde{E}\,\frac{p_{\perp}^{2}}{p_{\|}} }\,\mathrm{d}p_{\perp}\mathrm{d}p_{\|}\,.
\end{gathered}
\end{split}
\end{equation}
The last integral in (\ref{K_ava_RE_V2_appendix}) determines the avalanche runaway electron kinetic energy density and allows a semi-analytical simplification, meaning that the $p_{\perp}$-integration can be carried out and the remaining $p_{\|}$-integral has to be computed numerically. Thus, one once again utilizes the substitution (\ref{substitution_eta_j_par_appendix}) from subsection \ref{j_ava_RE_parallel_int_appendix_subsection}, which has been defined as: \vspace{-3mm}
\begin{equation}\label{substitution_eta_K_appendix}
\begin{split}
\begin{gathered}
\eta(p_{\perp})\hspace{-0.5mm}:=\hspace{-0.5mm}\dfrac{\tilde{E}}{p_{\|} }\hspace{-0.5mm}\left(\hspace{-0.5mm}1+p_{\|}^{2}+p_{\perp}^2\hspace{-0.5mm}\right)\,;\;\dfrac{\mathrm{d}\eta}{\mathrm{d}p_{\perp}}\hspace{-0.5mm}=\hspace{-0.5mm}\dfrac{2\,\tilde{E}\,p_{\perp}}{p_{\|}}\;;
\\[4pt]
 \eta(p_{\perp}\hspace{-0.5mm}=\hspace{-0.5mm}0)\hspace{-0.5mm}=\hspace{-0.5mm}\dfrac{\tilde{E}}{p_{\|} }\hspace{-0.5mm}\left(\hspace{-0.5mm}1+p_{\|}^{2}\hspace{-0.5mm} \right)\hspace{0.5mm},\;\eta(p_{\perp}\hspace{-0.5mm}\rightarrow\infty)\hspace{-0.5mm}=\hspace{-0.5mm}\infty\,.
\end{gathered}
\end{split}
\end{equation}
Hence, the integral from (\ref{K_ava_RE_V2_appendix}) becomes:\vspace{-1mm}
\begin{equation}\label{K_ava_RE_V3_appendix}
\begin{split}
\begin{gathered}
K_{\,RE}^{\,\textup{ava}}\underset{(\ref{substitution_eta_K_appendix})}{\overset{(\ref{K_ava_RE_V2_appendix})}{=}}\hspace{-0.5mm} \dfrac{ c^2\, \cancel{\tilde{E}}\,\textup{e}^{ \frac{2\,(E-1)}{c_{Z} \ln{(\Lambda)}}\,\frac{t}{\tau}}}{ c_{Z} \ln{(\Lambda)}}\hspace{-3.5mm} \displaystyle{\int\limits_{p_{\|}=-\infty}^{\infty}\hspace{-0.5mm}\int\limits_{\eta=\frac{\tilde{E}}{p_{\|} } ( 1+p_{\|}^{2})}^{\infty}}\hspace{-4.5mm} \dfrac{\cancel{p_{\perp}}\hspace{-0.8mm}\left(\frac{p_{\|} }{\tilde{E}}\,\eta-1\right)}{\cancel{p_{\|}}\,\frac{p_{\|} }{\tilde{E}}\,\eta} \dfrac{\cancel{p_{\|}}\,\textup{e}^{  -\frac{p_{\|}}{c_{Z} \ln{(\Lambda)}}-\eta+\frac{\tilde{E}}{p_{\|}}( 1+p_{\|}^{2}) }}{2\,\cancel{\tilde{E}}\,\cancel{p_{\perp}}}\,\mathrm{d}\eta\,\mathrm{d}p_{\|}
\\[6pt]
=  \dfrac{ c^2 \,\textup{e}^{ \frac{2\,(E-1)}{c_{Z} \ln{(\Lambda)}}\,\frac{t}{\tau}}}{ 2\,c_{Z} \ln{(\Lambda)}}\hspace{-1.5mm} \displaystyle{\int\limits_{p_{\|}=-\infty}^{\infty}}\hspace{-3mm}\textup{e}^{  \left(\tilde{E}-\frac{1}{c_{Z} \ln{(\Lambda)}}\right)p_{\|}+\frac{\tilde{E}}{p_{\|}}}\hspace{-2.5mm}\displaystyle{\int\limits_{\eta=\frac{\tilde{E}}{p_{\|} } ( 1+p_{\|}^{2})}^{\infty}}\hspace{-2.5mm} \left(1-\frac{\tilde{E}}{p_{\|}\,\eta}\right) \textup{e}^{-\eta} \,\mathrm{d}\eta\,\mathrm{d}p_{\|}
\\[6pt]
=  \dfrac{ c^2 \,\textup{e}^{ \frac{2\,(E-1)}{c_{Z} \ln{(\Lambda)}}\,\frac{t}{\tau}}}{ 2\,c_{Z} \ln{(\Lambda)}}\hspace{-1mm} \displaystyle{\int\limits_{p_{\|}=-\infty}^{\infty}}\hspace{-3mm}\textup{e}^{  \left(\tilde{E}-\frac{1}{c_{Z} \ln{(\Lambda)}}\right)p_{\|}+\frac{\tilde{E}}{p_{\|}}}\hspace{-1mm}\left(\displaystyle{\int\limits_{\eta=\frac{\tilde{E}}{p_{\|} } ( 1+p_{\|}^{2})}^{\infty}} \hspace{-6.5mm} \textup{e}^{-\eta} \,\mathrm{d}\eta-\frac{\tilde{E}}{p_{\|}}\displaystyle{\int\limits_{\eta=\frac{\tilde{E}}{p_{\|} } ( 1+p_{\|}^{2})}^{\infty}}\hspace{-5.5mm}\eta^{-1} \textup{e}^{-\eta} \,\mathrm{d}\eta\hspace{-0.5mm}\right)\hspace{-0.5mm}\mathrm{d}p_{\|}\,.
\end{gathered}
\end{split}
\end{equation}
The first appearing integral in (\ref{K_ava_RE_V3_appendix}) with respect to $\eta$ can be solved analytically and the second integral is expressable in terms of the upper incomplete gamma function $\Gamma(z,\,a)$ \cite{incgammafunc}:
\vspace{-2mm}
\begin{equation}\label{K_ava_RE_V4_appendix}
\begin{split}
\begin{gathered}
K_{\,RE}^{\,\textup{ava}}\underset{ }{\overset{(\ref{K_ava_RE_V3_appendix})}{=}}\dfrac{ c^2 \,\textup{e}^{ \frac{2\,(E-1)}{c_{Z} \ln{(\Lambda)}}\,\frac{t}{\tau}}}{ 2\,c_{Z} \ln{(\Lambda)}}\hspace{-1mm} \displaystyle{\int\limits_{p_{\|}=-\infty}^{\infty}}\hspace{-3mm}\textup{e}^{  \left(\tilde{E}-\frac{1}{c_{Z} \ln{(\Lambda)}}\right)p_{\|}+\frac{\tilde{E}}{p_{\|}}}
\\[4pt]
 \times\,\Biggl\lgroup \, \underbrace{ \bigl[-\textup{e}^{-\eta}\bigr]_{\eta=\frac{\tilde{E}}{p_{\|} } ( 1+p_{\|}^{2})}^{\eta\rightarrow\infty}}_{=\,\textup{e}^{-\frac{\tilde{E}}{p_{\|} } ( 1+p_{\|}^{2})}}-\frac{\tilde{E}}{p_{\|}}\,\Gamma\hspace{-0.5mm}\Biggl(0,\,\frac{\tilde{E}}{p_{\|} } ( 1+p_{\|}^{2})\Biggr) \hspace{-0.5mm} \Biggr\rgroup\mathrm{d}p_{\|}
\\[6pt]
=\dfrac{ c^2 \,\textup{e}^{ \frac{2\,(E-1)}{c_{Z} \ln{(\Lambda)}}\,\frac{t}{\tau}}}{ 2\,c_{Z} \ln{(\Lambda)}}  \displaystyle{\int\limits_{p_{\|}-\infty}^{\infty}} \underbrace{\textup{e}^{ -\frac{p_{\|}}{c_{Z} \ln{(\Lambda)}}} - \frac{\tilde{E}}{p_{\|}} \,\Gamma\hspace{-0.5mm}\Biggl(0,\,\frac{\tilde{E}}{p_{\|} } ( 1+p_{\|}^{2})\Biggr)\textup{e}^{  \left(\tilde{E}-\frac{1}{c_{Z} \ln{(\Lambda)}}\right)p_{\|}+\frac{\tilde{E}}{p_{\|}}}}_{=:\,\textup{I}_{\,3}(p_{\|})}\,\mathrm{d}p_{\|}\,.
\end{gathered}
\end{split}
\end{equation}
Note, that a finite lower integration bound $p_{\|,min}$ with $ 0\hspace{-0.3mm}\leq\hspace{-0.3mm}p_{\|,min} \hspace{-0.3mm}<\hspace{-0.3mm}\infty$ is needed for the last integral in (\ref{K_ava_RE_V4_appendix}), so that the upper incomplete gamma function is defined and produces real values. This leads to a finite total integration result for $K_{\,RE}^{\,\textup{ava}}$, where the integrand $\textup{I}_{\,3}(p_{\|})$ is plotted in figure \ref{fig_K_ava_integrand_3_appendix} for the possible interval for $p_{\|}$ and verifies that the integral has to converge, if a numerical integration routine is applied. 
\vspace{4mm}
\begin{figure}[H]
\begin{center}
\includegraphics[trim=62 52 120 41,width=1\textwidth,clip]{K_ava_integrand_3.pdf}
\captionsetup{format=hang,indention=0cm}
\caption[Characteristic and representation of the limits at infinity for the integrand function $\textup{I}_{\,3}(p_{\|})$ from (\ref{K_ava_RE_V4_appendix}) for $c_{Z}\approx 2.49$, $\ln{(\Lambda)}=14.9$ and\linebreak$\tilde{E}=2.8$]{Characteristic and representation of the limits at infinity\protect\footnotemark{} for the integrand function $\textup{I}_{\,3}(p_{\|})$ from (\ref{K_ava_RE_V4_appendix}) for $c_{Z}\approx 2.49$, $\ln{(\Lambda)}=14.9$ and $\tilde{E}=2.8$}
\label{fig_K_ava_integrand_3_appendix}
\end{center}
\end{figure}\footnotetext{\label{fig_K_ava_integrand3_footnote_appendix} The diagram in figure \ref{fig_K_ava_integrand_3_appendix} was created with the \textsc{MATLAB}-file "K_ava_integrand_3.m",\\ \hspace*{8.7mm}which can be found in the digital appendix.}
\newpage\noindent
If the finite lower integration bound $p_{\|,min}$ is used, the last integral in (\ref{K_ava_RE_V4_appendix}) yields:
\vspace{-1.5mm}
\begin{equation}\label{K_ava_RE_V5_appendix}
\begin{split}
\begin{gathered}
\underline{\underline{K_{\,RE}^{\,\textup{ava}}}}\underset{ }{\overset{(\ref{K_ava_RE_V4_appendix})}{=}} \dfrac{ c^2 \,\textup{e}^{ \frac{2\,(E-1)}{c_{Z} \ln{(\Lambda)}}\,\frac{t}{\tau}}}{ 2\,c_{Z} \ln{(\Lambda)}} \Biggl(\;\, \displaystyle{\int\limits_{p_{\|}=p_{\|,min}}^{\infty}}\hspace{-3mm}\textup{e}^{ -\frac{p_{\|}}{c_{Z} \ln{(\Lambda)}}}\,\mathrm{d}p_{\|}   
\\[6pt]
  -\,\tilde{E}\underbrace{\displaystyle{\int\limits_{p_{\|}=p_{\|,min}}^{\infty}}\hspace{-1mm} \frac{1}{p_{\|}} \,\Gamma\hspace{-0.5mm}\Biggl(0,\,\frac{\tilde{E}}{p_{\|} } ( 1+p_{\|}^{2})\Biggr)\textup{e}^{  \left(\tilde{E}-\frac{1}{c_{Z} \ln{(\Lambda)}}\right)p_{\|}+\frac{\tilde{E}}{p_{\|}}}\mathrm{d}p_{\|}}_{:=\,\textup{I}_{\,\textup{num,ava}}^{\,K,\,1\textup{D}}}\;\Biggr)
 \\[6pt]
= \dfrac{ c^2 \,\textup{e}^{ \frac{2\,(E-1)}{c_{Z} \ln{(\Lambda)}}\,\frac{t}{\tau}}}{ 2\,c_{Z} \ln{(\Lambda)}} \Biggl(\;\underbrace{ \left[-c_{Z} \ln{(\Lambda)}\,\textup{e}^{ -\frac{p_{\|}}{c_{Z} \ln{(\Lambda)}}}\right]_{p_{\|}=p_{\|,min}}^{p_{\|}\rightarrow\infty}}_{=\,c_{Z} \ln{(\Lambda)}\,\textup{e}^{ -\frac{p_{\|,min}}{c_{Z} \ln{(\Lambda)}}}}\hspace{-1.5mm} - \; \tilde{E}\cdot \textup{I}_{\,\textup{num,ava}}^{\,K,\,1\textup{D}}\Biggr)
\\[5pt]
=\underline{\underline{ \dfrac{ c^2 \,\textup{e}^{ \frac{2\,(E-1)}{c_{Z} \ln{(\Lambda)}}\,\frac{t}{\tau}}}{ 2\,c_{Z} \ln{(\Lambda)}} \biggl( c_{Z} \ln{(\Lambda)}\,\textup{e}^{ -\frac{p_{\|,min}}{c_{Z} \ln{(\Lambda)}}}  - \; \tilde{E}\cdot \textup{I}_{\,\textup{num,ava}}^{\,K,\,1\textup{D}}\biggr)}}\,.
\end{gathered}
\end{split}
\end{equation}
The boundaries of the integral $\textup{I}_{\,\textup{num,ava}}^{\,K,\,1\textup{D}}$ from (\ref{K_ava_RE_V5_appendix}) can be transformed, so that they are finite. Therefore, one recapitulates the substitution from (\ref{substitution1_RE_j_para_appendix}):\vspace{-1.5mm}
\begin{equation}\label{substitution1again_RE_K_appendix}
\begin{split}
\begin{gathered}
p_{\|}=p_{\|,min}+\dfrac{w}{1-w} \;;\;\dfrac{\mathrm{d}p_{\|}}{\mathrm{d}w}= \dfrac{1}{(1-w)^2}\;;\; w(p_{\|}=p_{\|,min})=0\;,\;w(p_{\|}\rightarrow\infty)=1\,.
\end{gathered}
\end{split}
\end{equation} 
Together with (\ref{K_ava_RE_V5_appendix}) it leads to:
\vspace{-1.5mm}
\begin{equation}\label{K_ava_RE_V6_appendix}
\begin{split}
\begin{gathered}
\underline{\underline{K_{\,RE}^{\,\textup{ava}}}}\underset{ }{\overset{(\ref{substitution1again_RE_K_appendix})}{=}} \dfrac{ c^2 \,\textup{e}^{ \frac{2\,(E-1)}{c_{Z} \ln{(\Lambda)}}\,\frac{t}{\tau}}}{ 2\,c_{Z} \ln{(\Lambda)}} \Biggl(c_{Z} \ln{(\Lambda)}\,\textup{e}^{ -\frac{p_{\|,min}}{c_{Z} \ln{(\Lambda)}}}   
\\[6pt]
 -\,\tilde{E} \displaystyle{\int\limits_{w=0}^{1}}  \frac{\Gamma\hspace{-0.5mm}\left(\hspace{-0.5mm}0,\,\frac{\tilde{E}\,(1-w)}{p_{\|,min}\,(1-w)+w}\left(\hspace{-0.5mm} 1\hspace{-0.5mm}+\hspace{-0.5mm}\left(\hspace{-0.5mm}p_{\|,min}\hspace{-0.5mm}+\hspace{-0.5mm}\frac{w}{1-w}\hspace{-0.5mm}\right)^{2} \right)\hspace{-0.5mm}\right)\,}{(p_{\|,min}\,(1-w)+w)(1-w)} 
 \\[7pt]
 \times\,\exp{ \left( \left(\tilde{E}-\frac{1}{c_{Z} \ln{(\Lambda)}}\right)\left(p_{\|,min}+\frac{w}{1-w}\right)+\frac{\tilde{E}\,(1-w)}{p_{\|,min}\,(1-w)+w}\right)}\,\mathrm{d}w \;\Biggr)
  \\[9pt]
=:\underline{\underline{ \dfrac{ c^2 \,\textup{e}^{ \frac{2\,(E-1)}{c_{Z} \ln{(\Lambda)}}\,\frac{t}{\tau}}}{ 2\,c_{Z} \ln{(\Lambda)}} \biggl( c_{Z} \ln{(\Lambda)}\,\textup{e}^{ -\frac{p_{\|,min}}{c_{Z} \ln{(\Lambda)}}}  - \; \tilde{E}\cdot \textup{I}_{\,\textup{num,ava}}^{\,K,\,1\textup{D}}\biggr)}}\,.
\end{gathered}
\end{split}
\end{equation}
Consequently, one can compute the mean mass-related kinetic energy density of an avalanche runaway electron population $K_{\,RE}^{\,\textup{ava}}$ from a two-dimensional numerical integration, whereby the rules (\ref{K_ava_RE_V3_num2D}) and (\ref{K_ava_RE_V3_num2D_sustituted}) are usable or from a one-dimensional numerical integration, utilizing the rules stated in (\ref{K_ava_RE_V5_appendix}) and (\ref{K_ava_RE_V6_appendix}).

\clearpage

\section{Numerical calculations}

\subsection{Runtime-based efficiency analysis for the computation of the moments of the avalanche runaway electron distribution function}\label{subsection_num_efficiency_appendix}

The runtime-based evaluation of the efficiency of the computation methods for the moments of the avalanche runaway distribution function is carried out with the help of the \textsc{MATLAB}-script "efficiency_analysis_for_moment_calculations.m", which is stored in the digital appendix.

The corresponding output for the values of the integrals and the parallel component of the current density for a certain set of parameters is shown for $p_{\|,min}=0$ in listing \ref{outMATLABeffcalcanalysisAppendix}. However, the runtime duration for each computation rule was calculated as the mean value of the runtime of $729$ computations for different parameter combinations with $\tilde{E}\in[5,\,45]$, $c_{Z}\in[5,\,45]$ and $\ln{(\Lambda)}\in[5,\,45]$, in order to achieve reliable results.\vspace{1mm}
\begin{lstlisting}[language=Matlab,keywordstyle=\empty,frame=single, caption={Output of the \textsc{MATLAB}-script "efficiency_analysis_for_moment_\\calculations.m" \vspace{1mm}},label={outMATLABeffcalcanalysisAppendix} ]
set of parameters:

EoverEcrit = 150; n_e = 1e+20 m^-3; k_B*T_e = 100 eV; Z_eff = 1.5;

calculated quantities:

lnLambda = 12.59741; E_crit = 0.06424 V/m; p_crit = 0.08192;
E_D = 328.24256 V/m; E_sa = 70.24391 V/m; E = 9.63532 V/m; 

I_num_ava_nRE = 1.00000002 
(2D-integration with "integral2", from 0 to 1, runtime benchmark, abs_tol~10^-6);
mean runtime duration = 0.0094 s (for 729 parameter combinations);

I_num_ava_nRE = 0.99998290 
(2D-integration with "trapz", from 0 to 1, runtime benchmark, 10^6 grid points);
mean runtime duration = 0.0165 s (for 729 parameter combinations);

I_num_ava_j_par_1D = 5.5701 (1D, from p_par_min to infinity);
mean runtime duration = 0.2654 s (for 729 parameter combinations);

I_num_ava_j_par_1D = 5.5701 (1D, from 0 to 1);
mean runtime duration = 0.2653 s (for 729 parameter combinations);

I_num_ava_j_par = 0.5102 (2D, from p_par_min to infinity);
mean runtime duration = 0.0383 s (for 729 parameter combinations);

I_num_ava_j_par = 0.5102 (2D, from 0 to 1);
mean runtime duration = 0.0063 s (for 729 parameter combinations);

I_num_ava_j_perp = 0.0224 (2D, from p_par_min to infinity);
mean runtime duration = 0.0365 s (for 729 parameter combinations);

I_num_ava_j_perp = 0.0224 (2D, from 0 to 1);
mean runtime duration = 0.0043 s (for 729 parameter combinations);


j_RE_ava_par_1D/(c*e*n_RE*exp(2*(E-1)/(c_Z*lnLambda)*t/tau)) 
		= u_RE_ava_par/c = -0.9688 (from 1D-integration);

j_RE_ava_par/(c*e*n_RE*exp(2*(E-1)/(c_Z*lnLambda)*t/tau)) 
		= u_RE_ava_par/c = -0.9688 (from 2D-integration);

j_RE_ava_perp/(c*e*n_RE*exp(2*(E-1)/(c_Z*lnLambda)*t/tau)) 
		= u_RE_ava_perp/c = -0.0426;

j_RE_ava/(c*e*n_RE*exp(2*(E-1)/(c_Z*lnLambda)*t/tau)) 
		= u_RE_ava/c = 0.9698; (check: 0.9698)

I_num_ava_K = 0.5026 (2D, from p_par_min to infinity);
mean runtime duration = 0.0404 s (for 729 parameter combinations);

I_num_ava_K = 0.5026 (2D, from 0 to 1);
mean runtime duration = 0.0097 s (for 729 parameter combinations);

I_num_ava_K_p_par_1D = 0.0480 (1D, from p_par_min to infinity);
mean runtime duration = 0.1058 s (for 729 parameter combinations);

I_num_ava_K_p_par_1D = 0.0480 (1D, from 0 to 1);
mean runtime duration = 0.1121 s (for 729 parameter combinations);

K_RE_ava/((c^2/2)*exp(2*(E-1)/(c_Z*lnLambda)*t/tau)) = 0.9544 (from 1D-integration);

K_RE_ava/((c^2/2)*exp(2*(E-1)/(c_Z*lnLambda)*t/tau)) = 0.9544 (from 2D-integration);
\end{lstlisting}

\clearpage

\subsection{Range of the physical quantities from the computation of the moments of the avalanche runaway electron distribution function}\label{subsection_generated_data}

The computation of the absolute values of the normalized parallel and orthogonal component $|u_{\,\|,RE}^{\,\textup{ava}}|/c$ and $u_{\,\perp,RE}^{\,\textup{ava}}|/c$ as well as the normalized magnitude $|u_{\,RE}^{\,\textup{ava}}/c$ of the mean velocity $\mathbf{u}_{\,RE}^{\,\mathrm{ava}}$ and the normalized mean mass-related kinetic energy density $K_{\,RE}^{\,\textup{ava}}/(c^2/2)$ of avalanche runaway electron populations with different electron temperatures is carried out with the \textsc{MATLAB}-codes "generate_num_data_10eV.m", "generate_ num_data_50eV.m", "generate_num_data_100eV.m" and "generate_num_\linebreak ~data_1000eV.m", which are stored in the digital appendix. The \textsc{MATLAB}-scripts "plot_ num_data_ 10eV.m", "plot_num_data_50eV.m", "plot_num_data_100eV.m" and "plot_num_data_1000eV.m" are then used produce graphical depiction of the computed results. Additionally, the console output allows to observe the minimum and maximum value of the physical quantities representing the results and is therefore shown in listing \ref{outMATLABgeneratedataAppendix}. It should be remarked that the mean value of the quantity $|u_{\,\perp,RE}^{\,\textup{ava}}|/|u_{\,\|,RE}^{\,\textup{ava}}|\cdot 100\,\%$ is only comparable for the electron temperatures\linebreak\mbox{$k_{B}T_{e}\in\lbrace 10\,\mathrm{eV},\,100\,\mathrm{eV},\,1000\,\mathrm{eV}\rbrace$}, because for those case the same parameter space:
\begin{equation}\label{EoverEcrit_range_appendix}
\lbrace n_{e}\rbrace\,\times\,\lbrace\vert E_{\|}\vert/E_{c}\rbrace\,=\,[1.01,\,1.1\cdot E_{D}/E{c}]\,\times\,[10^{19}\,\mathrm{m}^{-3},\,10^{21}\,\mathrm{m}^{-3}]
\end{equation}
from (\ref{EoverEcrit_range}) with $400\,\times\,400$ grid points were used. However, for the electron temperature $k_{B}T_{e}=50\,\mathrm{eV}$ the interval $\vert E_{\|}\vert/E_{c}\in[1.0001,\,6]$ was chosen for the normalized electric field with the same interval for the electron density and the same grid resolution of $400\,\times\,400$ grid points.\vspace{2mm}
\begin{lstlisting}[language=Matlab,keywordstyle=\empty,frame=single, caption={Combined output of the \textsc{MATLAB}-scripts "plot_num_data_10eV.m", \mbox{"plot_num_data_50eV.m", "plot_num_data_100eV.m" and "plot_}\\num_data_1000eV.m" \vspace{1mm}},,mathescape=true,label={outMATLABgeneratedataAppendix} ]
k_BT_e = 10 eV; Z_eff = 1.50;

min_u_RE_ava_par_over_c = 0.262320; max_u_RE_ava_par_over_c = 0.967382;

min_u_RE_ava_perp_over_c = 0.002288; max_u_RE_ava_perp_over_c = 0.937723;

min_u_RE_ava_perp_over_c = 0.24 $\%$; max_u_RE_ava_perp_over_c = 357.47 $\%$;
mean_u_perp_over_u_par = 1.33 $\%$; 

min_u_RE_ava_over_c = 0.959233; max_u_RE_ava_over_c = 0.973723;

min_K_RE_ava_over_csqhalf = 0.937932; max_K_RE_ava_over_csqhalf = 0.997979;


k_BT_e = 100 eV; Z_eff = 1.50;

min_u_RE_ava_par_over_c = 0.286065; max_u_RE_ava_par_over_c = 0.972566;

min_u_RE_ava_perp_over_c = 0.006721; max_u_RE_ava_perp_over_c = 0.928002;

min_u_RE_ava_perp_over_c = 0.69 $\%$; max_u_RE_ava_perp_over_c = 324.40 $\%$;
mean_u_perp_over_u_par = 2.16 $\%$; 

min_u_RE_ava_over_c = 0.963197; max_u_RE_ava_over_c = 0.972590;

min_K_RE_ava_over_csqhalf = 0.949604; max_K_RE_ava_over_csqhalf = 0.998287;


k_BT_e = 1000 eV; Z_eff = 1.50;

min_u_RE_ava_par_over_c = 0.306469; max_u_RE_ava_par_over_c = 0.981472;

min_u_RE_ava_perp_over_c = 0.019940; max_u_RE_ava_perp_over_c = 0.919063;

min_u_RE_ava_perp_over_c = 2.04 $\%$; max_u_RE_ava_perp_over_c = 299.89 $\%$;
mean_u_perp_over_u_par = 4.75 $\%$; 

min_u_RE_ava_over_c = 0.952208; max_u_RE_ava_over_c = 0.981948;

min_K_RE_ava_over_csqhalf = 0.957657; max_K_RE_ava_over_csqhalf = 0.998511;


k_BT_e = 50 eV; Z_eff = 1.50;

min_u_RE_ava_par_over_c = 0.035314; max_u_RE_ava_par_over_c = 0.938324;

min_u_RE_ava_perp_over_c = 0.211296; max_u_RE_ava_perp_over_c = 0.998115;

min_u_RE_ava_perp_over_c = 22.52 $\%$; max_u_RE_ava_perp_over_c = 2826.42 $\%$;
mean_u_perp_over_u_par = 47.11 $\%$; 

min_u_RE_ava_over_c = 0.935601; max_u_RE_ava_over_c = 0.998740;

min_K_RE_ava_over_csqhalf = 0.957416; max_K_RE_ava_over_csqhalf = 0.999952;
\end{lstlisting}

\clearpage

\newpage\thispagestyle{empty}\mbox{}\newpage

\end{document}